\documentclass[usenatbib]{mn2e}
\usepackage{graphicx,ifthen,url,float,setspace}
\def\ltsima{$\; \buildrel < \over \sim \;$}
\def\simlt{\lower.5ex\hbox{\ltsima}}
\def\gtsima{$\; \buildrel > \over \sim \;$}
\def\simgt{\lower.5ex\hbox{\gtsima}}
\newcommand {\scuba}{{\sc Scuba}}
\newcommand {\scubaii}{{\sc Scuba-2}}
\newcommand {\herschel}{{\it Herschel}}
\newcommand {\herschellong}{{\it Herschel Space Observatory}}
\newcommand {\uJy}{$\mu$Jy}
\newcommand {\um}{$\mu$m}

\def\um     {$\mu$m}
\def\ts     {\thinspace}
\def\kms    {\ifmmode{{\rm \ts km\ts s}^{-1}}\else{\ts km\ts s$^{-1}$}\fi}
\def\msol   {\ifmmode{{\rm M}_{\odot}}\else{M$_{\odot}$}\fi}
\def\lsol   {\ifmmode{{\rm L}_{\odot}}\else{L$_{\odot}$}\fi}
\def\zsol   {\ifmmode{{\rm Z}_{\odot}}\else{Z$_{\odot}$}\fi}
\def\etal   {{\rm et\ts al.}}
\def\aco    {\ifmmode{^{12}{\rm CO}(J\!=\!1\! \to \!0)}\else{$^{12}${\rm CO}($J$=1$\to$0)}\fi}
\def\bco    {\ifmmode{^{12}{\rm CO}(J\!=\!2\! \to \!1)}\else{$^{12}${\rm CO}($J$=2$\to$1)}\fi}
\def\cco    {\ifmmode{^{12}{\rm CO}(J\!=\!3\! \to \!2)}\else{$^{12}${\rm CO}($J$=3$\to$2)}\fi}
\def\dco    {\ifmmode{^{12}{\rm CO}(J\!=\!4\! \to \!3)}\else{$^{12}${\rm CO}($J$=4$\to$3)}\fi}
\def\gco    {\ifmmode{^{12}{\rm CO}(J\!=\!7\! \to \!6)}\else{$^{12}${\rm CO}($J$=7$\to$6)}\fi}

\def\ci     {\ifmmode{{\rm C}{\rm \small I}}\else{C\ts {\scriptsize I}}\fi}
\def\hi     {\ifmmode{{\rm H}{\rm \small I}}\else{H\ts {\scriptsize I}}\fi}
\def\hh     {\ifmmode{{\rm H}_2}\else{H$_2$}\fi}
\def\cone {\ifmmode{{\rm C}{\rm \small I}(^3\!P_1\!\to^3\!P_0)}
     \else{C\ts {\scriptsize I}{\small$(^3\!P_1\!\to\,^3\!P_0)$}}\fi}
\def\ctwo {\ifmmode{{\rm C}{\rm \small I}(^3\!P_2\!\to\,^3\!P_1)}
     \else{C\ts {\scriptsize I}{\small$(^3\!P_2\!\to\,^3\!P_1)$}}\fi}
\def\cij {\ifmmode{{\rm C}{\rm \small I}\,(^3P_i\to^3P_j)}\else{C\ts {\scriptsize I}\,{\small$(^3P_i\to^3P_j)$}}\fi}
\def\cii    {\ifmmode{{\rm C}{\rm \small II}}\else{C\ts {\scriptsize II}}\fi}
\def\tex {\ifmmode{{T}_{\rm ex}}\else{$T_{\rm ex}$}\fi}
\def\tmb {\ifmmode{{T}_{\rm mb}}\else{$T_{\rm mb}$}\fi}
\def\tkin {\ifmmode{{T}_{\rm kin}}\else{$T_{\rm kin}$}\fi}
\def\microns {\ifmmode{\mu{\rm m}}\else{$\mu$m}\fi}
\def\nhh   {\ifmmode{n({\rm H}_2)}\else{$n$(H$_2$)}\fi}

\newcommand{\ltaraw}{$\; \buildrel < \over \sim \;$}
\newcommand{\lta}{\lower.5ex\hbox{\ltaraw}}
\newcommand{\gtaraw}{$\; \buildrel > \over \sim \;$}
\newcommand{\gta}{\lower.5ex\hbox{\gtaraw}}

\newcommand{\msun}{{\rm\,M_\odot}}
\newcommand{\sfr}{{\rm\,M_\odot\,yr^{-1}}}
\newcommand{\lsun}{{\rm\,L_\odot}}

\loadboldmathitalic 
\title [\scubaii\ Galaxies in COSMOS]
{Characterisation of \scubaii\ 450\um\ and 850\um-selected Galaxies in the COSMOS Field}
\author[C.~M. Casey et al.]
{
Caitlin~M. Casey$^1$\thanks{Hubble Fellow; cmcasey@ifa.hawaii.edu}, Chian-Chou Chen$^1$, Lennox L. Cowie$^1$, Amy J. Barger$^{1,2,3}$, 
\newauthor Peter Capak$^4$, Olivier Ilbert$^5$, Michael Koss$^1$, Nicholas Lee$^1$, Emeric Le Floc'h$^6$, 
\newauthor  David B. Sanders$^1$, Jonathan P. Williams$^1$\\
$^1$ Institute for Astronomy, University of Hawai'i, 2680 Woodlawn Dr, Honolulu, HI 96822, USA \\
$^2$ Department of Astronomy, University of Wisconsin-Madison, 475 North Charter Street, Madison, WI 53706, USA \\
$^3$ Department of Physics and Astronomy, University of Hawai'i, 2505 Correa Road, Honolulu, HI 96822, USA \\
$^4$ Spitzer Science Center, California Institute of Technology, 1200 E. California Blvd, Pasadena, CA, 91125, USA \\
$^5$ Laboratoire d'Astrophysique de Marseille, 38 rue Frederic Joliot Curie, 13388 Marseille, France \\
$^6$ CEA-Saclay, Orme des Merisiers, Bat. 709, 91191 Gif-sur-Yvette, France \\
}
\date{Submitted \today.}
\begin{document} 
\maketitle 

\begin{abstract}
We present deep 450\um\ and 850\um\ observations of a large, uniformly
covered 394\,arcmin$^{2}$ area in the COSMOS field obtained with the
\scubaii\ instrument on the James Clerk Maxwell Telescope (JCMT).  We
achieve root-mean-square noise values of $\sigma_{\rm 450}$=4.13\,mJy
and $\sigma_{\rm 850}$=0.80\,mJy.  The differential and cumulative
number counts are presented and compared to similar previous works.
Individual point sources are identified at $>$3.6$\sigma$
significance, a threshold corresponding to a 3--5\%\ sample
contamination rate.  We identify 78 sources at 450\um\ and 99 at
850\um, with flux densities $S_{\rm 450}=$13--37\,mJy and $S_{\rm
  850}=$2--16\,mJy.  Only 62--76\%\ of 450\um\ sources are
850\um\ detected and 61--81\%\ of 850\um\ sources are
450\um\ detected.  The positional uncertainties at 450\um\ are
small (1--2.5\arcsec) and therefore allow a precise identification of
multiwavelength counterparts without reliance on detection at
24\um\ or radio wavelengths; we find that only 44\%\ of 450\um-sources
and 60\%\ of 850\um-sources have 24\um\ or radio counterparts.
450\um-selected galaxies peak at $\langle z\rangle=1.95\pm0.19$ and
850\um-selected galaxies peak at $\langle z\rangle=2.16\pm0.11$.  The
two samples occupy similar parameter space in redshift and luminosity,
while their median SED peak wavelengths differ by
$\sim$20--50\um\ (translating to $\Delta T_{\rm dust}=8-12$\,K, where
450\um-selected galaxies are warmer).  The similarities of the
450\um\ and 850\um\ populations, yet lack of direct overlap between
them, suggests that submillimeter surveys conducted at any single
far-infrared wavelength will be significantly incomplete (\simgt30\%)
at censusing infrared-luminous star formation at high-$z$.
\end{abstract}

\section{Introduction}\label{introduction}

\begin{figure}
\centering
\includegraphics[width=0.99\columnwidth]{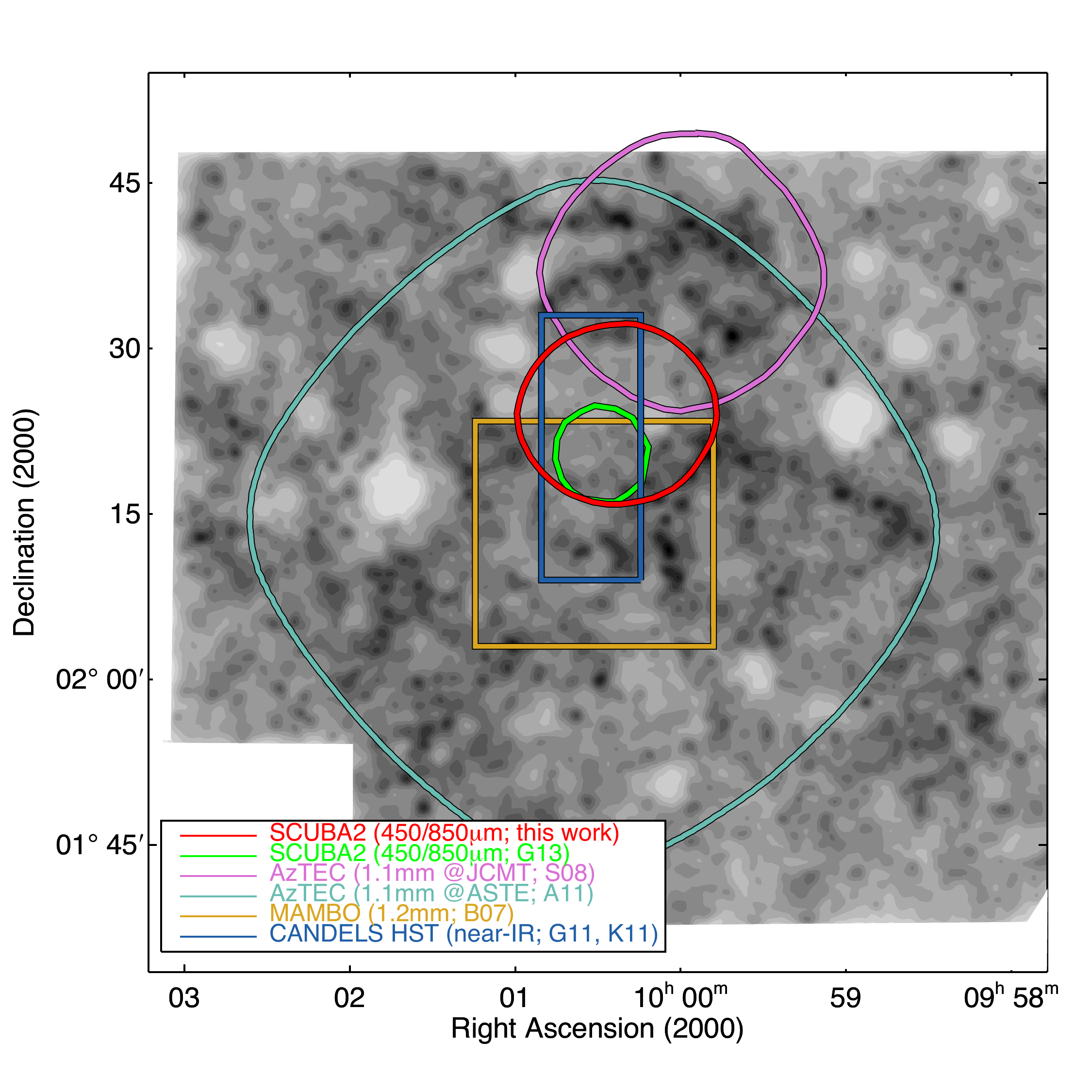}
\caption{ The surface density of galaxies showing large scale
  structure in the COSMOS field (darker shades indicate the most dense
  areas).  This uses the most recent compilation in
  \citet{scoville13a}.  
Overplotted are the areas surveyed in the submillimeter at
  1.2\,mm \citep[with MAMBO, $gold$;][]{bertoldi07a}, at 1.1mm
  \citep[with AzTEC on JCMT, $lavender$;][]{scott08a}, at 1.1mm
  \citep[with AzTEC on ASTE, $teal$;][]{aretxaga11a}.  Our
  450/850\um\ \scubaii\ coverage area is shown in $red$, and the
  \scubaii\ coverage of \citet{geach13a} and \citet{roseboom13a} is shown in $green$.  The
  CANDELS deep near-infrared HST coverage area is shown in $blue$.}
\label{fig:cosmos}
\end{figure}

The arrival of the Submillimeter Common User Bolometer Array
\citep[\scuba;][]{holland99a} on the James Clerk Maxwell Telescope
(JCMT) fifteen years ago ushered in a new age of galaxy evolution
studies by highlighting the importance of distant infrared-bright,
ultraluminous galaxies to the buildup of the Universe's stellar mass
and the formation of massive, local elliptical galaxies.
Submillimeter galaxies \citep[SMGs, canonically selected at
  850\um\ with $S_{\rm
    850}$\simgt2--5\,mJy;][]{smail97a,barger98a,hughes98a,eales99a}
have since been shown to have a peak volume density at
$z\approx2.2-2.5$ \citep{chapman03a,chapman05a,yun12a}.

Follow-up studies of their molecular and ionised gas properties
\citep{swinbank04a,neri03a,greve05a,tacconi06a,tacconi08a,bothwell10a,engel10a,banerji11a,alaghband-zadeh12a}
indicate that most star formation in bright SMGs is triggered by major
mergers of gas-rich disk galaxies, similar to local (ultra-) luminous
infrared galaxies, (U)LIRGs
\citep[e.g.][]{sanders88a,sanders96a,armus09a,u12a}.  Furthermore,
several detailed studies of the X-ray and mid-infrared properties of
SMGs have led to the conclusion that the majority of the SMGs'
bolometric energy is star formation dominated
\citep{alexander05a,pope08a,coppin08a,menendez-delmestre09a,coppin10a,laird10a}.
However, several recent studies have pointed out that, despite their
extreme star formation rates \simgt200$\sfr$, SMGs and high-$z$ ULIRGs
are not homogeneously described by major mergers
\citep[e.g.][]{daddi09a,bothwell10a,elbaz11a,rodighiero11a,alaghband-zadeh12a,hayward12a,targett11a}.
It is clear that the driving physical mechanisms for SMGs remains a
puzzle at high-$z$ despite these detailed multiwavelength studies.

The limiting factors in the analysis and characterisation of SMGs
is small number statistics and the lack of secure counterpart
identifications.  The vast majority of the studies mentioned
above$-$although thorough$-$are based on small samples of SMGs,
anywhere from 2-75 galaxies (but mostly on samples with less than
ten sources).  Some of these SMGs could have been mischaracterised due
to incorrect counterpart identifications, which may be a consequence
of the large beamsize of submm single-dish observations (the JCMT
beamsize at 850\um\ is $\sim$15\arcsec).  The problem of limited
statistics can be alleviated by wide-field submillimetre mapping.
Mapping large regions of sky at long wavelengths is technically
challenging, particularly if the maps are intended to be sufficiently
deep and sensitive enough to detect unlensed galaxies in the early
Universe.  New submillimeter/far-infrared (FIR) facilities, such as
AzTEC \citep{wilson08a} and primarily
\herschellong\ \citep{pilbratt10a}, have mapped larger sky areas in
recent years. \herschel\ has mapped hundreds of square degrees at
250--500\um \citep{eales10a,oliver12a}.  However, its surveys are more
limited in sensitivity than SCUBA, primarily due to the smaller
primary dish, which results in large beam sizes and much higher
confusion limits.

The new Submillimeter Common User Bolometric Array-2
\citep[\scubaii][]{holland13a} instrument on JCMT presents exciting
new avenues to explore infrared-bright starbursts in the early
Universe.  Besides the four-fold increase in bolometer arrays and
large field-of-view scans, which make 850\um\ mapping more efficient
than it was with \scuba, the \scubaii\ 450\um\ bolometers provide the
first opportunity to identify 450\um-bright sources with remarkable
resolution in the FIR; the JCMT beamsize at 450\um\ is $\sim$7\arcsec,
compared to $\sim$36\arcsec\ at 500\um\ on {\it Herschel}.  Direct
detection in the FIR with a small beamsize is a large step forward,
since multiwavelength counterpart identification has been a key
limiting factor in the interpretation of submillimeter-bright sources
over the past decade.  Deep \scubaii\ maps, which are much less
limited by confusion noise, are the scientific complement to the
large-scale mapping done with \herschel.

\begin{figure*}
\centering
\includegraphics[width=1.85\columnwidth]{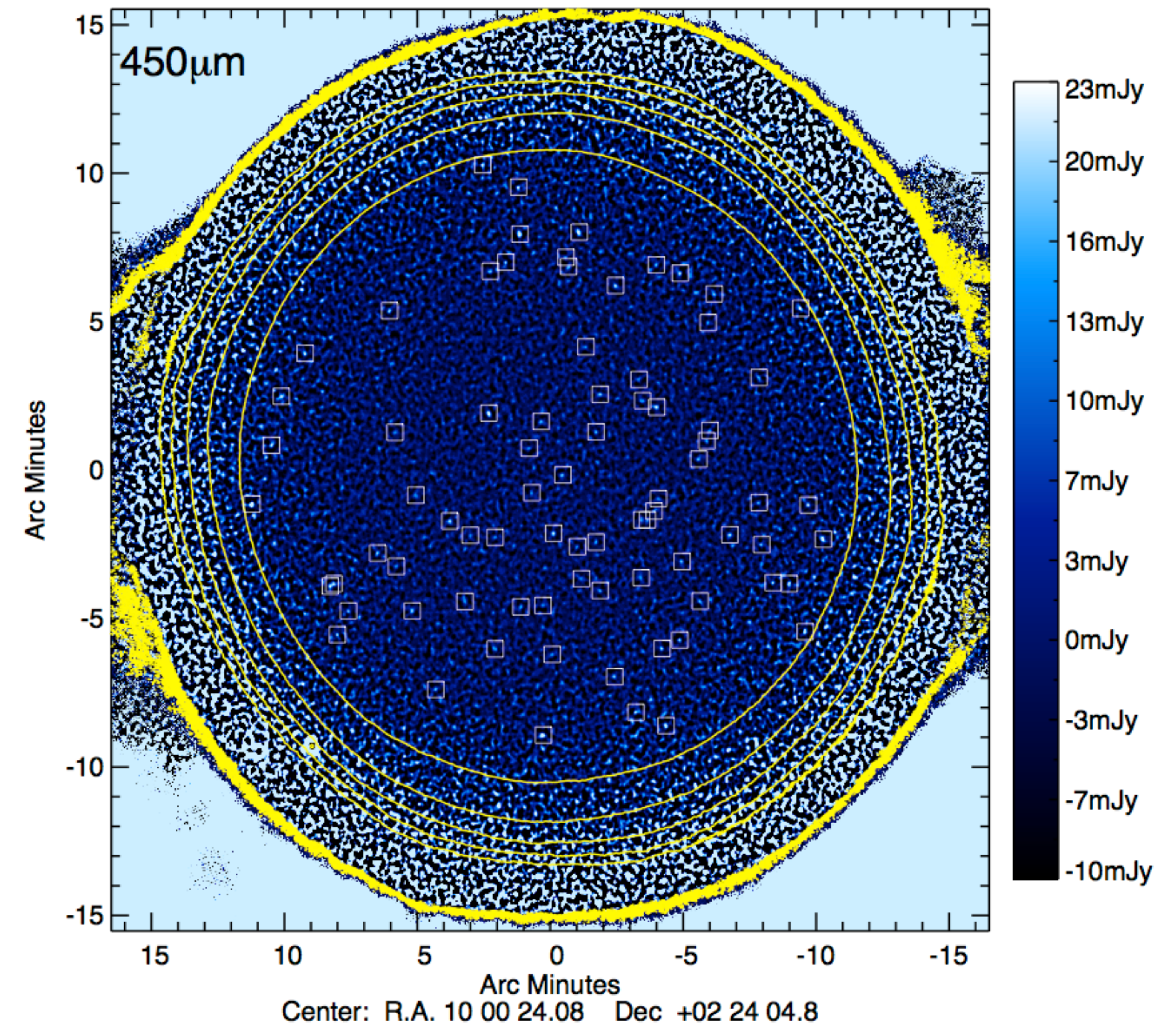}\\
\caption{ The COSMOS 450\um\ \scubaii\ map.  The colour scale is linear
  with upper and lower ranges of -10$<S_{\rm 450}<$23\,mJy/beam.
  Contours mark the 2, 3, 4, 5, and 6$\times$ the central RMS value,
  measured to be 4.13\,mJy/beam at 450\um.  We search for sources
  which are detected with RMS less than double the central value
  (roughly corresponding to the inner most contour).  The area probed
  by this cut has a radius $\sim$11.2\,arcmin and an area
  $\sim$394\,arcmin$^{2}$.  The 450\um--identified $>$3.6$\sigma$
  sources are identified with white boxes (a justification of the
  3.6$\sigma$ cut can be found in the text).}
\label{fig:map450}
\end{figure*}

\begin{figure*}
\centering
\includegraphics[width=1.85\columnwidth]{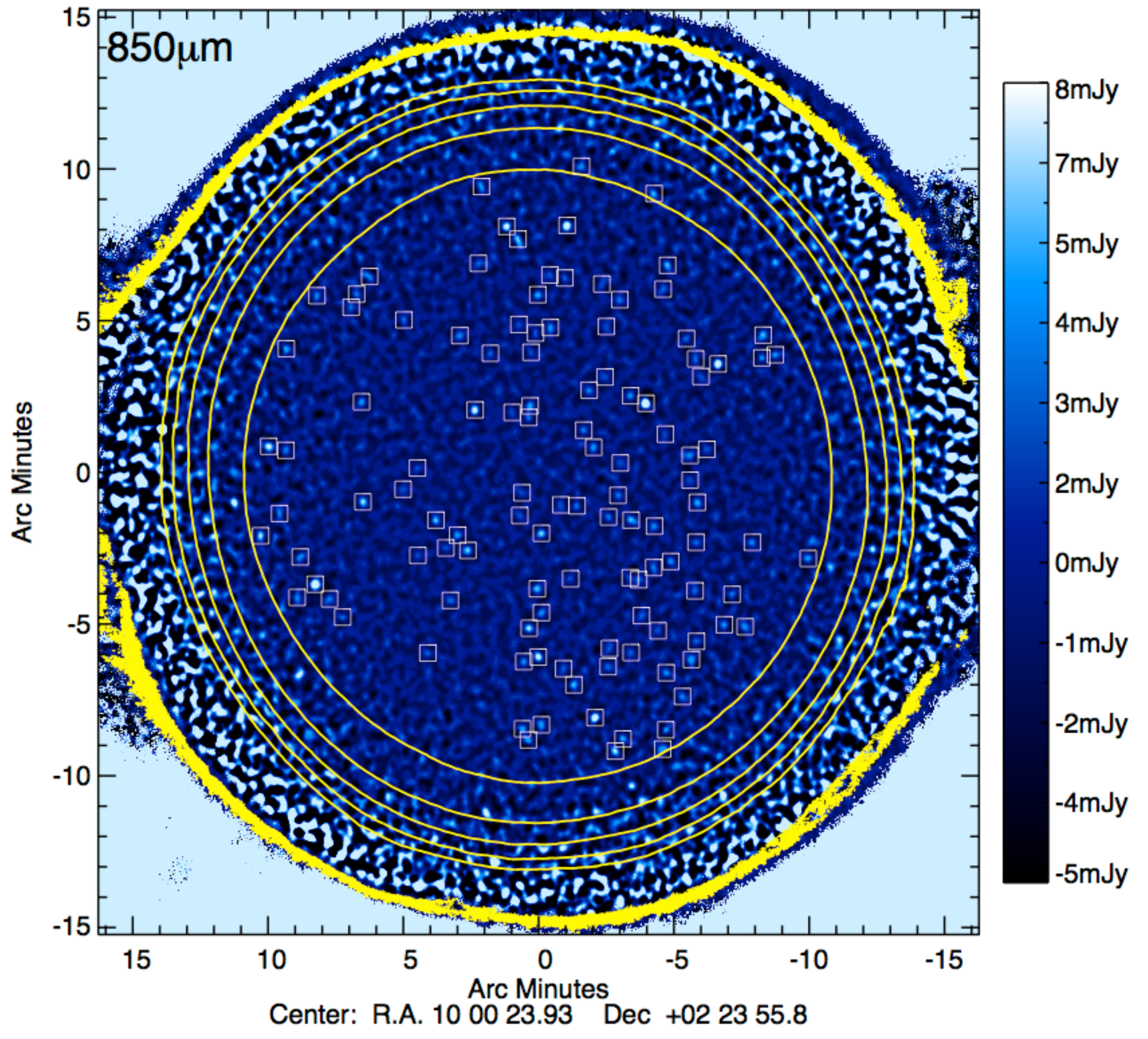}\\
\caption{The COSMOS 850\um\ \scubaii\ map, plotted in the same fashion
  as Figure~\ref{fig:map450}.  The colour scale is linear and runs from
  -5$<S_{\rm 850}<$8\,mJy/beam, and the measured central RMS is
  0.80\,mJy/beam.  All $>$3.6$\sigma$ sources are marked with white
  boxes.  Some sources appear outside the nominal 2$\times$RMS
  contour, but locally have noise that dips below the 2$\times$RMS
  threshold.}
\label{fig:map850}
\end{figure*}

This paper presents 450\um\ and 850\um\ observations from \scubaii\ in
a central region of the Cosmic Evolution Survey (COSMOS) field
\citep{scoville07a} surrounding the Cosmic Assembly Near-infrared Deep
Extragalactic Legacy Survey (CANDELS) area.  Recently,
\citet{geach13a} presented deep \scubaii\ 450\um\ observations in the
central $\sim$100\,arcmin$^2$ of this area.  Our observations have a
wider, more uniform coverage than the data presented in
\citet{geach13a}, with the motivation of selecting a homogeneous
sample, while the \citet{geach13a} data are deeper and patterned to
find fainter sources.  The pointing on the CANDELS area is motivated
by the perceived high correlation of submillimeter emission with
near-infrared emission \citep[e.g.][]{wang06a,serjeant08a} and the
ability to morphologically characterise SMG counterparts using the
{\it Hubble Space Telescope}'s high spatial resolution.  However, we
leave the detailed morphological analysis of these galaxies to a
future work.  This work is also complemented by the deeper, yet
smaller area \scubaii\ coverage and analysis in the lensed cluster
field A\,370 by \citet{chen13a}, which currently provides the deepest
450\um\ counts.

Section~\ref{sec:observations} describes our \scubaii\ observations
and data reduction as well as ancillary data used,
section~\ref{sec:pointsources} discusses the identification and flux
density measurements of point sources in the maps and differential and
cumulative number counts, section~\ref{sec:derived} describes the
multiwavelength properties of the \scubaii-selected galaxies, and
section \ref{sec:discussion} discusses their implications, relation to similar
starburst galaxy populations, and implications for future, deeper
\scubaii\ observations.  Throughout we assume a $\Lambda$ {\sc CDM}
cosmology with $H_{\rm 0}$=71\,km\,s$^{-1}$\,Mpc$^{-1}$ and
$\Omega_{\rm m}$=0.27 \citep{hinshaw09a}.

\section{Observations \&\ Data}\label{sec:observations}

Observations were taken with the \scubaii\ instrument on the JCMT on
2011-Dec-26, 2011-Dec-28, 2012-Feb-04, 2012-Feb-05, 2012-Feb-07,
2012-Apr-09, 2012-Apr-30, 2012-May-01, 2012-May-02, and 2012-Dec-21
under programs M11BH11A, M12AH11A and M12BH21A.  Conditions were
optimum, with the optical depth varying from 0.02$<\tau_{\rm
  225GHz}<0.05$ and averaging $\tau_{\rm 225GHz}\approx0.04$.  At the
wavelengths of our observations, the optical depths were approximately
$\tau_{450\mu\!m}\approx0.55$ and $\tau_{850\mu\!m}\approx0.16$.  The
integration times on-field per night were 3.5\,hrs, 3.6\,hrs,
3.1\,hrs, 5.5\,hrs, 5.5\,hrs, 4.5\,hrs, 2.9\,hrs, 3.8\,hrs, 0.6\,hrs
and 5.0\,hrs respectively, totalling 38.0\,hrs on-field.  We centred
our map at position 10:00:28.0, +02:24:00.  See
Figure~\ref{fig:cosmos} for a map of the galaxy density in the
surrounding area along with areas covered at other submm--mm
wavelengths.  The {\sc pong}-900 mapping pattern was used to achieve a
uniform RMS over a large area $\approx$15$\times$15\,arcmin\ in
preference to the Daisy mapping which has substantial patterning and
non-uniformity and covers a smaller area.

Raw data were downloaded via the Canadian Astronomy Data Centre (CADC)
JCMT Science Archive and processed with the \scubaii\ software
packages {\sc Smurf}, the Sub-Millimetre User Reduction Facility, and
{\sc Picard}, Pipeline for Combining and Analyzing Reduced Data.  Data
were split up by night, wavelength, and sub-array for processing.
{\sc Smurf} processing was first executed on each night's sub-array
data (which is equivalent to combining each sub-array per Minimum
Scheduable Block (MSB), and then combining all MSBs).  Partial maps
were constructed using the {\sc makemap} routine \citep{chapin13a} in
{\sc Smurf} with the default deep extragalactic configuration file
optimising background subtraction.  The {\sc makemap} uncalibrated
maps were combined at each stage (by sub-array, then nights) using the
{\sc Picard} recipe {\sc mosaic\_jcmt\_images}.  Each of the four
sub-arrays at 450\um\ and 850\um\ were combined for each night's worth
of data, and then the nights were combined to produce the unprocessed,
raw maps at 450\um\ and 850\um\ in units of pW.

We applied the matched-filter {\sc Picard} recipe to suppress pattern
noise and increase the signal from individual sources, which are
generally unresolved.  At 450\um, the map RMS noise drops
substantially from 14.08\,mJy to 4.13\,mJy.  A similar yet less
dramatic drop is seen at 850\um, from 2.18\,mJy to 0.80\,mJy.  Flux
calibration is done by applying the measured flux conversion factors
(FCF) from the first eight months of calibrator data, primarily from
Uranus and Mars (FCF$_{\rm
  850}$=556$\pm$45\,Jy\,pW$^{-1}$\,beam$^{-1}$ and FCF$_{\rm
  450}$=606$\pm$55\,Jy\,pW$^{-1}$\,beam$^{-1}$).  These flux
conversions factors differ from the current canonical values
\citep{dempsey13a} because we used an earlier version of {\sc Picard}.
Calibration data taken during the observations for this project agree
with these FCF values to $\sim$10\%.  Final reduced maps, in units of
mJy/beam, are shown in Figures~\ref{fig:map450} and \ref{fig:map850}.
The units are such that the flux density, in mJy, for an unresolved
source is simply the sources' peak value in the map.  Signal-to-noise
maps are constructed using the {\sc Picard} recipe {\sc makesnr};
these are not shown here but are quite similar to the final reduced
maps within the central area.  We found no evidence for a systematic
astrometric offset necessary after inspection of the radio data
described in the next section \citep[following the astrometric
  calibration procedure of][]{chen13a}.  Signal-to-noise maps are used
for the identification of point sources.

To avoid contamination from the less sensitive edges of the map in
calculating an overall characteristic map noise, we first measured the
effective area of the sensitive region of the map, where the RMS was
less than or equal to a factor of two times the RMS in the centre.  We
determined this boundary by measuring the RMS in concentric annuli
5\arcsec\ wide from a jackknife map, having subtracted out the signal
from point sources.  The jackknife map represents the instrumental noise of our
observations with sources removed and is constructed by subtracting
one half of our data from the other, then scaling the noise by the
square root of the integration time.  This 2$\times$RMS boundary lies
$\sim$11.2' from the map centre in both the 450\um\ and 850\um\ maps,
corresponding to an area of 394\,arcmin$^{2}$, slightly larger
than the anticipated 15'$\times$15' {\sc Pong}-900 target area.

\begin{figure*}
\includegraphics[width=0.99\columnwidth]{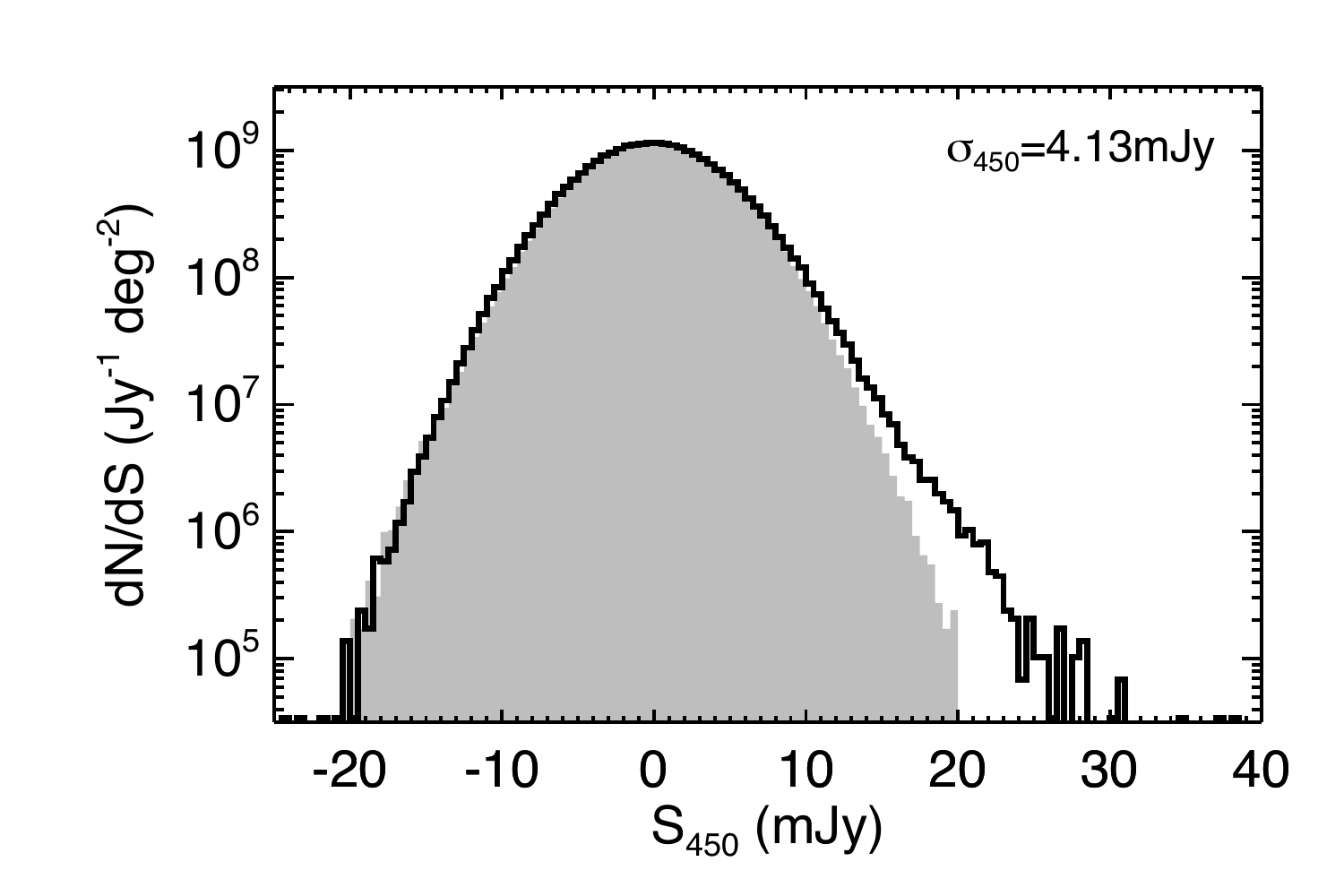}
\includegraphics[width=0.99\columnwidth]{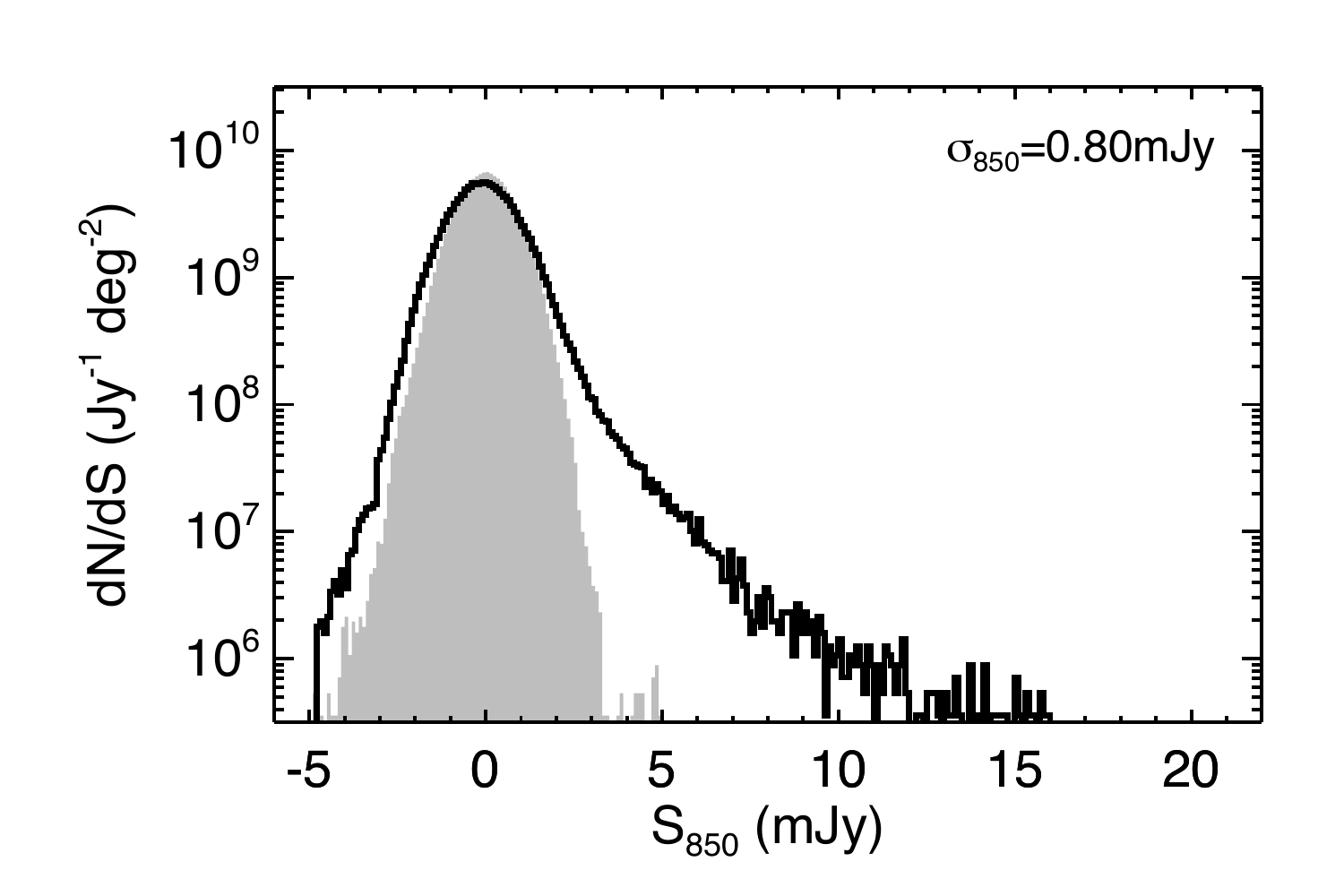}
\caption{ Pixel flux distributions (per unit area) within the central
  $\sim$400\,arcmin$^2$ area at 450\um\ and 850\um.  The solid black
  lines indicate the distributions of the data, showing excesses at
  high flux densities where point sources contribute to the
  distributions.  At lower flux densities, the distributions are
  dominated by instrument noise. The negative excesses are due to
  residual troughs around bright sources (a characteristic of the
  matched filter technique). The gray shaded histograms show the pixel
  noise distributions within the same region for the jackknife maps.
  As discussed in the text, the jackknife maps were constructed by
  differencing two halves of the data, which removes point sources and
  leaves only residual noise.}
\label{fig:differential}
\end{figure*}

The pixel flux distributions for the $\sim$400\,arcmin$^2$ central
region are shown in Figure~\ref{fig:differential} with the
distributions of the jackknife pure-noise maps, in the same
$\sim$400\,arcmin$^2$ area, highlighted in gray.  The positive excess
above the jackknife noise map is attributed to the detection of real
sources.  The negative flux excesses (prominent at 850\um) is due to
troughs around high-S/N sources and is a characteristic of the removal
of pattern noise and is exaggerated by the matched-filter technique.
In other words, in order to boost the signal in real sources,
especially at S/N ratios $>$5, matched-filter creates rings or troughs
of negative flux surrounding the source.  Since there are 44
850\um\ sources at S/N$>$5 and only 8 450\um\ sources at S/N$>$5, the
effect of a negative excess is more pronounced at 850\um.

\subsection{COSMOS Ancillary Data}

\begin{figure*}
\includegraphics[width=1.75\columnwidth]{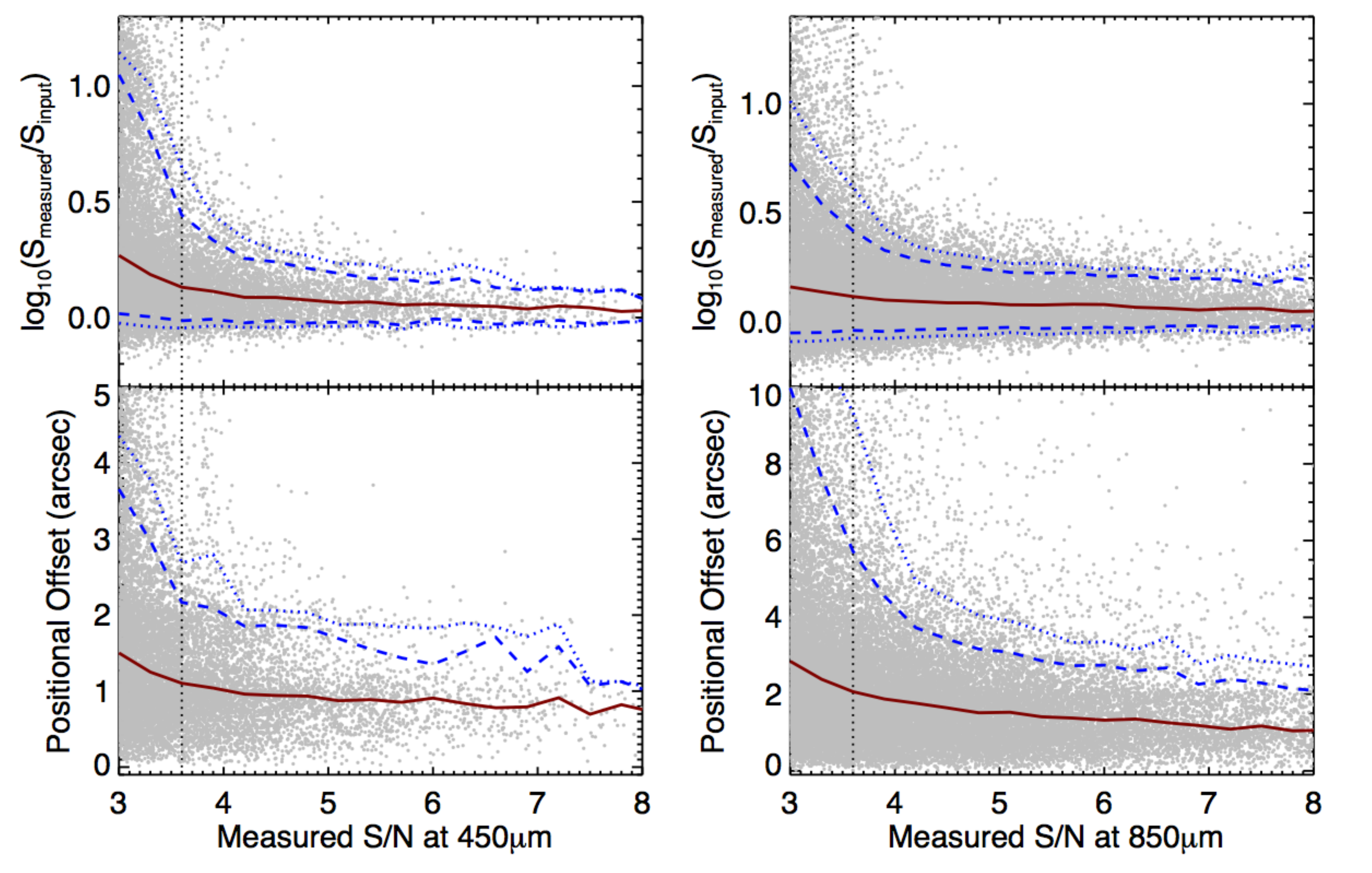}
\caption{ Our results from Monte Carlo simulations, showing more than
  10000 source realisations and the measured boosting factor and
  positional offset for each source (gray points) at 450\um\ (left)
  and 850\um\ (right).  The boosting factor is measured as the ratio
  of output measured flux density over input flux density.  We deboost
  our sources' flux densities using the median boosting factors (red
  lines).  The 90\%\ confidence intervals (dashed blue lines) and
  95\%\ confidence intervals (dotted blue lines) are overplotted.  We
  assign a positional uncertainty to sources according to the
  90\%\ confidence interval. The 3.6$\sigma$ signal-to-noise threshold
  is marked by dotted vertical lines. }
\label{fig:boosting}
\end{figure*}

We use the vast ancillary data available in the COSMOS field
\citep{scoville07a} to investigate the physical nature of the
\scubaii\ submillimeter-bright sources and compare with larger samples
of submillimeter sources previously analysed in the literature.
The imaging and associated data reduction for
{\it Spitzer Space Telescope} IRAC and MIPS data is described in
\citet{sanders07a}, \citet{le-floch09a} and \citet{frayer09a}.

Photometric redshifts were fit using the {\sc Le
  Phare}\footnote{http://www.cfht.hawaii.edu/$\sim$arnouts/LEPHARE/cfht\_lephare/
  lephare.html} code \citep{ilbert09a,ilbert10a} to multi-band
ultraviolet, optical, near-infrared (NIR) and mid-infrared (MIR)
photometry.  Stellar population templates
\citep{polletta07a,bruzual03a} were assumed as input to {\it Le
  Phare}.  Extinction is a free parameter of the fit and is given a
maximum value of $E(B-V)=0.5$, where galaxies redder than Sb have no
extinction (i.e. assuming older stellar populations rather than
dustier systems).  Not allowing a wider range of extinctions could
impact the quality of photometric redshifts in our sample, despite the
fact that 30+ photometric bands are used to measure redshift; this is
an issue we will investigate in a future work.

In this paper we quote the measured {\sc Le Phare} output
parameters of stellar mass and star formation rate from the
\citep{ilbert09a} catalogue for contrast.  Both are measured from
stellar template matching to UV/optical/near-infrared photometry only,
and in this paper, we refer to the template star formation rate as
``SFR$_{\rm UV}$.''  While this SED fitting method is widely used for
large, catalogue-based approaches to galaxy characterisation, we note
that it differs from the method of using the ultraviolet slope,
$\beta$, for measuring dust attenuation (and therefore dust-corrected
total star formation rates).  The latter method is used by
\citet{meurer99a} and \citet{reddy12a}, among others, and is shown to
more accurately predict dust attenuation in luminous infrared
galaxies.  We discuss the contrasting SFR measures later in
section~\ref{sec:normal}.

\section{Identification of Point Sources}\label{sec:pointsources}

Point sources are extracted in each map by isolating high
signal-to-noise pixel groups (e.g.  adjacent pixels) in the
signal-to-noise maps.  Within each pixel group, the highest
signal-to-noise pixel marks the object centre and point at which the
raw flux density is measured.  The raw flux densities are measured as
the peak flux density of the source in mJy/beam, assuming the sources
are unresolved (a safe assumption for high-redshift infrared
galaxies).  

Here we describe our use of Monte Carlo simulations to determine an
appropriate detection threshold for reporting detections, and for
measuring number counts, boosting factors and completeness and
contamination rates.

\subsection{Monte Carlo Simulations}\label{sec:simulations}

The estimation of cumulative number counts at both 450\um\ and 850\um,
along with the completeness and contamination rates of our samples,
deboosting factors, and adequate detection thresholds requires the use
of Monte Carlo simulations.  This is an iterative process which sheds
light on the observational impact and limitations on the true number
counts \citep[see ][for a thorough description of the
  process]{coppin05a}.  Monte Carlo testing is done by injecting fake
sources into a pure-noise, or jackknife, map.  Injecting sources into
our map with known flux densities and positions, we can measure the
accuracy by which we recover those sources using the above extraction
method.

Injecting fake sources into a noise map requires an $a\ priori$
assumption of intrinsic number counts; at 450\um, we vary the fit
parameters from \citet{chen13a} and \citet{geach13a}, which use a
double-power law and Schechter form respectively until our measured
raw number counts match those found in our Monte Carlo simulations.
At 850\um, we vary the parameters around those used in
\citet{knudsen08a}.  We use the Knudsen \etal\ initial variables to
take advantage of data at the faintest flux densities \citep[which are
  in agreement with previous blank-field number counts
  work][]{coppin06a}.
To generate substantial statistical samples of input and extracted
sources, we generate 200 maps at both wavelengths (which generates
$\sim$10,000 extracted sources).  Boosting, positional offset,
completeness and contamination are all estimated using the known input
sample, position and flux density, and the measured output sample,
position, flux density and signal-to-noise.

\begin{figure}
\includegraphics[width=0.89\columnwidth]{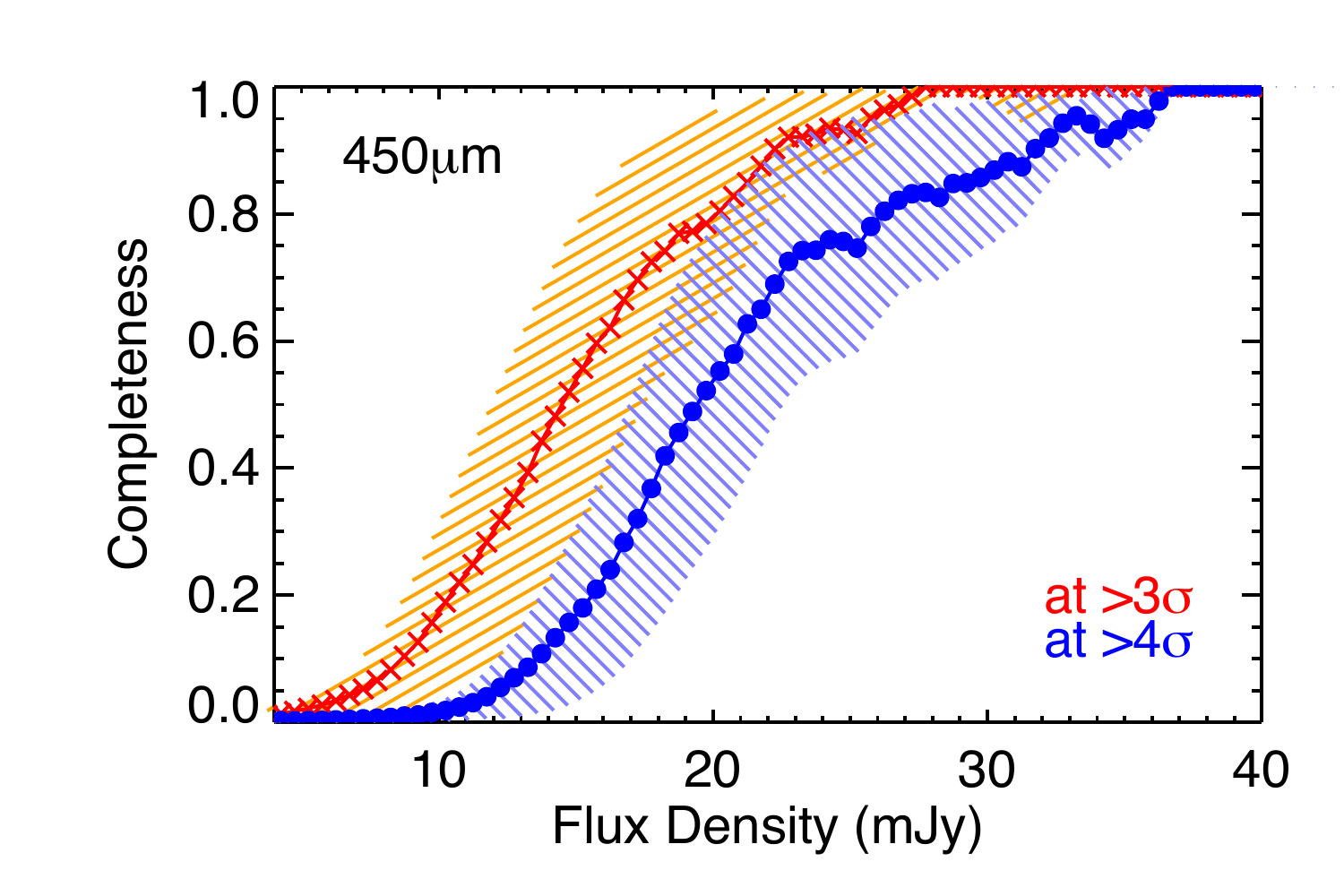}
\includegraphics[width=0.89\columnwidth]{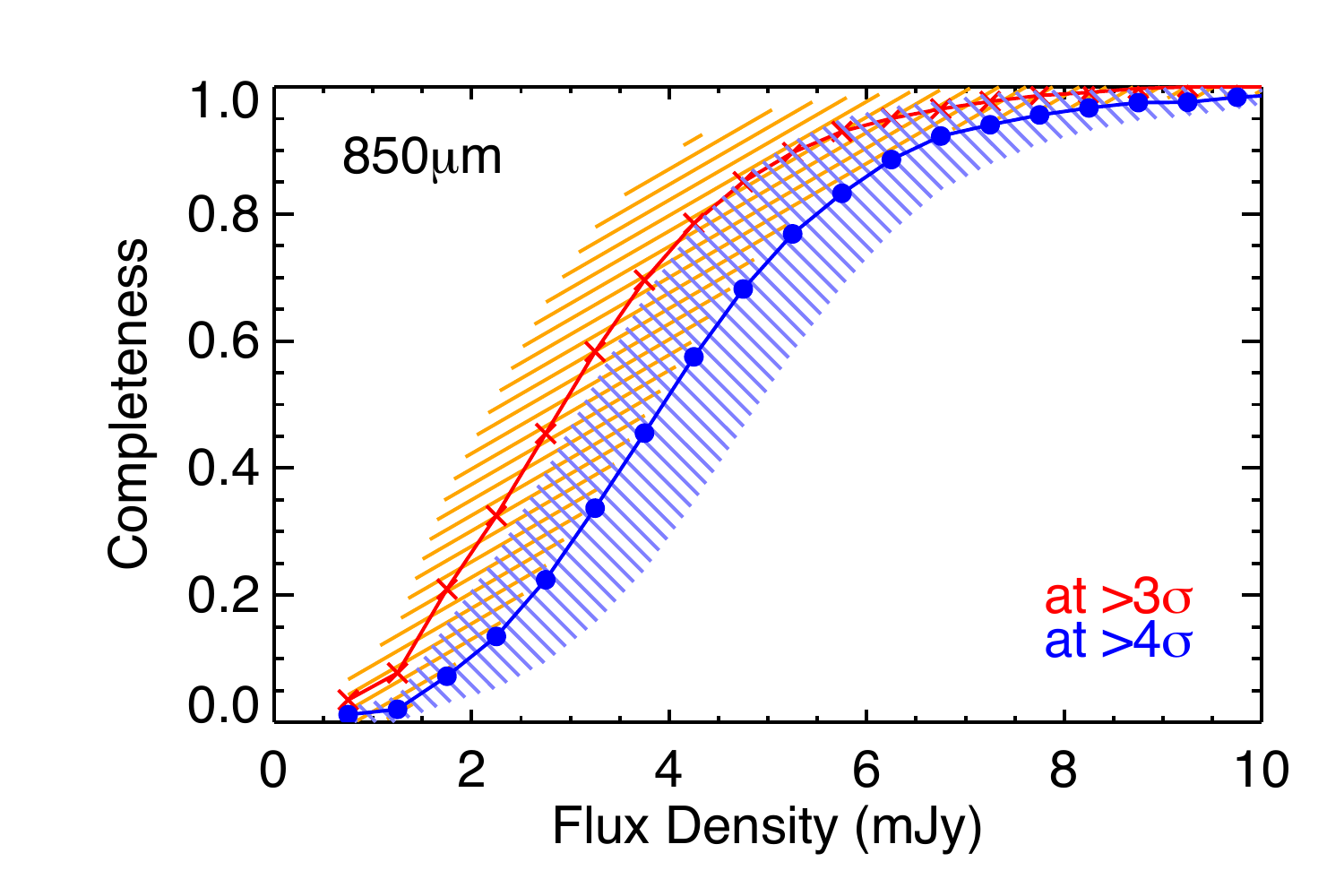}\\
\caption{The completeness curves at 450\um\ and 850\um.  Completeness
  is measured as a function of input flux density as the number of
  sources in a given flux bin which are recovered at the
  signal-to-noise threshold (3$\sigma$ and 4$\sigma$ are plotted
  here).  The uncertainty on the completeness curve is generated from
  a bimodal uncertainty.}
\label{fig:completeness}
\end{figure}

The flux densities of sources we measure from our maps need correction
for flux boosting. Sources' flux densities are expected to be boosted
from both Eddington boosting \citep{eddington13a} and confusion boosting.
We have estimated the
magnitude of these two boosting factors together using our simulations
results, which we show as a function of detection S/N in the top
panels of Figure~\ref{fig:boosting}.
Figure~\ref{fig:boosting} also shows the median positional offset
between the measured output position of a 450\um- or 850\um- source
and its input position in the bottom panels (a maximum search radius
of of 7\arcsec\ and 14\arcsec\ were used, respectively).  The positional
uncertainty quoted in Tables~\ref{tab:sources450} and
~\ref{tab:sources850} is the 90\%\ confidence interval, which we then
use for matching to radio, 24\um, and optical counterparts as
described in sections~\ref{sec:counterparts}.  Higher signal-to-noise
detections can yield very small positional uncertainties.

Before measuring the cumulative number counts or deciding on an
appropriate S/N detection threshold, it is essential to understand our
sample's completeness and contamination rate.  Completeness
deteriorates at low flux densities.  We estimate completeness from our
Monte Carlo simulations as the fraction of sources recovered
$>$3$\sigma$ or $>$4$\sigma$ in our simulated maps and is shown in
Figure~\ref{fig:completeness}.  As is expected, a more conservative
(4$\sigma$) cut on the sample results in a more incomplete sample at
moderate flux densities than a more liberal cut (3$\sigma$).  We will
take these completeness curves into consideration with the calculation
of the cumulative number counts.

\begin{figure}
\centering
\includegraphics[width=0.49\columnwidth]{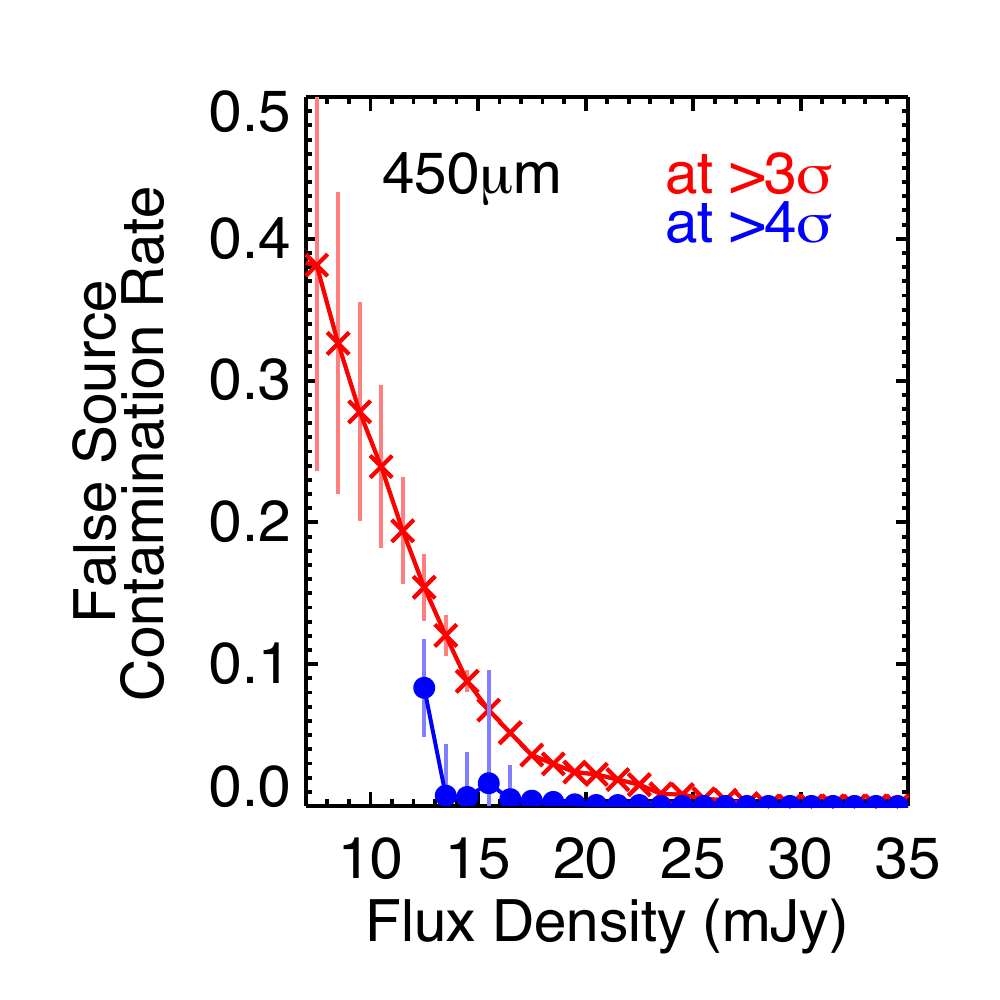}
\includegraphics[width=0.49\columnwidth]{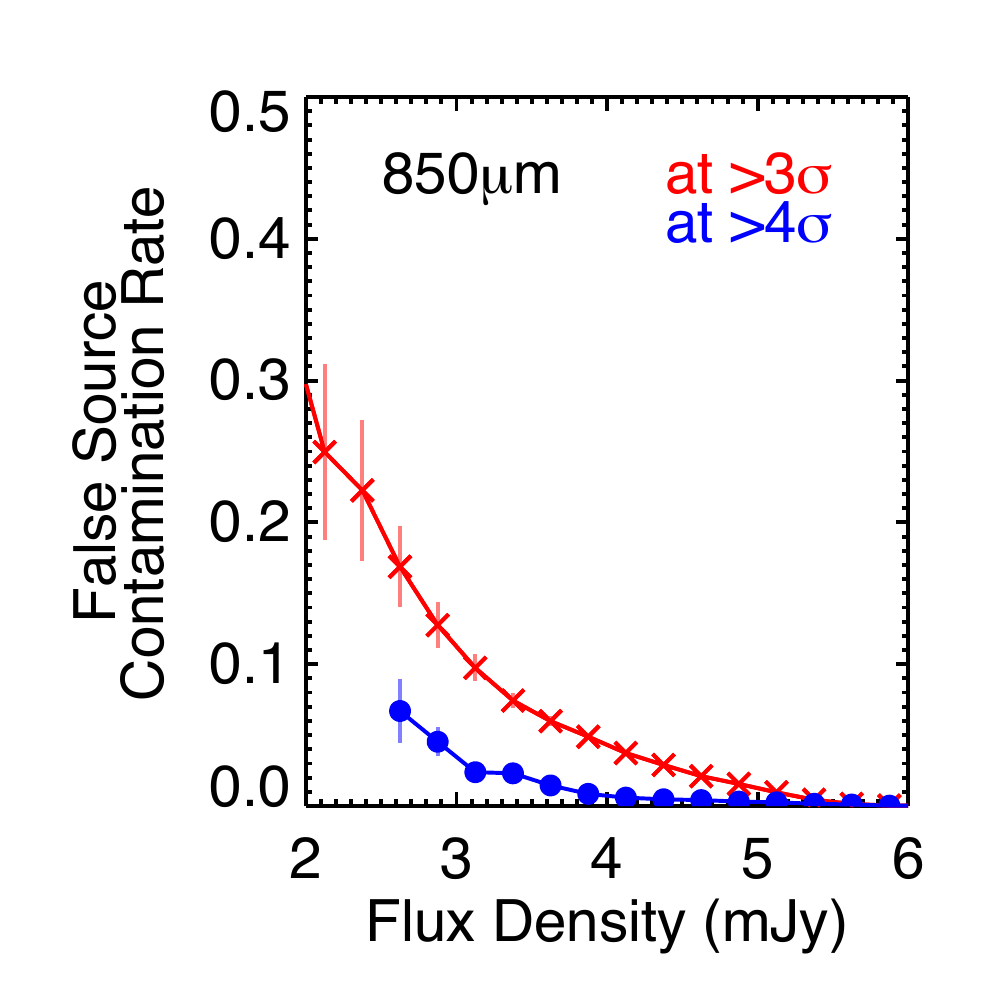}\\
\includegraphics[width=0.59\columnwidth]{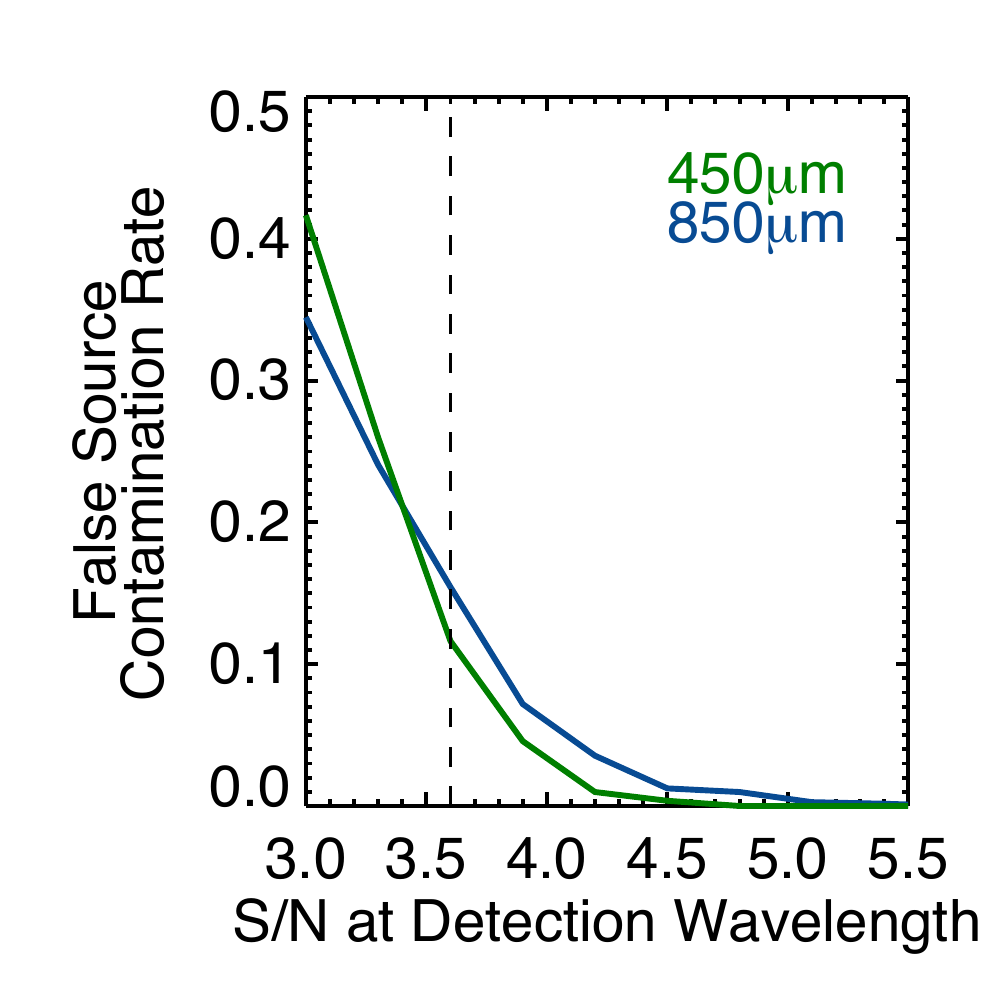}
\caption{The contamination rates at 450\um\ and 850\um\ as a function
  of flux density (top two panels) and signal-to-noise (bottom panel).
  The contamination rate is computed as the number of sources which
  are extracted from Monte Carlo simulations which are (a) truly
  spurious, caused by noise fluctuations, or (b) at flux densities
  lower than the nominal 3$\sigma$ detection limit, after correction
  for deboosting.  The sharp upturn in contamination at $\sim$12\,mJy
  at 450\um\ is due to small number counts in our simulations (despite
  $>$10K realisations, there are very few $>$4$\sigma$ sources at
  $\sim$12\,mJy).  We use the measured contamination rate curves in
  the bottom panel, as a function of signal-to-noise, to infer the
  likelihood that a source of a given signal-to-noise is a
  contaminant.  We plot our final signal-to-noise threshold of
  3.6$\sigma$ on the bottom panel, which shows that the maximum
  individual source contamination likelihood is $\sim$15\%\ in both
  450\um\ (green) and 850\um\ (blue) samples.}
\label{fig:contam}
\end{figure}

Deciding on an appropriate detection threshold requires an
understanding of our sample's contaminants.  We measure contamination
rate as a function of detection signal-to-noise.  At a given detection
S/N, the contaminating fraction is the number of sources which are
generated by fluctuations in the map due to noise or from sources well
below the nominal detection limit.  Since our maps are roughly
uniform, we fix this limiting flux density at the lowest deboosted
flux density of our 3$\sigma$ sample.  Any source with input flux
density below this threshold would not be expected to be measured in
our data even after Eddington and confusion boosting, so here is
considered a contaminant.  The contamination rate as a function of
both flux density and signal-to-noise is shown in
Figure~\ref{fig:contam}.

Figure~\ref{fig:contam} tells us the probability that a given source
with a given signal-to-noise ratio is spurious.  Accordingly, we are
able to assign a ``probability of contamination'' ($p_{\rm contam}$)
for each source.  We use this source-by-source probability to estimate
total contamination rates for samples defined by different
signal-to-noise detection thresholds.  
In a sample sized $N$, the probability of having no
contaminants is equal to $\prod_{i=1}^{N}(1-p_{i})$, the probability
of having one contaminant is equal to $\sum_{i=1}^{N}p_{i}\prod_{k\ne
  i}^{N}(1-p_{k})$, and so on.
For example, there are 37
sources in a 4$\sigma$ cut in our 450\um\ map.  We estimate that the
4$\sigma$ sample has a 76\%\ likelihood of having {\it no}
contaminants, and a 97\%\ chance of having a contamination rate of
$<$3\%.  
This cut is
quite conservative, so we push the detection threshold lower until we
estimate contamination rates of 3--5\%.  This is our target
contamination rate since it is low enough not to significantly impact
the global properties of the sample, while it allows for a far larger
sample size strengthening the analysis of the population.  At both
450\um\ and 850\um, we estimate that a 3.6$\sigma$ detection threshold
will yield a 3--5\%\ contamination rate.  As is seen in
Figure~\ref{fig:contam}, these limits correspond to similar `worst'
individual source contamination rates of $\sim$0.14 at 450\um\ and
$\sim$0.17 at 850\um\ (i.e. the lowest S/N sources in the 3.6$\sigma$
samples have a $\approx$14 or 17\%\ likelihood of being spurious).

\begin{table*}
\centering
\caption{450\um--identified $>$3.6$\sigma$ point sources in COSMOS.}
{\tiny \begin{tabular}{c@{ }c@{ }cccccc@{ }c@{ }c@{ }c@{ }c@{ }c}
\hline\hline
Name & Short Name & RA$_{\rm 450}$ & DEC$_{\rm 450}$ & S/N & S$_{\rm 450}$ & S$_{\rm 450}$ & $\Delta$($\alpha$,$\delta$) & $P_{\rm contam}$ & 850\um-  & Offset & S$_{\rm 850}$ & S$_{\rm 850}$   \\
     &            &                &                 &     &   {\sc Raw} & {\sc Deboosted} &     &                  & {\sc Source}  &  & {\sc Raw} & {\sc Deboosted} \\
     &            &                &                 &     &   [mJy] & [mJy] &  [\arcsec]   &                       &               & [\arcsec] & [mJy]     & [mJy] \\
\hline\hline
SMM\,J100028.6+023201... & 450.00 & 10:00:28.6 & 02:32:01 & 7.80 & 40.58$\pm$5.20 & 37.54$\pm$6.58 & 1.06 & $<$0.01 & 850.07 & 0.6   & 9.94$\pm$1.06 & 9.21$\pm$1.45 \\ 
SMM\,J100033.3+022559... & 450.01 & 10:00:33.3 & 02:25:59 & 6.89 & 28.43$\pm$4.13 & 25.62$\pm$4.96 & 1.51 & $<$0.01 & 850.02 & 0.6   & 9.47$\pm$0.80 & 8.22$\pm$0.70 \\ 
SMM\,J100109.0+022255... & 450.02 & 10:01:09.0 & 02:22:55 & 6.15 & 47.20$\pm$7.68 & 41.70$\pm$9.00 & 1.48 & $<$0.01 &        &       & 9.67$\pm$1.71 & \\ 
SMM\,J100008.0+022611... & 450.03 & 10:00:08.0 & 02:26:11 & 5.73 & 23.47$\pm$4.10 & 20.46$\pm$4.78 & 1.45 & $<$0.01 & 850.00 & 0.6   & 16.15$\pm$0.80 & 10.86$\pm$0.54 \\ 
SMM\,J100019.7+023205... & 450.04 & 10:00:19.7 & 02:32:05 & 5.62 & 29.15$\pm$5.19 & 25.35$\pm$6.04 & 1.48 & $<$0.01 & 850.03 & 1.5   & 12.11$\pm$1.03 & 10.54$\pm$0.90 \\ 
SMM\,J100023.6+022156... & 450.05 & 10:00:23.6 & 02:21:56 & 5.62 & 23.14$\pm$4.12 & 20.12$\pm$4.80 & 1.48 & $<$0.01 & 850.08 & 1.9   & 7.19$\pm$0.80 & 6.60$\pm$1.12 \\ 
SMM\,J100039.2+022221... & 450.06 & 10:00:39.2 & 02:22:21 & 5.43 & 22.29$\pm$4.11 & 19.32$\pm$4.76 & 1.55 & $<$0.01 & 850.12 & 1.3   & 6.49$\pm$0.80 & 5.83$\pm$1.13 \\ 
SMM\,J100014.2+023017... & 450.07 & 10:00:14.2 & 02:30:17 & 5.32 & 23.94$\pm$4.50 & 20.65$\pm$5.19 & 1.59 & $<$0.01 &        &       & 2.64$\pm$0.89 & \\ 
SMM\,J100025.1+021508... & 450.08 & 10:00:25.1 & 02:15:08 & 5.15 & 31.25$\pm$6.06 & 26.71$\pm$6.93 & 1.67 & $<$0.01 & (850.46) &       & 4.27$\pm$1.15 & \\ 
SMM\,J100016.6+022637... & 450.09 & 10:00:16.6 & 02:26:37 & 4.94 & 20.41$\pm$4.14 & 17.26$\pm$4.68 & 1.75 & $<$0.01 & 850.34 & 2.3   & 4.41$\pm$0.80 & 3.68$\pm$1.01 \\ 
SMM\,J100022.2+022354... & 450.10 & 10:00:22.2 & 02:23:54 & 4.85 & 20.09$\pm$4.14 & 16.91$\pm$4.67 & 1.78 & $<$0.01 & 850.135 & 5.5  & 2.60$\pm$0.80 & 1.87$\pm$1.01 \\ 
SMM\,J100025.4+022542... & 450.11 & 10:00:25.4 & 02:25:42 & 4.77 & 19.57$\pm$4.11 & 16.40$\pm$4.62 & 1.80 & $<$0.01 & 850.41 & 2.0   & 4.12$\pm$0.80 & 3.43$\pm$0.98 \\ 
SMM\,J095957.0+022153... & 450.12 & 09:59:57.0 & 02:21:53 & 4.76 & 22.03$\pm$4.63 & 18.45$\pm$5.21 & 1.81 & $<$0.01 &        &       & 1.71$\pm$0.89 & \\ 
SMM\,J100001.7+022426... & 450.13 & 10:00:01.7 & 02:24:26 & 4.75 & 19.96$\pm$4.20 & 16.72$\pm$4.73 & 1.81 & $<$0.01 & (850.32)&       & 3.81$\pm$0.82 & \\ 
SMM\,J100044.9+021919... & 450.14 & 10:00:44.9 & 02:19:19 & 4.74 & 21.73$\pm$4.59 & 18.18$\pm$5.17 & 1.81 & $<$0.01 &        &       & 1.28$\pm$0.91 & \\ 
SMM\,J100004.2+022059... & 450.15 & 10:00:04.2 & 02:20:59 & 4.60 & 19.92$\pm$4.33 & 16.53$\pm$4.88 & 1.83 & $<$0.01 & 850.28 & 1.8   & 4.86$\pm$0.84 & 4.07$\pm$1.08 \\ 
SMM\,J100057.2+022009... & 450.16 & 10:00:57.2 & 02:20:09 & 4.59 & 25.18$\pm$5.48 & 20.89$\pm$6.19 & 1.84 & $<$0.01 & (850.05) &       & 8.99$\pm$1.14 & \\ 
SMM\,J100017.2+022521... & 450.17 & 10:00:17.2 & 02:25:21 & 4.59 & 19.01$\pm$4.14 & 15.76$\pm$4.67 & 1.84 & $<$0.01 & 850.42 & 2.6   & 4.12$\pm$0.80 & 3.43$\pm$0.99 \\ 
SMM\,J100007.2+021803... & 450.18 & 10:00:07.2 & 02:18:03 & 4.49 & 22.63$\pm$5.04 & 18.66$\pm$5.68 & 1.85 & $<$0.01 &        &       & -0.76$\pm$0.95 &\\ 
SMM\,J095948.1+022014... & 450.19 & 09:59:48.1 & 02:20:14 & 4.47 & 28.25$\pm$6.33 & 23.24$\pm$7.14 & 1.86 & $<$0.01 &        &       & 0.77$\pm$1.16 & \\ 
SMM\,J100004.5+023042... & 450.20 & 10:00:04.5 & 02:30:42 & 4.44 & 23.28$\pm$5.24 & 19.11$\pm$5.91 & 1.87 & $<$0.01 & (850.24) &       & 4.62$\pm$1.01 & \\ 
SMM\,J100017.2+022138... & 450.21 & 10:00:17.2 & 02:21:38 & 4.40 & 18.14$\pm$4.12 & 14.83$\pm$4.64 & 1.88 & $<$0.01 &        &       & 0.72$\pm$0.80 & \\ 
SMM\,J100030.8+023104... & 450.22 & 10:00:30.8 & 02:31:04 & 4.34 & 21.10$\pm$4.86 & 17.15$\pm$5.47 & 1.90 & $<$0.01 & 850.149 & 4.1  & 3.04$\pm$0.96 & 2.16$\pm$1.21 \\ 
SMM\,J100004.6+021820... & 450.23 & 10:00:04.6 & 02:18:20 & 4.33 & 21.95$\pm$5.07 & 17.81$\pm$5.71 & 1.90 & $<$0.01 &        &       & -0.36$\pm$0.96 &\\ 
SMM\,J100050.1+022116... & 450.24 & 10:00:50.1 & 02:21:16 & 4.30 & 19.72$\pm$4.58 & 15.97$\pm$5.16 & 1.91 & $<$0.01 &        &       & 1.29$\pm$0.91 & \\ 
SMM\,J100028.5+021927... & 450.25 & 10:00:28.5 & 02:19:27 & 4.30 & 18.04$\pm$4.20 & 14.60$\pm$4.73 & 1.91 & $<$0.01 &        &       & -0.82$\pm$0.82 &\\ 
SMM\,J100016.6+022000... & 450.26 & 10:00:16.6 & 02:20:00 & 4.27 & 17.81$\pm$4.17 & 14.38$\pm$4.69 & 1.92 & $<$0.01 &        &       & -0.60$\pm$0.81 &\\ 
SMM\,J095942.9+022144... & 450.27 & 09:59:42.9 & 02:21:44 & 4.26 & 29.66$\pm$6.97 & 23.91$\pm$7.84 & 1.92 & $<$0.01 & 850.115 & 7.3  & 4.26$\pm$1.26 & 3.14$\pm$1.56 \\ 
SMM\,J100056.7+022014... & 450.28 & 10:00:56.7 & 02:20:14 & 4.24 & 22.76$\pm$5.37 & 18.32$\pm$6.04 & 1.93 & $<$0.01 & (850.05) &       & 10.85$\pm$1.12 &\\ 
SMM\,J100000.5+022503... & 450.29 & 10:00:00.5 & 02:25:03 & 4.22 & 18.01$\pm$4.27 & 14.46$\pm$4.80 & 1.93 & $<$0.01 &        &       & 0.94$\pm$0.83 & \\ 
SMM\,J100048.3+022926... & 450.30 & 10:00:48.3 & 02:29:26 & 4.16 & 20.99$\pm$5.05 & 16.76$\pm$5.69 & 1.95 & $<$0.01 &        &       & 2.23$\pm$1.04 & \\ 
SMM\,J100008.4+022241... & 450.31 & 10:00:08.4 & 02:22:41 & 4.11 & 16.96$\pm$4.13 & 13.46$\pm$4.69 & 1.97 & 0.02 &        &       & 1.17$\pm$0.80 & \\ 
SMM\,J100010.2+022624... & 450.32 & 10:00:10.2 & 02:26:24 & 4.10 & 16.86$\pm$4.11 & 13.37$\pm$4.67 & 1.97 & 0.02 & 850.26 & 3.0   & 4.87$\pm$0.80 & 4.12$\pm$1.06 \\ 
SMM\,J100021.3+023055... & 450.33 & 10:00:21.3 & 02:30:55 & 4.04 & 18.93$\pm$4.69 & 14.90$\pm$5.36 & 1.99 & 0.03 &        &       & 0.59$\pm$0.93 & \\ 
SMM\,J095945.2+022253... & 450.34 & 09:59:45.2 & 02:22:53 & 4.04 & 24.73$\pm$6.13 & 19.46$\pm$7.01 & 1.99 & 0.03 &        &       & 3.30$\pm$1.13 & \\ 
SMM\,J100014.4+021706... & 450.35 & 10:00:14.4 & 02:17:06 & 4.01 & 20.52$\pm$5.12 & 16.10$\pm$5.87 & 2.00 & 0.03 &        &       & -0.55$\pm$0.97 &\\ 
SMM\,J100101.0+022800... & 450.36 & 10:01:01.0 & 02:28:00 & 4.00 & 24.27$\pm$6.07 & 19.03$\pm$6.96 & 2.00 & 0.03 & 850.58 & 3.8   & 5.67$\pm$1.31 & 4.60$\pm$1.57 \\ 
SMM\,J100001.5+021939... & 450.37 & 10:00:01.5 & 02:19:39 & 4.00 & 19.20$\pm$4.80 & 15.05$\pm$5.51 & 2.00 & 0.03 &        &       & -0.98$\pm$0.91 &\\ 
SMM\,J100010.3+022223... & 450.38 & 10:00:10.3 & 02:22:23 & 4.00 & 16.47$\pm$4.12 & 12.91$\pm$4.73 & 2.01 & 0.03 & 850.14 & 2.8   & 6.25$\pm$0.80 & 5.55$\pm$1.11 \\ 
SMM\,J100011.2+021554... & 450.39 & 10:00:11.2 & 02:15:54 & 3.99 & 23.76$\pm$5.96 & 18.60$\pm$6.84 & 2.01 & 0.04 &        &       & -0.70$\pm$1.11 &\\ 
SMM\,J100033.1+023046... & 450.40 & 10:00:33.1 & 02:30:46 & 3.98 & 19.15$\pm$4.81 & 14.98$\pm$5.53 & 2.01 & 0.04 & 850.90 & 5.0   & 3.52$\pm$0.96 & 2.71$\pm$1.16 \\ 
SMM\,J100006.6+021527... & 450.41 & 10:00:06.6 & 02:15:27 & 3.96 & 26.42$\pm$6.67 & 20.62$\pm$7.67 & 2.02 & 0.04 &        &       & 1.39$\pm$1.25 & \\ 
SMM\,J100036.9+021938... & 450.42 & 10:00:36.9 & 02:19:38 & 3.94 & 16.71$\pm$4.24 & 13.01$\pm$4.89 & 2.02 & 0.04 & 850.48 & 3.6   & 3.99$\pm$0.83 & 3.30$\pm$1.00 \\ 
SMM\,J100028.8+023336... & 450.43 & 10:00:28.8 & 02:33:36 & 3.88 & 24.62$\pm$6.34 & 19.00$\pm$7.36 & 2.06 & 0.05 &        &       & 1.62$\pm$1.28 & \\ 
SMM\,J100054.5+021919... & 450.44 & 10:00:54.5 & 02:19:19 & 3.88 & 20.99$\pm$5.41 & 16.18$\pm$6.28 & 2.06 & 0.05 &        &       & -0.15$\pm$1.12 &\\ 
SMM\,J100056.1+021831... & 450.45 & 10:00:56.1 & 02:18:31 & 3.82 & 23.57$\pm$6.16 & 17.92$\pm$7.22 & 2.13 & 0.06 &        &       & -1.21$\pm$1.30 &\\ 
SMM\,J100032.4+022148... & 450.46 & 10:00:32.4 & 02:21:48 & 3.81 & 15.72$\pm$4.13 & 11.90$\pm$4.86 & 2.16 & 0.07 &        &       & 0.11$\pm$0.80 & \\ 
SMM\,J100025.2+021930... & 450.47 & 10:00:25.2 & 02:19:30 & 3.80 & 15.90$\pm$4.18 & 12.03$\pm$4.91 & 2.16 & 0.07 &        &       & 1.60$\pm$0.81 & \\ 
SMM\,J100009.4+022223... & 450.48 & 10:00:09.4 & 02:22:23 & 3.80 & 15.72$\pm$4.13 & 11.89$\pm$4.86 & 2.16 & 0.07 &        &       & 2.73$\pm$0.80 & \\ 
SMM\,J100000.0+022524... & 450.49 & 10:00:00.0 & 02:25:24 & 3.79 & 16.29$\pm$4.30 & 12.27$\pm$5.07 & 2.18 & 0.07 &        &       & -0.72$\pm$0.84 &\\ 
SMM\,J095952.2+022133... & 450.50 & 09:59:52.2 & 02:21:33 & 3.78 & 19.82$\pm$5.25 & 14.90$\pm$6.19 & 2.19 & 0.07 & 850.47 & 3.6   & 4.74$\pm$0.98 & 3.92$\pm$1.19 \\ 
SMM\,J100044.3+022313... & 450.51 & 10:00:44.3 & 02:23:13 & 3.77 & 15.64$\pm$4.15 & 11.74$\pm$4.90 & 2.20 & 0.08 &        &       & 1.27$\pm$0.81 & \\ 
SMM\,J100007.8+022306... & 450.52 & 10:00:07.8 & 02:23:06 & 3.77 & 15.57$\pm$4.13 & 11.68$\pm$4.88 & 2.20 & 0.08 &        &       & -0.46$\pm$0.80 &\\ 
SMM\,J100010.4+022026... & 450.53 & 10:00:10.4 & 02:20:26 & 3.74 & 15.69$\pm$4.19 & 11.69$\pm$4.97 & 2.23 & 0.08 & 850.109 & 0.6   & 2.84$\pm$0.82 & 2.13$\pm$1.00 \\ 
SMM\,J100026.8+022318... & 450.54 & 10:00:26.8 & 02:23:18 & 3.74 & 15.46$\pm$4.13 & 11.51$\pm$4.90 & 2.23 & 0.08 & 850.96 &  8.1  & 2.93$\pm$0.80 & 2.24$\pm$0.97 \\ 
SMM\,J100023.8+021751... & 450.55 & 10:00:23.8 & 02:17:51 & 3.74 & 17.06$\pm$4.57 & 12.70$\pm$5.42 & 2.24 & 0.08 & 850.06 & 3.1   & 9.30$\pm$0.89 & 8.42$\pm$0.92 \\ 
SMM\,J095959.3+023000... & 450.56 & 09:59:59.3 & 02:30:00 & 3.73 & 20.17$\pm$5.40 & 14.98$\pm$6.42 & 2.25 & 0.09 &        &       & -0.89$\pm$1.03 &\\ 
SMM\,J100032.4+021802... & 450.57 & 10:00:32.4 & 02:18:02 & 3.73 & 16.75$\pm$4.49 & 12.44$\pm$5.34 & 2.25 & 0.09 &        &       & 0.56$\pm$0.88 & \\ 
SMM\,J100036.1+022152... & 450.58 & 10:00:36.1 & 02:21:52 & 3.72 & 15.32$\pm$4.12 & 11.35$\pm$4.90 & 2.26 & 0.09 & 850.21 & 3.6   & 5.29$\pm$0.80 & 4.60$\pm$1.08 \\ 
SMM\,J100019.4+022024... & 450.59 & 10:00:19.4 & 02:20:24 & 3.71 & 15.33$\pm$4.13 & 11.33$\pm$4.92 & 2.27 & 0.09 & 850.101 & 3.3   & 2.87$\pm$0.80 & 2.19$\pm$0.98 \\ 
SMM\,J100008.1+023059... & 450.60 & 10:00:08.1 & 02:30:59 & 3.71 & 18.82$\pm$5.07 & 13.90$\pm$6.04 & 2.27 & 0.09 &        &       & 1.81$\pm$0.99 & \\ 
SMM\,J100018.7+022813... & 450.61 & 10:00:18.7 & 02:28:13 & 3.70 & 15.48$\pm$4.18 & 11.42$\pm$4.98 & 2.28 & 0.09 &        &       & 1.24$\pm$0.81 & \\ 
SMM\,J100047.3+022049... & 450.62 & 10:00:47.3 & 02:20:49 & 3.70 & 16.48$\pm$4.45 & 12.15$\pm$5.30 & 2.28 & 0.09 &        &       & -0.18$\pm$0.89 &\\ 
SMM\,J100020.0+022129... & 450.63 & 10:00:20.0 & 02:21:29 & 3.70 & 15.23$\pm$4.12 & 11.23$\pm$4.90 & 2.29 & 0.09 &        &       & 0.05$\pm$0.80 & \\ 
SMM\,J100010.7+022707... & 450.64 & 10:00:10.7 & 02:27:07 & 3.68 & 15.18$\pm$4.12 & 11.14$\pm$4.91 & 2.31 & 0.10 &        &       & 0.86$\pm$0.80 & \\ 
SMM\,J100027.2+022448... & 450.65 & 10:00:27.2 & 02:24:48 & 3.68 & 15.13$\pm$4.12 & 11.08$\pm$4.92 & 2.32 & 0.10 &        &       & 0.00$\pm$0.80 & \\ 
SMM\,J100104.6+022633... & 450.66 & 10:01:04.6 & 02:26:33 & 3.66 & 23.50$\pm$6.42 & 17.15$\pm$7.68 & 2.34 & 0.10 & 850.146 & 4.1   & 4.29$\pm$1.36 & 3.06$\pm$1.70 \\ 
SMM\,J100000.3+022902... & 450.67 & 10:00:00.3 & 02:29:02 & 3.65 & 17.79$\pm$4.87 & 12.96$\pm$5.83 & 2.34 & 0.10 &        &       & 1.08$\pm$0.94 & \\ 
SMM\,J095945.7+021837... & 450.68 & 09:59:45.7 & 02:18:37 & 3.65 & 28.64$\pm$7.84 & 20.86$\pm$9.38 & 2.35 & 0.10 &        &       & -0.56$\pm$1.46 &\\ 
SMM\,J100047.5+022520... & 450.69 & 10:00:47.5 & 02:25:20 & 3.64 & 15.52$\pm$4.26 & 11.27$\pm$5.11 & 2.36 & 0.11 &        &       & 0.72$\pm$0.84 & \\ 
SMM\,J095952.6+022711... & 450.70 & 09:59:52.6 & 02:27:11 & 3.64 & 18.86$\pm$5.18 & 13.68$\pm$6.21 & 2.36 & 0.11 &        &       & 1.28$\pm$0.99 & \\ 
SMM\,J095946.3+022931... & 450.71 & 09:59:46.3 & 02:29:31 & 3.64 & 28.16$\pm$7.74 & 20.42$\pm$9.28 & 2.37 & 0.11 &        &       & 1.44$\pm$1.45 & \\ 
SMM\,J095950.5+022016... & 450.72 & 09:59:50.5 & 02:20:16 & 3.63 & 21.50$\pm$5.92 & 15.58$\pm$7.09 & 2.37 & 0.11 &        &       & 0.12$\pm$1.09 & \\ 
SMM\,J100106.1+022454... & 450.73 & 10:01:06.1 & 02:24:54 & 3.63 & 23.49$\pm$6.48 & 17.00$\pm$7.76 & 2.38 & 0.11 &        &       & -1.96$\pm$1.39 &\\ 
SMM\,J100041.3+021640... & 450.74 & 10:00:41.3 & 02:16:40 & 3.61 & 20.01$\pm$5.54 & 14.44$\pm$6.64 & 2.39 & 0.11 &        &       & -0.38$\pm$1.10 &\\ 
SMM\,J100034.2+023421... & 450.75 & 10:00:34.2 & 02:34:21 & 3.61 & 28.05$\pm$7.76 & 20.23$\pm$9.31 & 2.39 & 0.11 &        &       & -1.80$\pm$1.55 &\\ 
SMM\,J100021.7+023114... & 450.76 & 10:00:21.7 & 02:31:14 & 3.61 & 17.34$\pm$4.80 & 12.50$\pm$5.76 & 2.39 & 0.11 &        &       & 0.64$\pm$0.95 & \\ 
SMM\,J095952.6+022258... & 450.77 & 09:59:52.6 & 02:22:58 & 3.61 & 17.85$\pm$4.95 & 12.85$\pm$5.94 & 2.40 & 0.11 &        &       & 0.11$\pm$0.95 & \\ 
\hline\hline
\end{tabular}
}

\label{tab:sources450}
\begin{spacing}{0.7}
{\scriptsize {\bf Table Notes.} The $>$3.6$\sigma$ 450\um--detected
  sources we extract within the central 394\arcmin$^2$ of our COSMOS
  map.  The 3.6$\sigma$ detection threshold is chosen based on an
  estimated 3--5\%\ contamination rate. The list is ordered by
  detection signal-to-noise ratio (S/N).  The ``{\sc Raw}'' flux
  densities are those measured directly from our map. The ``{\sc
    Deboosted}'' flux densities are those given after correction for
  confusion and Eddington boosting as a function of detection S/N, as
  described in section~\ref{sec:simulations}.  We also measure a
  90\%\ confidence interval for positional uncertainties and estimated
  probability of contamination, $P_{\rm contam}$, from the results of
  our Monte Carlo tests as functions of detection S/N.  The last four
  columns contain details on the corresponding 850\um\ counterparts if
  they exist (their short name, offset from 450\um\ position, raw and
  deboosted flux densities).  If a 450\um\ source is not detected at
  850\um\ at $>$3$\sigma$ then the flux density is measured at the
  450\um\ position.  Also see Table~\ref{tab:sources850} for more
  details on 850\um\ sources.}
\end{spacing}
\end{table*}

\subsection{Number Counts}

\begin{figure*}
\includegraphics[width=0.8\columnwidth]{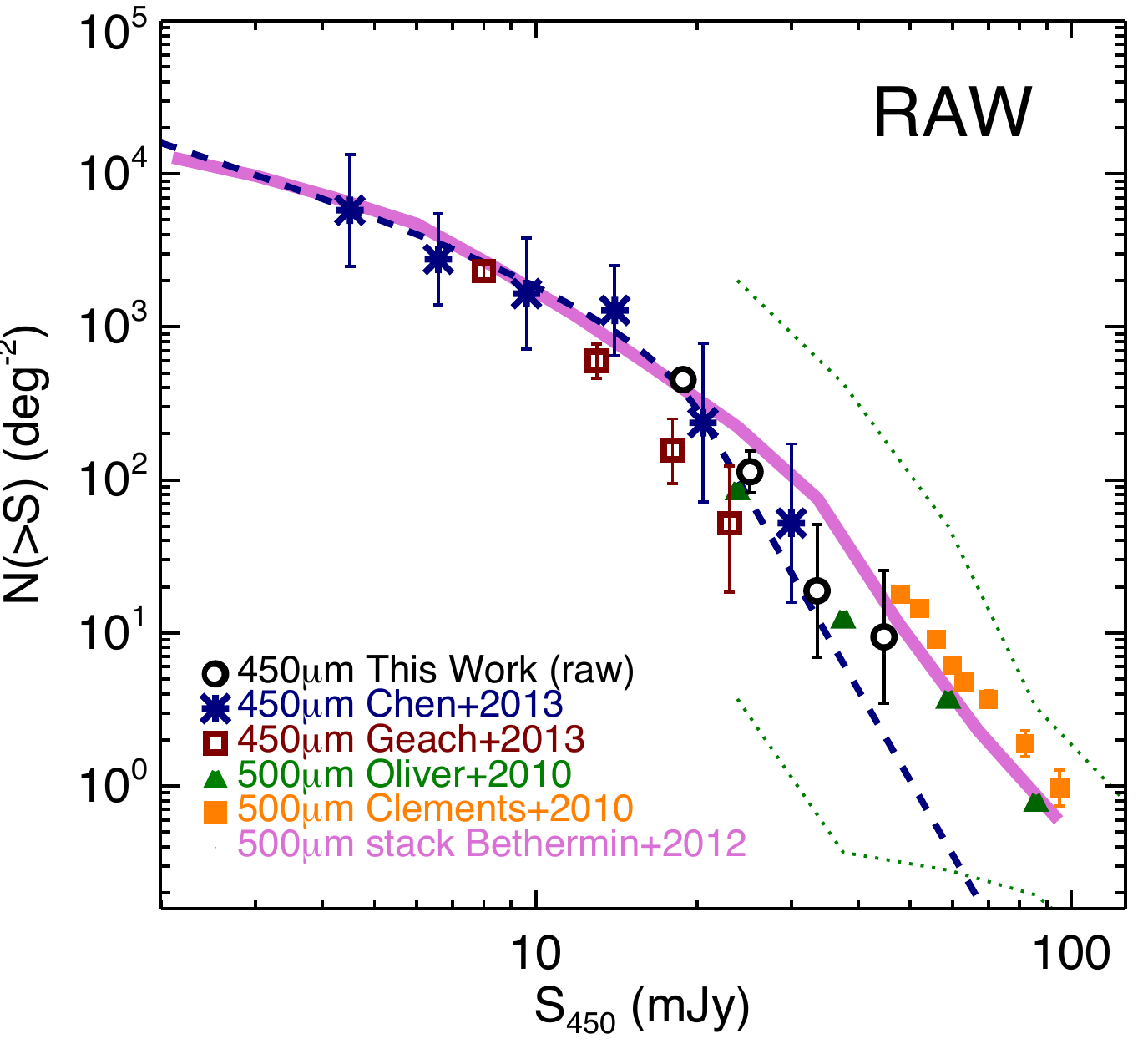}
\includegraphics[width=0.8\columnwidth]{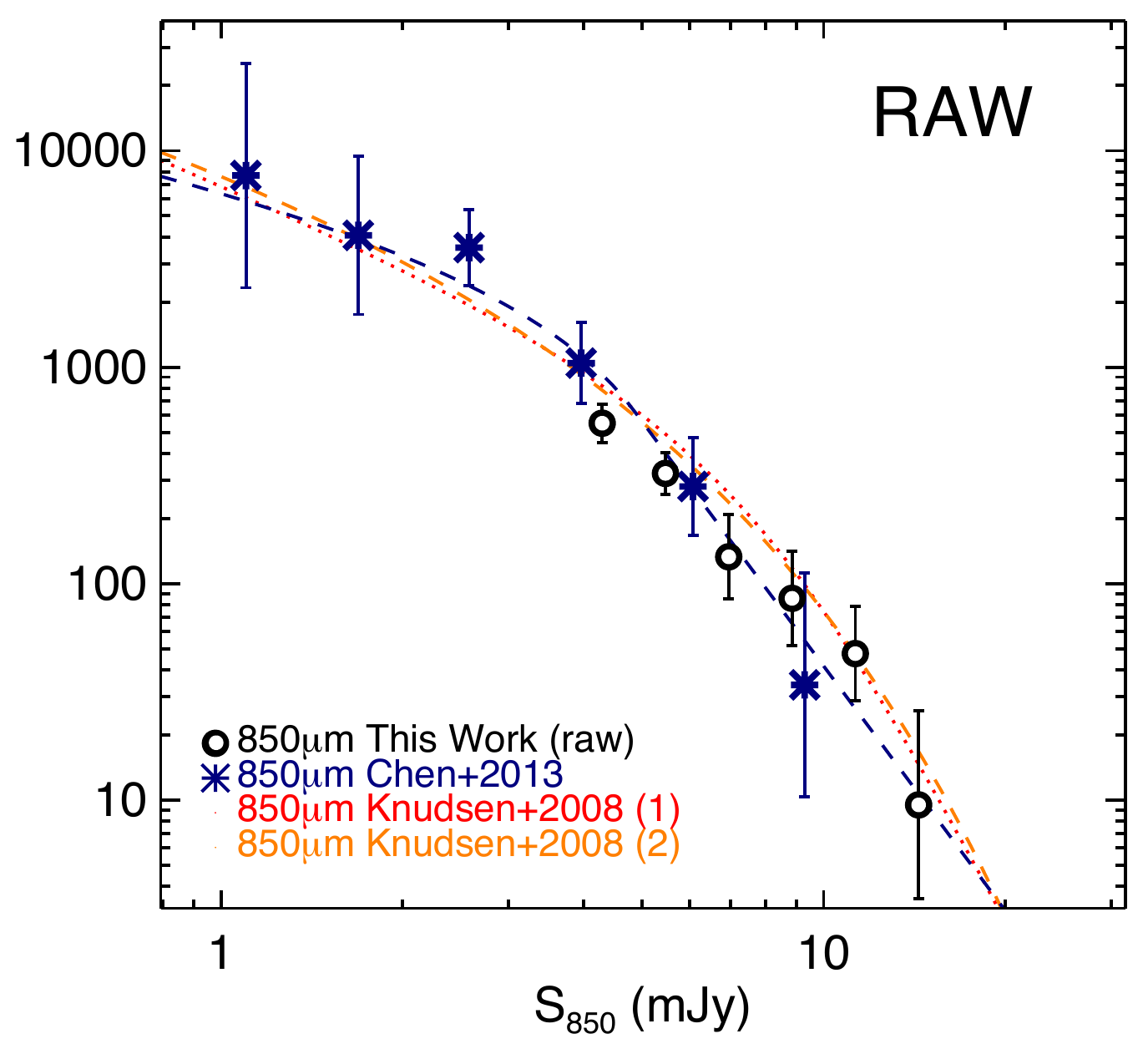}\\
\includegraphics[width=0.8\columnwidth]{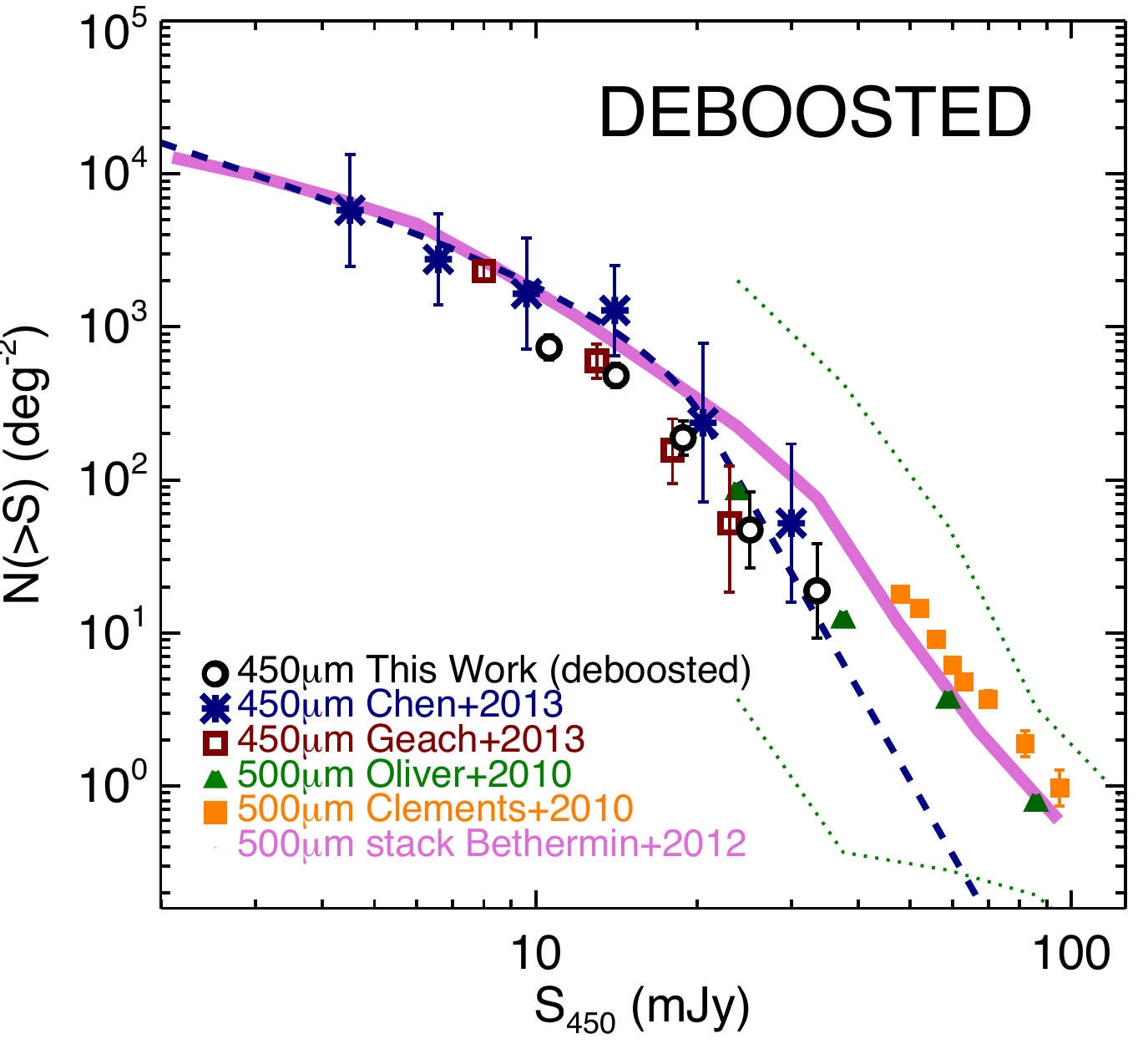}
\includegraphics[width=0.8\columnwidth]{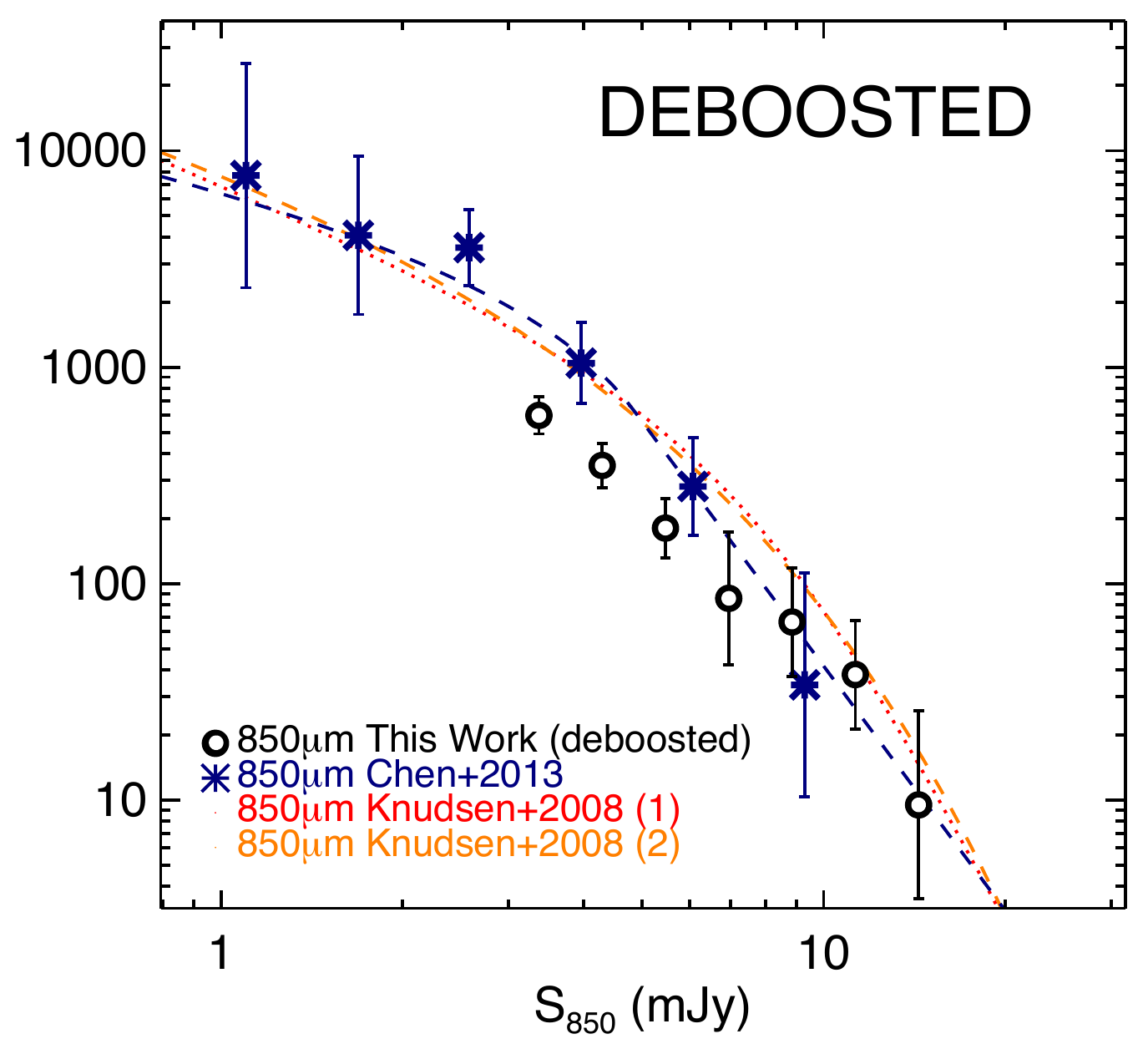}\\
\includegraphics[width=0.8\columnwidth]{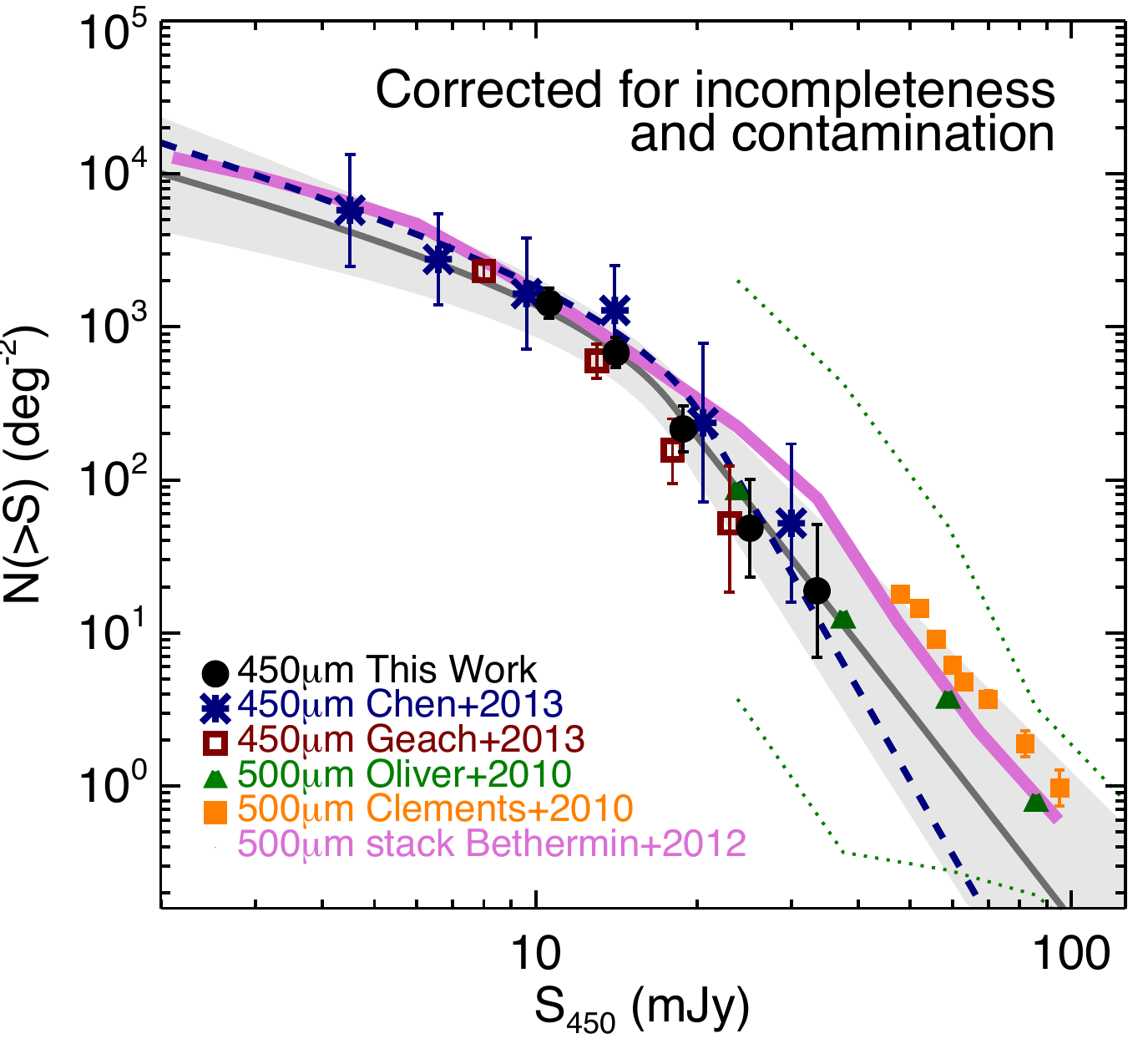}
\includegraphics[width=0.8\columnwidth]{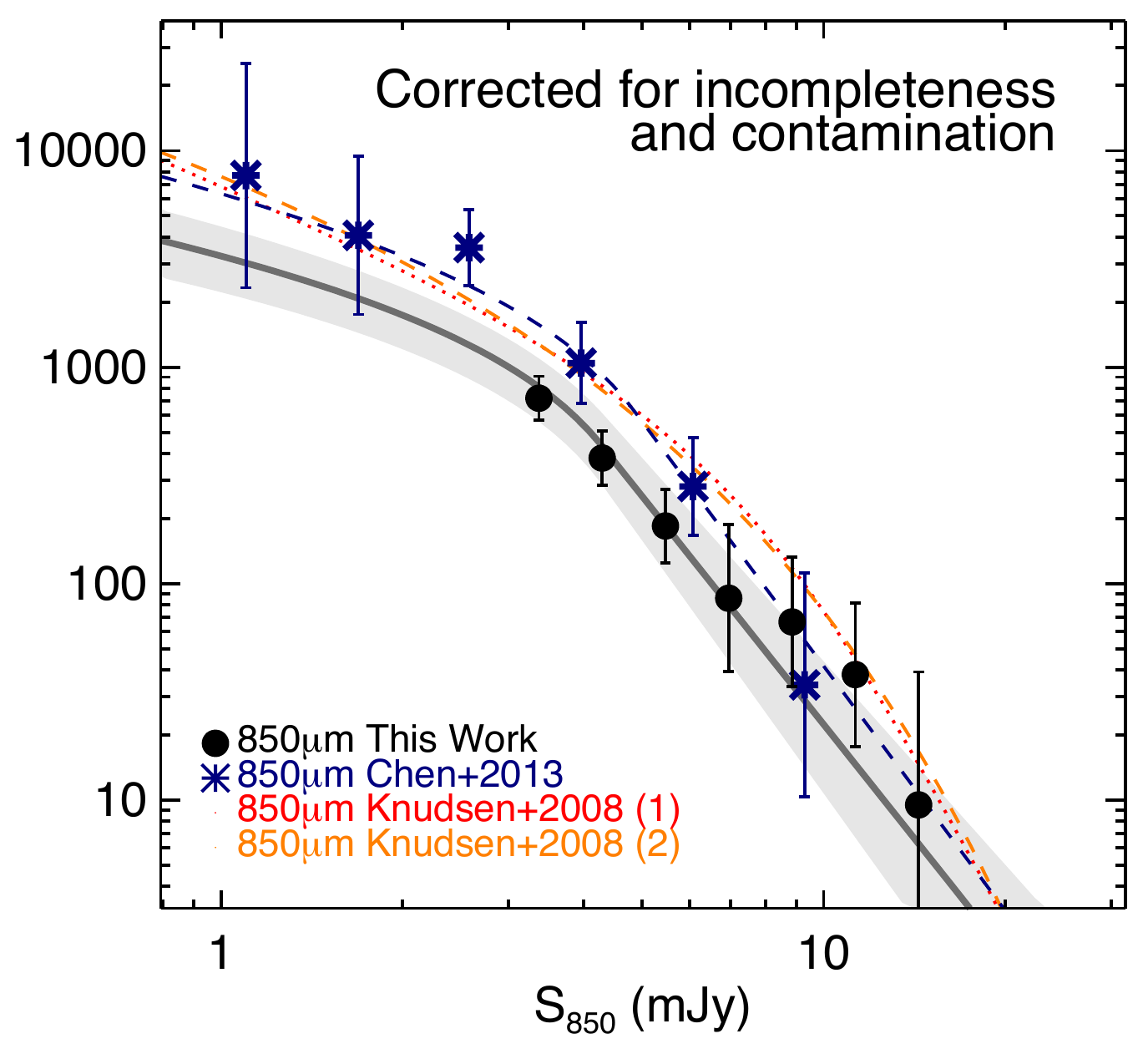}
\caption{The cumulative number counts at 450\um\ (left column) and
  850\um\ (right column) from our COSMOS data.  The top row shows the
  raw number counts, with extracted flux densities from our map, the
  middle row corrects the sources' flux densities for confusion and
  Eddington boosting, and the bottom row corrects the deboosted number
  counts for incompleteness and contamination (as discussed in the
  text).  The best-fit double power law parameterisation of the number
  counts are shown in gray in both bottom panels, with associated
  light gray uncertainties.  We compare our measured counts (black
  open and closed circles) to those in the literature, all corrected
  for incompleteness and contamination.  At 450\um, we compare to the
  recent results of \citet{chen13a} (blue asterisks, blue dashed line)
  and \citet{geach13a} (red squares), and we note that the latter work
  is data taken in the same sky region as our COSMOS map although
  covering only $\sim$1/4 the area.  We also compare 450\um\ data to
  500\um\ data from the {\it Herschel}-{\sc Spire} instrument from
  \citet{oliver10a} (green triangles) and \citet{clements10b} (orange
  squares).  The \citet{bethermin12a} (pink curve) work is not from
  direct 500\um\ extractions but rather stacks on 24\um\ sources at
  500\um.  At 850\um, we overplot the \citet{chen13a} results and the
  best-fit models from Eq. 3 \& 4 in \citet{knudsen08a} (red-orange
  curves), called (1) and (2) here respectively; we omit many data
  measurements taken at 850\um\ to avoid clutter on the plot. }
\label{fig:numbercounts}
\end{figure*}

We calculate the cumulative number counts \citep[as has been done at
  different wavelengths in the
  literature:][]{barger99a,smail02a,cowie02a,scott02a,webb03a,borys03a,barnard04a,coppin06a,knudsen08a,oliver10a,chen13a}
for our sample in three stages and plot the results in
Figure~\ref{fig:numbercounts}.  The first is straightforward, as it is
the cumulative number counts of the raw flux densities measured from
our maps above $>$3.6$\sigma$ at 450\um\ and above $>$3.6$\sigma$ at
850\um\ (first panel on Fig~\ref{fig:numbercounts}, labelled `{\sc
  raw}').  The second calculation incorporates the correction for flux
boosting, as shown in Figure~\ref{fig:boosting}.  Here we correct the
flux density of each individual source according to its detection
signal-to-noise ratio.  The uncertainty in the deboosting factor is
propagated to the new uncertainty in the sources' deboosted flux
density.  As a result, the sources' ratio of deboosted flux density to
its uncertainty is lower than the sources' measured signal-to-noise
ratios. The net effect of deboosting is a shift towards lower flux
densities (in the $x$-direction on the number counts plot), seen in
the middle panels of Figure~\ref{fig:numbercounts}.

The final effect which needs to be accounted for in our computation of
the cumulative number counts is incompleteness and contamination in
our samples.  As shown in Figure~\ref{fig:completeness} and
Figure~\ref{fig:contam}, our samples are not complete down to our
nominal $\sim$15\,mJy$\approx$3.6$\sigma_{\rm 450}$ or
$\sim$2.9\,mJy$\approx$3.6$\sigma_{\rm 850}$ detection threshold.
Furthermore, 3--5\%\ of our sources are likely contaminants.  To
correct for sources missing and contaminant sources we construct a
correction factor, $c_{i}$, for each source, based on its flux density
and signal-to-noise, such that $c_{i}=(1-f_{contam})/f_{complete}$.
We correct the number counts accordingly.  Not surprisingly, this
correction for incompleteness and contamination has an effect on the
low flux density bins and a minimal effect on the high flux density
bins.  Our deboosted and corrected number counts are given in
Table~\ref{tab:numbercounts}.

In functional form, we can assume that the differential number counts
follow a few different forms.  The types we adopt in this paper are a
broken double power law \citep[as most recently used in][]{chen13a},
taking the form:
\begin{equation}
\frac{dN}{dS}\,=\,\left\{
\begin{array}{lr}
\frac{N_{\rm 0}}{S_{\rm 0}}(\frac{S}{S_{0}})^{-\alpha} & : S\le S_{0} \\
\frac{N_{\rm 0}}{S_{\rm 0}}(\frac{S}{S_{0}})^{-\beta}  & : S>S_{0} \\
\end{array}
\right.
\label{eq:powerlaw}
\end{equation}
where S$_{\rm 0}$ represents the turnover flux density, N$_{\rm 0}$ a
normalisation for the function in source density per mJy, $\alpha$ is
the faint end slope and $\beta$ is the bright end slope (presumed to
be steeper).  Alternatively we can assume the number counts follow a
Schechter function, as used in \citet{geach13a}\footnote{Note that
  Equation 1 of Geach et al. should be written $dN/dS =
  (N^{\prime}/S^{\prime})(S/S^{\prime})^{1-\alpha}exp(-S/S^{\prime})$,
  and the best-fit parameter $N^{\prime}$ from their paper should be
  quoted as $N^{\prime}=4900\pm1040\,$deg$^{-2}$\,mJy$^{-1}$ rather
  than $N^{\prime}=490\pm104\,$deg$^{-2}$. Also note we use an
  exponent of $-\alpha$ instead of $1-\alpha$.  By our equation, the
  parameters of the Geach et al. data are: N$_{\rm
    0}=4900\,$deg$^{-2}$\,mJy$^{-1}$, S$_{\rm 0}=$10\,mJy, and
  $\alpha=$2.0.}:
\begin{equation}
\frac{dN}{dS}\,=\,\frac{N_{0}}{S_{0}}\left(\frac{S}{S_{0}}\right)^{-\alpha}e^{-\left(\frac{S}{S_{0}}\right)}
\label{eq:schechter}
\end{equation}
We perform a $\chi^2$ optimisation to determine best-fit parameter
values at 450\um\ and 850\um\ respectively in
equations~\ref{eq:powerlaw} and ~\ref{eq:schechter}, which are given
in Table~\ref{tab:numbercounts}.  These best-fit parameter values are
tested for consistency with the Monte Carlo simulations; in other
words, we adjust the fit parameters, populate the jackknife maps
accordingly, then extract sources and re-measure the raw number
counts.  The parameters which reproduce the raw number counts most
accurately are used in the Monte Carlo tests for boosting factor,
positional uncertainty and estimating contamination and completeness
rates, as given in the previous section.  Once we arrive at the {\it
  corrected} number counts, we then remeasure our best-fit parameters
with a $\chi^2$ test which agrees with our initial injected counts
within uncertainty.  
We find the power-law fit produces a better fit than a Schechter fit
to both datasets (with marginally better reduced $\chi^2$, 7 versus 12
at 450\um\ and 3 versus 5 at 850\um), so we use the double power-law
Monte Carlo predictions throughout.

Noting a deficit in our 850\um\ number counts relative to many
literature measurements
\citep{smail02a,cowie02a,scott02a,webb03a,borys03a,knudsen08a,chen13a},
we investigate the strength of this deficit.  To do this we compare
our best-fit double powerlaw function to the Knudsen \etal\ double
powerlaw, integrated over the flux density range where our survey is
sensitive (2.5--18\,mJy).  This comparison indicates that our survey
is 2.1$^{+1.0}_{-0.6}$ times less dense than the nominal
850\um\ source density.  We also test to see if the deficit exists at
450\um, but we find our 450\um\ source density is consistent with the
only other independent measurement at this wavelength from
\citet{chen13a}.  

\begin{table}
\centering
\caption{ True Number Counts at 450\um, 850\um, and best-fit double power-law/Schechter function parameters.}
{\scriptsize \begin{tabular}{c@{ }c@{ }c@{ }cc@{ }c@{ }c@{ }c}
\hline\hline
\multicolumn{4}{l}{\underline{\bf 450\um-Number Counts}} & \multicolumn{4}{l}{\underline{\bf 850\um-Number Counts}} \\
$S_{\rm 450}$ & N & dN/dS & N($>S$)  & $S_{\rm 850}$ & N & dN/dS & N($>S$) \\
    (mJy)     &   & {\tiny (mJy$^{-1}$\,deg$^{-2}$)} & (deg$^{-2}$)    & (mJy) & & {\tiny (mJy$^{-1}$\,deg$^{-2}$)} & (deg$^{-2}$) \\
\hline\hline
10.59 & 15 & 154$^{+51}_{-38}$      & 1430$^{+370}_{-290}$    & 2.65  & 20 & 579$^{+157}_{-124}$    & 1040$^{+300}_{-230}$ \\
14.13 & 27 & 129$^{+31}_{-25}$      & 679$^{+175}_{-139}$     & 3.37  & 23 & 412$^{+104}_{-83}$     & 719$^{+189}_{-150}$ \\
18.84 & 29 & 66.1$^{+15.0}_{-12.2}$ & 215$^{+88}_{-62}$       & 4.29  & 27 & 306$^{+70}_{-57}$      & 380$^{+127}_{-95}$ \\
25.12 &  5 & 6.87$^{+4.13}_{-2.58}$ & 48$^{+52}_{-25}$        & 5.46  & 19 & 78.2$^{+31.0}_{-22.2}$ & 185$^{+88}_{-60}$ \\
33.50 &  1 & 0.97$^{+1.80}_{-0.63}$ & 18$^{+32}_{-12}$        & 6.96  &  7 & 40.4$^{+19.7}_{-13.2}$ & 85.6$^{+101.5}_{-46.4}$ \\
      &    &                        &                         & 8.86  &  3 & 13.3$^{+11.1}_{-6.0}$  & 66.5$^{+66.1}_{-33.2}$ \\
      &    &                        &                         & 11.28 &  4 & 13.9$^{+9.6}_{-5.7}$   & 38.0$^{+43.6}_{-20.3}$ \\
      &    &                        &                         & 14.37 &  1 & 2.7$^{+5.1}_{-1.8}$    & 9.5$^{+29.6}_{-7.2}$\\
\hline\hline
\multicolumn{8}{c}{\underline{\bf Best-Fit Parameters to Equation~\ref{eq:powerlaw}}} \\
\multicolumn{4}{c}{\underline{450\um\ Parameters}} & \multicolumn{4}{c}{\underline{850\um\ Parameters}} \\
$N_{0}=$   & \multicolumn{3}{c}{(1.4$\pm$0.4)$\times$10$^{3}$}      & \multicolumn{4}{c}{(1.5$\pm$0.4)$\times$10$^{3}$} \\
$S_{0}=$   & \multicolumn{3}{c}{18$\pm$8}       & \multicolumn{4}{c}{4.3$\pm$1.8} \\
$\alpha=$  & \multicolumn{3}{c}{1.91$\pm$0.40}  & \multicolumn{4}{c}{1.34$\pm$0.09} \\
$\beta=$   & \multicolumn{3}{c}{5.5$\pm$1.0}    & \multicolumn{4}{c}{4.50$\pm$0.38} \\
\hline\hline
\multicolumn{8}{c}{\underline{\bf Best-Fit Parameters to Equation~\ref{eq:schechter}}} \\
\multicolumn{4}{c}{\underline{450\um\ Parameters}} & \multicolumn{4}{c}{\underline{850\um\ Parameters}} \\
$N_{0}=$   & \multicolumn{3}{c}{(1.1$\pm$0.4)$\times$10$^{4}$}  & \multicolumn{4}{c}{(3.3$\pm$1.4)$\times$10$^{3}$} \\
$S_{0}=$   & \multicolumn{3}{c}{6.1$\pm$1.0}                    & \multicolumn{4}{c}{3.7$\pm$1.1} \\
$\alpha=$  & \multicolumn{3}{c}{0.5$\pm$1.3}                    & \multicolumn{4}{c}{1.4$\pm$1.1} \\
\hline\hline 
\end{tabular}
}
\label{tab:numbercounts}
{\small {\bf Table Notes.}  Number counts are given in differential
  and cumulative form.  The number of sources contributing to each bin
  is given as the number ``N.''  The best-fit parameters to the number
  counts are given at bottom and are fit to the differential counts.
}
\end{table}

\subsection{Source Lists}

The 450\um\ sources detected $>$3.6$\sigma$ are summarised in
Table~\ref{tab:sources450} and the 850\um\ sources detected at
$>$3.6$\sigma$ are summarised in Table~\ref{tab:sources850} (given at
the end of the paper due to its length).  Both raw and deboosted flux
densities are given in the tables, along with a 90\%\ confidence
interval on positional uncertainty, $\Delta(\alpha,\delta)$, the
probability of contamination or spuriousness for that source ($P_{\rm
  contam}$), and the corresponding source detected at the opposite
wavelength if it exists.  The search for counterparts at the opposite
wavelength is done within a radius corresponding to the sum in
quadrature of the positional uncertainties, $\Delta(\alpha,\delta)$,
at both 450\um\ and 850\um\ (a function of the detection S/N for each
source).  If no counterpart is found, the measured flux density at the
opposite wavelength is extracted at the detected source's position and
quoted as a 3$\sigma$ upper limit.  Its flux density is not deboosted.

Of the 78 450\um\ sources, 19 have 850\um\ counterparts detected above
3.6$\sigma$ (a further 6 sources are detected at 850\um\ above
3$\sigma$, and 5 others are close [$<$12\arcsec] but not formally
matched).  Such a low fraction of 450\um\ corresponding to
850\um\ sources (\simlt(19--30)/78$\approx$24--38\%) suggests that
indeed, the 450\um\ population is intrinsically different than the
850\um\ population, either probing a different redshift regime,
luminosity regime, or temperature regime than the 850\um\ population.
Similarly, only 19 of 99 850\um\ sources are 450\um\ detected
$>3.6\sigma$, with an additional 16 which are marginal $>3\sigma$
detections and 4 nearby sources ($<$12\arcsec) which are not formally
matched.  This translates to a 450\um-detection rate of
850\um\ sources of 19--39\%.  We investigate the reasons for such
little overlap in the next section.

\citet{chen13a} showed that sources detected at both 450\um\ and
850\um\ at low-S/N (e.g. 3--4$\sigma$) have effective S/N higher than
the individual S/N in either band, in other words, that a source which
is 3$\sigma$ at 450\um\ and 3$\sigma$ at 850\um\ is effectively a
4$\sigma$ detection.  We use this technique to extract 450\um\ sources
below our nominal detection threshold.  
The additional 3$<\sigma<$3.6 450\um\ sources we detect with
850\um\ counterparts are given in Table~\ref{tab:extrasources450}.  We
exclude these sources from our 450\um\ number count computation due to
the unquantified bias which requiring a 850\um\ counterpart enforces
on the sample, but we can use this sample for comparisons to the
independently-detected $>$3.6$\sigma$ 450\um\ sample in redshift,
luminosity and temperature space.

\begin{table*}
\centering
\caption{Marginal 3$<\sigma<$3.6 450\um--identified sources with 850\um\ $>$3$\sigma$ Counterparts.}
{\scriptsize \begin{tabular}{c@{ }ll@{ }c@{ }ccc@{ }c@{ }ccc@{ }c@{ }c}
\hline\hline
Name  & 850\um--       & 450\um--   & Offset & RA$_{\rm 450}$ & DEC$_{\rm 450}$ & S/N$_{\rm 850}$ & S/N$_{\rm 450}$ & S$_{\rm 450}$ & S$_{\rm 450}$   & S$_{\rm 850}$ & S$_{\rm 850}$   & $\Delta$($\alpha$,$\delta$) \\
      & Short          & Short      &           &            &                 &                 &                 & {\sc Raw}     & {\sc Deboosted} & {\sc Raw}     & {\sc Deboosted} &          \\
      &      Name      &   Name     & [\arcsec] &            &                 &                 &                 &    [mJy]      &      [mJy]      &    [mJy]      &      [mJy]      &  [\arcsec] \\
\hline\hline
SMM\,J100031.0+022751... & 850.62 & 450.78 & 0.9   & 10:00:31.0 & +02:27:51 & 4.29 & 3.60 & 14.94$\pm$4.15 & 10.72$\pm$4.99 & 3.46$\pm$0.81 & 2.80$\pm$0.96 & 2.42 \\ 
SMM\,J100034.3+022121... & 850.10 & 450.81 & 0.8   & 10:00:34.3 & +02:21:21 & 8.60 & 3.58 & 14.71$\pm$4.11 & 10.47$\pm$4.98 & 6.90$\pm$0.80 & 6.24$\pm$1.13 & 2.45 \\ 
SMM\,J095950.8+022745... & 850.55 & 450.86 & 3.6   & 09:59:50.8 & +02:27:45 & 4.51 & 3.55 & 19.80$\pm$5.58 & 13.95$\pm$6.80 & 4.77$\pm$1.06 & 3.89$\pm$1.27 & 2.50 \\ 
SMM\,J100025.1+021847... & 850.09 & 450.87 & 2.8   & 10:00:25.1 & +02:18:47 & 8.89 & 3.55 & 15.24$\pm$4.30 & 10.74$\pm$5.24 & 7.47$\pm$0.84 & 6.83$\pm$1.17 & 2.50 \\ 
SMM\,J100022.2+022842... & 850.23 & 450.94 & 0.6   & 10:00:22.2 & +02:28:42 & 6.24 & 3.50 & 14.67$\pm$4.19 & 10.14$\pm$5.17 & 5.06$\pm$0.81 & 4.31$\pm$1.07 & 2.58 \\ 
SMM\,J100052.8+021906... & 850.103 & 450.95 & 3.1   & 10:00:52.8 & +02:19:06 & 3.54 & 3.50 & 18.59$\pm$5.31 & 12.83$\pm$6.56 & 3.88$\pm$1.10 & 2.94$\pm$1.34 & 2.59 \\ 
SMM\,J100049.8+022448... & 850.133 & 450.96 & 8.2   & 10:00:49.8 & +02:24:48 & 3.25 & 3.49 & 15.25$\pm$4.36 & 10.50$\pm$5.39 & 2.85$\pm$0.88 & 2.06$\pm$1.09 & 2.60 \\ 
SMM\,J095950.7+022823... & 850.33 & 450.99 & 3.7   & 09:59:50.7 & +02:28:23 & 5.54 & 3.47 & 20.15$\pm$5.80 & 13.76$\pm$7.20 & 6.20$\pm$1.12 & 5.18$\pm$1.41 & 2.63 \\ 
SMM\,J100025.0+022757... & 850.85 & 450.105 & 4.5   & 10:00:25.0 & +02:27:57 & 3.73 & 3.44 & 14.24$\pm$4.14 & 9.60$\pm$5.16 & 2.99$\pm$0.80 & 2.32$\pm$0.97 & 2.69 \\ 
SMM\,J100041.8+022358... & 850.92 & 450.106 & 5.6   & 10:00:41.8 & +02:23:58 & 3.67 & 3.44 & 14.19$\pm$4.13 & 9.57$\pm$5.15 & 2.94$\pm$0.80 & 2.27$\pm$0.97 & 2.69 \\ 
SMM\,J100025.6+023051... & 850.159 & 450.126 & 12.3   & 10:00:25.6 & +02:30:51 & 3.05 & 3.36 & 15.79$\pm$4.70 & 10.32$\pm$5.91 & 2.89$\pm$0.95 & 2.02$\pm$1.19 & 2.83 \\ 
SMM\,J100029.5+022131... & 850.131 & 450.133 & 2.6   & 10:00:29.5 & +02:21:31 & 3.27 & 3.32 & 13.67$\pm$4.12 & 8.78$\pm$5.19 & 2.64$\pm$0.81 & 1.91$\pm$1.00 & 2.90 \\ 
SMM\,J100011.8+022935... & 850.52 & 450.134 & 1.7   & 10:00:11.8 & +02:29:35 & 4.73 & 3.31 & 14.45$\pm$4.36 & 9.26$\pm$5.50 & 4.09$\pm$0.86 & 3.37$\pm$1.04 & 2.91 \\ 
SMM\,J095953.0+022641... & 850.163 & 450.135 & 3.5   & 09:59:53.0 & +02:26:41 & 3.00 & 3.31 & 16.78$\pm$5.07 & 10.76$\pm$6.39 & 2.91$\pm$0.97 & 2.02$\pm$1.22 & 2.91 \\ 
SMM\,J100010.1+021758... & 850.83 & 450.166 & 1.8   & 10:00:10.1 & +02:17:58 & 3.77 & 3.24 & 15.87$\pm$4.90 & 9.79$\pm$6.20 & 3.52$\pm$0.93 & 2.74$\pm$1.12 & 3.08 \\ 
SMM\,J100005.4+022516... & 850.104 & 450.173 & 6.4   & 10:00:05.4 & +02:25:16 & 3.51 & 3.22 & 13.31$\pm$4.13 & 8.11$\pm$5.23 & 2.82$\pm$0.80 & 2.12$\pm$0.98 & 3.13 \\ 
SMM\,J100022.6+023023... & 850.88 & 450.179 & 6.8   & 10:00:22.6 & +02:30:23 & 3.69 & 3.21 & 14.51$\pm$4.52 & 8.80$\pm$5.73 & 3.31$\pm$0.90 & 2.56$\pm$1.08 & 3.15 \\ 
SMM\,J100014.1+022836... & 850.114 & 450.189 & 8.2   & 10:00:14.1 & +02:28:36 & 3.40 & 3.17 & 13.37$\pm$4.22 & 7.92$\pm$5.33 & 2.81$\pm$0.83 & 2.08$\pm$1.02 & 3.25 \\ 
SMM\,J100005.0+021718... & 850.49 & 450.193 & 0.6   & 10:00:05.0 & +02:17:18 & 4.79 & 3.16 & 17.40$\pm$5.50 & 10.27$\pm$6.95 & 4.91$\pm$1.02 & 4.05$\pm$1.24 & 3.27 \\ 
SMM\,J100011.1+021507... & 850.53 & 450.206 & 4.8   & 10:00:11.1 & +02:15:07 & 4.68 & 3.14 & 20.69$\pm$6.60 & 12.04$\pm$8.32 & 5.72$\pm$1.22 & 4.70$\pm$1.47 & 3.33 \\ 
SMM\,J100023.6+021916... & 850.27 & 450.215 & 2.7   & 10:00:23.6 & +02:19:16 & 6.08 & 3.11 & 13.05$\pm$4.20 & 7.48$\pm$5.28 & 5.00$\pm$0.82 & 4.23$\pm$1.08 & 3.39 \\ 
SMM\,J100032.0+023324... & 850.67 & 450.240 & 5.9   & 10:00:32.0 & +02:33:24 & 4.07 & 3.06 & 19.01$\pm$6.21 & 10.61$\pm$7.75 & 5.15$\pm$1.26 & 4.12$\pm$1.50 & 3.51 \\ 
SMM\,J100041.3+022534... & 850.151 & 450.247 & 11.5   & 10:00:41.3 & +02:25:34 & 3.12 & 3.05 & 12.58$\pm$4.13 & 6.97$\pm$5.14 & 2.52$\pm$0.81 & 1.79$\pm$1.02 & 3.54 \\ 
SMM\,J100013.4+022224... & 850.40 & 450.252 & 3.7   & 10:00:13.4 & +02:22:24 & 5.20 & 3.04 & 12.57$\pm$4.14 & 6.91$\pm$5.15 & 4.16$\pm$0.80 & 3.47$\pm$0.99 & 3.57 \\ 
\hline\hline
\end{tabular}
}

\label{tab:extrasources450}
\begin{spacing}{0.7}
{\scriptsize {\bf Table Notes.}  Sources extracted down to 3$\sigma$
  in the 450\um\ map using 850\um\ $>$3$\sigma$ positional priors.
  The 450\um\ $>$3$\sigma$ list (with 274 sources) was checked against
  the 850\um\ $>$3$\sigma$ list (with 164 sources) and 25 sources were
  found.  The positions here are given by their 450\um\ detection with
  associated positional uncertainty and the Offset is the offset
  between 450\um\ and 850\um\ centroid.  The short names are numbered
  in order of decreasing S/N for all sources extracted in the map,
  e.g. from 450.00 at S/N$=7.80$ to to 450.273 with S/N$=3.00$, hence
  the high, non-sequential numbers given at both 450\um\ and 850\um.}
\end{spacing}
\end{table*}

\subsection{Counterpart Matching}\label{sec:counterparts}

Determining the optical/near-infrared counterparts to submillimeter
sources is complex, but thanks to the low positional uncertainty on
our 450\um\ sources, it is a much more straightforward
process than in other submillimeter maps.  A handful of sources in our sample (five
sources at 450\um, eight at 850\um) have interferometric observations
either with the SMA \citep{younger07a,younger09a} or the Plateau de
Bure Interferometer \citep{smolcic12a} where the counterparts to submm
observations are known, but the rest of the sample needs careful
counterpart analysis.  
Our counterpart matching procedure at
450\um\ and 850\um\ is the same at both wavelengths, but we note that
the added uncertainty at 850\um\ generates many more potential
counterparts and is, therefore, naturally more uncertain.  

All optical ($i$-band), near-infrared (3.6\um), 24\um, and 1.4\,GHz
radio sources which fall within the 90\%\ confidence positional
uncertainty are considered as possible counterparts.  Optical sources
are taken from the Ilbert \etal\ catalogue, which identifies
individual sources down to $i\sim$27.  Although we consider sources
within the nominal positional uncertainty found by our simulations (as
shown in Figure~\ref{fig:boosting}), additional accommodation must be
made for uncertainty in the positions measured at other wavelengths or
source size.  Since \scubaii\ 450\um\ positional uncertainties are on
the order of 1--2\arcsec, a source's size or positional uncertainty at
other wavelengths is not negligible.  At optical and near-infrared
wavelengths we are able to estimate sources' `positional uncertainty'
by considering its measured size; we fix the positional uncertainty
for each source individually by taking the area (measured in
arcseconds squared) and elongation (both measured using {\sc
  SExtractor}\footnote{{\sc SExtractor}: a Source Extractor program
  developed in \citet{bertin96a}.}) to convert to a semi-major axis.
Most optical/near-infrared sources have approximated semi-major axes
$\sim$0.5\arcsec, with 90\%\ of sources being $<$1\arcsec.  Therefore,
when seeking potential matches to 450\um\ sources, we search within a
radius which is the 450\um\ source's positional
uncertainty and the optical/near-infrared sources' approximated
semi-major axis summed in quadrature.

\begin{table*}
\centering
\caption{Identification of \scubaii\ Sources at 870\um--1.2\,mm}
{\scriptsize
\begin{tabular}{llccccll}
\hline\hline
450\um- & 850\um- & {\sc Alternate} & $\lambda_{\rm obs}$ & S$_{\lambda}$ & {\tiny {\sc Interfer-}} & {\sc Inst./} & {\sc Reference} \\
{\sc Name }& {\sc Name} & {\sc Name}     &    [mm]             & [mJy]    & {\tiny {\sc ometric?}}  & {\sc Tel} & \\
\hline\hline
450.00 & 850.07  & {\tiny AzTEC\,J100028.94+023200.3}  & 1.1  & 3.8$\pm$0.9   &   & AzTEC & \citet{scott08a} \\
450.01 & 850.02  & AzTEC/C80                           & 1.1  & 4.1$\pm$0.9   &   & AzTEC  & \citet{aretxaga11a} \\
       &         & COSLA-47                            & 0.87 & 9.0$\pm$2.8   &   & LABOCA  & Navarette \etal\ in prep \\
       &         & COSLA-47                            & 1.3  & 3.11$\pm$0.59 & Y & PdBI  & \citet{smolcic12a} \\
450.03 & 850.00  & {\tiny AzTEC\,J100008.03+022612.1}  & 1.1  & 8.3$\pm$1.1   &   & AzTEC  & \citet{scott08a} \\
       &         & AzTEC-2                             & 0.89 & 12.4$\pm$1.0  & Y & SMA  & \citet{younger07a,younger09a} \\
       &         & AzTEC/C3                            & 1.1  & 10.5$\pm$1.0  &   & AzTEC  & \citet{aretxaga11a} \\
450.04 & 850.03  & {\tiny AzTEC\,J100019.73+023206.0}  & 1.1  & 6.5$\pm$1.3   &   & AzTEC  & \citet{scott08a} \\
       &         & AzTEC-5                             & 0.89 & 9.3$\pm$1.3   & Y & SMA  & \citet{younger07a,younger09a} \\
       &         & AzTEC/C42                           & 1.1  & 4.8$\pm$1.1   &   & AzTEC  & \citet{aretxaga11a} \\
450.05 & 850.08  & AzTEC/C38                           & 1.1  & 5.1$\pm$1.0   &   & AzTEC  & \citet{aretxaga11a} \\
       &         & COSLA-35                            & 0.87 & 8.2$\pm$2.2   &   & LABOCA  & Navarette \etal\ in prep \\
       &         & COSLA-35                            & 1.3  & 2.15$\pm$0.51 & Y & PdBI  & \citet{smolcic12a} \\
450.07 & 850.98  & AzTEC/C169                          & 1.1  & 3.1$\pm$0.8   &   & AzTEC  & \citet{aretxaga11a} \\
450.08 & 850.46  & COSLA-8                             & 0.87 & 6.9$\pm$1.6   &   & LABOCA  & Navarette \etal\ in prep \\
       &         & COSLA-8                             & 1.3  & 2.65$\pm$0.62 & Y & PdBI  & \citet{smolcic12a} \\
450.16 &         & COSBO3$**$                          & 1.2  & 7.4$\pm$1.1   &   & MAMBO  & \citet{bertoldi07a} \\
       &         & AzTEC/C6$**$                        & 1.1  & 9.6$\pm$1.0   &   & AzTEC  & \citet{aretxaga11a} \\
450.20 & 850.24  & {\tiny AzTEC\,J100004.54+023040.1}  & 1.1  & 3.3$\pm$0.8   &   & AzTEC  & \citet{scott08a} \\
       &         & AzTEC/C150                          & 1.1  & 3.3$\pm$1.2   &   & AzTEC  & \citet{aretxaga11a} \\
450.27 & 850.115 & AzTEC/C65                           & 1.1  & 4.4$\pm$1.0   &   & AzTEC  & \citet{aretxaga11a} \\
450.28 & 850.05  & COSBO3$**$                          & 1.2  & 7.4$\pm$1.1   &   & MAMBO  & \citet{bertoldi07a} \\
       &         & AzTEC/C6$**$                        & 1.1  & 9.6$\pm$1.0   &   & AzTEC  & \citet{aretxaga11a} \\
450.38 & 850.14  & AzTEC/C24                           & 1.1  & 5.7$\pm$1.0   &   & AzTEC  & \citet{aretxaga11a} \\
450.55 & 850.06  & COSBO7                              & 1.2  & 5.0$\pm$0.9   &   & MAMBO  & \citet{bertoldi07a} \\
       &         & AzTEC/C160                          & 1.1  & 3.1$\pm$1.2   &   & AzTEC  & \citet{aretxaga11a} \\
450.66 & 850.146 & AzTEC/C66                           & 1.1  & 4.3$\pm$0.9   &   & AzTEC  & \citet{aretxaga11a} \\
\hline
        & 850.01 & {\tiny AzTEC\,J095957.22+022729.3}  & 1.1  & 5.8$\pm$1.0   &   & AzTEC  & \citet{scott08a} \\
        &        & AzTEC-9                             & 0.89 & 13.5$\pm$1.8  & Y & SMA  & \citet{younger07a,younger09a} \\
        &        & AzTEC/C18                           & 1.1  & 7.9$\pm$1.5   &   & AzTEC  & \citet{aretxaga11a} \\
        & 850.04 & COSBO1                              & 1.2  & 6.2$\pm$0.9   &   & MAMBO  & \citet{bertoldi07a} \\
        &        & AzTEC/C7                            & 1.1  & 8.9$\pm$1.1   &   & AzTEC  & \citet{aretxaga11a} \\
        & 850.13 & AzTEC/C114                          & 1.1  & 3.7$\pm$0.9   &   & AzTEC  & \citet{aretxaga11a} \\
        & 850.15 & {\tiny AzTEC\,J100025.23+022608.0}  & 1.1  & 1.9$\pm$0.6   &   & AzTEC  & \citet{scott08a} \\
        &        & AzTEC/C30                           & 1.1  & 5.5$\pm$1.1   &   & AzTEC  & \citet{aretxaga11a} \\
        & 850.18 & {\tiny AzTEC\,J100023.98+022950.0}  & 1.1  & 2.6$\pm$0.7   &   & AzTEC  & \citet{scott08a} \\
        & 850.20 & {\tiny AzTEC\,J100026.68+023128.1}  & 1.1  & 2.8$\pm$0.8   &   & AzTEC  & \citet{scott08a} \\
        & 850.22 & COSLA-50                            & 0.87 & 5.6$\pm$1.6   &   & LABOCA  & Navarette \etal\ in prep \\
        &        & COSLA-50                            & 1.3  & ...           & Y & PdBI  & \citet{smolcic12a} \\
        &        & AzTEC/C33                           & 1.1  & 5.3$\pm$1.1   &   & AzTEC  & \citet{aretxaga11a} \\
        & 850.25 & COSBO19                             & 1.2  & 3.0$\pm$0.8   &   & MAMBO  & \citet{bertoldi07a} \\
        &        & COSLA-38                            & 0.87 & 5.8$\pm$1.6   &   & LABOCA  & Navarette \etal\ in prep \\
        &        & COSLA-38                            & 1.3  & 8.19$\pm$1.85 & Y & PdBI  & \citet{smolcic12a} \\
450.99  & 850.33 & {\tiny AzTEC\,J095950.69+022829.5}  & 1.1  & 3.6$\pm$0.9   &   & AzTEC  & \citet{scott08a} \\
        & 850.35 & AzTEC/C74                           & 1.1  & 4.2$\pm$0.9   &   & AzTEC  & \citet{aretxaga11a} \\
        & 850.50 & AzTEC/C35                           & 1.1  & 5.2$\pm$1.0   &   & AzTEC  & \citet{aretxaga11a} \\
        & 850.57 & AzTEC/C45                           & 1.1  & 4.8$\pm$1.0   &   & AzTEC  & \citet{aretxaga11a} \\
450.78  & 850.62 & {\tiny AzTEC\,J100031.06+022751.5}  & 1.1  & 2.7$\pm$0.8   &   & AzTEC  & \citet{scott08a} \\
        & 850.63 & COSBO36                             & 1.2  & 5.7$\pm$1.3   &   & MAMBO  & \citet{bertoldi07a} \\
        &        & AzTEC/C71                           & 1.1  & 4.3$\pm$1.1   &   & AzTEC  & \citet{aretxaga11a} \\
        & 850.94 & COSBO29                             & 1.2  & 3.1$\pm$1.0   &   & MAMBO  & \citet{bertoldi07a} \\
        &        & AzTEC/C162                          & 1.1  & 3.1$\pm$1.2   &   & AzTEC  & \citet{aretxaga11a} \\
\hline\hline
\end{tabular}
}
\label{tab:1mm}
\begin{spacing}{0.7}
{\scriptsize {\bf Table Notes.}  Counterparts of \scubaii\ sources at
  870\um--1.2\,mm as identified by LABOCA, PdBI, AzTEC, SMA and MAMBO
  with given references. Sources are divided into
  450\um\ $>$3.6$\sigma$ and 850\um\ $>$3.6$\sigma$ samples.  None of
  the marginal 3$<\sigma<$3.6 450\um/850\um\ sources were reported as
  detections in the given references.  Sources are searched within a
  10\arcsec\ (450\um\ sources) or 12\arcsec\ (850\um\ sources) search
  radius; this is approximately one beamsize at $\sim$850\um--1.2\,mm
  and represents a regime where chance coincidence between submm
  sources is very low ($p-value<0.01$).  Flux densities ($S_\lambda$) are
  deboosted and not raw.  Sources marked with ** correspond to
  multiple \scubaii\ sources in the Table, i.e., 450.16, 450.28 and
  850.05 all correspond to COSBO3/AzTEC/C6.}
\end{spacing}
\end{table*}

The positional uncertainty of 24\um\ and 1.4\,GHz sources is estimated
with a more statistical method (since the resolution is not adequate
to resolve source sizes) by measuring the global offset of 24\um\ and
1.4\,GHz sources to $K$-band and $i$-band positions.  We find that
90\%\ of 24\um\ sources lie within 1.05\arcsec\ of their associated
optical/near-infrared counterpart and 90\%\ of 1.4\,GHz sources lie
within 0.85\,\arcsec.  Therefore, we add 1.05\arcsec\ and
0.85\arcsec\ (in quadrature) to our search radius for 24\um\ and
1.4\,GHz counterparts, respectively.  We observe many instances where
24\um\ and 1.4\,GHz counterparts fall outside the 450\um\ positional
uncertainty but lie $<$1\arcsec\ beyond the positional uncertainty
(e.g. 450.01).  The median search radii at 450\um\ and 850\um\ were
2.2\arcsec\ and 4.2\arcsec, respectively.

The advantage of identifying radio counterparts comes from the
well-known FIR/radio correlation for starburst galaxies
\citep{helou85a,condon92a}, whereby a radio-detected galaxy is likely
to be FIR-bright, and thus correspond to the source generating the FIR
emission detected in our \scubaii\ maps
\citep{barger00a,chapman03a,chapman05a}.  MIPS 24\um-emission also
correlates with FIR emission since both are seen in dusty galaxies,
although the relationship is more complex due to the variation of SED
types in the mid-infrared \citep[PAH emission features drifting in and
  out of the band, and AGN generated power law
  emission][]{le-floch05a,lee10a}.  The existence of a
mid-infrared or radio counterpart within the beamsize of
450\um\ observations can help distinguish that source from others as
the likely counterpart, since the likelihood of random coincidence is
quite low \citep[this likelihood is often called the
  $p$-value;][]{downes86a}\footnote{We calculate the $p$-value
  (defined by $p=1-exp(-\pi n \theta^2)$ where $n$ is the source
  density, $\theta$ the angular offset or search area) by assuming the
  following source densities which we measure in this area of the
  COSMOS field: 885,900\,deg$^{-2}$ at $i$-band; 151,900\,deg$^{-2}$
  at 3.6\,\um; 19,400\,deg$^{-2}$ at 24\,\um; and 2,300\,deg$^{-2}$ at
  1.4\,GHz.}.

A summary of counterparts is found in Table~\ref{tab:counterparts450}
for 450\um\ sources, Table~\ref{tab:counterpartsextra450} for marginal
450\um\ sources, and Table~\ref{tab:counterparts850} for
850\um\ sources which are not 450\um-detected.  Photometric redshifts
\citep{ilbert10a} are quoted where available.  Also, the handful of
sources which have interferometric observations have their known
counterparts in bold in Tables~\ref{tab:counterparts450},
\ref{tab:counterpartsextra450} and \ref{tab:counterparts850}.  The
interferometrically-observed subset of \scubaii\ sources (along with
all \scubaii\ sources which were detected with other submillimeter
instruments) is given in Table~\ref{tab:1mm}.  The best-guess
counterpart is the source with the lowest $p$-value;
generally, a $p$-value $<$0.05 can be regarded as confident, while
larger $p$-values are more tentative.  Sources with $p$-values $>$0.5
are removed from
Tables~\ref{tab:counterparts450}--\ref{tab:counterparts850} since they
are more likely not truly-associated counterparts.
Figure~\ref{fig:cutouts450} (given at the end of this paper) show
12\arcsec$\times$12\arcsec\ postage stamp cutouts of the
450\um\ sources in the optical ($Biz$ tricolor) and near-infrared
(IRAC 3.6\um) with radio and 24\um\ counterparts identified for
reference.

A significant fraction of sources, both at 450\um\ and 850\um\ lack
24\um\ or radio counterparts.  At 450\um, only 44\%\ (=34/78) of
sources have 24\um\ or radio counterparts and at 850\um, only
60\%\ \citep[=59/99, which is consistent with previous rates at
  850\um;][]{coppin05a}.  These fractions should improve with deeper
radio data \citep[e.g. as in][]{barger12a}, which should be available
in the coming years in the COSMOS field, but it would not fix the bias
introduced in submillimeter samples which are reliant on these
counterparts.  Studies which are reliant on multiwavelength
counterpart identification, yet which have a significant number of
galaxies lacking counterparts, will be biased.  However, note that the
majority of sources without 24\um\ or radio counterparts {\it do} have
IRAC near-infrared counterparts ($\sim$33/44=75\%), and of those that
do not, 7 sources have no counterparts at all (i.e. also lack any
optical counterparts).  This leaves only 4 objects (in both
450\um\ and 850\um\ samples) which rely on matching directly to
optical counterparts$-$the most uncertain method generating the
highest $p$-values.  The median $p$-values for the different
counterpart matching methods at 450\um\ are the following: 0.0009
(1.4\,GHz, 16 galaxies), 0.02 (24\um, 18 galaxies), 0.09 (3.6\um, 33
galaxies), and 0.30 (optical/$i$-band, 4 galaxies).

\subsection{Reliability of Counterpart Identifications}\label{sec:purity}

Identifying the correct multiwavelength counterpart(s) for each
submillimeter source is the most important, but also the most
difficult aspect of characterising the submillimetre galaxy
population.  Recently, it has become clear that direct, far-infrared
interferometry is the only guaranteed method of identifying
counterparts correctly \citep[e.g. recent results from SMA, PdBI and
  ALMA targeting 850\um--1.1\,mm
  sources;][]{younger08a,wang11a,barger12a,smolcic12a,karim13a,hodge13a}.
For 850\um\ sources detected with a $\sim$15\arcsec\ beamsize,
interferometric work suggests that best-guess counterpart matching to
radio and/or 24\um\ counterparts will fail $\sim$30\%\ of the time
\citep[][and Smail, private communication]{hodge13a}, dependent on
ancillary field depth and robustness of FIR/radio and FIR/mid-infrared
correlations.  This provides a good estimate to the accuracy of the
counterparts given in Table~\ref{tab:sources850} for our
850\um\ sample, but what is the reliability of our 450\um\ source
counterparts?

For the five 450\um\ sources which have interferometric observations,
four would have been traditionally matched with their truthful
counterpart using our method; two of those sit within the
450\um\ positional uncertainty $<$1.5\arcsec, two are outside but
still $<$2\arcsec\ away, and one source, 450.08, is 6.8\arcsec\ away
from our \scubaii\ position.  Unfortunately, small number statistics
on interferometric observations limit interpretation.

In the absence of interferometry, we can speculate that the reduced
beamsize with respect to 850\um\ would see a reduction in counterpart
contamination proportional to the reduction in sky area searched for
potential counterparts.  The positional uncertainties of 450\um\ range
from 1.0--2.4\arcsec\ at $>$3.6$\sigma$ (with a median of
$\sim$2.1\arcsec), while 850\um\ positional uncertainties vary from
2--6\arcsec\ at $>$3.6$\sigma$ (median of 3.6\arcsec).  This suggests
the same counterpart identification strategy (when applied to
450\um\ samples) should fail only $\sim$5\%\ of the time.

While this thought experiment gives a good ballpark figure for
counterpart correctness, we can also use the individual source
probabilities of correctness, given by the $p$-value to compute the
likely number of incorrect counterparts.  The likelihood of all 71
450\um\ counterparts being correct (note that 7 sources lack any
counterparts) is $\sim$0.4\%, the probability of one contaminant is
2.7\%, the probability of two is 8\%, three is 15\%, four is 20\%,
five is 20\%, six is 15\%, seven is 8\%, and higher than seven
contaminants is 11\%\ likely.  Therefore, we anticipate that
5--10\%\ of our 450\um\ sample has incorrect counterpart matches; it
is most likely the sources with incorrect counterparts are those with
the highest $p$-values.  While this emphasises the importance of
securing the correct counterparts for detailed follow-up on individual
sources, this will not impact the bulk result of our analysis on the
450\um\ population (in terms of redshift distribution, luminosities,
etc) significantly.

\begin{figure}
\centering
\includegraphics[width=0.60\columnwidth]{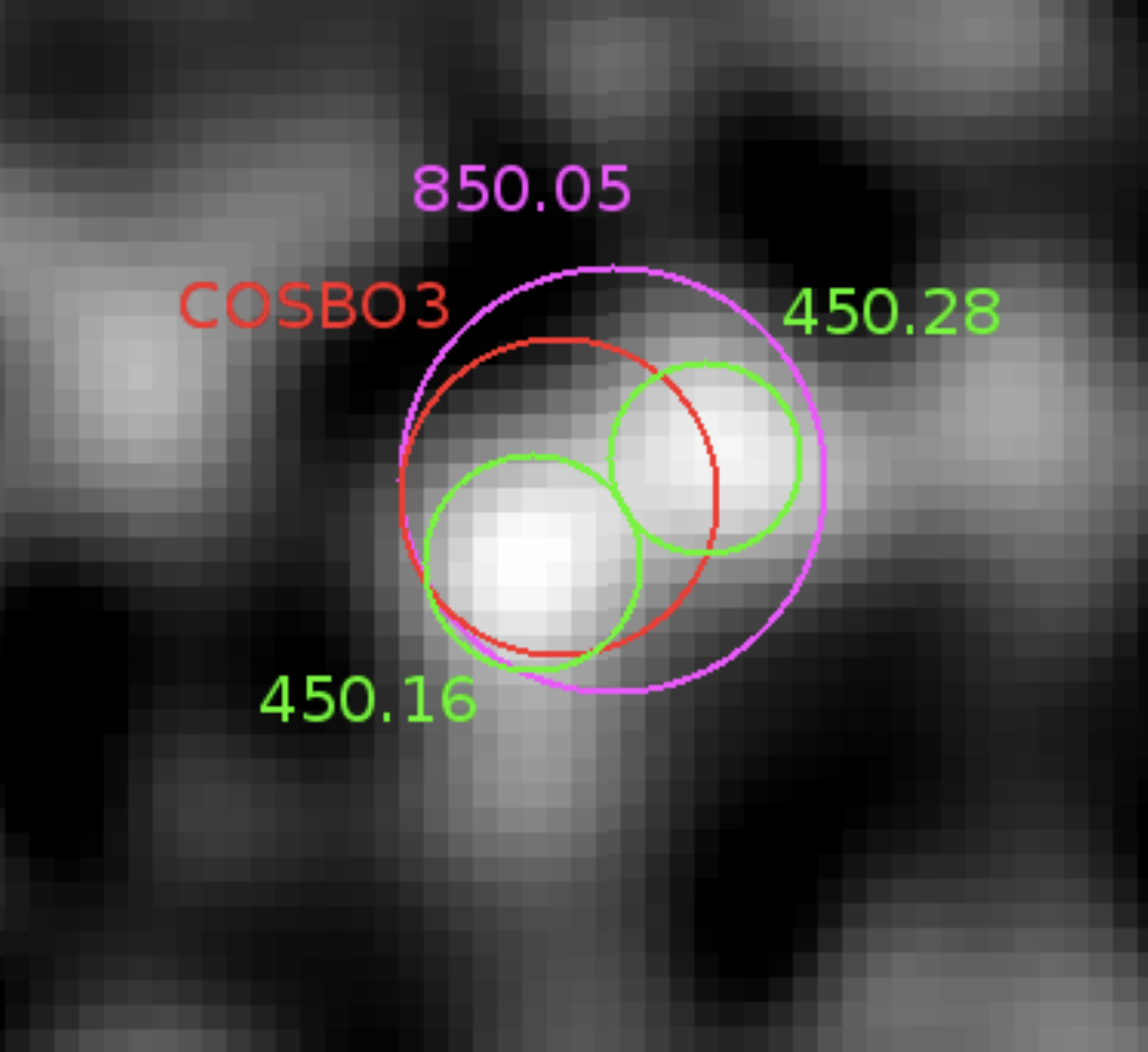}
\caption{A 35$\times$40\arcsec\ 450\um\ signal-to-noise map cutout of
  450.16 and 450.28.  The 450\um\ sources are circled and labelled in
  green, while COSBO3, the MAMBO-detected 1.2\,mm source is shown in
  red and the 850\um\ \scubaii\ source, 850.05, is shown in magenta.
  The beamsizes at 450\um, 850\um, and 1.2mm are 7\arcsec, 15\arcsec,
  and 11\arcsec\ respectively.}
\label{fig:multiple}
\end{figure}

\subsection{Multiplicity}

The improved resolution from interferometry makes it clear that a
potentially large fraction (20--50\%) of bright ($>$10\,mJy) SMGs are
actually multiples \citep{wang11a,barger12a,hodge13a}.
Although \scubaii\ is a bolometer array and not an interferometer, the
high resolution at 450\um\ can detect multiple sources when
longer-wavelength observations only detect one SMG.  This phenomenon
is found once at high S/N in our sample, surrounding sources 450.16
and 450.28; the system is illustrated in Figure~\ref{fig:multiple}.
There is only one long-wavelength submillimeter source detected in
this area: 850.05 in our sample, as well as COSBO3, detected at
1.2\,mm \citep{bertoldi07a}, and AzTEC/C6, detected at 1.1\,mm
\citep{aretxaga11a}.  However, we find two 450\um\ sources within the
\scubaii\ 850\um\ beam: 450.16 with a S/N of $\sim$4.6 and 450.28 with
a S/N of $\sim$4.2.  This is the best unequivocal identification of a
multiple at 450\um\ in our sample.

This demonstrates that shorter-wavelength, smaller beamsize bolometers
can be used to probe submillimeter galaxy multiplicity in blank-field
surveys at potentially lower observational cost than interferometric
follow-up (e.g. future large-field surveys from LMT and CCAT).

\begin{figure}
\centering
\includegraphics[width=0.99\columnwidth]{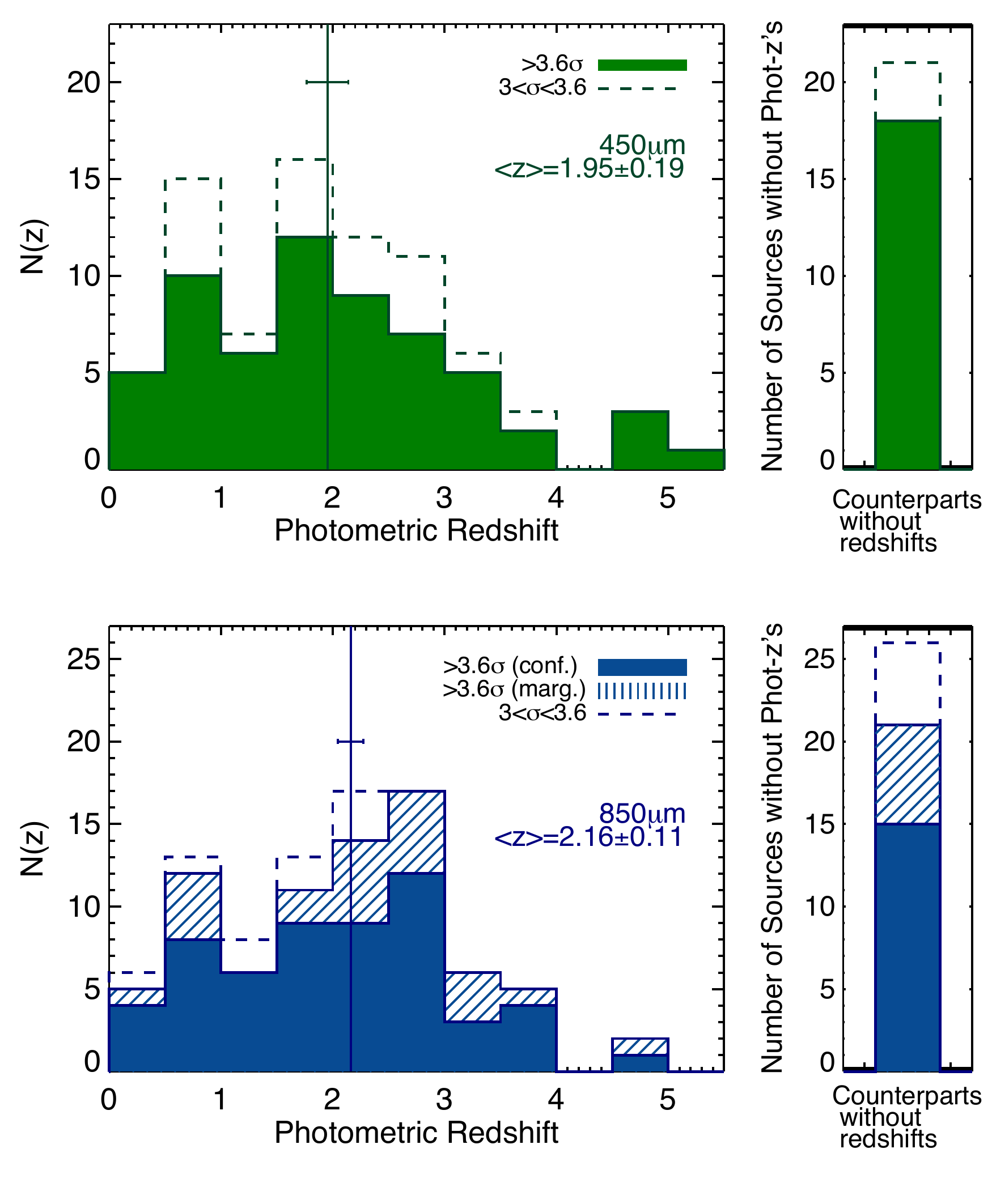}
\caption{The distribution in photometric redshift for
  \scubaii\ sources identified at 450\um\ ($green$) and
  850\um\ ($blue$).  The median redshifts of the two samples are
  $\langle z\rangle=1.95$ and $\langle z\rangle=2.16$ respectively.
  At 450\um, we illustrate two distributions, the first consisting of
  just $>$3.6$\sigma$ 450\um\ sources (solid green shaded region), and
  the second of the 3$<\sigma<$3.6 sample (dashed green line).  We
  also indicate the number of sources without photometric redshifts at
  right.  Due to the larger beamsize at 850\um, our counterpart
  matching is done first using 24\um, radio, or 450\um\ identification
  (solid blue shaded region).  The remaining 850\um\ SMGs without
  24\um/radio/450\um\ counterparts are shown in the hashed region.
  The dashed blue line corresponds to the 3$<\sigma<$3.6 sample. The
  median redshifts are marked with vertical lines and 1$\sigma$
  bootstrap uncertainty on the median. }
\label{fig:nz}
\end{figure}

\begin{figure}
\centering
\includegraphics[width=0.99\columnwidth]{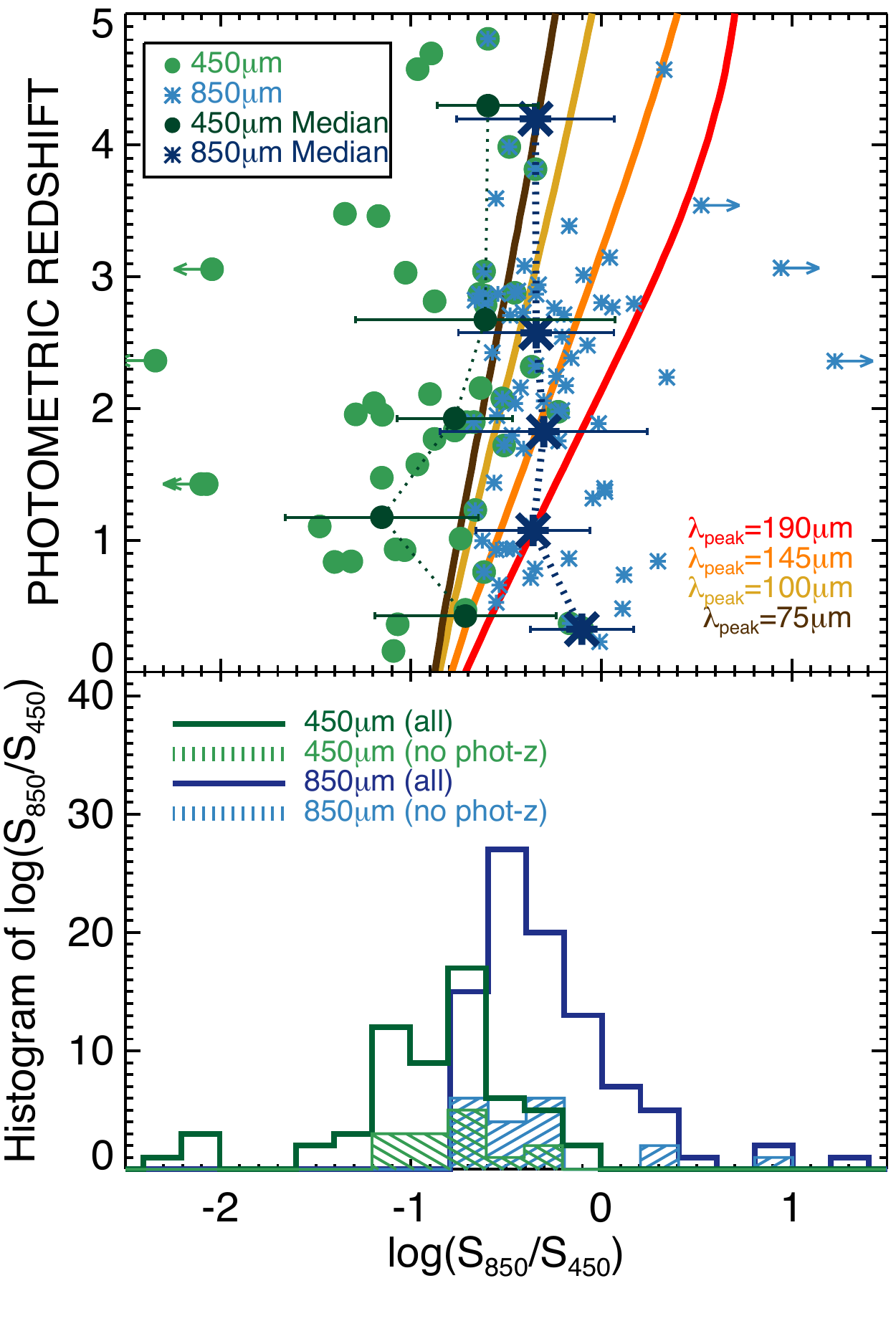}
\caption{ The FIR colour properties of the \scubaii\ sample, measured
  as $\log(S_{\rm 850}/S_{\rm 450})$.  At top, we plot against the
  optical/near-infrared photometric redshift.  Sources with arrows
  represent galaxies which have flux densities close to zero in one
  band or the other; sources with negative flux densities are not
  included here.  The 450\um\ sample (green) has, on average,
  450\um\ flux densities which exceed 850\um\ flux densities by a
  factor of 5, while the 850\um\ sample (blue) has naturally `redder'
  colours.  Median values, binned by redshift, are shown in dark
  green/blue with deviations in the population indicated by error bars
  (the uncertainties in the median are much smaller
  $\sim$0.05--0.08\,dex).  Slight redshift evolution is seen in the
  median.  We overplot evolution of constant temperature (or constant
  SED peak wavelength) and note that our samples are inconsistent with
  non-evolving SED shape (i.e. the dust temperatures of high redshift,
  more luminous sources are much higher than they are at low
  redshift).  At bottom, we plot the distribution in FIR colour for
  both samples with the subset of the samples which do not have any
  photometric redshift.  The distribution of colours for the samples
  without redshifts resembles the parent population, indicating no
  obvious redshift-colour bias.  }
\label{fig:fircolor}
\end{figure}

\section{Derived Source Characteristics}\label{sec:derived}

\subsection{Redshift Distributions}

Figure~\ref{fig:nz} shows the distributions of optical/near-infrared
photometric redshifts \citep{ilbert10a} for the best-guess
counterparts identified in Tables~\ref{tab:counterparts450} and
\ref{tab:counterparts850}.  Photometric redshifts span $0<z<5.1$, with
most sources at $z<3$.  The median redshift for the 450\um\ sample is
$z=1.95\pm0.19$ while the median for the 850\um\ population is
$z=2.16\pm0.11$, in agreement with previous work at
850\um\ \citep{chapman05a,wardlow11a}.  The uncertainties on the
median are generated via bootstraping methods.

At 850\um, we split our sample by its potential biases.  The increased
beamsize of 850\um\ observations means that 850\um\ sources which lack
24\um\ or radio identifications have more uncertain counterparts.  We
illustrate the distribution of redshifts for the 24\um/radio
identified subset, then those without (dashed area) and finally those
from the $3<\sigma<3.6$ marginal catalogue.  Note that the number of
sources lacking redshift identifications is $\approx$18--21 (amounting
to 21--23\%) of either sample.

Are those sources without photometric redshifts intrinsically
different than those with photometric redshifts?  We investigate this
by seeing if the FIR properties, i.e. $S_{\rm 850}$ to $S_{\rm 450}$
colour, (a) evolves with redshift, or (b) differs between the sample
of galaxies with photometric redshifts and without.
Figure~\ref{fig:fircolor} plots the FIR colour of both $>$3.6$\sigma$
samples against photometric redshift and in histogram form.  As
expected, 450\um\ sources are much brighter at 450\um\ than
850\um\ and vice versa.  There is some shallow evolution of the
$\log(S_{\rm 850}$/$S_{\rm 450})$ colour with redshift, although the
statistical variation within subsamples dominates.  
As we find that sources display a wide range of FIR properties
irrespective of redshift, this does not necessarily bode well for work
which makes use of FIR photometric redshifts which assume a certain
FIR template with fixed temperature and use FIR photometry to estimate
the redshift more precisely than $\Delta z\sim2$ \citep[some work
  using this technique include][, but often under specific luminosity
  restrictions, not broadly applicable to a wide range of high-$z$
  ULIRGs]{roseboom12a,barger12a,chen13a}.  Furthermore, sources which
do not have photometric redshifts show roughly the same distribution
in FIR colour as the parent population of galaxies they are selected
from.  This suggests that the systems lacking photometric redshifts
are not intrinsically biased with FIR properties.  While there is
significant variation in SED shape, the differences between the
measured medians for 450\um\ and 850\um\ samples (measured by
uncertainty of the median) are statistically significant at
$>$5$\sigma$ at $z$\simlt$2$ and $\sim$3$\sigma$ at $z\sim3$.

It is difficult to say if sources without photometric redshifts differ
significantly from those with redshift estimates.  \citet{wardlow11a}
estimate the redshift distribution of 870\um\ sources in CDFS without
photometric redshifts using the density of IRAC 3.6\um\ emitters, and
guess that they largely sit at $z\sim2.5$, near the expected peak of
the whole population.  The SCUBA2 population is likely similar to the
870\um\ galaxies from \citet{wardlow11a}, although only spectroscopic
confirmation will unequivocally reveal their true nature.

\subsection{Infrared luminosities \&\ dust properties}\label{sec:sed}

\begin{figure*}
\centering
\includegraphics[width=0.99\columnwidth]{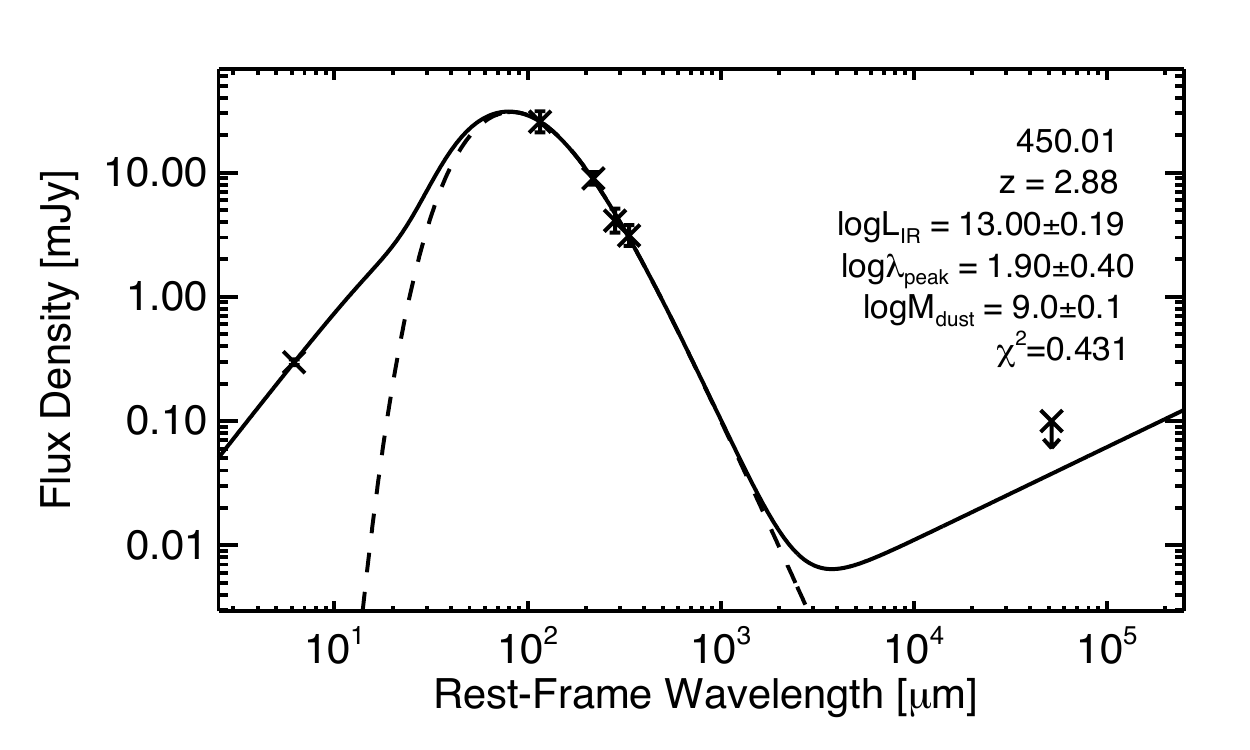}
\includegraphics[width=0.99\columnwidth]{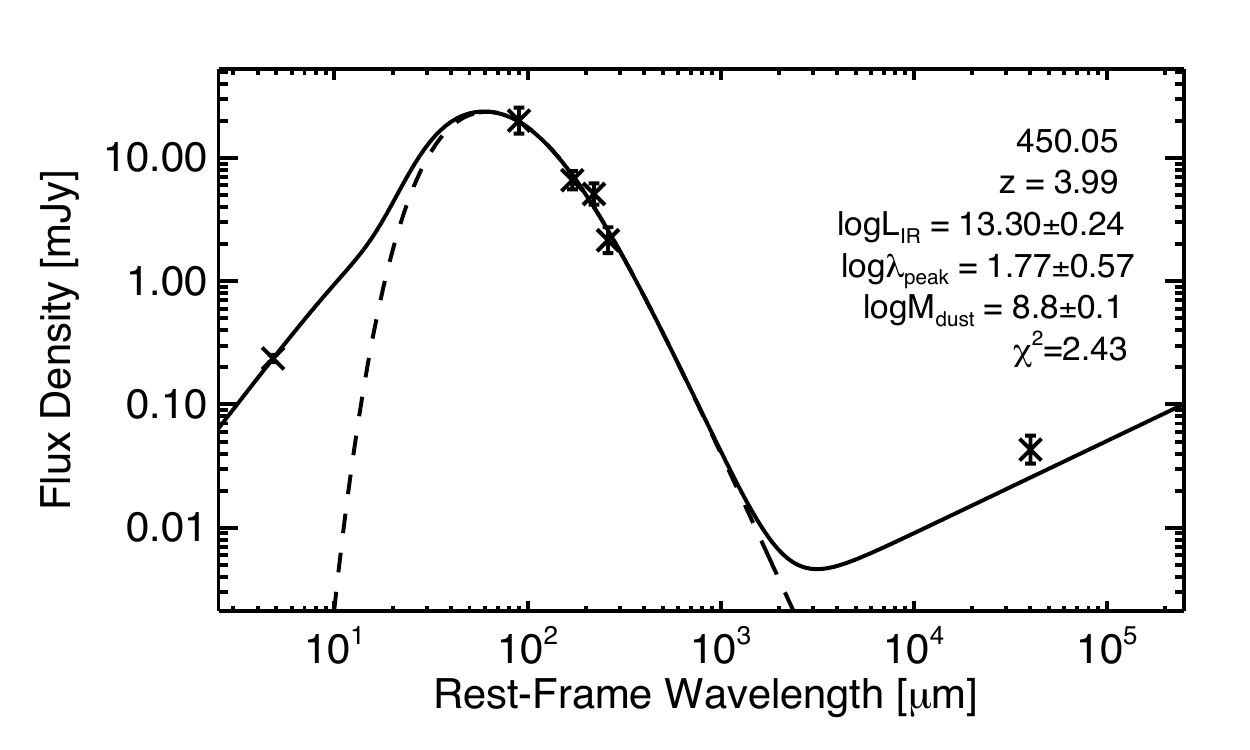}\\
\includegraphics[width=0.99\columnwidth]{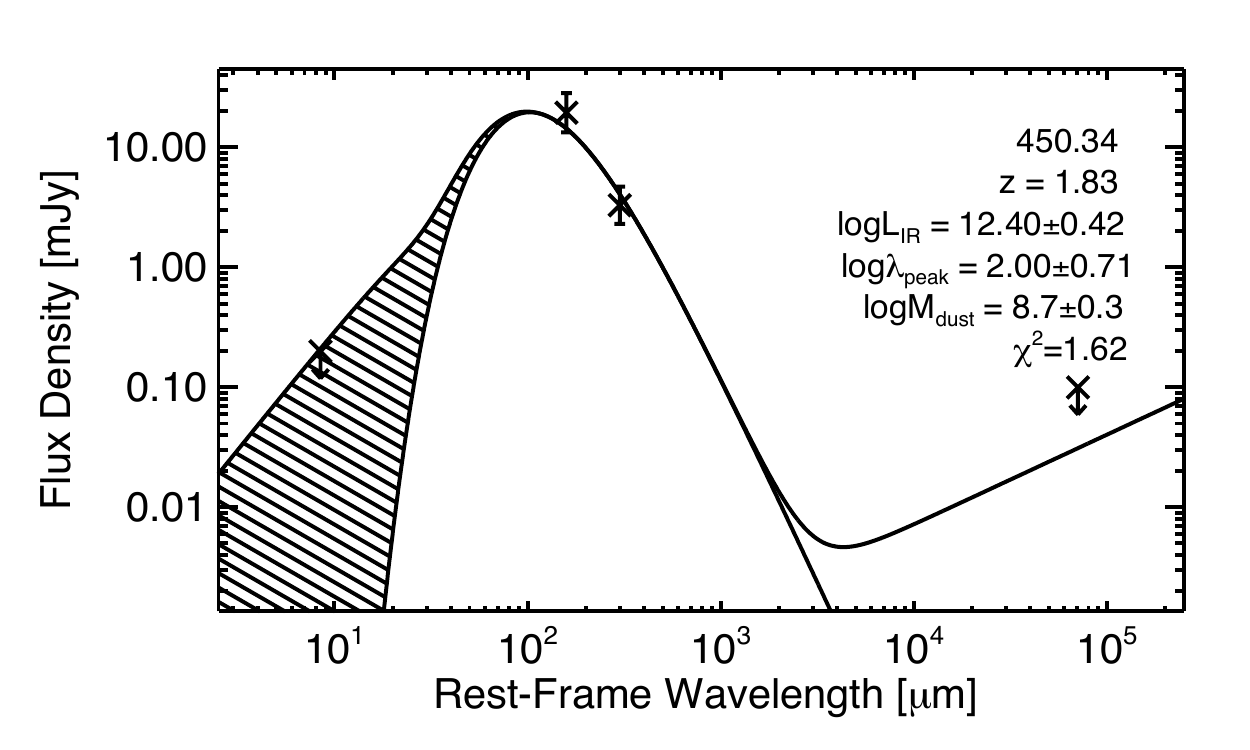}
\includegraphics[width=0.99\columnwidth]{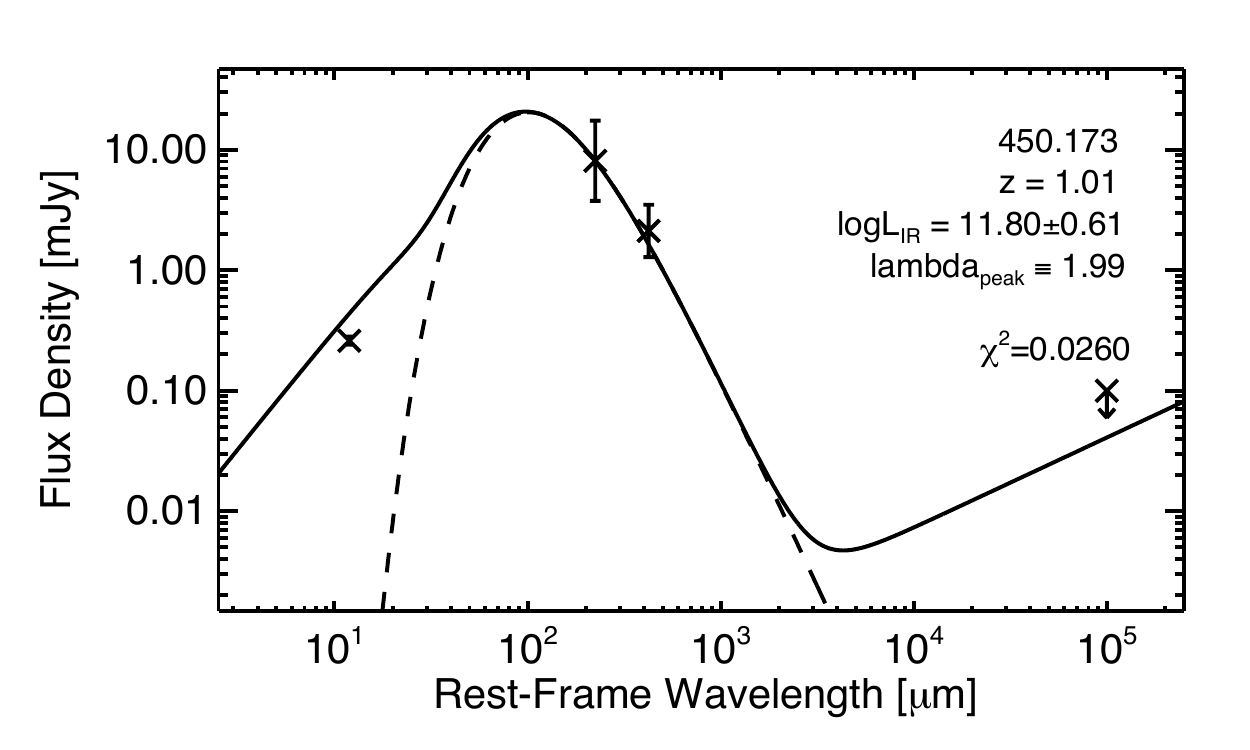}\\
\includegraphics[width=0.99\columnwidth]{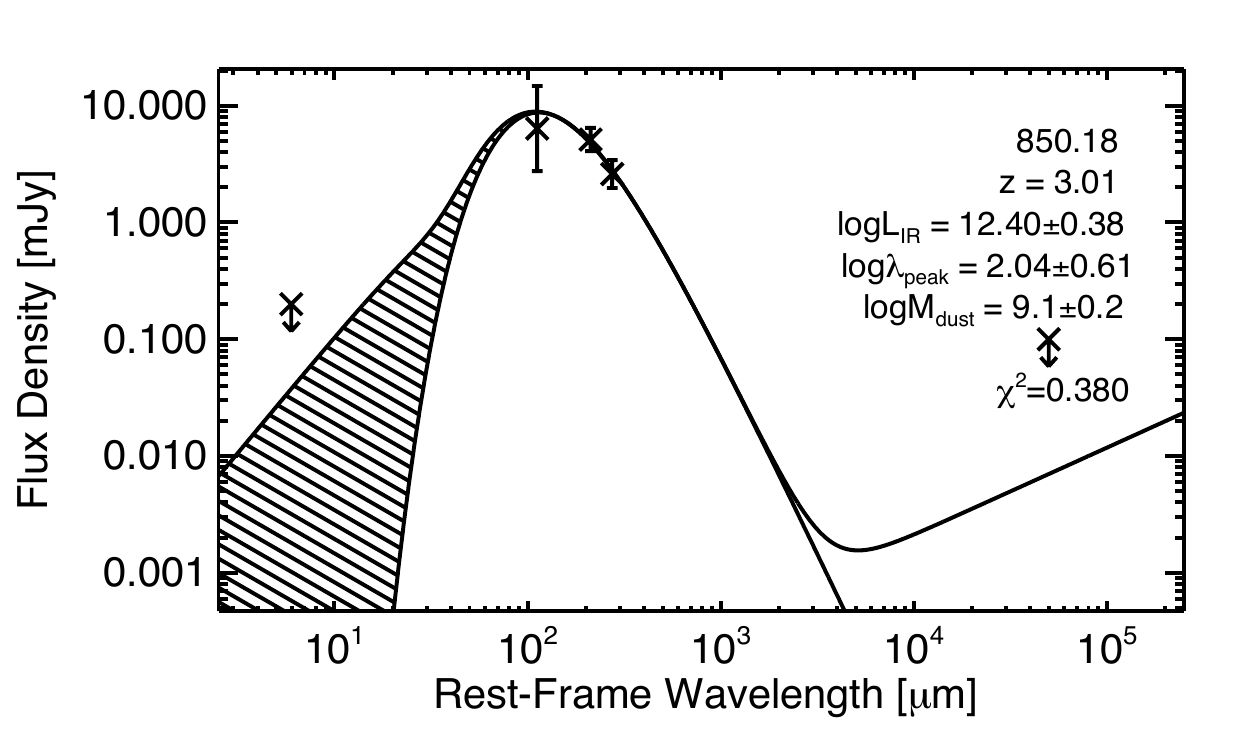}
\includegraphics[width=0.99\columnwidth]{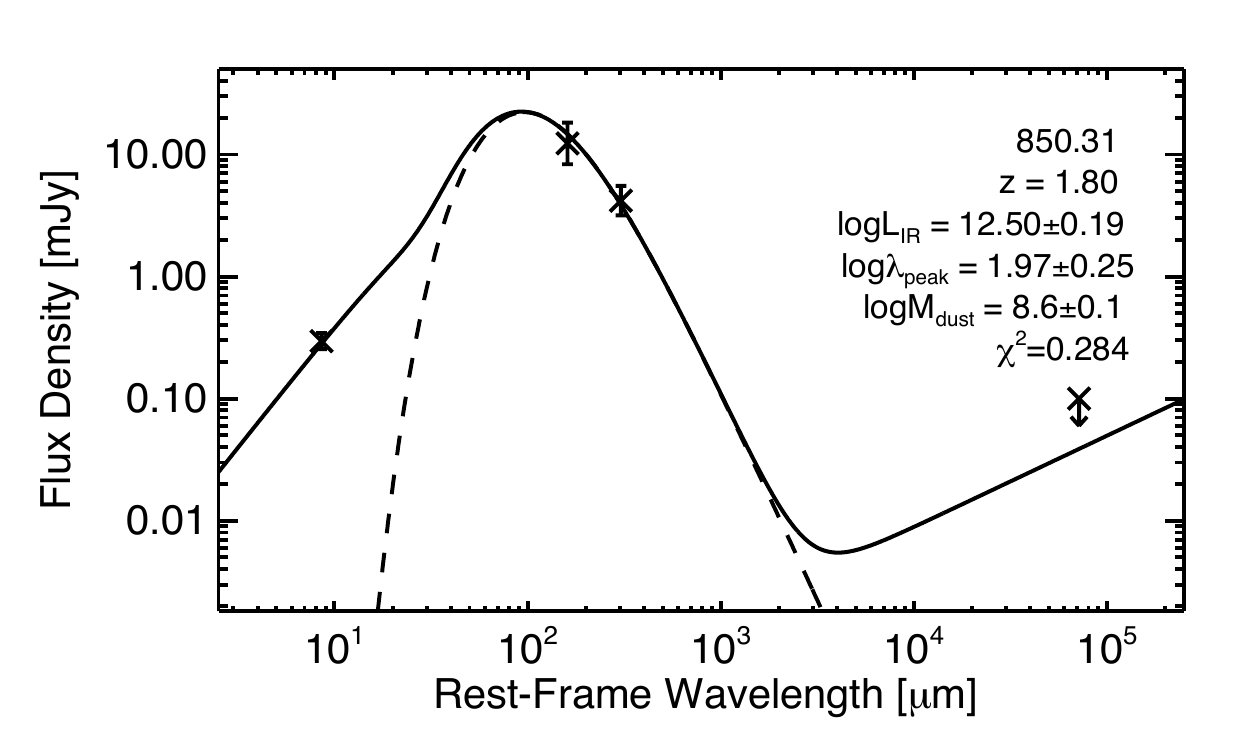}\\
\caption{ Example far-infrared SEDs for select \scubaii\ sources, some
  450\um-selected and some 850\um-selected.  The assumed
  optical/near-infrared photometric redshift is specified in the upper
  right corner of each panel, as well as the measured $L_{\rm IR}$,
  $\lambda_{\rm peak}$, $M_{\rm dust}$, and $\chi^2$ goodness of fit.
  If the SED fit used a fixed temperature due to a lack of $>$2
  infrared photometric detections, then no dust mass is measured.  In
  cases where there is no 24\um\ detection, the mid-infrared portion
  of the SED is shaded to indicate its uncertainty.  The uncertainty
  in $L_{\rm IR}$ reflects this uncertainty.  The radio portion of the
  SEDs are extrapolated from assuming the FIR/radio correlation holds
  \citep[and evolves as measured in][]{ivison10a,ivison10b} and are
  not fit to the radio data.
}
\label{fig:seds}
\end{figure*}

We can go beyond basic FIR colour comparisons with our \scubaii\ data
and use SED fits to measure infrared luminosities, temperatures and in
some cases dust masses.  Simple greybody SED fitting combined with a
mid-infrared powerlaw gives us this information with a minimum number
of free parameters or model assumptions.  We fit all available
infrared photometry, from 24\um\ MIPS, 70\um\ MIPS, 450\um\ \scubaii,
850\um\ \scubaii, 870\um\ LABOCA, 1.1\,mm AzTEC, 1.2\,mm MAMBO, and
1.3\,mm PdBI.

The combination of a greybody and mid-infrared powerlaw accounts for
both galaxy-wide cold dust emission and smaller-scale warm dust
emission \citep{blain03a,kovacs06a}.  The fitting method is described in full in
\citet{casey12a} and is represented by the equation
\begin{equation}
S_{\nu} = N_{\rm bb}\frac{(1-e^{-(\nu/\nu^{0})^{\beta}})\nu^{3}}{e^{hc\nu/kT} - 1} + N_{\rm pl} (c/\nu)^\alpha e^{-(\nu_{c}/\nu)^{2}}
\end{equation}
where $S_{\nu}$ is flux density (in units of mJy), $T$ is greybody
dust temperature, $\beta$ is the emissivity index of the greybody,
$\alpha$ is the mid-infrared powerlaw slope, and $N_{\rm bb}$ and
$N_{\rm pl}$ are the normalisations of the greybody and powerlaw,
respectively ($N_{\rm pl}$ is a fixed function of $N_{\rm bb}$, $T$,
and $\alpha$).  Note however that in this work we do not quote our
best-fit temperatures since they are heavily dependent on the assumed
opacity and emissivity model.  For example, an SED peaking at
100\um\ can be described as 29\,K (blackbody), 31\,K (optically thin
greybody), 44\,K (greybody with $\tau=1$ at 100\um), or 46\,K
(greybody with $\tau=1$ at 200\um).  All of these models have been
used in the literature to derive dust temperature despite the fact
that they are not directly comparable
\citep[e.g.][]{blain03a,kovacs10a,casey09b,coppin08a,hwang10a}; see
\citet{casey12a}, Figure 2, for more details on the impact of model
assumption on measured dust temperature.  Instead of estimating
temperatures, we estimate the rest-frame SED peak wavelength of
$S_{\nu}$ called $\lambda_{\rm peak}$ ($\propto 1/T_{\rm dust}$),
which is much more easily constrained by the data.

Since the number of far-infrared photometric data points is limited to
2--5 (from {\it Spitzer}-MIPS, \scubaii, AzTEC and MAMBO
measurements), we only fit SEDs with two free parameters, $L_{\rm
  IR}$, and $\lambda_{\rm peak}$ and do not attempt to constrain
emissivity or mid-infrared powerlaw index\footnote{Although for
  sources with $\ge$4 photometric points at $>$3$\sigma$, we allow
  variation in $\alpha$; this removes some reliance on the 24\um\ flux
  to map directly to the FIR peak.}.  The emissivity index,
$\beta=1.5$ is fixed \citep[e.g.][]{pope08a,younger09a,chapin10a}, as
is the mid-infrared powerlaw index, $\alpha=2.0$
\citep[e.g.][]{blain03a,casey12a}.  A representative sample of
best-fit SEDs, including both high and low signal-to-noise sources, is
illustrated in Figure~\ref{fig:seds}.

Table~\ref{tab:physical} gives the basic derived properties from the
SED fits, including $\chi^2$ goodness of fit, $L_{\rm IR}$, the
implied SFR from $L_{\rm IR}$ to one significant
figure\footnote{Assuming the \citet{Kennicutt98a} $L_{\rm IR}$-SFR
  scaling and a Salpeter IMF.}, and the SED peak wavelength and dust
masses, $\lambda_{\rm peak}$ and $M_{\rm dust}$.  Note that all of the
flux densities used in the fits do factor into $\chi^2$, even
non-detections; instead of treating non-detections as upper limits, we
treat them as flux density measurements with very low significance
(given by the extracted flux density from the map at the extracted
position).  The uncertainties on $L_{\rm IR}$, $\lambda_{\rm peak}$
and $M_{\rm dust}$ are derived from the photometric uncertainties in
the FIR data and average $\sim$0.2-0.4\,dex, $\sim$0.2--0.4\,dex and
$\sim$0.4--0.5\,dex respectively.  These uncertainties \textit{do not}
account for uncertainty in the IMF or dust absorption coefficient
which impact the SFR and $M_{\rm dust}$ respectively.  Sources with
just two FIR photometric points (and upper limits at other bands) have
SEDs fixed to the mean dust $\lambda_{\rm peak}$ of their parent
sample, either 450\um-selected galaxies or 850\um-selected galaxies.
In the next section we measure these peak wavelengths as
$\log\lambda_{\rm peak}=2.05\pm0.04$ and $\log\lambda_{\rm
  peak}=2.12\pm0.03$, respectively.  We do not attempt to constrain
the SEDs of sources without photometric redshifts.

\begin{figure*}
  \centering
  \includegraphics[width=0.75\columnwidth]{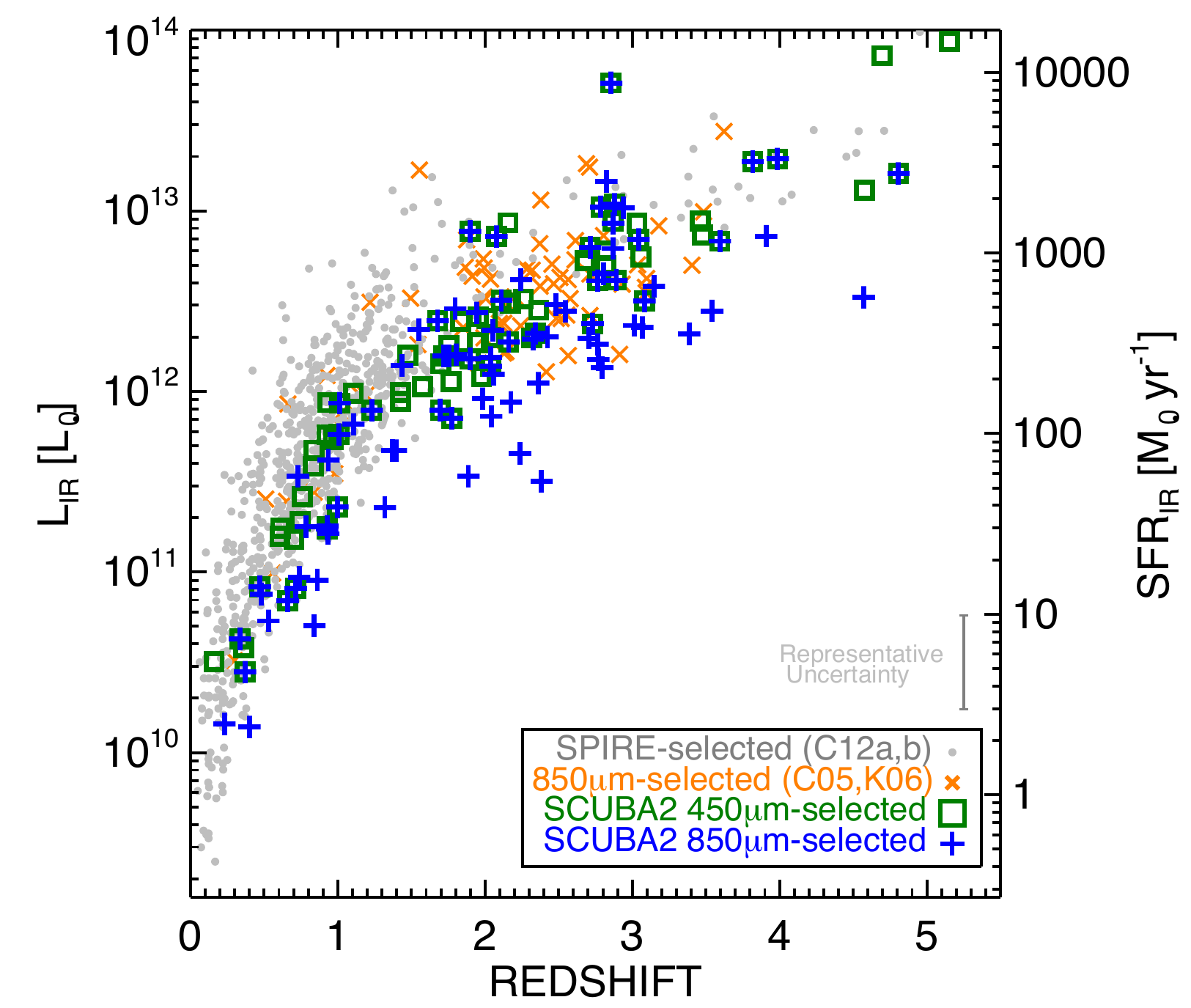}
  \includegraphics[width=0.59\columnwidth]{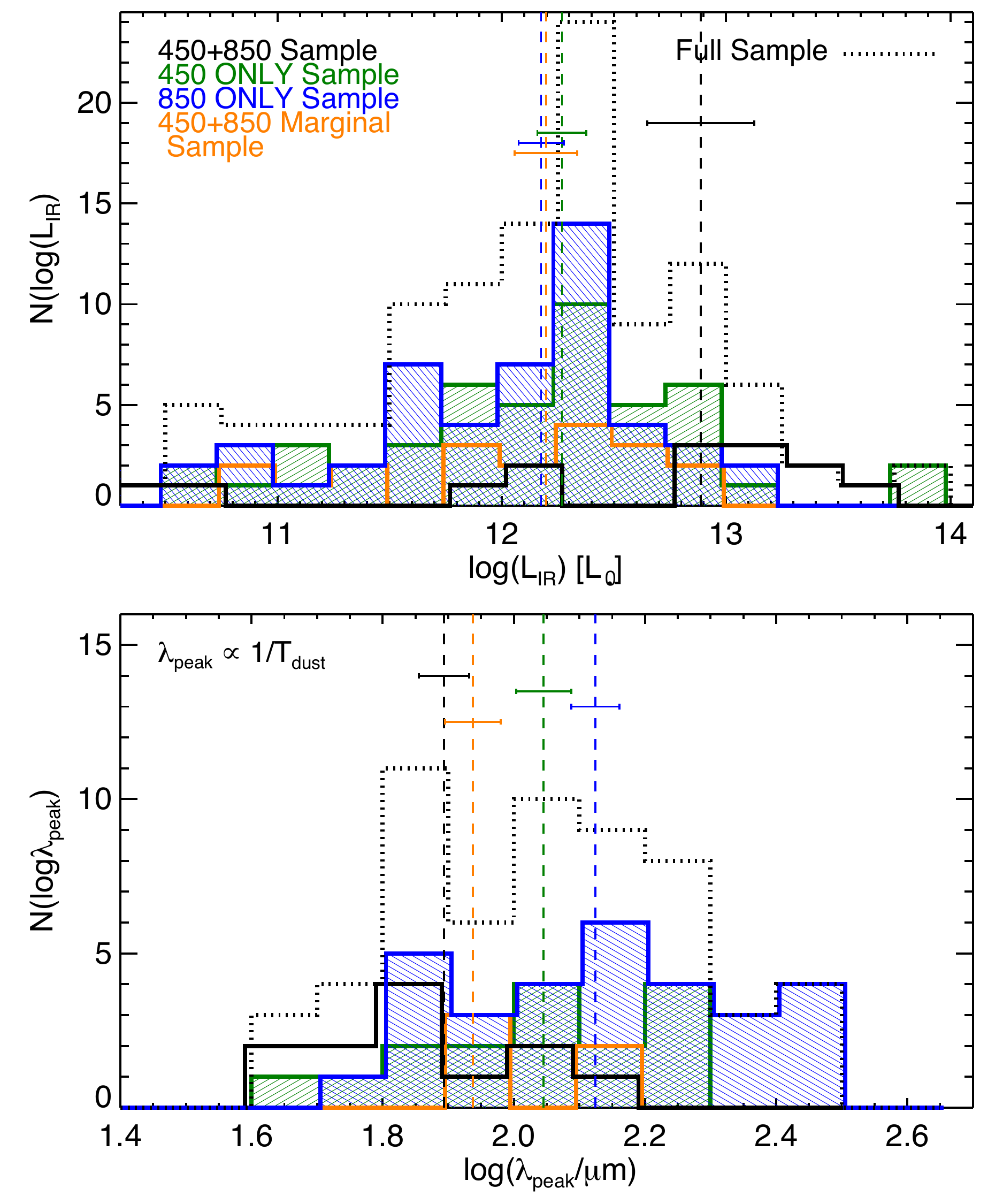}
  \includegraphics[width=0.65\columnwidth]{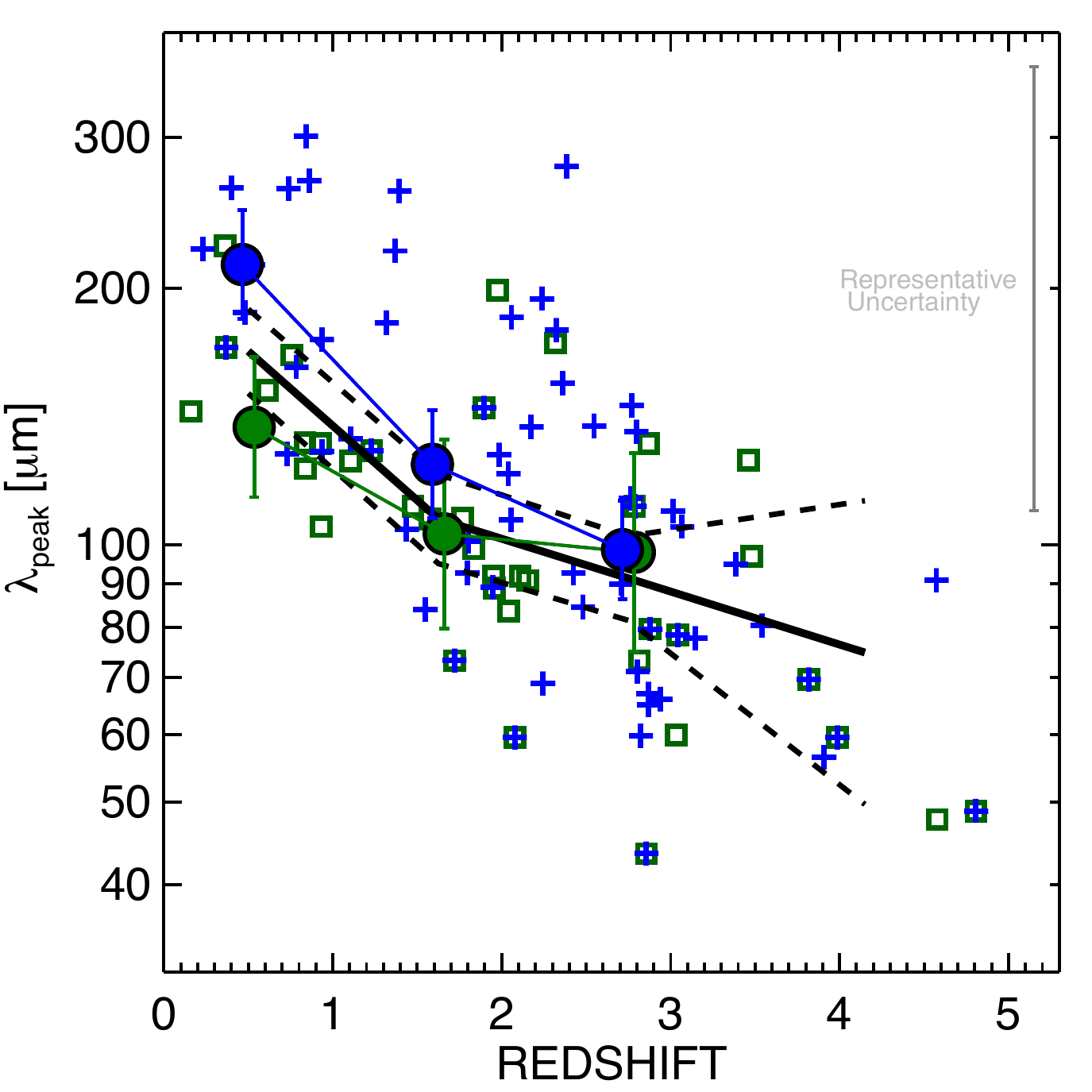}
  \caption{{\bf LEFT:} Infrared luminosity against photometric
    redshift for 450\um\ (green squares) and 850\um-selected (blue
    crosses) sources.  For context, the 850\um-selected SCUBA sources
    \citep[orange crosses;][]{chapman05a,kovacs06a} and {\it
      Herschel}-SPIRE-selected, spectroscopically confirmed sources
    \citep[gray circles;][]{casey12b,casey12c} are overplotted.  The
    mean redshift of the SCUBA2 sample is similar to the original
    SCUBA-selected sources and higher than SPIRE sources while the
    \scubaii\ luminosities, at a given redshift, probe to slightly lower
    luminosities ($\sim$0.1--0.4\,dex) than SPIRE sources.
{\bf MIDDLE:} Histograms of infrared luminosity and SED peak
wavelength (inversely proportional to dust temperature), for various
subsamples of \scubaii\ sources: the whole sample (dotted lines), the
overlapping $>$3.6$\sigma$ 450 and 850 sources (detected in both
bands, black), 450\um\ $>$3.6$\sigma$ sources not detected at
850\um\ (green), 850\um\ $>$3.6$\sigma$ sources not detected at
450\um\ (blue), and the marginal 3$<\sigma<$3.6 450 and 850 source
catalogue (orange).  Vertical lines mark the median value for each
subset and 1$\sigma$ statistical variation on the median determined by
bootstrapping.
{\bf RIGHT:} SED peak wavelength against redshift for sources which
have peak wavelength constraints.  The median SED peak wavelength for
all \scubaii\ galaxies as a function of redshift is shown in black
with dashed 67\%\ intervals and error bars representing uncertainty on
the median.  The median dust temperature evolution of 450\um-only
detected galaxies is shown as green circles, while 850\um-only
detected galaxies are in blue.  At low redshifts ($z\le2$) the two
populations have statistically distinct SED shapes, echoing the
results from Figure~\ref{fig:fircolor}.
}
  \label{fig:lirz}
\end{figure*}

\subsection{Bulk Infrared Properties of the Population}

Figure~\ref{fig:lirz} illustrates infrared luminosities against
photometric redshift for both 450\um- and 850\um-selected galaxies.
Both 450\um\ and 850\um\ samples by and large sit at $1<z<4$ with
$10^{12}<L_{\rm IR}<10^{13.2}$\,$\lsun$.  This is very similar to the
luminosity and redshift range probed by the original
\scuba\ 850\um\ surveys; the distribution of the \citet{chapman05a}
spectroscopically-confirmed sample \citep[with improved luminosity
  estimates using photometry from][]{kovacs06a} is shown for
comparison.  The parameter space probed by $\sim$1\,mm-selected
populations \citep[e.g.][]{yun12a} is similar.  We also compare to the
recent spectroscopically-confirmed sample of {\it Herschel}-{\sc
  spire} sources analysed in \citet{casey12b} and \citet{casey12c};
the {\sc spire} sources (selected at 250--500\um) peak in density at
lower redshifts and lower luminosities, although at a given redshift,
the \scubaii\ sources probe $\sim$0.1--0.5\,dex fainter sources than
SPIRE.

The distributions in infrared luminosity and peak SED wavelength (the
two free parameters of our SED fitting technique) are plotted in the
middle of Figure~\ref{fig:lirz} and can be used to assess basic
differences between subsets of the population.  What we find is that
the subset detected at both 450\um\ and 850\um\ are more
luminous than the marginal sample and those detected at only one
wavelength.  The median SED peak wavelength of the 850\um-only
sample
is longer than those detected at both wavelengths
(translating to cooler temperatures).  The subsets which are
450\um-only detected and marginal 3$<\sigma<$3.6 sources have median
SED peak wavelengths which lie between.  Uncertainties on the median
values in luminosity and SED peak wavelength are determined through
bootstrapping.

Figure~\ref{fig:lirz} also shows the change in SED peak wavelength
with redshift, where higher redshift sources tend to have hotter SEDs
peaking at shorter rest-frame wavelengths.  This increase in
temperature (decrease in SED peak wavelength) also correlates with
$L_{\rm IR}$, whereby the higher redshift sources are also the most
luminous and hotter, which is a well-documented correlation
\citep[e.g.][]{soifer87a,soifer89a,chapman03a,chapin09a,hwang10a}.
Figure~\ref{fig:lirz} shows us that the sources detected only at
450\um\ are statistically warmer (peak at shorter wavelengths) than
average and 850\um\ are statistically cooler (peak at longer
wavelengths), and that the difference in median peak wavelengths is
$\Delta\log(\lambda_{\rm peak})=0.16$\,dex, which is $\sim$50\um, at
$z$\simlt$1$ and 0.10\,dex at $1<z<2.5$, which is $\sim$20\um.

These results suggest that, by and large, \scubaii\ 450\um\ and
850\um\ galaxies are not significantly dissimilar in their redshifts,
luminosities, or SED peak wavelengths/temperatures, but that systems
detected in only one of the two bands can be primarily distinguished
based on their SED peak wavelengths (or temperatures) rather than
their luminosities or redshifts.

\subsection{Galaxies' distribution in $z$--L$_{\rm IR}$--$\lambda_{\rm peak}$ space}\label{sec:predictive}

To better understand the relative selection differences between
450\um\ and 850\um, we use Monte Carlo methods to test for
detectability as a function of the primary SED parameters, $z$,
$L_{\rm IR}$, and $\lambda_{\rm peak}$.  For a given redshift, luminosity
and SED peak wavelength, we generate a number of SEDs satisfying those
characteristics with varying emissivity, opacities, and mid-infrared
powerlaw slopes and then compute the relative probability that a
galaxy of those characteristics ($z$, $L_{\rm IR}$, and $\lambda_{\rm
  peak}$) is detectable in our survey.  Instead of setting a strict
flux density threshold which is representative of our sample
(e.g. $S_{\rm 450}>14.9$\,mJy or $S_{\rm 850}>2.9$\,mJy, at
$>$3.6$\sigma$), we inject galaxies of these given SED types into our
jackknife maps and measure the $>3.6\sigma$ detectability.

We then assume an underlying distribution or density of galaxies in
$z$--L$_{\rm IR}$--$\lambda_{\rm peak}$ space, as given by
integrated-infrared luminosity functions
\citep{le-floch05a,caputi07a,magnelli11a,casey12b,gruppioni13a}.
However, the luminosity functions only constrain the distribution of
galaxies in $z$ and L$_{\rm IR}$, therefore we must assume some
distribution of sources in SED peak wavelength, or temperature.  The
simplest assumption is that any galaxy SED, regardless of luminosity
or redshift, should peak at 100$\pm$40\um.  This model of the
population fails since it predicts that 450\um-detected galaxies peak
at $\langle z\rangle=0.8$ while 850\um-detected galaxies peak at
$\langle z\rangle=2.8$.  It also predicts that the populations have
indistinguishable SED shapes and that nearly all 450\um\ sources will
be 850\um\ detected.  Both the predicted redshift and SED peak
wavelength distributions and fraction of overlap between populations
disagree significantly with our observations.

Adjusting the distribution of sources in SED peak wavelength produces
results which are consistent with our observations.  In other words,
by invoking a correlation whereby L$_{\rm IR}\propto$T$_{\rm
  dust}\propto 1/\lambda_{\rm peak}$, we can predict $\langle
z\rangle=$1.5--2.0 for 450\um\ sources and $\langle z\rangle=$2.3-2.6
for 850\um\ sources.  Perhaps most pertinent, we can recover
approximately the same overlap fraction of the population, namely that
only $\sim$25--40\%\ of the 450\um\ galaxies are 850\um\ detected and
vice versa.  The predicted SED peak wavelength distributions also
differ between 450\um\ and 850\um\ with this model.  Unfortunately,
our data are not significant enough to place meaningful constraints on
the slope of the underlying correlation between luminosity and peak
wavelength, or provide better estimates to overlap fraction than
$\sim$25\%.  However, future large samples (N$\approx$5000) selected
at even more wavelengths across the submillimeter will provide
meaningful constraints on the distribution of galaxies in $z$--L$_{\rm
  IR}$--$\lambda_{\rm peak}$ space.

These simulations can also be used to test the reliability of our SED
fits.  For instance, does SED fitting at 450\um\ and 850\um\ (vs. a
more fully-sampled range of wavelengths) have a systematic effect on
measured SED peak wavelength?  Having generated sets of SEDs which
describe a galaxy of known $z$, $L_{\rm IR}$ and $\lambda_{\rm peak}$,
we can re-measure the SED characteristics by refitting the noise-added
450\um\ and 850\um\ photometry (and using an additional constraint at
observed 24\um).  At high signal-to-noise, where galaxies are luminous
enough to be detected at both wavelengths, the scatter in the $\log$
of measured SED peak wavelength is 0.05\,dex with no systematic
offset.  At lower luminosities, where galaxies are more likely to be
detected in only one of the two bands, the scatter in measured SED
peak wavelength increases to 0.11\,dex although there is still no
systematic offset.  We attribute the lack of systematic offset in
measured peak wavelengths to the \scubaii\ selection wavelengths,
which sample the SED in significantly different regimes, e.g. the
SED's peak and the SED's Rayleigh-Jeans tail.  Even galaxies which are
not formally detected in one of the two bands still has a flux density
constraint at both wavelengths which, on average, provides an accurate
estimate to the SED peak wavelength.

\section{Discussion}\label{sec:discussion}

The ultimate goal of this work is the characterisation of galaxies
which emit at 450\um\ and 850\um, how they relate to other
similarly-selected galaxy populations, how they relate to the more
extensively-studied optical/UV-selected populations, and how important
they are in the context of total star formation in the Universe.
Follow-up studies can pursue source characterisation in more detail.
These initial observations from \scubaii\ are also incredibly valuable
in providing first insight of high-resolution bolometer observations
which will become more commonplace with completion of the Large
Millimeter Telescope (LMT) and the Cornell-Caltech Atacama Telescope
(CCAT).

\subsection{Are 450\um\ and 850\um\ populations different?}

When contrasting the physical characteristics of 450\um\ with
850\um\ galaxies, like redshift distribution, luminosity, and dust
temperature, this work suggests that the two populations are quite
similar.  
Are these similarities between 450\um\ and 850\um\ galaxies
what we expect, given the lack of direct overlap of the samples?

If you na\"{i}vely assume there is little evolution or luminosity
dependence of SMG SEDs, then selecting at shorter wavelengths than
850\um\ would select a lower redshift population.  If you take the
mean redshift for 850\um-selected SMGs as $\langle z\rangle=2.2$ and
assume they peak at rest-frame 100\um, then no change in SED type
would predict the mean redshift for 450\um-selected galaxies is
$\langle z\rangle\approx$0.5.  This assumption fails primarily since
it does not consider that 850\um\ selects galaxies almost exclusively
on the Raleigh-Jeans tail of dust emission versus the peak.  Adjusting
this to reflect that 450\um\ will select more galaxies near their peak
(a factor of $\sim$2 in wavelength) implies that 450\um\ galaxies
should peak near $\langle z\rangle\approx1$ (this test is similar to that in section~\ref{sec:predictive}).  Still, this is
inconsistent with our data which predicts that both populations peak
at around $z\sim2$.

\citet{geach13a} and \citet{roseboom13a} find a 
450\um\ redshift distribution which averages $\langle z\rangle=1.3$,
which is statistically different than our finding of $\langle
z\rangle=2.0\pm0.2$.  Note however that the median distribution of the
Geach \etal\ and Roseboom \etal\ sample is $z\sim1.6$, significantly
different than the average, and closer to the median value we measure.
The offset between median redshifts between the two samples is likely
due to the difference in depths of coverage between our surveys; the
Geach \etal\ and Roseboom \etal\ work is a factor of
$\approx$3$\times$ deeper at its centre, while ours is a factor of
$\approx$4$\times$ larger, so our sample naturally picks up more
luminous sources that might sit at higher redshifts than the fainter,
lower redshift 450\um-population.

Despite the overlap in redshift distribution with 850\um\ sources,
62--76\%\ of 450\um\ sources are not 850\um\ detected.  Similarly,
61--81\%\ of 850\um\ sources are not 450\um-detected. If the
difference is not the sources' redshifts, then the SED properties
could be the cause of the lack of overlap between the galaxy
populations.  Dust temperature has been thought to cause a significant
selection difference between 850\um--1\,mm and $<$500\um\ populations
\citep{blain04a,chapman04a,casey09a}, where the former population is
biased towards colder temperature SEDs.  Before {\it Herschel} and
\scubaii, this population of warm-dust high-$z$ ULIRGs was called
Optically Faint Radio Galaxies (OFRGs) or Submm-faint, Star-Forming
Radio Galaxies (SFRGs), and was shown to exist in \citet{casey09a}.
In this paper, our finding that 850\um-only sources are statistically
cooler (at a fixed redshift or luminosity) than 450\um-only sources is
consistent with this hypothesis and is the most prominent difference
between 450\um-only and 850\um-only populations, even though the
average SED peak wavelength offset is only $\sim$20\um\ at $z\sim2$
(information presented in Figure~\ref{fig:lirz}).

Certainly the lack of one-to-one overlap in the 450\um\ and
850\um-samples implies that systems selected at any one wavelength in
the FIR are {\it not} representative of {\it all} ultraluminous
infrared star formation at high redshift.  This becomes an issue when
studies try to compare the relative importance of normal galaxies to
infrared galaxies in the context of universal star formation and
completeness for interpretation of the star formation rate density.

\subsection{Relation to `Normal' Galaxies}\label{sec:normal}

Recent work has proposed that most submillimeter or infrared-luminous
systems are a different, more extreme class of galaxy than most
`normal' star forming galaxies in the Universe
\citep[e.g.][]{daddi09a,genzel10a,tacconi10a}. This
distinction is likely valid when speaking of the most luminous subset
of SMGs. SMGs at $\approx$10$^{13}\lsun$ are nearly all found to be
major merger-driven starbursts, much more extreme and much more rare than
normal, more modest star forming galaxies at comparable redshifts
\citep[e.g.][]{engel10a}.

Although many SMGs exhibit extreme properties not seen in typical star
forming systems, this is not true of the entire infrared-luminous
population.  Here we measure a mean infrared luminosity of
\scubaii\ galaxies as $\approx$1--2$\times$10$^{12}\lsun$, and the
average star formation rates in our sample are $\approx$200$\sfr$,
which is typical of high mass ($>$10$^{10}\msun$) `normal' galaxies at
$z\approx2$.  Most normal galaxies at this epoch have specific star
formation rates (sSFR) of 1--10\,Gyr$^{-1}$ while \scubaii\ galaxies
span sSFRs from 1--100\,Gyr$^{-1}$, averaging 10\,Gyr$^{-1}$ at the
upper end of sSFRs for normal galaxies.  This is consistent with the
idea that SMGs, although rare and extreme systems, can exhibit
both properties of `normal' star forming galaxies and extreme
starbursts during different stages of evolution \citep[as suggested in
  the models of][]{hayward13a}.

Comparing the star formation rates from the UV/optical/near-infrared
SED fits (dubbed ``SFR$_{\rm UV}$'') directly with the star formation
rates in the far-infrared, there is a substantial discrepancy.
Despite the fact that SFR$_{\rm UV}$ is supposedly corrected for the
effects of dust extinction, a cap on the maximum extinction
introduced in the {\sc Le Phare} SED fitting procedure implies that
very dust obscured galaxies will have much higher SFR$_{\rm IR}$ than
SFR$_{\rm UV}$.  By our measure, SFR$_{\rm UV}$ is underestimating the
star formation rate by a factor of $\simgt$13 (where the total SFR can
be approximated as SFR$_{\rm IR}$ in infrared-luminous galaxies).

Table~\ref{tab:physical} gives the {\sc Le Phare}-estimated quantities
for star formation rate and stellar mass, as derived from the UV
through to the near-infrared.
These \scubaii\ sources are known to be unusual in that the {\it are}
directly detected in the infrared, so the discrepancy between star
formation rate measurements is not unexpected or new
\citep{rosa-gonzalez02a,dye08a,yun12a}.  Nevertheless, the discrepancy
should emphasise the need to treat extinction carefully in model fits,
especially for infrared-luminous galaxies. This potentially will alter
the interpretation of galaxies' specific star formation rates.  A
future work, which will summarise our progress in confirming redshifts
spectroscopically, will explore the total energy output of these
galaxies in more depth and provide more context for comparing the
sample to `normal' galaxies.

\subsection{Contribution to Universal Star Formation}

\begin{figure}
\centering
\includegraphics[width=0.99\columnwidth]{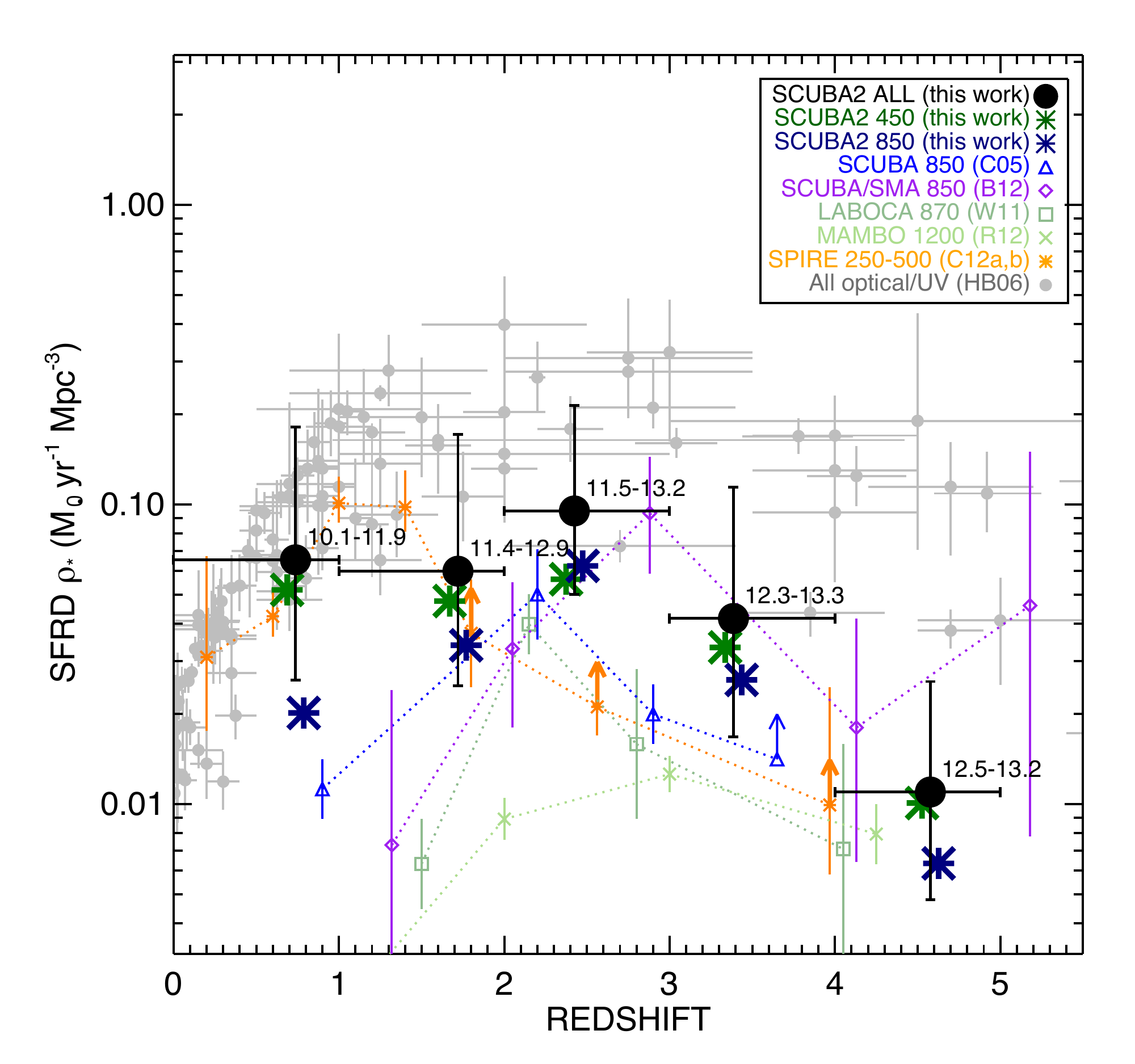}
\caption{The estimated contribution of \scubaii\ sources in this paper
  to the cosmic star formation rate density (SFRD).  We split up the
  contribution estimates by selection wavelength, showing both the
  contribution from 450\um-selected (dark green) and 850\um-selected
  sources (navy), and we compare the total (black points) to
  literature SFRD values for similarly selected populations.  Note
  that the total SFRD estimate is incomplete and only represents
  sources characterised at $>$3.6$\sigma$ in this paper and does not
  extrapolate to fainter luminosities; thus they should be regarded as
  lower limits.  The range of luminosities represented by each black
  point is given adjacent to the point, in logarithmic solar
  luminosities.  Galaxies with $\log L_{\rm IR}>13.6$ are excluded
  from this estimate due to potential lensing bias.  Literature
  comparisons are \citet{chapman05a} 850\um-selected SCUBA sources
  (blue), \citet{barger12a} SCUBA sources (purple), \citet{wardlow11a}
  870\um-selected LABOCA sources (light sea green), the
  \citet{roseboom12a} 1.2mm-selected MAMBO sources (light green), and
  the \citet{casey12b,casey12c} {\it Herschel}-{\sc spire}
  250--500\um-selected sources (orange).  }
\label{fig:sfrd}
\end{figure}

Here we measure the net contribution of \scubaii\ galaxies to the star
formation rate density (SFRD) of the Universe, or in other words, how
significantly they impact net star formation in the Universe at any
given epoch.

We compute SFRD estimates for the \scubaii\ population in this paper
using the 1/$V_{\rm max}$ method.  Within a redshift bin, we sum each
galaxy's star formation rate divided by accessible volume.  Our survey
area is roughly uniform over 394\,arcmin$^2$, although individual
sources' accessible volume is computed as a function of the local map
noise.
Note that we avoid estimating the luminosity functions directly given
the limited statistics of this sample, where \simgt500 galaxies would
be necessary to place accurate constraints in $L_{\rm IR}$ and $z$.
The maximum accessible volume for a source is computed given that
sources' selection wavelength, flux density, and peak wavelength
(either measured explicitly or assumed to be $\log\lambda_{\rm
  peak}=2.05$ [450\um] or $2.12$ [850\um] if unconstrained).  We split
the sample into five redshift bins spanning the range of our data:
$0<z<1$, $1<z<2$, $2<z<3$, $3<z<4$, and $4<z<5$.  We exclude the three
individual sources which have luminosities $\log L_{\rm IR}>13.6$, as
they might be contaminated by AGN or could be lensed.
Figure~\ref{fig:sfrd} shows the results.  The black
points$-$representing the total contribution from 450\um\ and
850\um\ sources$-$should be regarded as lower limits as they only
constitute narrow dynamic range in luminosity where our survey is
sensitive ($>3.6\sigma$).

Figure~\ref{fig:sfrd} shows us that (a) work done to-date on measuring
the SFRD from infrared/submm samples has been incomplete and is highly
dependent on survey depth, but (b) with more multi-wavelength
submillimeter surveys$-$like the \scubaii\ survey presented
herein$-$we are beginning to see more of the infrared contribution to
cosmic star formation which we had not previously seen.  This is
particularly evident at $z\sim1-2$, where 450\um-selected galaxies are
dominant, while at $z\sim2-3$ 450\um\ and 850\um-selected galaxies
contribute equally, and 850\um-selected galaxies might have a more
prominent contribution to the SFRD at $z\sim3$.  This survey is
conducted at similar depths to previous submillimeter surveys, however
it probes more of cosmic infrared-based star formation due to the
multiwavelength submillimeter approach.  Although deeper, wider
surveys are necessary for complete surveys, probing a larger dynamic
range in luminosity, there is no doubt that the multi-wavelength
selection approach is also necessary, as these \scubaii\ samples
represent a more complete subset of ultraluminous activity at these
epochs than single-wavelength selected populations at 850\um--1.2\,mm.

The contribution from infrared-luminous galaxies appears 1--10 times
less significant than the SFRD contribution from more normal, optical
and UV selected galaxies \citep*{hopkins06a}.  However, here the
estimates from Hopkins \&\ Beacom are corrected for dust extinction,
in that they are scaled up by a factor proportional to the rest-frame
UV slope.  As addressed earlier by the discrepancy between optical
SED-derived SFR and infrared SFR, there might be cause for concern if
the dust extinction is not properly handled for some of the
potentially dustier optically-selected galaxies.

\section{Conclusions}

With simultaneous mapping at 450\um\ and 850\um, the
\scubaii\ instrument is making substantial headway in the detection and
characterisation of infrared emission in the distant Universe.  The
$\sim$7\arcsec\ high-resolution blank-field mapping at 450\um\ opens
up a unique parameter space in high-$z$ submillimeter science, and the
efficient mapping at 850\um\ means that the confusion limit is reached
in far less time than previous observations.  This paper has presented
initial \scubaii\ mapping of a uniformly-covered 394\,arcmin$^2$ area
in the COSMOS field at both 450\um\ and 850\um.  We reach the
following conclusions about our maps and the identification of point sources:
\begin{itemize}
  \item The root-mean-square (RMS) noise values of our maps are
    $\sigma_{\rm 450}=$4.13\,mJy and $\sigma_{\rm 850}=$0.80\,mJy at
    our map centres and find largely uniform RMS over a
    22.4\,arcmin\ diameter (394\,arcmin$^2$), the threshold where the
    RMS exceeds two times the central RMS value.
  \item We extract point sources down to a 3.6$\sigma$ detection
    threshold at 450\um\ and 850\um, where we find 78 and 99 sources
    respectively.  These extraction limits are determined via Monte
    Carlo simulations and an expected sample contamination rates of
    3--5\%.
  \item Source number counts at 450\um\ and 850\um\ are measured and
    we compare them to previous results at both wavelengths.  We
    provide best-fit double-power law and Schechter function fits.
  \item Positional uncertainties of 1--2.5\,\arcsec\ at 450\um\ and
    2--6\,\arcsec\ at 850\um\ are estimated using Monte Carlo tests;
    we use this uncertainty to match both 450\um\ and 850\um\ sources
    to multiwavelength counterparts.  The low positional uncertainties
    at 450\um\ allow direct matching to counterparts in the
    optical/near-infrared, without relying on detection at 24\um\ or
    radio wavelengths.  Our 850\um\ counterpart matching is done first
    by identifying 24\um\ and radio counterparts, then if absent,
    taken as the nearest neighbour optical source.  
  \item 56\%\ of all 450\um\ galaxies and 40\%\ of all
    850\um\ galaxies lack both 24\um\ and radio counterparts down to
    the deep field detection limits ($S_{\rm 24}\approx80\,$\uJy\ and
    $S_{\rm 1.4}\approx\,$50\,\uJy).  This suggests that submillimeter
    samples relying on identification at either wavelength could be
    significantly biased, producing up to $\sim$1/2
    mis-identifications.
\end{itemize}

After analysis of field number counts and counterpart matching
techniques, we also analyse the population characteristics of
450\um\ and 850\um\ galaxies, how they differ from one another and how
they relate to other high-redshift galaxy populations.  We reach the
following conclusions on their physical attributes:
\begin{itemize}
  \item The redshift distributions for both populations is measured
    using the extensive COSMOS ancillary optical and near-infrared
    data.  The 450\um\ population peaks at $\langle z\rangle
    =1.95\pm0.19$ while the 850\um\ population peaks at $\langle
    z\rangle =2.16\pm0.11$.
  \item The far-infrared colour of \scubaii\ sources (measured as
    $\log({\rm S}_{\rm 850}/{\rm S}_{\rm 450})$) is found not to
    evolve significantly with redshift, although 450\um\ identified
    galaxies have statistically `bluer' far-infrared SEDs than
    850\um\ identified galaxies.  Both 450\um\ and 850\um\ populations
    have colours which are inconsistent with a fixed SED peak
    wavelength (i.e. fixed temperature) across $0<z<5$.
  \item 850\um-detected sources which are 450\um-dropouts peak at
    $z\approx2$ and do not sit at higher redshifts as would be
    suspected if SED shape were not evolving or changing with infrared
    luminosity.  Similarly 450\um-detected sources which are
    850\um-dropouts also sit at $z\approx2$ and not at lower
    redshifts.  This is consistent with model populations where there
    is correlation between luminosity and dust temperatures, although
    whether or not that correlation evolves with $z$ is unconstrained
    by this data.
  \item Infrared luminosities, SED peak wavelengths and dust masses
    are estimated for both populations, and we determine that
    450\um\ galaxies and 850\um\ galaxies are equally luminous and
    peak at similar rest-frame wavelengths.  The most significant
    distinction is the SED peak wavelength of 450\um-only or
    850\um-only detected galaxies, which differs by $\sim$20--50\um\ (or
    8--12\,K, with 450\um-sources being the warmer subset).
  \item The star formation rates measured directly in the far-infrared
    exceed those predicted from UV/optical/near-infrared photometry by
    $\sim$13 times if restrictions are placed on extinction when
    generating best-fit stellar population fits.
  \item The contribution of these \scubaii\ sources to the cosmic
    infrared luminosity density, or the star formation rate density,
    is measured to be higher than previous submm surveys, not due to
    depth, but due to multi-wavelength selection.  The 450\um\ and
    850\um\ samples complement one another in that they only overlap
    by $\sim$30\%, so together they constitute a more complete census
    of luminous infrared activity than either subset.  The total SFRD
    contribution we measure here is a factor of 2--3\,times higher
    than previous single-wavelength $\sim$850\um\ selection at the
    same depth.  This demonstrates that multiple selection wavelengths
    are necessary for a complete census of infrared luminous star
    formation.
\end{itemize}
\scubaii\ high-resolution 450\um\ mapping has allowed the first
detailed look at an infrared-luminous population not surveyed before.
This work has revealed the necessity to exercise caution when
identifying multiwavelength counterparts of submillimeter sources
(especially those with a large beamsize) and the biases and
limitations of previously analysed 850\um--1\,mm samples.  Direct
far-infrared interferometry still proves the best method for
unequivocally identifying multiwavelength counterparts, however future
work from CCAT will enable large field-of-view mapping with resolution
significantly improved over JCMT, both at 850\um\ and 450\um.  This
work provides a stepping stone to link infrared-luminous systems to
the more ubiquitous, `normal' star forming galaxies across a wide
range of epochs, shedding light on galaxy evolution and the nature of
cosmic star formation.

\section*{Acknowledgements}

We would like to thank the referee for a very thoughtful and helpful
report in reviewing this manuscript.  CMC is generously supported by a
Hubble Fellowship from Space Telescope Science Institute, grant
HST-HF-51268.01-A. CCC and LC are generously supported by NSF grant
AST 0709356.  AB thanks the University of Wisconsin Research Committee
with funds granted by the Wisconsin Alumni Research Foundation, and
the David and Lucile Packard Foundation. CMC and DBS would like to
thank the Aspen Center for Physics and the NSF grant 1066293 for many
fruitful conversations with the community regarding this work during
``The Obscured Universe'' summer workshop.
The James Clerk Maxwell Telescope is operated by the Joint Astronomy
Centre on behalf of the Science and Technology Facilities Council of
the United Kingdom, the National Research Council of Canada, and
(until 31 March 2013) the Netherlands Organisation for Scientific
Research.  Additional funds for the construction of \scubaii\ were
provided by the Canada Foundation for Innovation.  The authors also
wish to recognise and acknowledge the very significant cultural role
and reverence that the summit of Mauna Kea has always had within the
indigenous Hawaiian community.  We are most fortunate to have the
opportunity to conduct observations from this mountain.

\bibliography{caitlin-bibdesk}

\begin{thebibliography}{118}
\expandafter\ifx\csname natexlab\endcsname\relax\def\natexlab#1{#1}\fi

\bibitem[{{Alaghband-Zadeh} {et~al}\mbox{.}(2012){Alaghband-Zadeh}, {Chapman},
  {Swinbank}, {Smail}, {Harrison}, {Alexander}, {Casey}, {Dav{\'e}},
  {Narayanan}, {Tamura}, \& {Umehata}}]{alaghband-zadeh12a}
{Alaghband-Zadeh} S. {et~al.}, 2012, \mnras, 3329

\bibitem[{{Alexander} {et~al}\mbox{.}(2005){Alexander}, {Bauer}, {Chapman},
  {Smail}, {Blain}, {Brandt}, \& {Ivison}}]{alexander05a}
{Alexander} D.~M., {Bauer} F.~E., {Chapman} S.~C., {Smail} I., {Blain} A.~W.,
  {Brandt} W.~N., {Ivison} R.~J., 2005, \apj, 632, 736

\bibitem[{{Aretxaga} {et~al}\mbox{.}(2011){Aretxaga}, {Wilson}, {Aguilar},
  {Alberts}, {Scott}, {Scoville}, {Yun}, {Austermann}, {Downes}, {Ezawa},
  {Hatsukade}, {Hughes}, {Kawabe}, {Kohno}, {Oshima}, {Perera}, {Tamura}, \&
  {Zeballos}}]{aretxaga11a}
{Aretxaga} I. {et~al.}, 2011, \mnras, 415, 3831

\bibitem[{{Armus} {et~al}\mbox{.}(2009){Armus}, {Mazzarella}, {Evans},
  {Surace}, {Sanders}, {Iwasawa}, {Frayer}, {Howell}, {Chan}, {Petric},
  {Vavilkin}, {Kim}, {Haan}, {Inami}, {Murphy}, {Appleton}, {Barnes}, {Bothun},
  {Bridge}, {Charmandaris}, {Jensen}, {Kewley}, {Lord}, {Madore}, {Marshall},
  {Melbourne}, {Rich}, {Satyapal}, {Schulz}, {Spoon}, {Sturm}, {U}, {Veilleux},
  \& {Xu}}]{armus09a}
{Armus} L. {et~al.}, 2009, \pasp, 121, 559

\bibitem[{{Banerji} {et~al}\mbox{.}(2011){Banerji}, {Chapman}, {Smail},
  {Alaghband-Zadeh}, {Swinbank}, {Dunlop}, {Ivison}, \& {Blain}}]{banerji11a}
{Banerji} M., {Chapman} S.~C., {Smail} I., {Alaghband-Zadeh} S., {Swinbank}
  A.~M., {Dunlop} J.~S., {Ivison} R.~J., {Blain} A.~W., 2011, \mnras, 418, 1071

\bibitem[{{Barger} {et~al}\mbox{.}(2000){Barger}, {Cowie}, \&
  {Richards}}]{barger00a}
{Barger} A.~J., {Cowie} L.~L., {Richards} E.~A., 2000, \aj, 119, 2092

\bibitem[{{Barger} {et~al}\mbox{.}(1998){Barger}, {Cowie}, {Sanders}, {Fulton},
  {Taniguchi}, {Sato}, {Kawara}, \& {Okuda}}]{barger98a}
{Barger} A.~J., {Cowie} L.~L., {Sanders} D.~B., {Fulton} E., {Taniguchi} Y.,
  {Sato} Y., {Kawara} K., {Okuda} H., 1998, \nat, 394, 248

\bibitem[{{Barger} {et~al}\mbox{.}(1999){Barger}, {Cowie}, {Smail}, {Ivison},
  {Blain}, \& {Kneib}}]{barger99a}
{Barger} A.~J., {Cowie} L.~L., {Smail} I., {Ivison} R.~J., {Blain} A.~W.,
  {Kneib} J., 1999, \aj, 117, 2656

\bibitem[{{Barger} {et~al}\mbox{.}(2012){Barger}, {Wang}, {Cowie}, {Owen},
  {Chen}, \& {Williams}}]{barger12a}
{Barger} A.~J., {Wang} W.-H., {Cowie} L.~L., {Owen} F.~N., {Chen} C.-C.,
  {Williams} J.~P., 2012, \apj, 761, 89

\bibitem[{{Barnard} {et~al}\mbox{.}(2004){Barnard}, {Vielva}, {Pierce-Price},
  {Blain}, {Barreiro}, {Richer}, \& {Qualtrough}}]{barnard04a}
{Barnard} V.~E., {Vielva} P., {Pierce-Price} D.~P.~I., {Blain} A.~W.,
  {Barreiro} R.~B., {Richer} J.~S., {Qualtrough} C., 2004, \mnras, 352, 961

\bibitem[{{Bertin} \& {Arnouts}(1996)}]{bertin96a}
{Bertin} E., {Arnouts} S., 1996, \aaps, 117, 393

\bibitem[{{Bertoldi} {et~al}\mbox{.}(2007){Bertoldi}, {Carilli}, {Aravena},
  {Schinnerer}, {Voss}, {Smolcic}, {Jahnke}, {Scoville}, {Blain}, {Menten},
  {Lutz}, {Brusa}, {Taniguchi}, {Capak}, {Mobasher}, {Lilly}, {Thompson},
  {Aussel}, {Kreysa}, {Hasinger}, {Aguirre}, {Schlaerth}, \&
  {Koekemoer}}]{bertoldi07a}
{Bertoldi} F. {et~al.}, 2007, \apjs, 172, 132

\bibitem[{{B{\'e}thermin} {et~al}\mbox{.}(2012){B{\'e}thermin}
  {et~al.}}]{bethermin12a}
{B{\'e}thermin} M., {et~al.}, 2012, A\&A, 542

\bibitem[{{Blain} {et~al}\mbox{.}(2003){Blain}, {Barnard}, \&
  {Chapman}}]{blain03a}
{Blain} A.~W., {Barnard} V.~E., {Chapman} S.~C., 2003, \mnras, 338, 733

\bibitem[{{Blain} {et~al}\mbox{.}(2004){Blain}, {Chapman}, {Smail}, \&
  {Ivison}}]{blain04a}
{Blain} A.~W., {Chapman} S.~C., {Smail} I., {Ivison} R., 2004, \apj, 611, 725

\bibitem[{{Borys} {et~al}\mbox{.}(2003){Borys}, {Chapman}, {Halpern}, \&
  {Scott}}]{borys03a}
{Borys} C., {Chapman} S., {Halpern} M., {Scott} D., 2003, \mnras, 344, 385

\bibitem[{{Bothwell} {et~al}\mbox{.}(2010){Bothwell}, {Chapman}, {Tacconi},
  {Smail}, {Ivison}, {Casey}, {Bertoldi}, {Beswick}, {Biggs}, {Blain}, {Cox},
  {Genzel}, {Greve}, {Kennicutt}, {Muxlow}, {Neri}, \& {Omont}}]{bothwell10a}
{Bothwell} M.~S. {et~al.}, 2010, \mnras, 405, 219

\bibitem[{{Bruzual} \& {Charlot}(2003)}]{bruzual03a}
{Bruzual} G., {Charlot} S., 2003, \mnras, 344, 1000

\bibitem[{{Caputi} {et~al}\mbox{.}(2007){Caputi}, {Lagache}, {Yan}, {Dole},
  {Bavouzet}, {Le Floc'h}, {Choi}, {Helou}, \& {Reddy}}]{caputi07a}
{Caputi} K.~I. {et~al.}, 2007, \apj, 660, 97

\bibitem[{{Casey}(2012)}]{casey12a}
{Casey} C.~M., 2012, \mnras, 425, 3094

\bibitem[{{Casey} {et~al}\mbox{.}(2012{\natexlab{a}}){Casey}, {Berta},
  {B{\'e}thermin}, {Bock}, {Bridge}, {Budynkiewicz}, {Burgarella}, {Chapin},
  {Chapman}, {Clements}, {Conley}, {Conselice}, {Cooray}, {Farrah},
  {Hatziminaoglou}, {Ivison}, {le Floc'h}, {Lutz}, {Magdis}, {Magnelli},
  {Oliver}, {Page}, {Pozzi}, {Rigopoulou}, {Riguccini}, {Roseboom}, {Sanders},
  {Scott}, {Seymour}, {Valtchanov}, {Vieira}, {Viero}, \& {Wardlow}}]{casey12b}
{Casey} C.~M. {et~al.}, 2012{\natexlab{a}}, \apj, 761, 140

\bibitem[{{Casey} {et~al}\mbox{.}(2012{\natexlab{b}}){Casey}, {Berta},
  {B{\'e}thermin}, {Bock}, {Bridge}, {Burgarella}, {Chapin}, {Chapman},
  {Clements}, {Conley}, {Conselice}, {Cooray}, {Farrah}, {Hatziminaoglou},
  {Ivison}, {le Floc'h}, {Lutz}, {Magdis}, {Magnelli}, {Oliver}, {Page},
  {Pozzi}, {Rigopoulou}, {Riguccini}, {Roseboom}, {Sanders}, {Scott},
  {Seymour}, {Valtchanov}, {Vieira}, {Viero}, \& {Wardlow}}]{casey12c}
{Casey} C.~M. {et~al.}, 2012{\natexlab{b}}, \apj, 761, 139

\bibitem[{{Casey} {et~al}\mbox{.}(2009{\natexlab{a}}){Casey}, {Chapman},
  {Beswick}, {Biggs}, {Blain}, {Hainline}, {Ivison}, {Muxlow}, \&
  {Smail}}]{casey09b}
{Casey} C.~M. {et~al.}, 2009{\natexlab{a}}, \mnras, 399, 121

\bibitem[{{Casey} {et~al}\mbox{.}(2009{\natexlab{b}}){Casey}, {Chapman},
  {Muxlow}, {Beswick}, {Alexander}, \& {Conselice}}]{casey09a}
{Casey} C.~M., {Chapman} S.~C., {Muxlow} T.~W.~B., {Beswick} R.~J., {Alexander}
  D.~M., {Conselice} C.~J., 2009{\natexlab{b}}, \mnras, 395, 1249

\bibitem[{{Chapin} {et~al}\mbox{.}(2010){Chapin}, {Chapman}, {Coppin},
  {Devlin}, {Dunlop}, {Greve}, {Halpern}, {Hasselfield}, {Hughes}, {Ivison},
  {Marsden}, {Moncelsi}, {Netterfield}, {Pascale}, {Scott}, {Smail}, {Viero},
  {Walter}, {Weiss}, \& {van der Werf}}]{chapin10a}
{Chapin} E.~L. {et~al.}, 2010, \mnras, 1682

\bibitem[{{Chapin} {et~al}\mbox{.}(2009){Chapin}, {Hughes}, \&
  {Aretxaga}}]{chapin09a}
{Chapin} E.~L., {Hughes} D.~H., {Aretxaga} I., 2009, \mnras, 393, 653

\bibitem[{{Chapin} {et~al}\mbox{.}(2013){Chapin} {et~al.}}]{chapin13a}
{Chapin} E.~L., {et~al.}, 2013, MNRAS, 430, 2545

\bibitem[{{Chapman} {et~al}\mbox{.}(2005){Chapman}, {Blain}, {Smail}, \&
  {Ivison}}]{chapman05a}
{Chapman} S.~C., {Blain} A.~W., {Smail} I., {Ivison} R.~J., 2005, \apj, 622,
  772

\bibitem[{{Chapman} {et~al}\mbox{.}(2003){Chapman}, {Helou}, {Lewis}, \&
  {Dale}}]{chapman03a}
{Chapman} S.~C., {Helou} G., {Lewis} G.~F., {Dale} D.~A., 2003, \apj, 588, 186

\bibitem[{{Chapman} {et~al}\mbox{.}(2004){Chapman}, {Smail}, {Blain}, \&
  {Ivison}}]{chapman04a}
{Chapman} S.~C., {Smail} I., {Blain} A.~W., {Ivison} R.~J., 2004, \apj, 614,
  671

\bibitem[{{Chen} {et~al}\mbox{.}(2013){Chen}, {Cowie}, {Barger}, {Casey},
  {Lee}, {Sanders}, {Wang}, \& {Williams}}]{chen13a}
{Chen} C.-C., {Cowie} L.~L., {Barger} A.~J., {Casey} C.~M., {Lee} N., {Sanders}
  D.~B., {Wang} W.-H., {Williams} J.~P., 2013, \apj, 762, 81

\bibitem[{{Clements} {et~al}\mbox{.}(2010){Clements}, {Rigby}, {Maddox},
  {Dunne}, {Mortier}, {Pearson}, {Amblard}, {Auld}, {Baes}, {Bonfield},
  {Burgarella}, {Buttiglione}, {Cava}, {Cooray}, {Dariush}, {de Zotti}, {Dye},
  {Eales}, {Frayer}, {Fritz}, {Gardner}, {Gonzalez-Nuevo}, {Herranz}, {Ibar},
  {Ivison}, {Jarvis}, {Lagache}, {Leeuw}, {Lopez-Caniego}, {Negrello},
  {Pascale}, {Pohlen}, {Rodighiero}, {Samui}, {Serjeant}, {Sibthorpe}, {Scott},
  {Smith}, {Temi}, {Thompson}, {Valtchanov}, {van der Werf}, \&
  {Verma}}]{clements10b}
{Clements} D.~L. {et~al.}, 2010, \aap, 518, L8

\bibitem[{{Condon}(1992)}]{condon92a}
{Condon} J.~J., 1992, \araa, 30, 575

\bibitem[{{Coppin} {et~al}\mbox{.}(2006){Coppin}, {Chapin}, {Mortier}, {Scott},
  {Borys}, {Dunlop}, {Halpern}, {Hughes}, {Pope}, {Scott}, {Serjeant}, {Wagg},
  {Alexander}, {Almaini}, {Aretxaga}, {Babbedge}, {Best}, {Blain}, {Chapman},
  {Clements}, {Crawford}, {Dunne}, {Eales}, {Edge}, {Farrah}, {Gazta{\~n}aga},
  {Gear}, {Granato}, {Greve}, {Fox}, {Ivison}, {Jarvis}, {Jenness}, {Lacey},
  {Lepage}, {Mann}, {Marsden}, {Martinez-Sansigre}, {Oliver}, {Page},
  {Peacock}, {Pearson}, {Percival}, {Priddey}, {Rawlings}, {Rowan-Robinson},
  {Savage}, {Seigar}, {Sekiguchi}, {Silva}, {Simpson}, {Smail}, {Stevens},
  {Takagi}, {Vaccari}, {van Kampen}, \& {Willott}}]{coppin06a}
{Coppin} K. {et~al.}, 2006, \mnras, 372, 1621

\bibitem[{{Coppin} {et~al}\mbox{.}(2005){Coppin}, {Halpern}, {Scott}, {Borys},
  \& {Chapman}}]{coppin05a}
{Coppin} K., {Halpern} M., {Scott} D., {Borys} C., {Chapman} S., 2005, \mnras,
  357, 1022

\bibitem[{{Coppin} {et~al}\mbox{.}(2010){Coppin}, {Pope},
  {Men{\'e}ndez-Delmestre}, {Alexander}, {Dunlop}, {Egami}, {Gabor}, {Ibar},
  {Ivison}, {Austermann}, {Blain}, {Chapman}, {Clements}, {Dunne}, {Dye},
  {Farrah}, {Hughes}, {Mortier}, {Page}, {Rowan-Robinson}, {Scott}, {Simpson},
  {Smail}, {Swinbank}, {Vaccari}, \& {Yun}}]{coppin10a}
{Coppin} K. {et~al.}, 2010, \apj, 713, 503

\bibitem[{{Coppin} {et~al}\mbox{.}(2008){Coppin}, {Swinbank}, {Neri}, {Cox},
  {Alexander}, {Smail}, {Page}, {Stevens}, {Knudsen}, {Ivison}, {Beelen},
  {Bertoldi}, \& {Omont}}]{coppin08a}
{Coppin} K.~E.~K. {et~al.}, 2008, \mnras, 389, 45

\bibitem[{{Cowie} {et~al}\mbox{.}(2002){Cowie}, {Barger}, \&
  {Kneib}}]{cowie02a}
{Cowie} L.~L., {Barger} A.~J., {Kneib} J.-P., 2002, \aj, 123, 2197

\bibitem[{{Daddi} {et~al}\mbox{.}(2009){Daddi}, {Dannerbauer}, {Stern},
  {Dickinson}, {Morrison}, {Elbaz}, {Giavalisco}, {Mancini}, {Pope}, \&
  {Spinrad}}]{daddi09a}
{Daddi} E. {et~al.}, 2009, \apj, 694, 1517

\bibitem[{{Dempsey} {et~al}\mbox{.}(2013){Dempsey} {et~al.}}]{dempsey13a}
{Dempsey} J.~T., {et~al.}, 2013, MNRAS, 430, 2534

\bibitem[{{Downes} {et~al}\mbox{.}(1986){Downes}, {Peacock}, {Savage}, \&
  {Carrie}}]{downes86a}
{Downes} A.~J.~B., {Peacock} J.~A., {Savage} A., {Carrie} D.~R., 1986, \mnras,
  218, 31

\bibitem[{{Dye} {et~al}\mbox{.}(2008){Dye}, {Eales}, {Aretxaga}, {Serjeant},
  {Dunlop}, {Babbedge}, {Chapman}, {Cirasuolo}, {Clements}, {Coppin}, {Dunne},
  {Egami}, {Farrah}, {Ivison}, {van Kampen}, {Pope}, {Priddey}, {Rieke},
  {Schael}, {Scott}, {Simpson}, {Takagi}, {Takata}, \& {Vaccari}}]{dye08a}
{Dye} S. {et~al.}, 2008, \mnras, 386, 1107

\bibitem[{{Eales} {et~al}\mbox{.}(2010){Eales}, {Dunne}, {Clements}, {Cooray},
  {de Zotti}, {Dye}, {Ivison}, {Jarvis}, {Lagache}, {Maddox}, {Negrello},
  {Serjeant}, {Thompson}, {van Kampen}, {Amblard}, {Andreani}, {Baes},
  {Beelen}, {Bendo}, {Benford}, {Bertoldi}, {Bock}, {Bonfield}, {Boselli},
  {Bridge}, {Buat}, {Burgarella}, {Carlberg}, {Cava}, {Chanial}, {Charlot},
  {Christopher}, {Coles}, {Cortese}, {Dariush}, {da Cunha}, {Dalton}, {Danese},
  {Dannerbauer}, {Driver}, {Dunlop}, {Fan}, {Farrah}, {Frayer}, {Frenk},
  {Geach}, {Gardner}, {Gomez}, {Gonz{\'a}lez-Nuevo}, {Gonz{\'a}lez-Solares},
  {Griffin}, {Hardcastle}, {Hatziminaoglou}, {Herranz}, {Hughes}, {Ibar},
  {Jeong}, {Lacey}, {Lapi}, {Lawrence}, {Lee}, {Leeuw}, {Liske},
  {L{\'o}pez-Caniego}, {M{\"u}ller}, {Nandra}, {Panuzzo}, {Papageorgiou},
  {Patanchon}, {Peacock}, {Pearson}, {Phillipps}, {Pohlen}, {Popescu},
  {Rawlings}, {Rigby}, {Rigopoulou}, {Robotham}, {Rodighiero}, {Sansom},
  {Schulz}, {Scott}, {Smith}, {Sibthorpe}, {Smail}, {Stevens}, {Sutherland},
  {Takeuchi}, {Tedds}, {Temi}, {Tuffs}, {Trichas}, {Vaccari}, {Valtchanov},
  {van der Werf}, {Verma}, {Vieria}, {Vlahakis}, \& {White}}]{eales10a}
{Eales} S. {et~al.}, 2010, \pasp, 122, 499

\bibitem[{{Eales} {et~al}\mbox{.}(1999){Eales}, {Lilly}, {Gear}, {Dunne},
  {Bond}, {Hammer}, {Le F{\`e}vre}, \& {Crampton}}]{eales99a}
{Eales} S., {Lilly} S., {Gear} W., {Dunne} L., {Bond} J.~R., {Hammer} F., {Le
  F{\`e}vre} O., {Crampton} D., 1999, \apj, 515, 518

\bibitem[{{Eddington}(1913)}]{eddington13a}
{Eddington} A.~S., 1913, \mnras, 73, 359

\bibitem[{{Elbaz} {et~al}\mbox{.}(2011){Elbaz}, {Dickinson}, {Hwang},
  {D{\'{\i}}az-Santos}, {Magdis}, {Magnelli}, {Le Borgne}, {Galliano},
  {Pannella}, {Chanial}, {Armus}, {Charmandaris}, {Daddi}, {Aussel}, {Popesso},
  {Kartaltepe}, {Altieri}, {Valtchanov}, {Coia}, {Dannerbauer}, {Dasyra},
  {Leiton}, {Mazzarella}, {Alexander}, {Buat}, {Burgarella}, {Chary}, {Gilli},
  {Ivison}, {Juneau}, {Le Floc'h}, {Lutz}, {Morrison}, {Mullaney}, {Murphy},
  {Pope}, {Scott}, {Brodwin}, {Calzetti}, {Cesarsky}, {Charlot}, {Dole},
  {Eisenhardt}, {Ferguson}, {F{\"o}rster Schreiber}, {Frayer}, {Giavalisco},
  {Huynh}, {Koekemoer}, {Papovich}, {Reddy}, {Surace}, {Teplitz}, {Yun}, \&
  {Wilson}}]{elbaz11a}
{Elbaz} D. {et~al.}, 2011, \aap, 533, A119+

\bibitem[{{Engel} {et~al}\mbox{.}(2010){Engel}, {Tacconi}, {Davies}, {Neri},
  {Smail}, {Chapman}, {Genzel}, {Cox}, {Greve}, {Ivison}, {Blain}, {Bertoldi},
  \& {Omont}}]{engel10a}
{Engel} H. {et~al.}, 2010, \apj, 724, 233

\bibitem[{{Frayer} {et~al}\mbox{.}(2009){Frayer}, {Sanders}, {Surace},
  {Aussel}, {Salvato}, {Le Floc'h}, {Huynh}, {Scoville}, {Afonso-Luis},
  {Bhattacharya}, {Capak}, {Fadda}, {Fu}, {Helou}, {Ilbert}, {Kartaltepe},
  {Koekemoer}, {Lee}, {Murphy}, {Sargent}, {Schinnerer}, {Sheth}, {Shopbell},
  {Shupe}, \& {Yan}}]{frayer09a}
{Frayer} D.~T. {et~al.}, 2009, \aj, 138, 1261

\bibitem[{{Geach} {et~al}\mbox{.}(2013){Geach}, {Chapin}, {Coppin}, {Dunlop},
  {Halpern}, {Smail}, {van der Werf}, {Serjeant}, {Farrah}, {Roseboom},
  {Targett}, {Arumugam}, {Asboth}, {Blain}, {Chrysostomou}, {Clarke}, {Ivison},
  {Jones}, {Karim}, {Mackenzie}, {Meijerink}, {Michalowski}, {Scott},
  {Simpson}, {Swinbank}, {Alexander}, {Almaini}, {Aretxaga}, {Best}, {Chapman},
  {Clements}, {Conselice}, {Danielson}, {Eales}, {Edge}, {Gibb}, {Hughes},
  {Jenness}, {Knudsen}, {Lacey}, {Marsden}, {McMahon}, {Oliver}, {Page},
  {Peacock}, {Rigopoulou}, {Robson}, {Spaans}, {Stevens}, {Webb}, {Willott},
  {Wilson}, \& {Zemcov}}]{geach13a}
{Geach} J.~E. {et~al.}, 2013, MNRAS, 432, 53

\bibitem[{{Genzel} {et~al}\mbox{.}(2010){Genzel}, {Tacconi}, {Gracia-Carpio},
  {Sternberg}, {Cooper}, {Shapiro}, {Bolatto}, {Bouch{\'e}}, {Bournaud},
  {Burkert}, {Combes}, {Comerford}, {Cox}, {Davis}, {Schreiber},
  {Garcia-Burillo}, {Lutz}, {Naab}, {Neri}, {Omont}, {Shapley}, \&
  {Weiner}}]{genzel10a}
{Genzel} R. {et~al.}, 2010, \mnras, 407, 2091

\bibitem[{{Greve} {et~al}\mbox{.}(2005){Greve}, {Bertoldi}, {Smail}, {Neri},
  {Chapman}, {Blain}, {Ivison}, {Genzel}, {Omont}, {Cox}, {Tacconi}, \&
  {Kneib}}]{greve05a}
{Greve} T.~R. {et~al.}, 2005, \mnras, 359, 1165

\bibitem[{{Gruppioni} {et~al}\mbox{.}(2013){Gruppioni} {et~al.}}]{gruppioni13a}
{Gruppioni} C., {et~al.}, 2013, MNRAS, 432, 23

\bibitem[{{Hayward} {et~al}\mbox{.}(2013){Hayward}, {Behroozi}, {Somerville},
  {Primack}, {Moreno}, \& {Wechsler}}]{hayward13a}
{Hayward} C.~C., {Behroozi} P.~S., {Somerville} R.~S., {Primack} J.~R.,
  {Moreno} J., {Wechsler} R.~H., 2013, MNRAS, 434, 2572

\bibitem[{{Hayward} {et~al}\mbox{.}(2012){Hayward}, {Jonsson}, {Kere{\v s}},
  {Magnelli}, {Hernquist}, \& {Cox}}]{hayward12a}
{Hayward} C.~C., {Jonsson} P., {Kere{\v s}} D., {Magnelli} B., {Hernquist} L.,
  {Cox} T.~J., 2012, \mnras, 424, 951

\bibitem[{{Helou} {et~al}\mbox{.}(1985){Helou}, {Soifer}, \&
  {Rowan-Robinson}}]{helou85a}
{Helou} G., {Soifer} B.~T., {Rowan-Robinson} M., 1985, \apjl, 298, L7

\bibitem[{{Hinshaw} {et~al}\mbox{.}(2009){Hinshaw}, {Weiland}, {Hill},
  {Odegard}, {Larson}, {Bennett}, {Dunkley}, {Gold}, {Greason}, {Jarosik},
  {Komatsu}, {Nolta}, {Page}, {Spergel}, {Wollack}, {Halpern}, {Kogut},
  {Limon}, {Meyer}, {Tucker}, \& {Wright}}]{hinshaw09a}
{Hinshaw} G. {et~al.}, 2009, \apjs, 180, 225

\bibitem[{{Hodge} {et~al}\mbox{.}(2013){Hodge}, {Karim}, {Smail}, {Swinbank},
  {Walter}, {Biggs}, {Ivison}, {Weiss}, {Alexander}, {Bertoldi}, {Brandt},
  {Chapman}, {Coppin}, {Cox}, {Danielson}, {Dannerbauer}, {De Breuck},
  {Decarli}, {Edge}, {Greve}, {Knudsen}, {Menten}, {Rix}, {Schinnerer},
  {Simpson}, {Wardlow}, \& {van der Werf}}]{hodge13a}
{Hodge} J.~A. {et~al.}, 2013, \apj, 768, 91

\bibitem[{{Holland} {et~al}\mbox{.}(1999){Holland}, {Robson}, {Gear},
  {Cunningham}, {Lightfoot}, {Jenness}, {Ivison}, {Stevens}, {Ade}, {Griffin},
  {Duncan}, {Murphy}, \& {Naylor}}]{holland99a}
{Holland} W.~S. {et~al.}, 1999, \mnras, 303, 659

\bibitem[{{Holland} {et~al}\mbox{.}(2013){Holland} {et~al.}}]{holland13a}
{Holland} W.~S., {et~al.}, 2013, MNRAS, 430, 2513

\bibitem[{{Hopkins} \& {Beacom}(2006)}]{hopkins06a}
{Hopkins} A.~M., {Beacom} J.~F., 2006, \apj, 651, 142

\bibitem[{{Hughes} {et~al}\mbox{.}(1998){Hughes}, {Serjeant}, {Dunlop},
  {Rowan-Robinson}, {Blain}, {Mann}, {Ivison}, {Peacock}, {Efstathiou}, {Gear},
  {Oliver}, {Lawrence}, {Longair}, {Goldschmidt}, \& {Jenness}}]{hughes98a}
{Hughes} D.~H. {et~al.}, 1998, \nat, 394, 241

\bibitem[{{Hwang} {et~al}\mbox{.}(2010){Hwang} {et~al.}}]{hwang10a}
{Hwang} H.~S., {et~al.}, 2010, MNRAS, 409

\bibitem[{{Ilbert} {et~al}\mbox{.}(2009){Ilbert}, {Capak}, {Salvato}, {Aussel},
  {McCracken}, {Sanders}, {Scoville}, {Kartaltepe}, {Arnouts}, {Le Floc'h},
  {Mobasher}, {Taniguchi}, {Lamareille}, {Leauthaud}, {Sasaki}, {Thompson},
  {Zamojski}, {Zamorani}, {Bardelli}, {Bolzonella}, {Bongiorno}, {Brusa},
  {Caputi}, {Carollo}, {Contini}, {Cook}, {Coppa}, {Cucciati}, {de la Torre},
  {de Ravel}, {Franzetti}, {Garilli}, {Hasinger}, {Iovino}, {Kampczyk},
  {Kneib}, {Knobel}, {Kovac}, {Le Borgne}, {Le Brun}, {F{\`e}vre}, {Lilly},
  {Looper}, {Maier}, {Mainieri}, {Mellier}, {Mignoli}, {Murayama}, {Pell{\`o}},
  {Peng}, {P{\'e}rez-Montero}, {Renzini}, {Ricciardelli}, {Schiminovich},
  {Scodeggio}, {Shioya}, {Silverman}, {Surace}, {Tanaka}, {Tasca}, {Tresse},
  {Vergani}, \& {Zucca}}]{ilbert09a}
{Ilbert} O. {et~al.}, 2009, \apj, 690, 1236

\bibitem[{{Ilbert} {et~al}\mbox{.}(2010){Ilbert}, {Salvato}, {Le Floc'h},
  {Aussel}, {Capak}, {McCracken}, {Mobasher}, {Kartaltepe}, {Scoville},
  {Sanders}, {Arnouts}, {Bundy}, {Cassata}, {Kneib}, {Koekemoer}, {Le
  F{\`e}vre}, {Lilly}, {Surace}, {Taniguchi}, {Tasca}, {Thompson}, {Tresse},
  {Zamojski}, {Zamorani}, \& {Zucca}}]{ilbert10a}
{Ilbert} O. {et~al.}, 2010, \apj, 709, 644

\bibitem[{{Ivison} {et~al}\mbox{.}(2010{\natexlab{a}}){Ivison}, {Alexander},
  {Biggs}, {Brandt}, {Chapin}, {Coppin}, {Devlin}, {Dickinson}, {Dunlop},
  {Dye}, {Eales}, {Frayer}, {Halpern}, {Hughes}, {Ibar}, {Kov{\'a}cs},
  {Marsden}, {Moncelsi}, {Netterfield}, {Pascale}, {Patanchon}, {Rafferty},
  {Rex}, {Schinnerer}, {Scott}, {Semisch}, {Smail}, {Swinbank}, {Truch},
  {Tucker}, {Viero}, {Walter}, {Wei{\ss}}, {Wiebe}, \& {Xue}}]{ivison10a}
{Ivison} R.~J. {et~al.}, 2010{\natexlab{a}}, \mnras, 402, 245

\bibitem[{{Ivison} {et~al}\mbox{.}(2010{\natexlab{b}}){Ivison}, {Magnelli},
  {Ibar}, {Andreani}, {Elbaz}, {Altieri}, {Amblard}, {Arumugam}, {Auld},
  {Aussel}, {Babbedge}, {Berta}, {Blain}, {Bock}, {Bongiovanni}, {Boselli},
  {Buat}, {Burgarella}, {Castro-Rodr{\'{\i}}guez}, {Cava}, {Cepa}, {Chanial},
  {Cimatti}, {Cirasuolo}, {Clements}, {Conley}, {Conversi}, {Cooray}, {Daddi},
  {Dominguez}, {Dowell}, {Dwek}, {Eales}, {Farrah}, {F{\"o}rster Schreiber},
  {Fox}, {Franceschini}, {Gear}, {Genzel}, {Glenn}, {Griffin}, {Gruppioni},
  {Halpern}, {Hatziminaoglou}, {Isaak}, {Lagache}, {Levenson}, {Lu}, {Lutz},
  {Madden}, {Maffei}, {Magdis}, {Mainetti}, {Maiolino}, {Marchetti},
  {Morrison}, {Mortier}, {Nguyen}, {Nordon}, {O'Halloran}, {Oliver}, {Omont},
  {Owen}, {Page}, {Panuzzo}, {Papageorgiou}, {Pearson}, {P{\'e}rez-Fournon},
  {P{\'e}rez Garc{\'{\i}}a}, {Poglitsch}, {Pohlen}, {Popesso}, {Pozzi},
  {Rawlings}, {Raymond}, {Rigopoulou}, {Riguccini}, {Rizzo}, {Rodighiero},
  {Roseboom}, {Rowan-Robinson}, {Saintonge}, {Sanchez Portal}, {Santini},
  {Schulz}, {Scott}, {Seymour}, {Shao}, {Shupe}, {Smith}, {Stevens}, {Sturm},
  {Symeonidis}, {Tacconi}, {Trichas}, {Tugwell}, {Vaccari}, {Valtchanov},
  {Vieira}, {Vigroux}, {Wang}, {Ward}, {Wright}, {Xu}, \& {Zemcov}}]{ivison10b}
{Ivison} R.~J. {et~al.}, 2010{\natexlab{b}}, \aap, 518, L31+

\bibitem[{{Karim} {et~al}\mbox{.}(2013){Karim}, {Swinbank}, {Hodge}, {Smail},
  {Walter}, {Biggs}, {Simpson}, {Danielson}, {Alexander}, {Bertoldi}, {de
  Breuck}, {Chapman}, {Coppin}, {Dannerbauer}, {Edge}, {Greve}, {Ivison},
  {Knudsen}, {Menten}, {Schinnerer}, {Wardlow}, {Wei{\ss}}, \& {van der
  Werf}}]{karim13a}
{Karim} A. {et~al.}, 2013, \mnras, 432, 2

\bibitem[{{Kennicutt}(1998{\natexlab{a}})}]{kennicutt98b}
{Kennicutt}, Jr. R.~C., 1998{\natexlab{a}}, \araa, 36, 189

\bibitem[{{Kennicutt}(1998{\natexlab{b}})}]{Kennicutt98a}
{Kennicutt}, Jr. R.~C., 1998{\natexlab{b}}, \apj, 498, 541

\bibitem[{{Knudsen} {et~al}\mbox{.}(2008){Knudsen}, {van der Werf}, \&
  {Kneib}}]{knudsen08a}
{Knudsen} K.~K., {van der Werf} P.~P., {Kneib} J.-P., 2008, \mnras, 384, 1611

\bibitem[{{Kov{\'a}cs} {et~al}\mbox{.}(2006){Kov{\'a}cs}, {Chapman}, {Dowell},
  {Blain}, {Ivison}, {Smail}, \& {Phillips}}]{kovacs06a}
{Kov{\'a}cs} A., {Chapman} S.~C., {Dowell} C.~D., {Blain} A.~W., {Ivison}
  R.~J., {Smail} I., {Phillips} T.~G., 2006, \apj, 650, 592

\bibitem[{{Kov{\'a}cs} {et~al}\mbox{.}(2010){Kov{\'a}cs}, {Omont}, {Beelen},
  {Lonsdale}, {Polletta}, {Fiolet}, {Greve}, {Borys}, {Cox}, {De Breuck},
  {Dole}, {Dowell}, {Farrah}, {Lagache}, {Menten}, {Bell}, \&
  {Owen}}]{kovacs10a}
{Kov{\'a}cs} A. {et~al.}, 2010, \apj, 717, 29

\bibitem[{{Laird} {et~al}\mbox{.}(2010){Laird}, {Nandra}, {Pope}, \&
  {Scott}}]{laird10a}
{Laird} E.~S., {Nandra} K., {Pope} A., {Scott} D., 2010, \mnras, 401, 2763

\bibitem[{{Le Floc'h} {et~al}\mbox{.}(2009){Le Floc'h}, {Aussel}, {Ilbert},
  {Riguccini}, {Frayer}, {Salvato}, {Arnouts}, {Surace}, {Feruglio},
  {Rodighiero}, {Capak}, {Kartaltepe}, {Heinis}, {Sheth}, {Yan}, {McCracken},
  {Thompson}, {Sanders}, {Scoville}, \& {Koekemoer}}]{le-floch09a}
{Le Floc'h} E. {et~al.}, 2009, \apj, 703, 222

\bibitem[{{Le Floc'h} {et~al}\mbox{.}(2005){Le Floc'h}, {Papovich}, {Dole},
  {Bell}, {Lagache}, {Rieke}, {Egami}, {P{\'e}rez-Gonz{\'a}lez},
  {Alonso-Herrero}, {Rieke}, {Blaylock}, {Engelbracht}, {Gordon}, {Hines},
  {Misselt}, {Morrison}, \& {Mould}}]{le-floch05a}
{Le Floc'h} E. {et~al.}, 2005, \apj, 632, 169

\bibitem[{{Lee} {et~al}\mbox{.}(2010){Lee}, {Le Floc'h}, {Sanders}, {Frayer},
  {Arnouts}, {Ilbert}, {Aussel}, {Salvato}, {Scoville}, \&
  {Kartaltepe}}]{lee10a}
{Lee} N. {et~al.}, 2010, \apj, 717, 175

\bibitem[{{Magnelli} {et~al}\mbox{.}(2011){Magnelli}, {Elbaz}, {Chary},
  {Dickinson}, {Le Borgne}, {Frayer}, \& {Willmer}}]{magnelli11a}
{Magnelli} B., {Elbaz} D., {Chary} R.~R., {Dickinson} M., {Le Borgne} D.,
  {Frayer} D.~T., {Willmer} C.~N.~A., 2011, \aap, 528, A35

\bibitem[{{Men{\'e}ndez-Delmestre}
  {et~al}\mbox{.}(2009){Men{\'e}ndez-Delmestre}, {Blain}, {Smail}, {Alexander},
  {Chapman}, {Armus}, {Frayer}, {Ivison}, \& {Teplitz}}]{menendez-delmestre09a}
{Men{\'e}ndez-Delmestre} K. {et~al.}, 2009, \apj, 699, 667

\bibitem[{{Meurer} {et~al}\mbox{.}(1999){Meurer}, {Heckman}, \&
  {Calzetti}}]{meurer99a}
{Meurer} G.~R., {Heckman} T.~M., {Calzetti} D., 1999, \apj, 521, 64

\bibitem[{{Neri} {et~al}\mbox{.}(2003){Neri}, {Genzel}, {Ivison}, {Bertoldi},
  {Blain}, {Chapman}, {Cox}, {Greve}, {Omont}, \& {Frayer}}]{neri03a}
{Neri} R. {et~al.}, 2003, \apjl, 597, L113

\bibitem[{{Oliver} {et~al}\mbox{.}(2010){Oliver}, {Frost}, {Farrah},
  {Gonzalez-Solares}, {Shupe}, {Henriques}, {Roseboom}, {Alfonso-Luis},
  {Babbedge}, {Frayer}, {Lencz}, {Lonsdale}, {Masci}, {Padgett}, {Polletta},
  {Rowan-Robinson}, {Siana}, {Smith}, {Surace}, \& {Vaccari}}]{oliver10a}
{Oliver} S. {et~al.}, 2010, \mnras, 405, 2279

\bibitem[{{Oliver} {et~al}\mbox{.}(2012){Oliver}, {Bock}, {Altieri}, {Amblard},
  {Arumugam}, {Aussel}, {Babbedge}, {Beelen}, {B{\'e}thermin}, {Blain},
  {Boselli}, {Bridge}, {Brisbin}, {Buat}, {Burgarella},
  {Castro-Rodr{\'{\i}}guez}, {Cava}, {Chanial}, {Cirasuolo}, {Clements},
  {Conley}, {Conversi}, {Cooray}, {Dowell}, {Dubois}, {Dwek}, {Dye}, {Eales},
  {Elbaz}, {Farrah}, {Feltre}, {Ferrero}, {Fiolet}, {Fox}, {Franceschini},
  {Gear}, {Giovannoli}, {Glenn}, {Gong}, {Gonz{\'a}lez Solares}, {Griffin},
  {Halpern}, {Harwit}, {Hatziminaoglou}, {Heinis}, {Hurley}, {Hwang}, {Hyde},
  {Ibar}, {Ilbert}, {Isaak}, {Ivison}, {Lagache}, {Le Floc'h}, {Levenson},
  {Faro}, {Lu}, {Madden}, {Maffei}, {Magdis}, {Mainetti}, {Marchetti},
  {Marsden}, {Marshall}, {Mortier}, {Nguyen}, {O'Halloran}, {Omont}, {Page},
  {Panuzzo}, {Papageorgiou}, {Patel}, {Pearson}, {P{\'e}rez-Fournon}, {Pohlen},
  {Rawlings}, {Raymond}, {Rigopoulou}, {Riguccini}, {Rizzo}, {Rodighiero},
  {Roseboom}, {Rowan-Robinson}, {S{\'a}nchez Portal}, {Schulz}, {Scott},
  {Seymour}, {Shupe}, {Smith}, {Stevens}, {Symeonidis}, {Trichas}, {Tugwell},
  {Vaccari}, {Valtchanov}, {Vieira}, {Viero}, {Vigroux}, {Wang}, {Ward},
  {Wardlow}, {Wright}, {Xu}, \& {Zemcov}}]{oliver12a}
{Oliver} S.~J. {et~al.}, 2012, \mnras, 424, 1614

\bibitem[{{Pilbratt} {et~al}\mbox{.}(2010){Pilbratt}, {Riedinger}, {Passvogel},
  {Crone}, {Doyle}, {Gageur}, {Heras}, {Jewell}, {Metcalfe}, {Ott}, \&
  {Schmidt}}]{pilbratt10a}
{Pilbratt} G.~L. {et~al.}, 2010, \aap, 518, L1

\bibitem[{{Polletta} {et~al}\mbox{.}(2007){Polletta}, {Tajer}, {Maraschi},
  {Trinchieri}, {Lonsdale}, {Chiappetti}, {Andreon}, {Pierre}, {Le F{\`e}vre},
  {Zamorani}, {Maccagni}, {Garcet}, {Surdej}, {Franceschini}, {Alloin},
  {Shupe}, {Surace}, {Fang}, {Rowan-Robinson}, {Smith}, \&
  {Tresse}}]{polletta07a}
{Polletta} M. {et~al.}, 2007, \apj, 663, 81

\bibitem[{{Pope} {et~al}\mbox{.}(2008){Pope}, {Chary}, {Alexander}, {Armus},
  {Dickinson}, {Elbaz}, {Frayer}, {Scott}, \& {Teplitz}}]{pope08a}
{Pope} A. {et~al.}, 2008, \apj, 675, 1171

\bibitem[{{Reddy} {et~al}\mbox{.}(2012){Reddy}, {Dickinson}, {Elbaz},
  {Morrison}, {Giavalisco}, {Ivison}, {Papovich}, {Scott}, {Buat},
  {Burgarella}, {Charmandaris}, {Daddi}, {Magdis}, {Murphy}, {Altieri},
  {Aussel}, {Dannerbauer}, {Dasyra}, {Hwang}, {Kartaltepe}, {Leiton},
  {Magnelli}, \& {Popesso}}]{reddy12a}
{Reddy} N. {et~al.}, 2012, \apj, 744, 154

\bibitem[{{Rodighiero} {et~al}\mbox{.}(2011){Rodighiero}, {Daddi},
  {Baronchelli}, {Cimatti}, {Renzini}, {Aussel}, {Popesso}, {Lutz}, {Andreani},
  {Berta}, {Cava}, {Elbaz}, {Feltre}, {Fontana}, {F{\"o}rster Schreiber},
  {Franceschini}, {Genzel}, {Grazian}, {Gruppioni}, {Ilbert}, {Le Floch},
  {Magdis}, {Magliocchetti}, {Magnelli}, {Maiolino}, {McCracken}, {Nordon},
  {Poglitsch}, {Santini}, {Pozzi}, {Riguccini}, {Tacconi}, {Wuyts}, \&
  {Zamorani}}]{rodighiero11a}
{Rodighiero} G. {et~al.}, 2011, \apjl, 739, L40

\bibitem[{{Rosa-Gonz{\'a}lez} {et~al}\mbox{.}(2002){Rosa-Gonz{\'a}lez},
  {Terlevich}, \& {Terlevich}}]{rosa-gonzalez02a}
{Rosa-Gonz{\'a}lez} D., {Terlevich} E., {Terlevich} R., 2002, \mnras, 332, 283

\bibitem[{{Roseboom} {et~al}\mbox{.}(2013){Roseboom}, {Dunlop}, {Cirasuolo},
  {Geach}, {Smail}, {Halpern}, {van der Werf}, {Almaini}, {Arumugam}, {Asboth},
  {Auld}, {Blain}, {Bremer}, {Bock}, {Bowler}, {Buitrago}, {Chapin}, {Chapman},
  {Chrysostomou}, {Clarke}, {Conley}, {Coppin}, {Danielson}, {Farrah}, {Glenn},
  {Hatziminaoglou}, {Ibar}, {Ivison}, {Jenness}, {van Kampen}, {Karim},
  {Mackenzie}, {Marsden}, {Meijerink}, {Micha{\l}owski}, {Oliver}, {Page},
  {Pearson}, {Scott}, {Simpson}, {Smith}, {Spaans}, {Swinbank}, {Symeonidis},
  {Targett}, {Valiante}, {Viero}, {Wang}, {Willott}, \& {Zemcov}}]{roseboom13a}
{Roseboom} I.~G. {et~al.}, 2013, ArXiv e-prints

\bibitem[{{Roseboom} {et~al}\mbox{.}(2012){Roseboom}, {Ivison}, {Greve},
  {Amblard}, {Arumugam}, {Auld}, {Aussel}, {Bethermin}, {Blain}, {Bock},
  {Boselli}, {Brisbin}, {Buat}, {Burgarella}, {Castro-Rodr{\'{\i}}guez},
  {Cava}, {Chanial}, {Chapin}, {Chapman}, {Clements}, {Conley}, {Conversi},
  {Cooray}, {Dowell}, {Dunlop}, {Dwek}, {Eales}, {Elbaz}, {Farrah},
  {Franceschini}, {Glenn}, {Griffin}, {Halpern}, {Hatziminaoglou}, {Ibar},
  {Isaak}, {Lagache}, {Levenson}, {Lu}, {Madden}, {Maffei}, {Mainetti},
  {Marchetti}, {Marsden}, {Morrison}, {Mortier}, {Nguyen}, {O'Halloran},
  {Oliver}, {Omont}, {Page}, {Panuzzo}, {Papageorgiou}, {Pearson},
  {P{\'e}rez-Fournon}, {Pohlen}, {Rawlings}, {Raymond}, {Rigopoulou}, {Rizzo},
  {Rodighiero}, {Rowan-Robinson}, {Schulz}, {Scott}, {Seymour}, {Shupe},
  {Smith}, {Stevens}, {Symeonidis}, {Trichas}, {Tugwell}, {Vaccari},
  {Valtchanov}, {Vieira}, {Viero}, {Vigroux}, {Wardlow}, {Wang}, {Wright},
  {Xu}, \& {Zemcov}}]{roseboom12a}
{Roseboom} I.~G. {et~al.}, 2012, \mnras, 419, 2758

\bibitem[{{Sanders} \& {Mirabel}(1996)}]{sanders96a}
{Sanders} D.~B., {Mirabel} I.~F., 1996, \araa, 34, 749

\bibitem[{{Sanders} {et~al}\mbox{.}(2007){Sanders}, {Salvato}, {Aussel},
  {Ilbert}, {Scoville}, {Surace}, {Frayer}, {Sheth}, {Helou}, {Brooke},
  {Bhattacharya}, {Yan}, {Kartaltepe}, {Barnes}, {Blain}, {Calzetti}, {Capak},
  {Carilli}, {Carollo}, {Comastri}, {Daddi}, {Ellis}, {Elvis}, {Fall},
  {Franceschini}, {Giavalisco}, {Hasinger}, {Impey}, {Koekemoer}, {Le
  F{\`e}vre}, {Lilly}, {Liu}, {McCracken}, {Mobasher}, {Renzini}, {Rich},
  {Schinnerer}, {Shopbell}, {Taniguchi}, {Thompson}, {Urry}, \&
  {Williams}}]{sanders07a}
{Sanders} D.~B. {et~al.}, 2007, \apjs, 172, 86

\bibitem[{{Sanders} {et~al}\mbox{.}(1988){Sanders}, {Soifer}, {Elias},
  {Neugebauer}, \& {Matthews}}]{sanders88a}
{Sanders} D.~B., {Soifer} B.~T., {Elias} J.~H., {Neugebauer} G., {Matthews} K.,
  1988, \apjl, 328, L35

\bibitem[{{Scott} {et~al}\mbox{.}(2008){Scott}, {Austermann}, {Perera},
  {Wilson}, {Aretxaga}, {Bock}, {Hughes}, {Kang}, {Kim}, {Mauskopf}, {Sanders},
  {Scoville}, \& {Yun}}]{scott08a}
{Scott} K.~S. {et~al.}, 2008, \mnras, 385, 2225

\bibitem[{{Scott} {et~al}\mbox{.}(2002){Scott}, {Fox}, {Dunlop}, {Serjeant},
  {Peacock}, {Ivison}, {Oliver}, {Mann}, {Lawrence}, {Efstathiou},
  {Rowan-Robinson}, {Hughes}, {Archibald}, {Blain}, \& {Longair}}]{scott02a}
{Scott} S.~E. {et~al.}, 2002, \mnras, 331, 817

\bibitem[{{Scoville} {et~al}\mbox{.}(2013){Scoville}, {Arnouts}, {Aussel},
  {Benson}, {Bongiorno}, {Bundy}, {Calvo}, {Capak}, {Carollo}, {Civano},
  {Dunlop}, {Elvis}, {Faisst}, {Finoguenov}, {Fu}, {Giavalisco}, {Guo},
  {Ilbert}, {Iovino}, {Kajisawa}, {Kartaltepe}, {Leauthaud}, {Le F{\`e}vre},
  {LeFloch}, {Lilly}, {Liu}, {Manohar}, {Massey}, {Masters}, {McCracken},
  {Mobasher}, {Peng}, {Renzini}, {Rhodes}, {Salvato}, {Sanders}, {Sarvestani},
  {Scarlata}, {Schinnerer}, {Sheth}, {Shopbell}, {Smol{\v c}i{\'c}},
  {Taniguchi}, {Taylor}, {White}, \& {Yan}}]{scoville13a}
{Scoville} N. {et~al.}, 2013, \apjs, 206, 3

\bibitem[{{Scoville} {et~al}\mbox{.}(2007){Scoville}, {Aussel}, {Brusa},
  {Capak}, {Carollo}, {Elvis}, {Giavalisco}, {Guzzo}, {Hasinger}, {Impey},
  {Kneib}, {LeFevre}, {Lilly}, {Mobasher}, {Renzini}, {Rich}, {Sanders},
  {Schinnerer}, {Schminovich}, {Shopbell}, {Taniguchi}, \&
  {Tyson}}]{scoville07a}
{Scoville} N. {et~al.}, 2007, \apjs, 172, 1

\bibitem[{{Serjeant} {et~al}\mbox{.}(2008){Serjeant}, {Dye}, {Mortier},
  {Peacock}, {Egami}, {Cirasuolo}, {Rieke}, {Borys}, {Chapman}, {Clements},
  {Coppin}, {Dunlop}, {Eales}, {Farrah}, {Halpern}, {Mauskopf}, {Pope},
  {Rowan-Robinson}, {Scott}, {Smail}, \& {Vaccari}}]{serjeant08a}
{Serjeant} S. {et~al.}, 2008, \mnras, 386, 1907

\bibitem[{{Smail} {et~al}\mbox{.}(1997){Smail}, {Ivison}, \&
  {Blain}}]{smail97a}
{Smail} I., {Ivison} R.~J., {Blain} A.~W., 1997, \apjl, 490, L5+

\bibitem[{{Smail} {et~al}\mbox{.}(2002){Smail}, {Ivison}, {Blain}, \&
  {Kneib}}]{smail02a}
{Smail} I., {Ivison} R.~J., {Blain} A.~W., {Kneib} J., 2002, \mnras, 331, 495

\bibitem[{{Smol{\v c}i{\'c}} {et~al}\mbox{.}(2012){Smol{\v c}i{\'c}},
  {Aravena}, {Navarrete}, {Schinnerer}, {Riechers}, {Bertoldi}, {Feruglio},
  {Finoguenov}, {Salvato}, {Sargent}, {McCracken}, {Albrecht}, {Karim},
  {Capak}, {Carilli}, {Cappelluti}, {Elvis}, {Ilbert}, {Kartaltepe}, {Lilly},
  {Sanders}, {Sheth}, {Scoville}, \& {Taniguchi}}]{smolcic12a}
{Smol{\v c}i{\'c}} V. {et~al.}, 2012, \aap, 548, A4

\bibitem[{{Soifer} {et~al}\mbox{.}(1989){Soifer}, {Boehmer}, {Neugebauer}, \&
  {Sanders}}]{soifer89a}
{Soifer} B.~T., {Boehmer} L., {Neugebauer} G., {Sanders} D.~B., 1989, \aj, 98,
  766

\bibitem[{{Soifer} {et~al}\mbox{.}(1987){Soifer}, {Sanders}, {Madore},
  {Neugebauer}, {Danielson}, {Elias}, {Lonsdale}, \& {Rice}}]{soifer87a}
{Soifer} B.~T., {Sanders} D.~B., {Madore} B.~F., {Neugebauer} G., {Danielson}
  G.~E., {Elias} J.~H., {Lonsdale} C.~J., {Rice} W.~L., 1987, \apj, 320, 238

\bibitem[{{Swinbank} {et~al}\mbox{.}(2004){Swinbank}, {Smail}, {Chapman},
  {Blain}, {Ivison}, \& {Keel}}]{swinbank04a}
{Swinbank} A.~M., {Smail} I., {Chapman} S.~C., {Blain} A.~W., {Ivison} R.~J.,
  {Keel} W.~C., 2004, \apj, 617, 64

\bibitem[{{Tacconi} {et~al}\mbox{.}(2010){Tacconi}, {Genzel}, {Neri}, {Cox},
  {Cooper}, {Shapiro}, {Bolatto}, {Bouch{\'e}}, {Bournaud}, {Burkert},
  {Combes}, {Comerford}, {Davis}, {Schreiber}, {Garcia-Burillo},
  {Gracia-Carpio}, {Lutz}, {Naab}, {Omont}, {Shapley}, {Sternberg}, \&
  {Weiner}}]{tacconi10a}
{Tacconi} L.~J. {et~al.}, 2010, \nat, 463, 781

\bibitem[{{Tacconi} {et~al}\mbox{.}(2006){Tacconi}, {Neri}, {Chapman},
  {Genzel}, {Smail}, {Ivison}, {Bertoldi}, {Blain}, {Cox}, {Greve}, \&
  {Omont}}]{tacconi06a}
{Tacconi} L.~J. {et~al.}, 2006, \apj, 640, 228

\bibitem[{{Tacconi} {et~al}\mbox{.}(2008){Tacconi} {et~al.}}]{tacconi08a}
{Tacconi} L.~J., {et~al.}, 2008, \apj, 680, 246

\bibitem[{{Targett}(2011)}]{targett11a}
{Targett} T.~A., 2011, MNRAS

\bibitem[{{U} {et~al}\mbox{.}(2012){U}, {Sanders}, {Mazzarella}, {Evans},
  {Howell}, {Surace}, {Armus}, {Iwasawa}, {Kim}, {Casey}, {Vavilkin},
  {Dufault}, {Larson}, {Barnes}, {Chan}, {Frayer}, {Haan}, {Inami}, {Ishida},
  {Kartaltepe}, {Melbourne}, \& {Petric}}]{u12a}
{U} V. {et~al.}, 2012, \apjs, 203, 9

\bibitem[{{Wang} {et~al}\mbox{.}(2006){Wang}, {Cowie}, \& {Barger}}]{wang06a}
{Wang} W.-H., {Cowie} L.~L., {Barger} A.~J., 2006, \apj, 647, 74

\bibitem[{{Wang} {et~al}\mbox{.}(2011){Wang}, {Cowie}, {Barger}, \&
  {Williams}}]{wang11a}
{Wang} W.-H., {Cowie} L.~L., {Barger} A.~J., {Williams} J.~P., 2011, \apjl,
  726, L18

\bibitem[{{Wardlow} {et~al}\mbox{.}(2011){Wardlow}, {Smail}, {Coppin},
  {Alexander}, {Brandt}, {Danielson}, {Luo}, {Swinbank}, {Walter}, {Wei{\ss}},
  {Xue}, {Zibetti}, {Bertoldi}, {Biggs}, {Chapman}, {Dannerbauer}, {Dunlop},
  {Gawiser}, {Ivison}, {Knudsen}, {Kov{\'a}cs}, {Lacey}, {Menten}, {Padilla},
  {Rix}, \& {van der Werf}}]{wardlow11a}
{Wardlow} J.~L. {et~al.}, 2011, \mnras, 415, 1479

\bibitem[{{Webb} {et~al}\mbox{.}(2003){Webb}, {Eales}, {Foucaud}, {Lilly},
  {McCracken}, {Adelberger}, {Steidel}, {Shapley}, {Clements}, {Dunne}, {Le
  F{\`e}vre}, {Brodwin}, \& {Gear}}]{webb03a}
{Webb} T.~M. {et~al.}, 2003, \apj, 582, 6

\bibitem[{{Wilson} {et~al}\mbox{.}(2008){Wilson}, {Petitpas}, {Iono}, {Baker},
  {Peck}, {Krips}, {Warren}, {Golding}, {Atkinson}, {Armus}, {Cox}, {Ho},
  {Juvela}, {Matsushita}, {Mihos}, {Pihlstrom}, \& {Yun}}]{wilson08a}
{Wilson} C.~D. {et~al.}, 2008, \apjs, 178, 189

\bibitem[{{Younger} {et~al}\mbox{.}(2007){Younger}, {Fazio}, {Huang}, {Yun},
  {Wilson}, {Ashby}, {Gurwell}, {Lai}, {Peck}, {Petitpas}, {Wilner}, {Iono},
  {Kohno}, {Kawabe}, {Hughes}, {Aretxaga}, {Webb}, {Mart{\'{\i}}nez-Sansigre},
  {Kim}, {Scott}, {Austermann}, {Perera}, {Lowenthal}, {Schinnerer}, \&
  {Smol{\v c}i{\'c}}}]{younger07a}
{Younger} J.~D. {et~al.}, 2007, \apj, 671, 1531

\bibitem[{{Younger} {et~al}\mbox{.}(2008){Younger}, {Fazio}, {Wilner}, {Ashby},
  {Blundell}, {Gurwell}, {Huang}, {Iono}, {Peck}, {Petitpas}, {Scott},
  {Wilson}, \& {Yun}}]{younger08a}
{Younger} J.~D. {et~al.}, 2008, \apj, 688, 59

\bibitem[{{Younger} {et~al}\mbox{.}(2009){Younger}, {Omont}, {Fiolet}, {Huang},
  {Fazio}, {Lai}, {Polletta}, {Rigopoulou}, \& {Zylka}}]{younger09a}
{Younger} J.~D. {et~al.}, 2009, \mnras, 394, 1685

\bibitem[{{Yun} {et~al}\mbox{.}(2012){Yun}, {Scott}, {Guo}, {Aretxaga},
  {Giavalisco}, {Austermann}, {Capak}, {Chen}, {Ezawa}, {Hatsukade}, {Hughes},
  {Iono}, {Johnson}, {Kawabe}, {Kohno}, {Lowenthal}, {Miller}, {Morrison},
  {Oshima}, {Perera}, {Salvato}, {Silverman}, {Tamura}, {Williams}, \&
  {Wilson}}]{yun12a}
{Yun} M.~S. {et~al.}, 2012, \mnras, 420, 957

\end{thebibliography}

\clearpage
\begin{figure*}
\centering
\includegraphics[width=0.33\columnwidth]{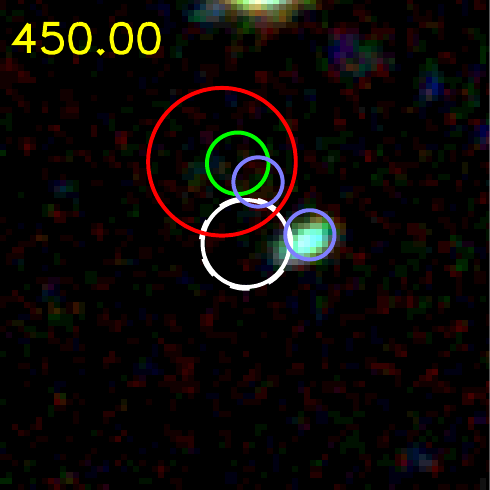}
\includegraphics[width=0.33\columnwidth]{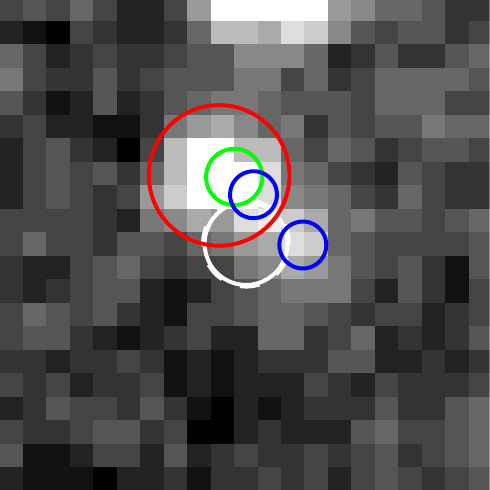}
\includegraphics[width=0.33\columnwidth]{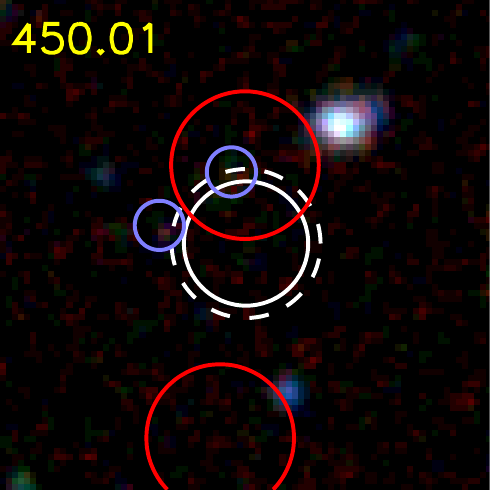}
\includegraphics[width=0.33\columnwidth]{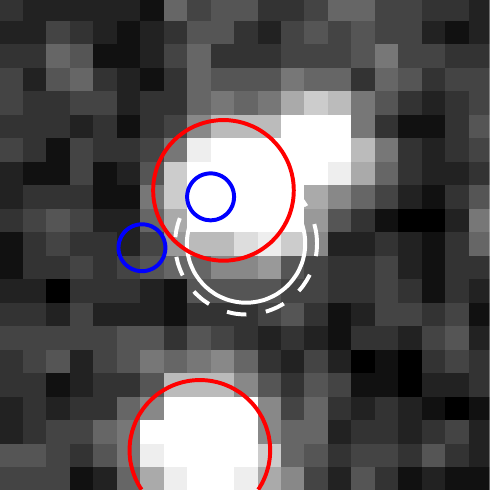}
\includegraphics[width=0.33\columnwidth]{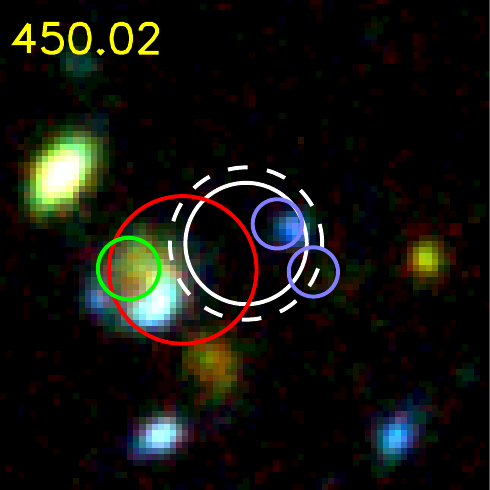}
\includegraphics[width=0.33\columnwidth]{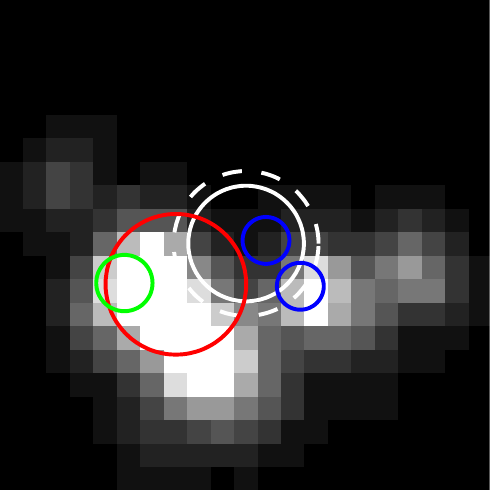}
\includegraphics[width=0.33\columnwidth]{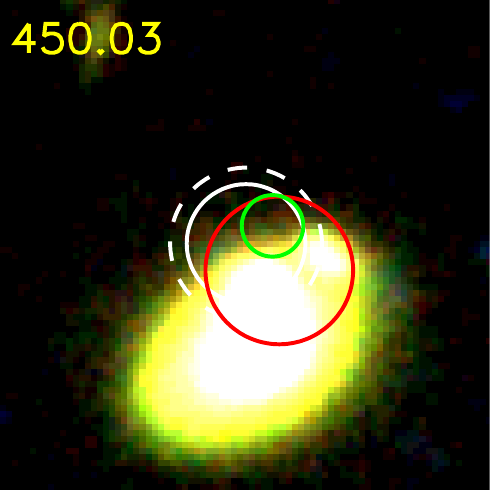}
\includegraphics[width=0.33\columnwidth]{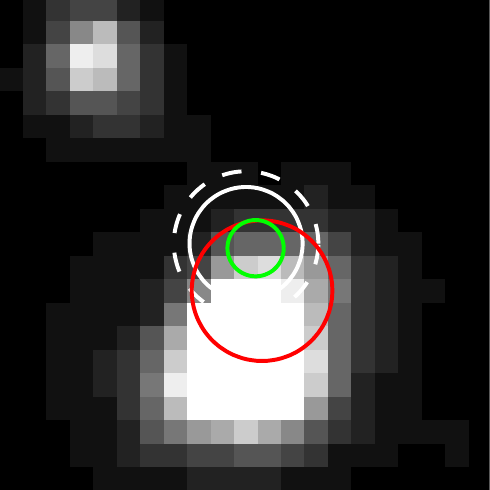}
\includegraphics[width=0.33\columnwidth]{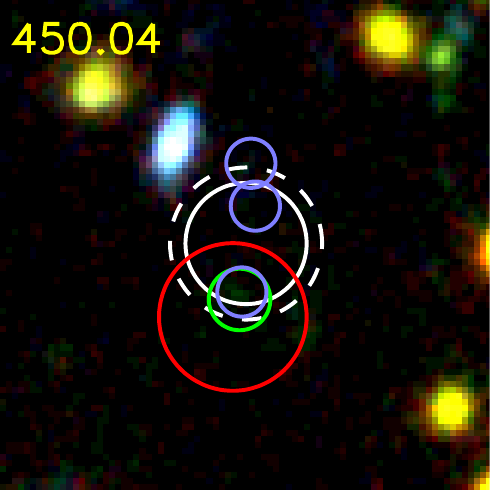}
\includegraphics[width=0.33\columnwidth]{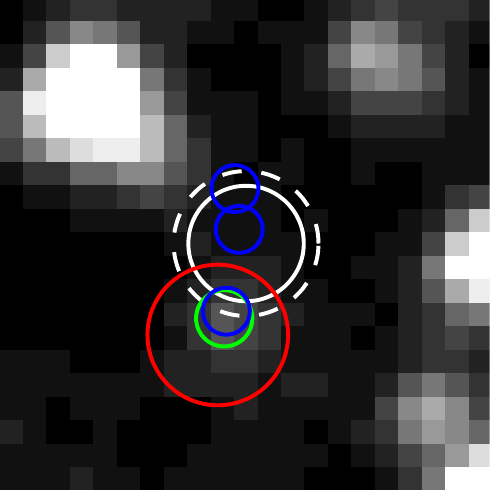}
\includegraphics[width=0.33\columnwidth]{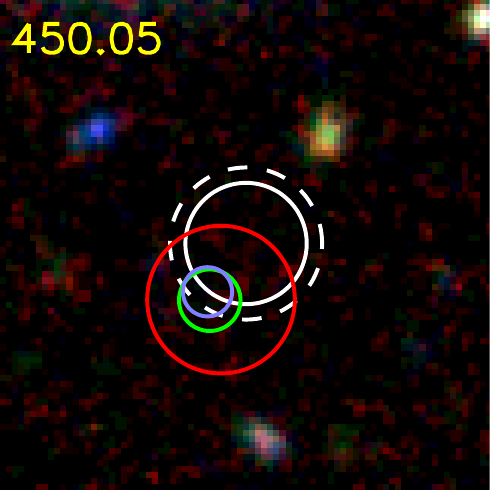}
\includegraphics[width=0.33\columnwidth]{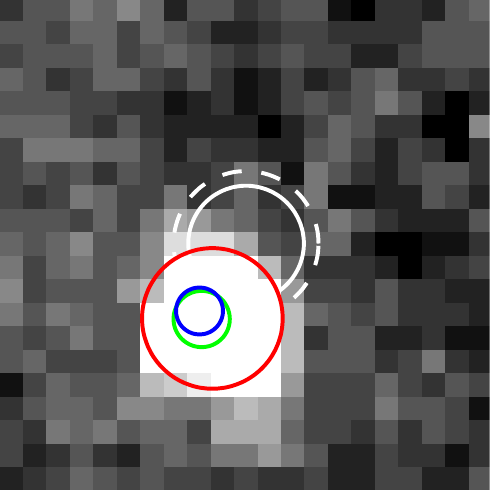}
\includegraphics[width=0.33\columnwidth]{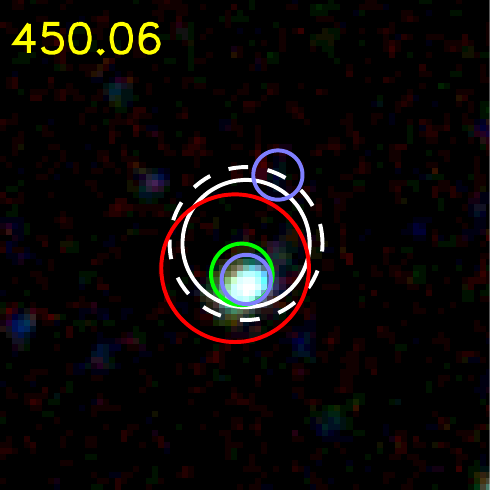}
\includegraphics[width=0.33\columnwidth]{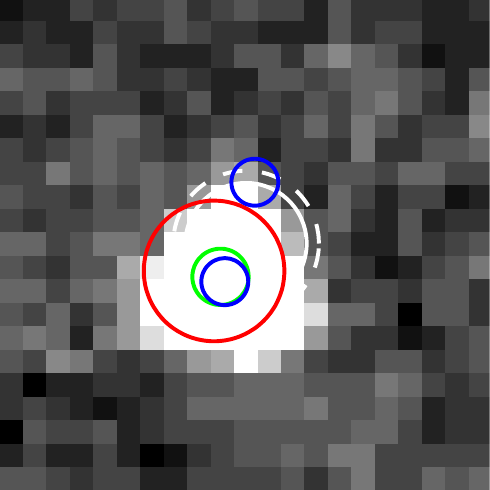}
\includegraphics[width=0.33\columnwidth]{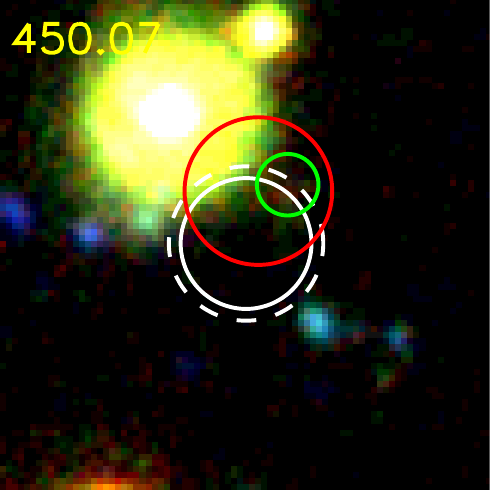}
\includegraphics[width=0.33\columnwidth]{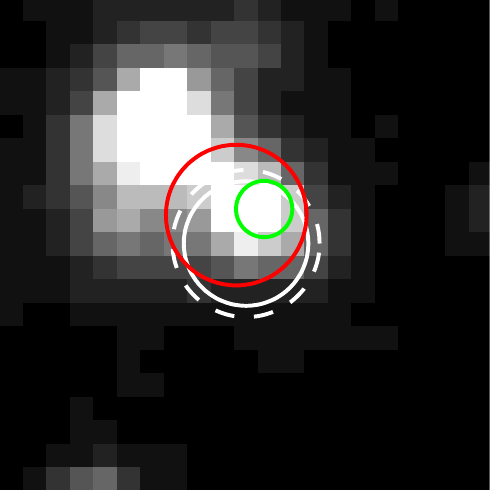}
\includegraphics[width=0.33\columnwidth]{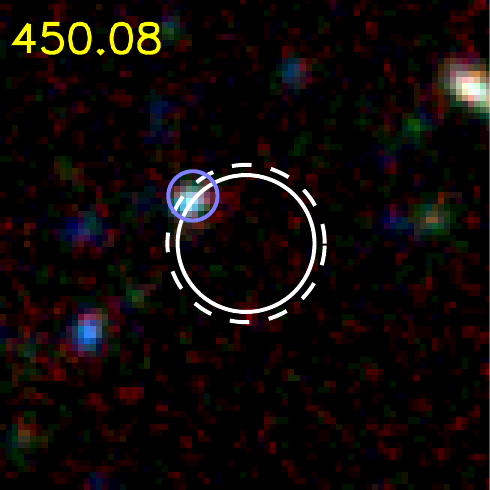}
\includegraphics[width=0.33\columnwidth]{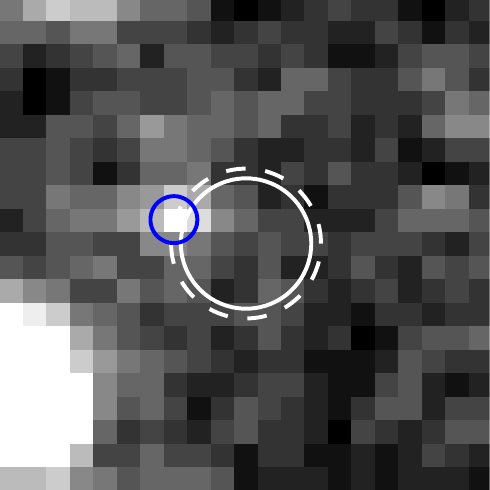}
\includegraphics[width=0.33\columnwidth]{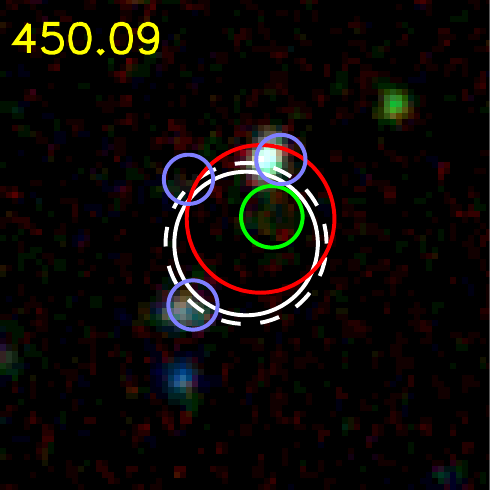}
\includegraphics[width=0.33\columnwidth]{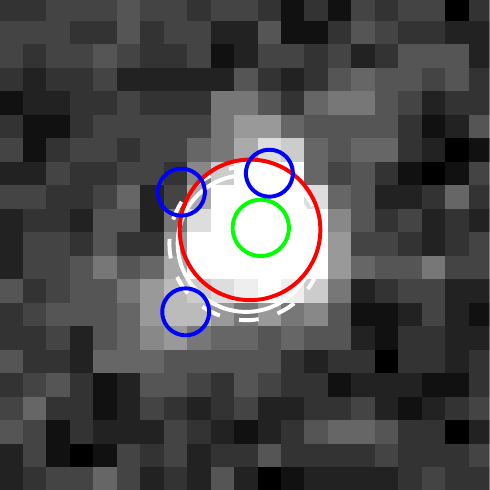}
\includegraphics[width=0.33\columnwidth]{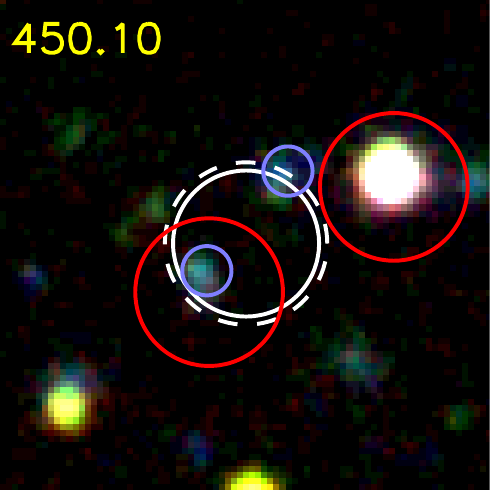}
\includegraphics[width=0.33\columnwidth]{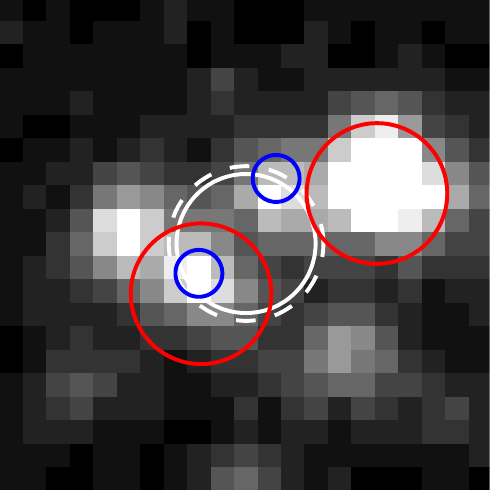}
\includegraphics[width=0.33\columnwidth]{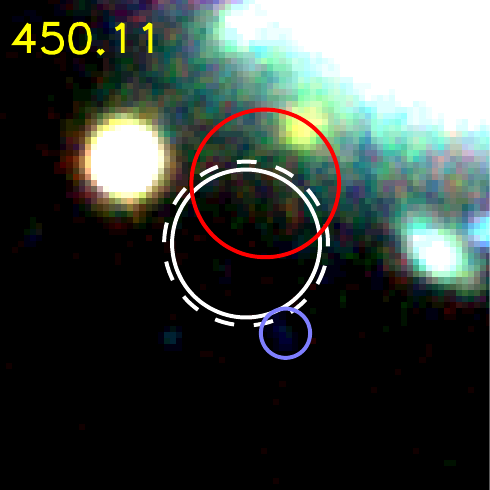}
\includegraphics[width=0.33\columnwidth]{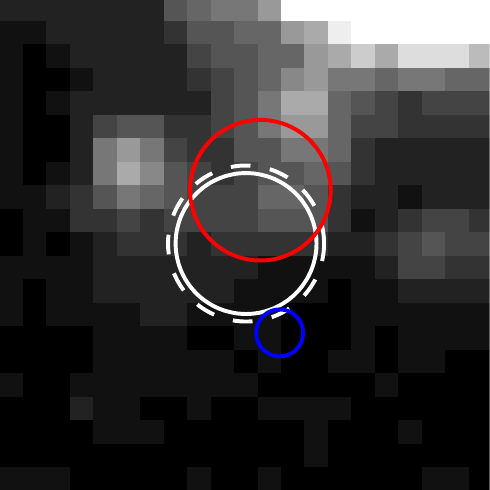}
\includegraphics[width=0.33\columnwidth]{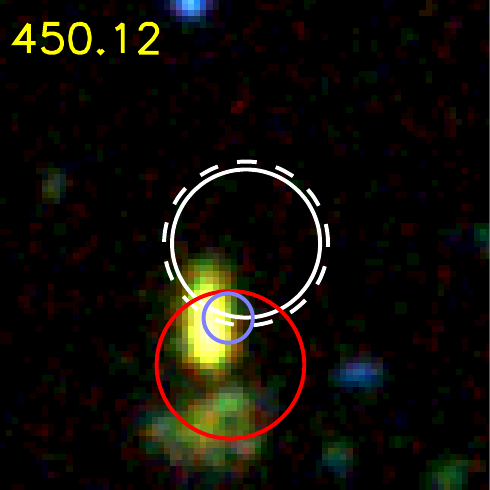}
\includegraphics[width=0.33\columnwidth]{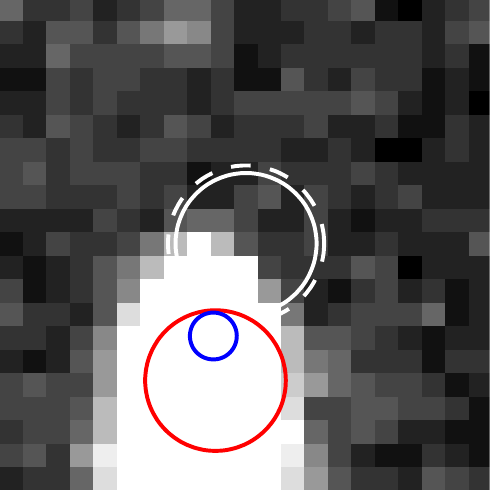}
\includegraphics[width=0.33\columnwidth]{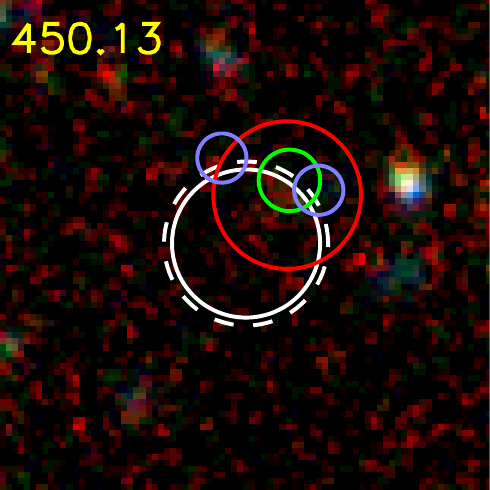}
\includegraphics[width=0.33\columnwidth]{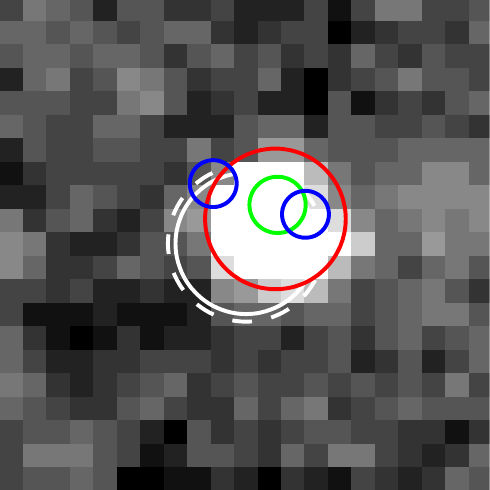}
\includegraphics[width=0.33\columnwidth]{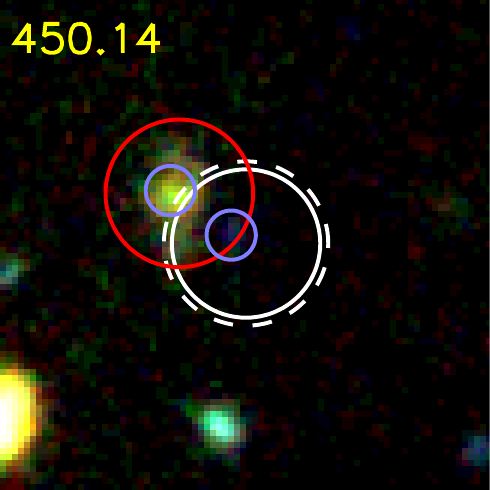}
\includegraphics[width=0.33\columnwidth]{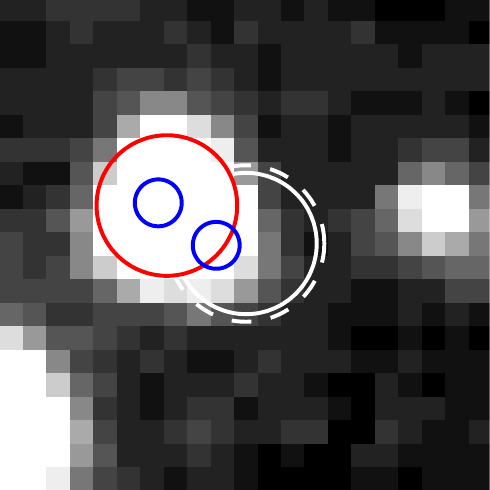}
\includegraphics[width=0.33\columnwidth]{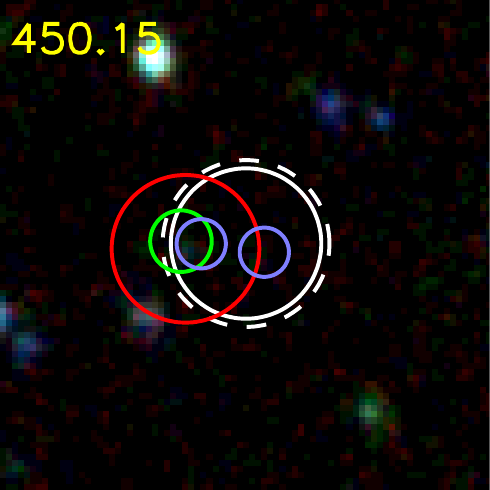}
\includegraphics[width=0.33\columnwidth]{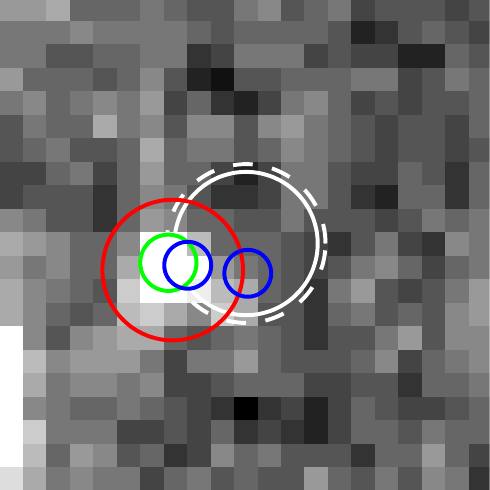}
\includegraphics[width=0.33\columnwidth]{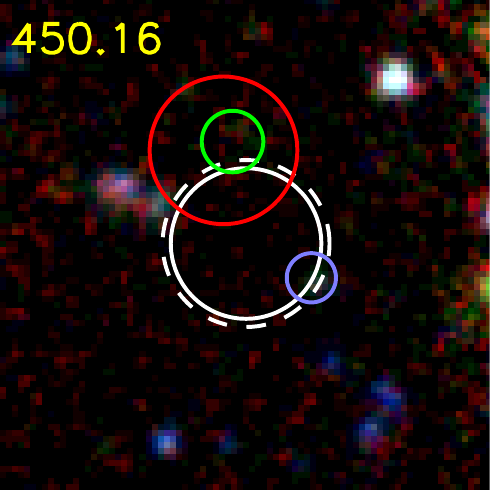}
\includegraphics[width=0.33\columnwidth]{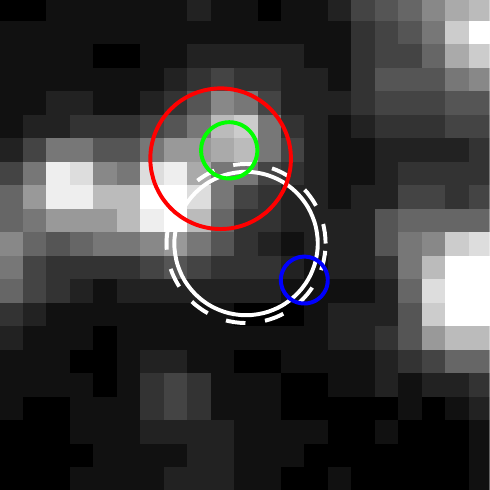}
\includegraphics[width=0.33\columnwidth]{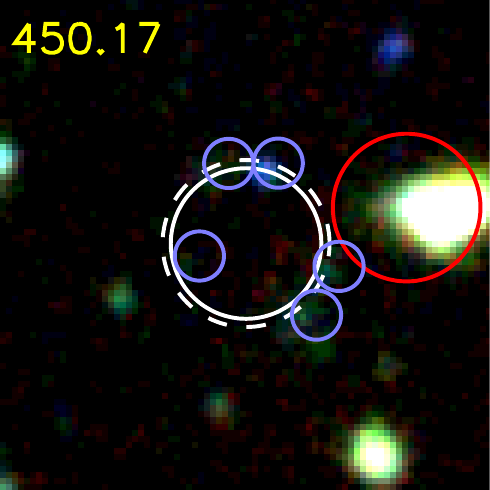}
\includegraphics[width=0.33\columnwidth]{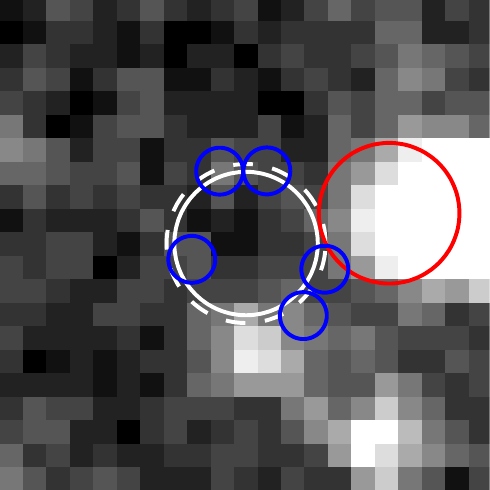}
\includegraphics[width=0.33\columnwidth]{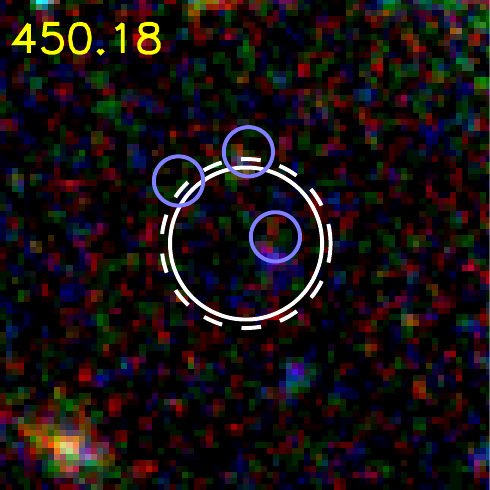}
\includegraphics[width=0.33\columnwidth]{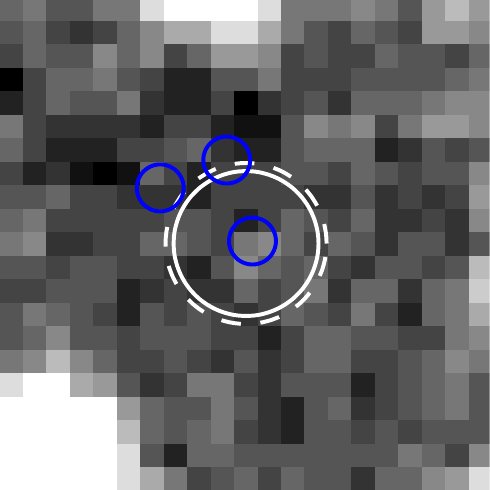}
\includegraphics[width=0.33\columnwidth]{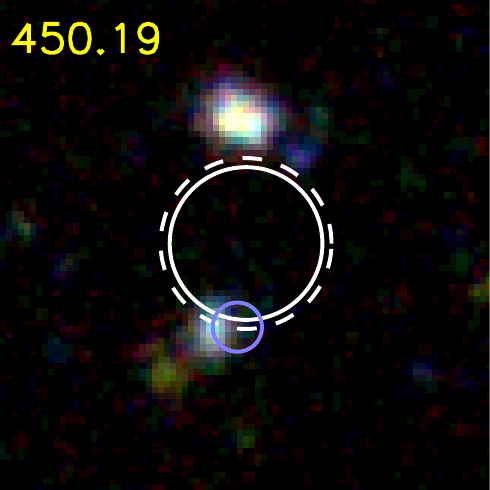}
\includegraphics[width=0.33\columnwidth]{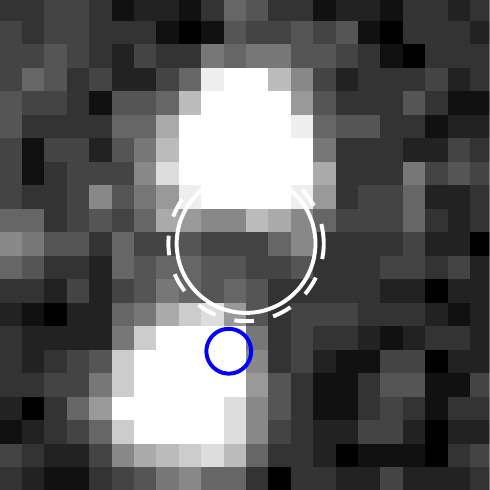}
\includegraphics[width=0.33\columnwidth]{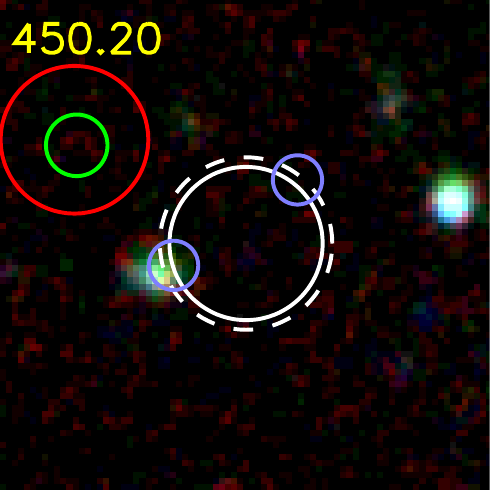}
\includegraphics[width=0.33\columnwidth]{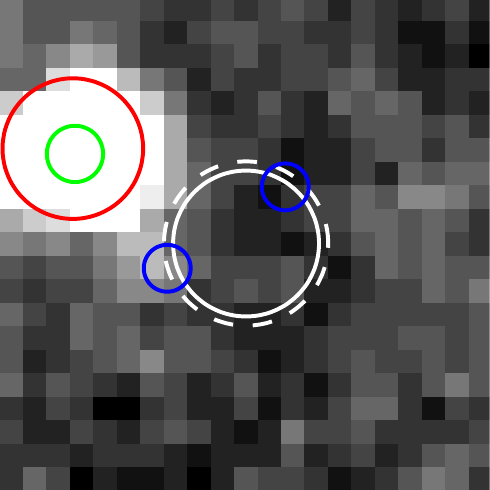}
\caption{ 12\arcsec$\times$12\arcsec\ cutouts for 450\um-detected
  sources in optical tri-colour (SUBARU B, I, and Z) and {\it
    Spitzer}-{\sc IRAC} 3.6\um.  White solid circles indicate the
  position of the 450\um\ source with associated positional
  uncertainty, varying 1.5--3\arcsec.  The outter ring (white dashed
  circle) corresponds to the 95\%\ confidence interval on position,
  while the inner is the 90\%\ interval.  {\it Spitzer}-{\sc Mips}
  24\um\ source positions are marked with red circles and 1.4\,GHz
  source positions are marked with green circles.  Optical
  counterparts are marked with blue circles.  }
\label{fig:cutouts450}
\end{figure*}
\clearpage
\begin{center}
\includegraphics[width=0.32\columnwidth]{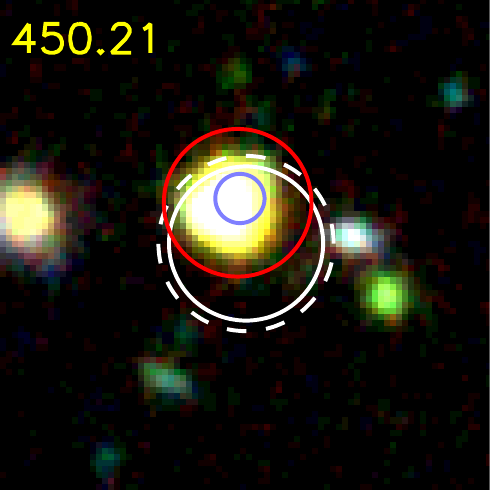}\includegraphics[width=0.32\columnwidth]{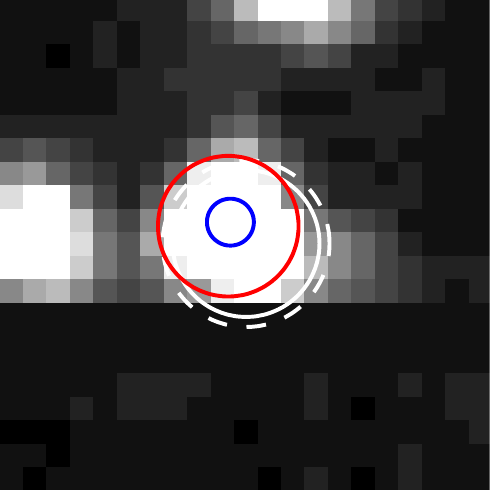}\includegraphics[width=0.32\columnwidth]{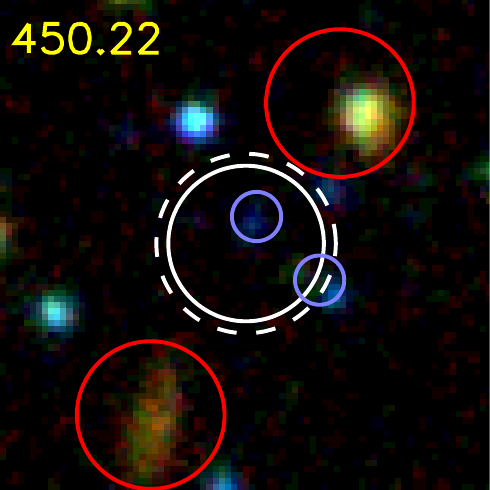}\includegraphics[width=0.32\columnwidth]{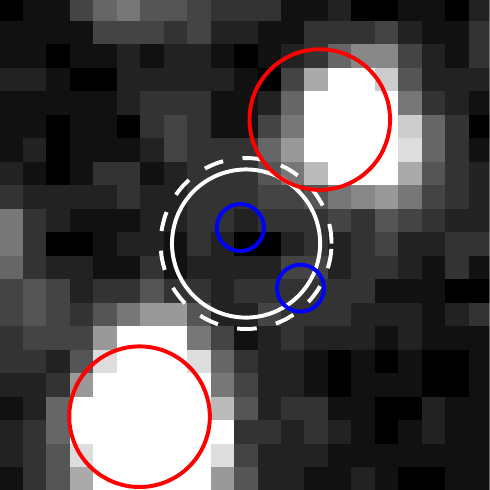}\includegraphics[width=0.32\columnwidth]{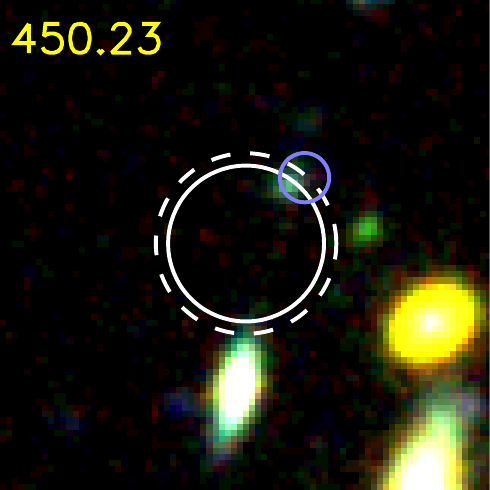}\includegraphics[width=0.32\columnwidth]{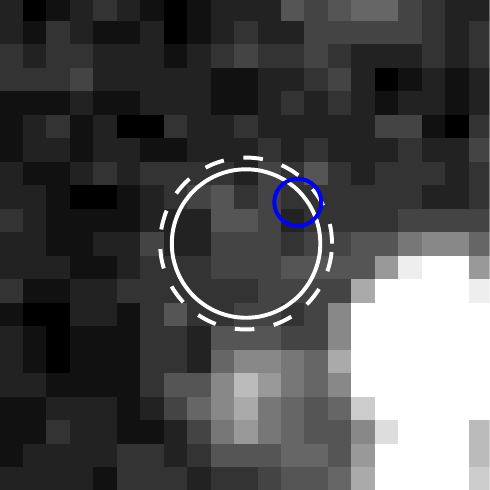}
\includegraphics[width=0.32\columnwidth]{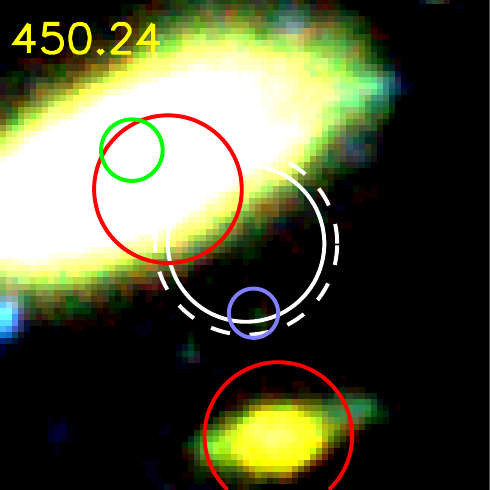}\includegraphics[width=0.32\columnwidth]{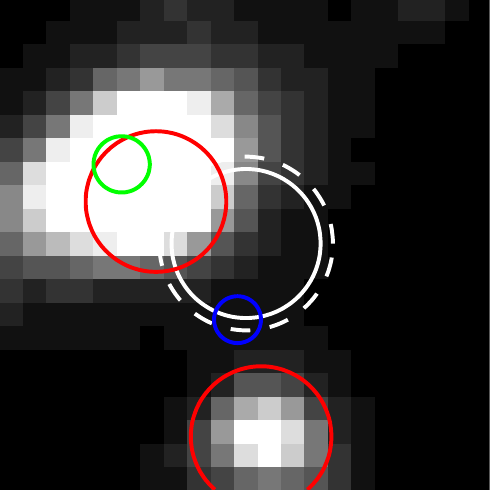}\includegraphics[width=0.32\columnwidth]{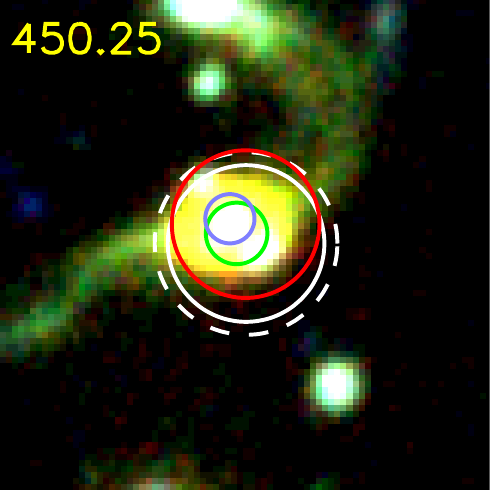}\includegraphics[width=0.32\columnwidth]{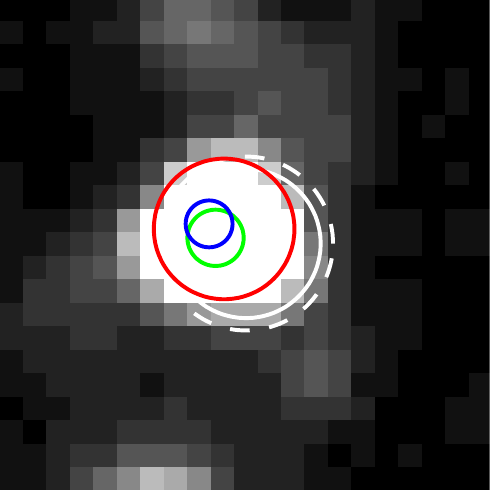}\includegraphics[width=0.32\columnwidth]{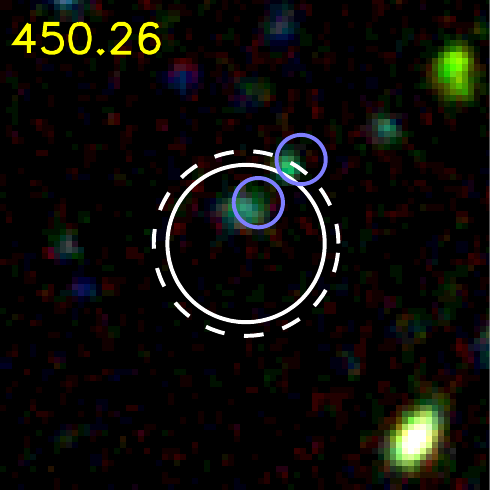}\includegraphics[width=0.32\columnwidth]{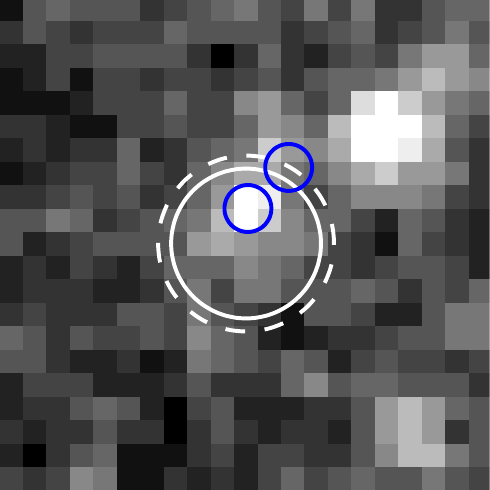}
\includegraphics[width=0.32\columnwidth]{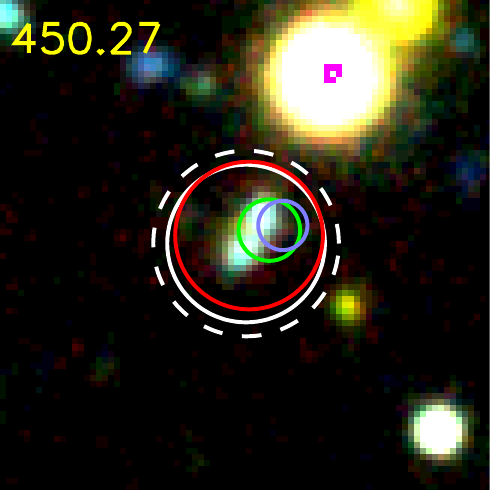}\includegraphics[width=0.32\columnwidth]{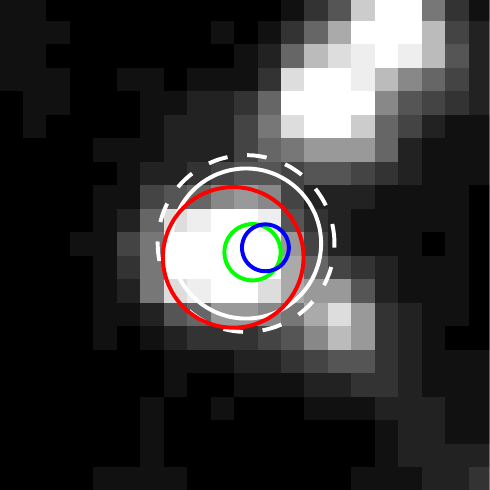}\includegraphics[width=0.32\columnwidth]{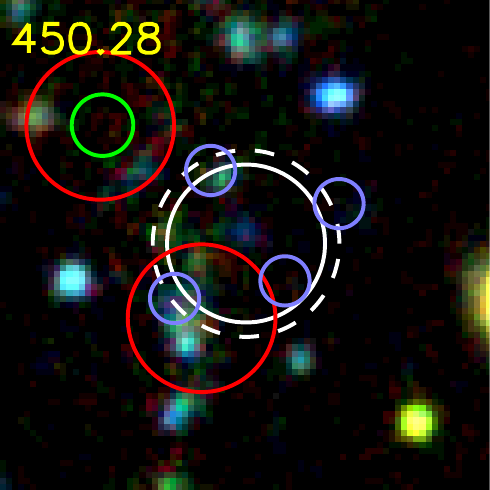}\includegraphics[width=0.32\columnwidth]{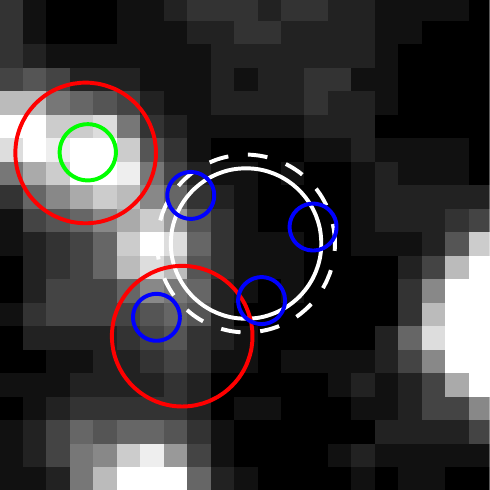}\includegraphics[width=0.32\columnwidth]{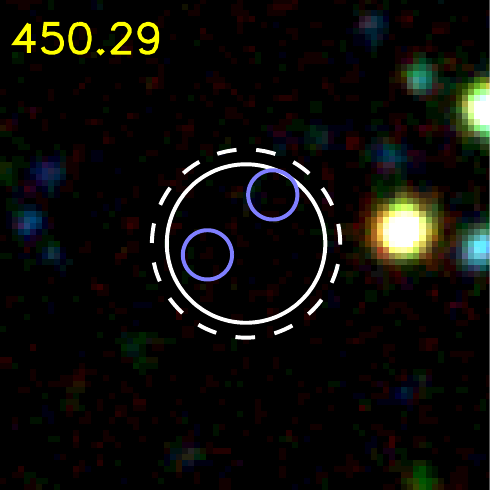}\includegraphics[width=0.32\columnwidth]{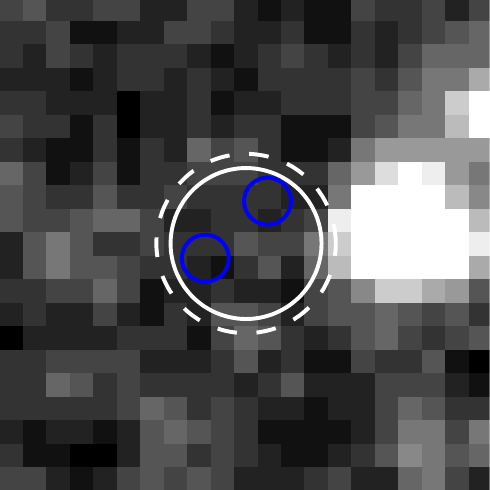}
\includegraphics[width=0.32\columnwidth]{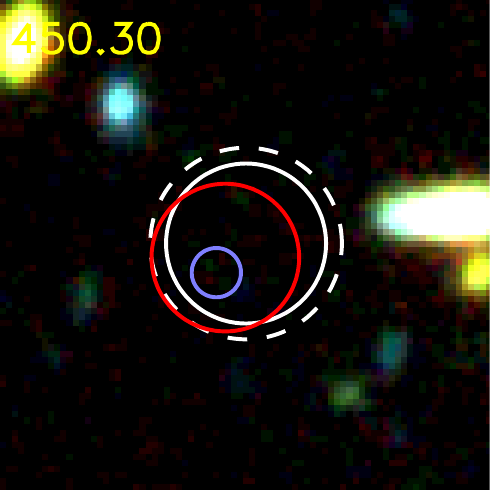}\includegraphics[width=0.32\columnwidth]{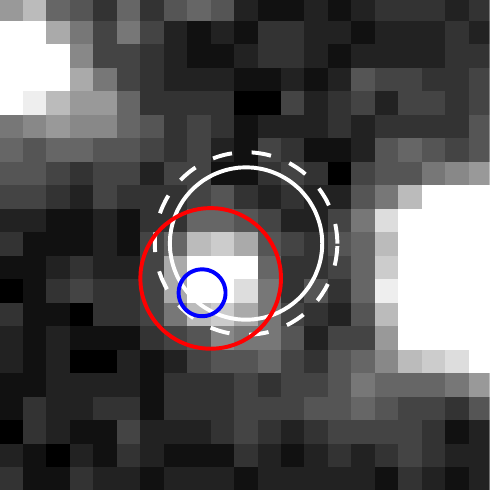}\includegraphics[width=0.32\columnwidth]{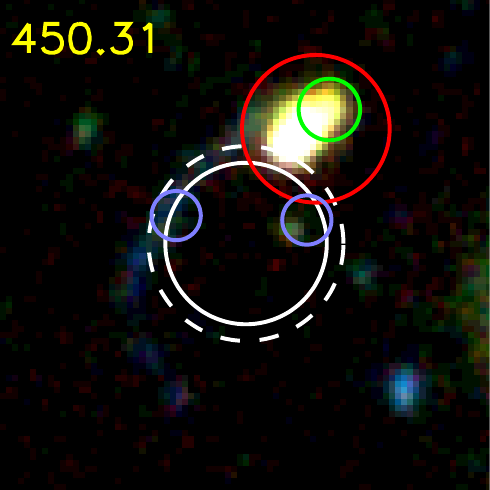}\includegraphics[width=0.32\columnwidth]{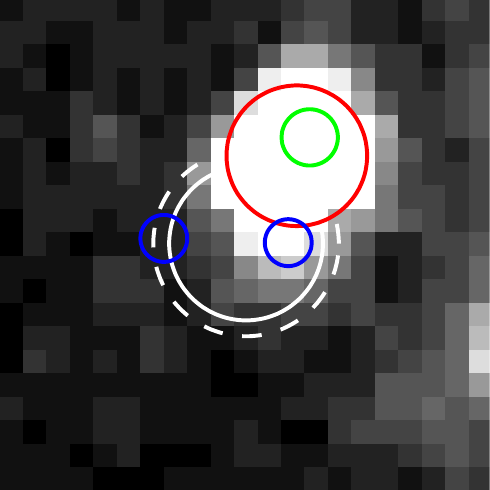}\includegraphics[width=0.32\columnwidth]{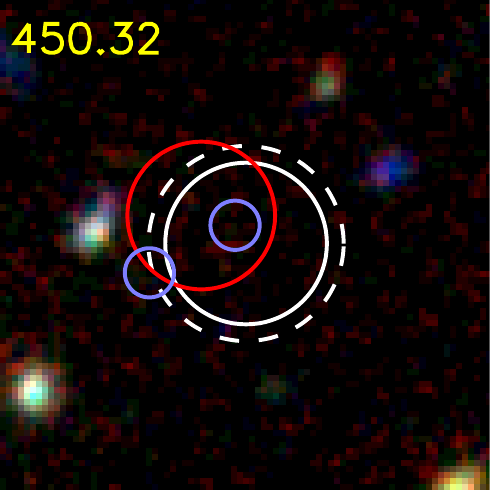}\includegraphics[width=0.32\columnwidth]{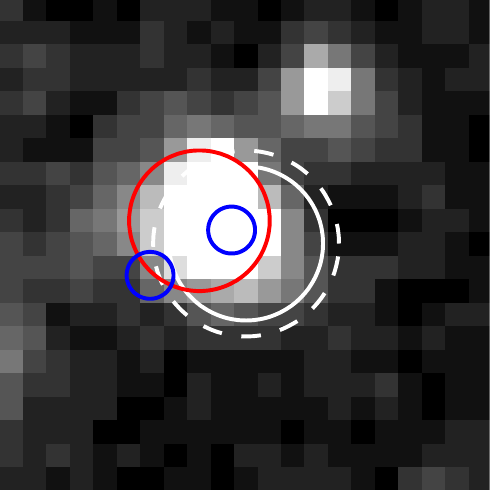}
\includegraphics[width=0.32\columnwidth]{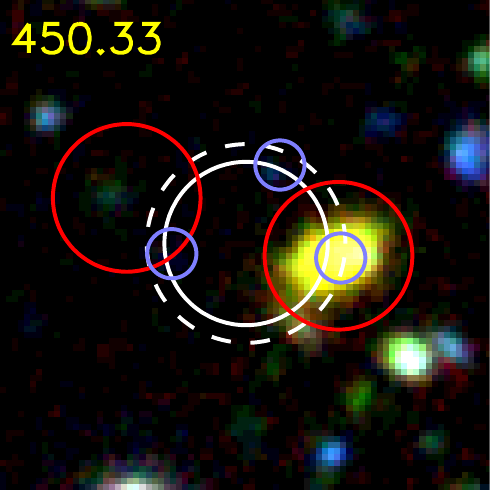}\includegraphics[width=0.32\columnwidth]{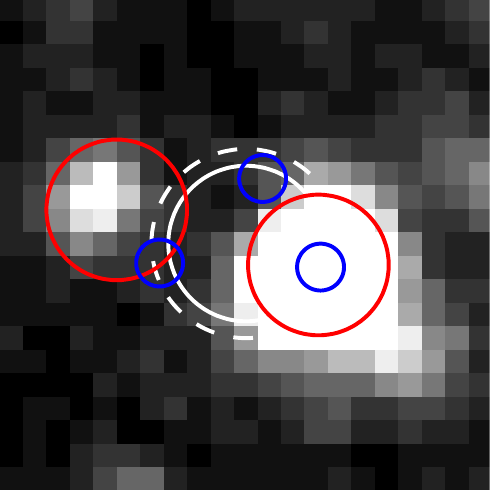}\includegraphics[width=0.32\columnwidth]{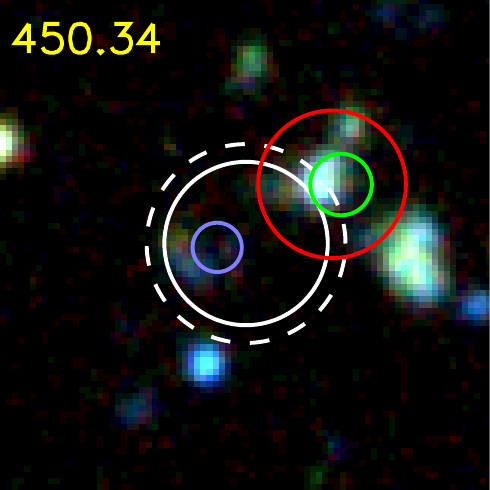}\includegraphics[width=0.32\columnwidth]{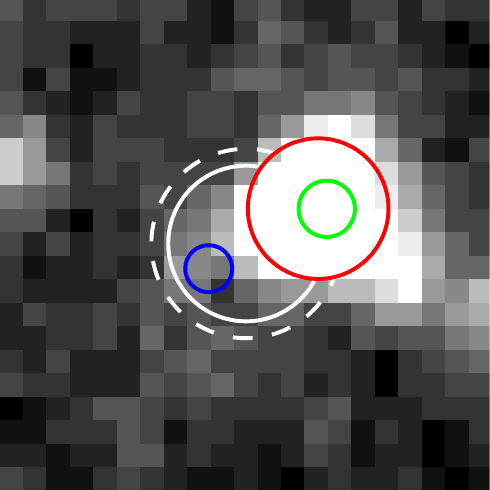}\includegraphics[width=0.32\columnwidth]{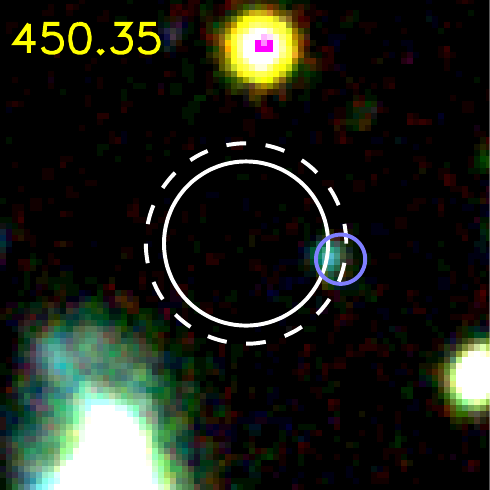}\includegraphics[width=0.32\columnwidth]{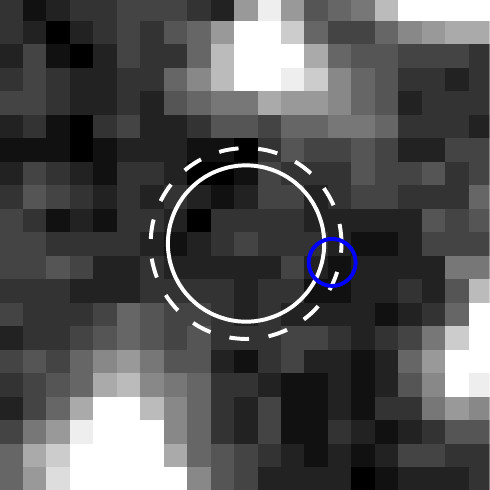}
\includegraphics[width=0.32\columnwidth]{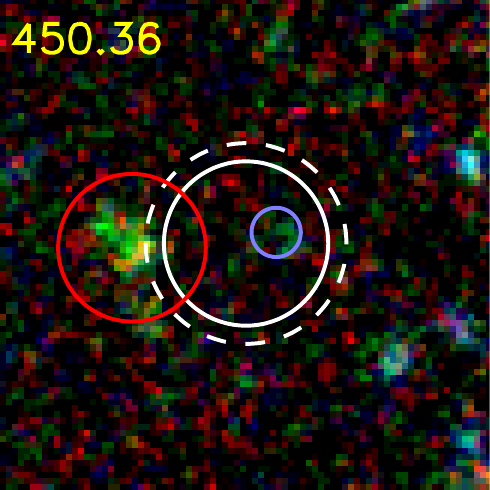}\includegraphics[width=0.32\columnwidth]{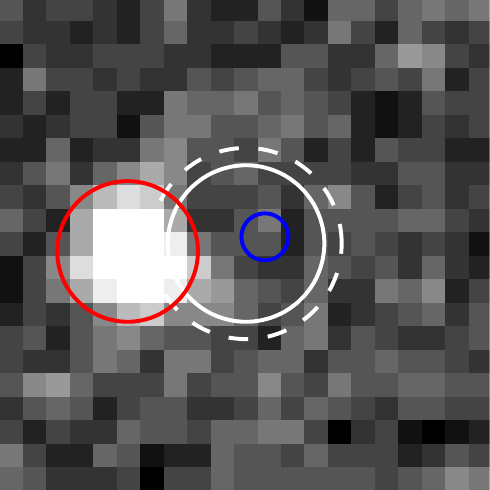}\includegraphics[width=0.32\columnwidth]{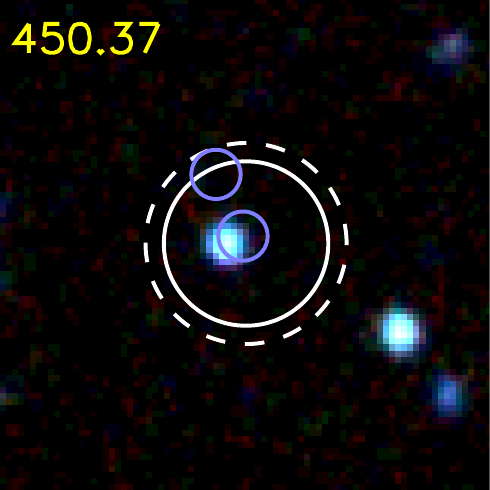}\includegraphics[width=0.32\columnwidth]{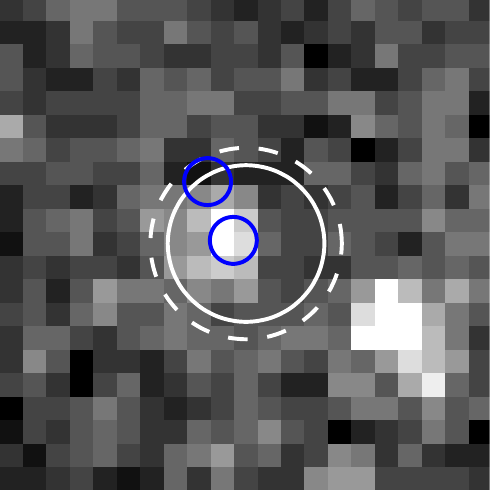}\includegraphics[width=0.32\columnwidth]{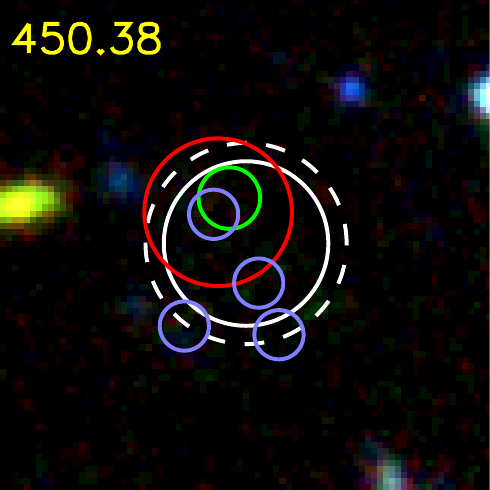}\includegraphics[width=0.32\columnwidth]{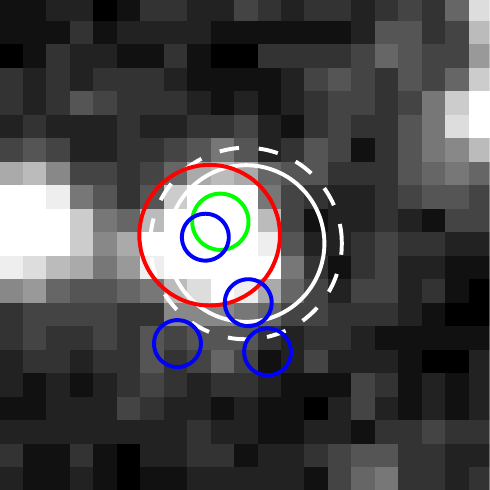}
\includegraphics[width=0.32\columnwidth]{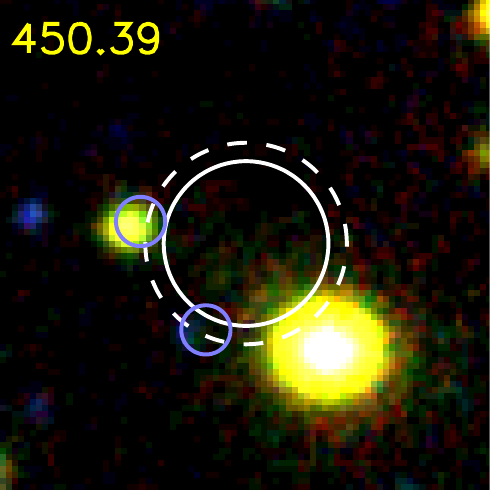}\includegraphics[width=0.32\columnwidth]{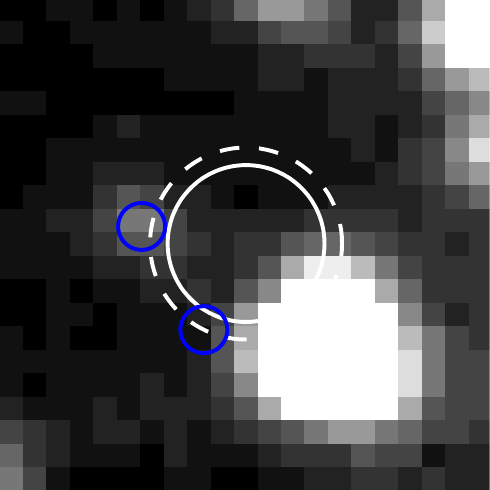}\includegraphics[width=0.32\columnwidth]{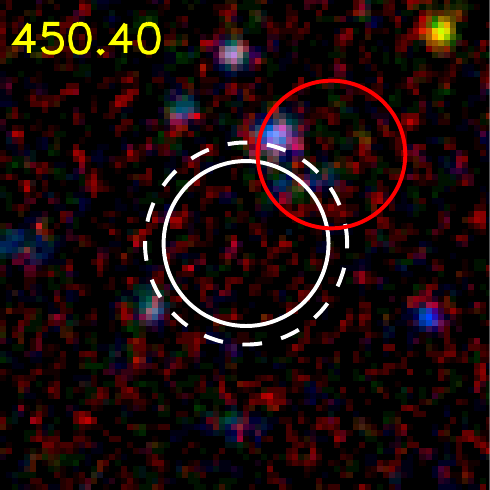}\includegraphics[width=0.32\columnwidth]{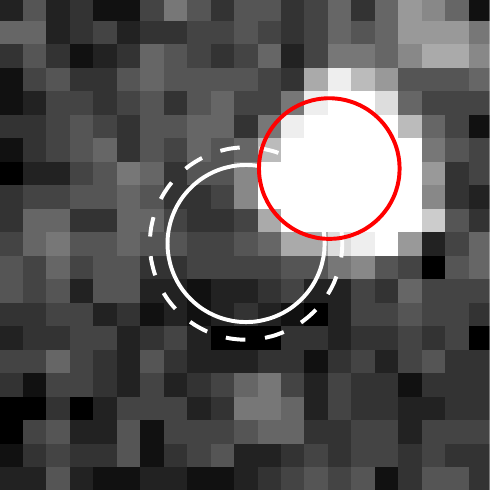}\includegraphics[width=0.32\columnwidth]{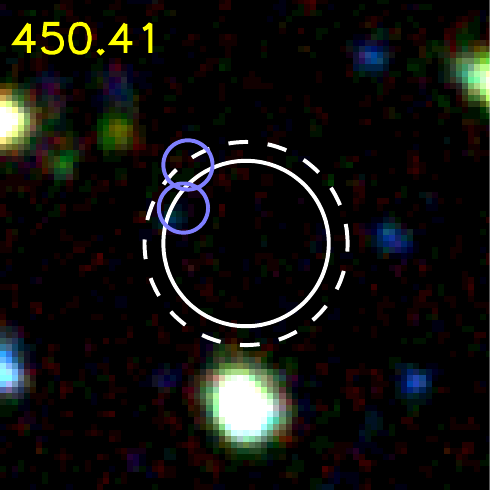}\includegraphics[width=0.32\columnwidth]{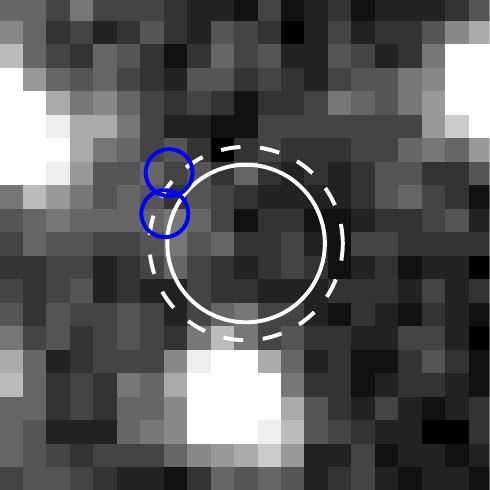}
\centerline{Figure~\ref{fig:cutouts450} -- continued.}\end{center}
\clearpage
\begin{center}
\includegraphics[width=0.32\columnwidth]{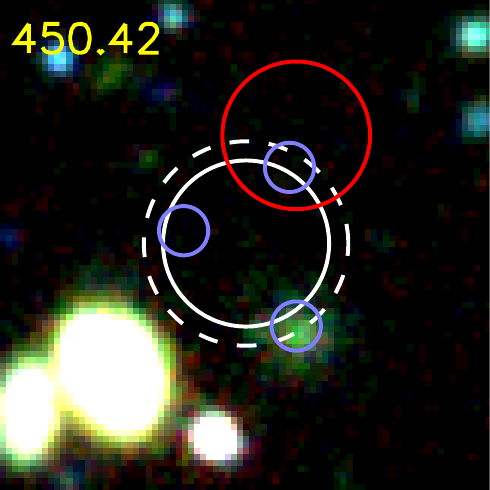}\includegraphics[width=0.32\columnwidth]{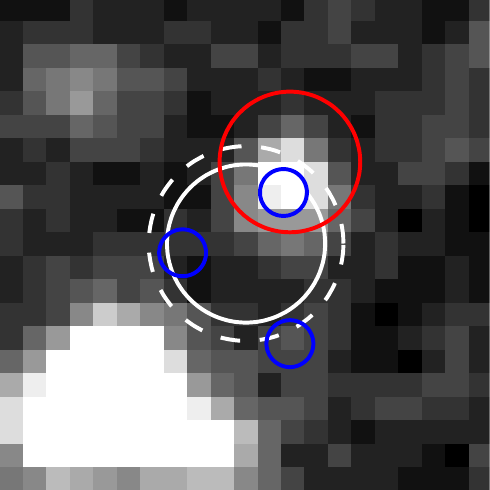}\includegraphics[width=0.32\columnwidth]{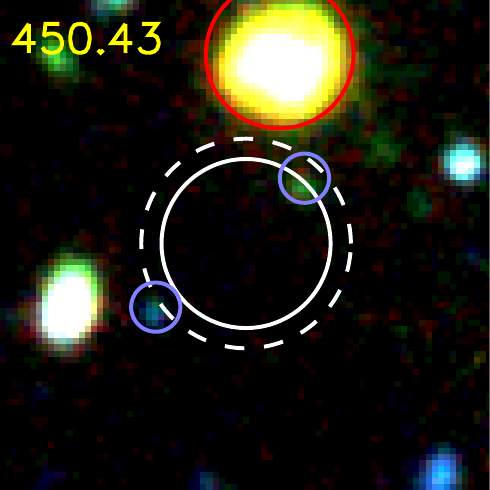}\includegraphics[width=0.32\columnwidth]{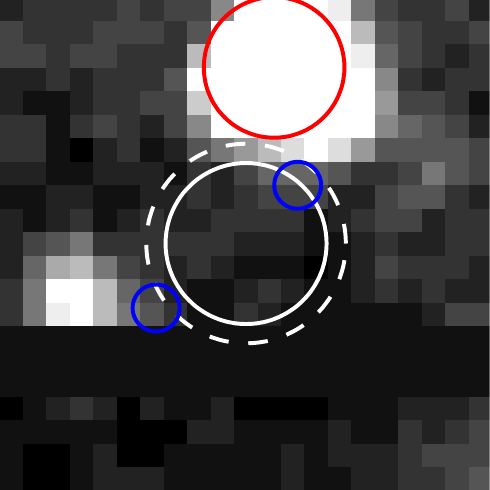}\includegraphics[width=0.32\columnwidth]{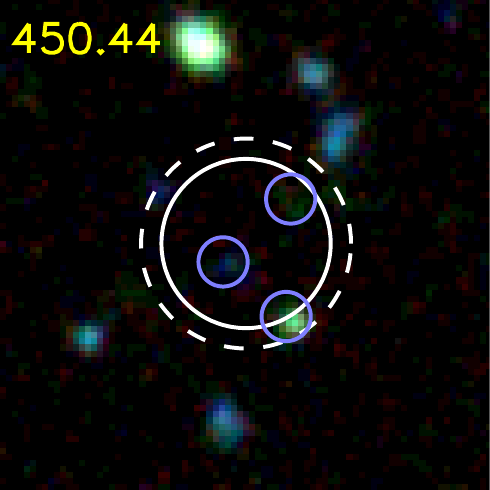}\includegraphics[width=0.32\columnwidth]{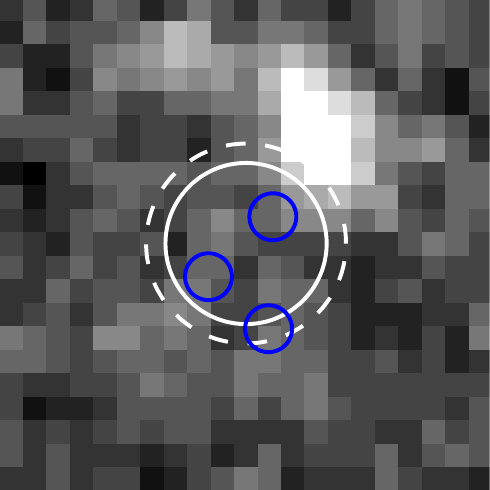}
\includegraphics[width=0.32\columnwidth]{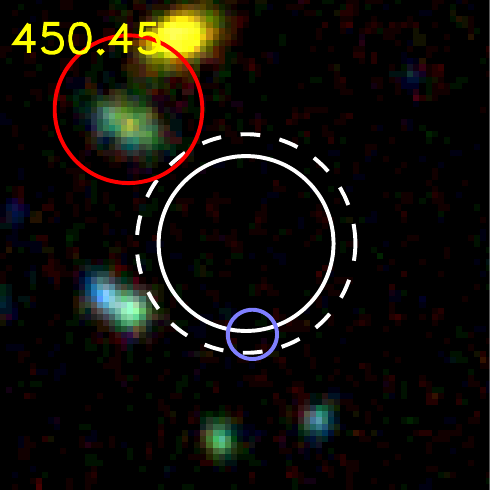}\includegraphics[width=0.32\columnwidth]{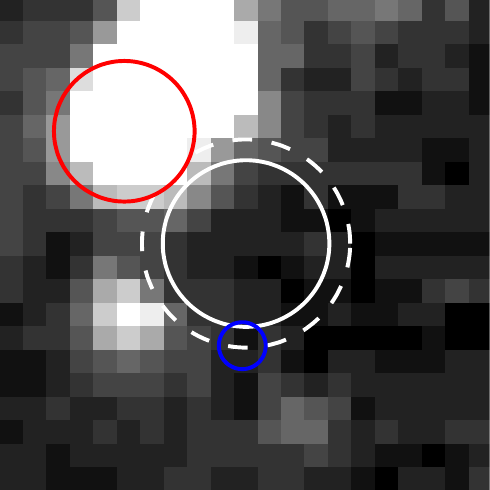}\includegraphics[width=0.32\columnwidth]{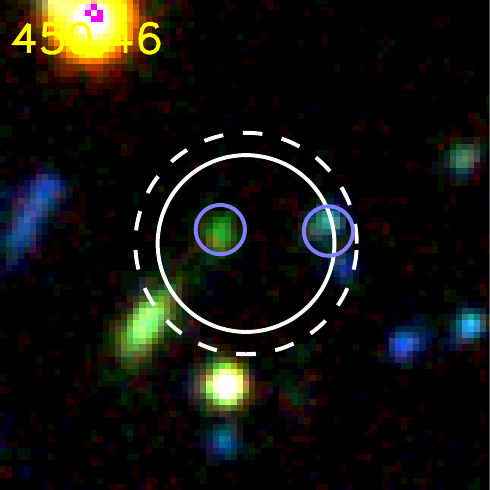}\includegraphics[width=0.32\columnwidth]{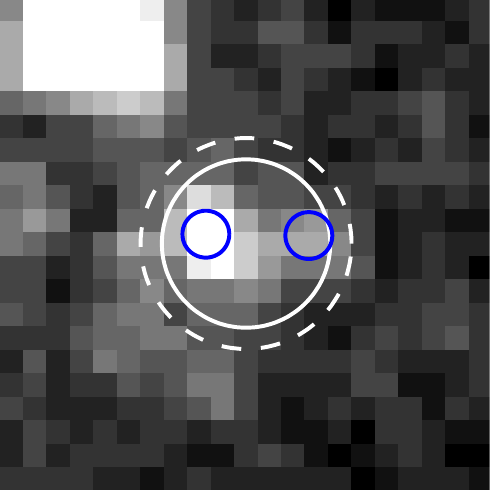}\includegraphics[width=0.32\columnwidth]{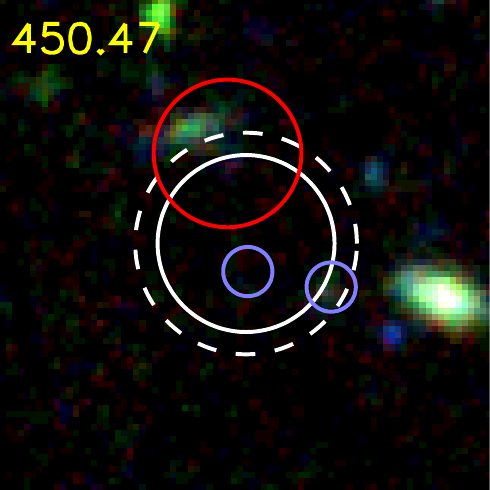}\includegraphics[width=0.32\columnwidth]{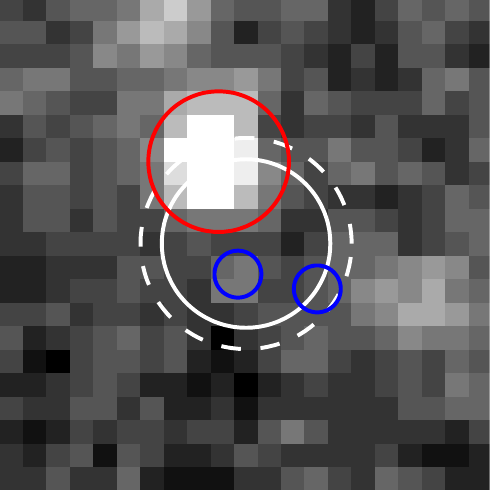}
\includegraphics[width=0.32\columnwidth]{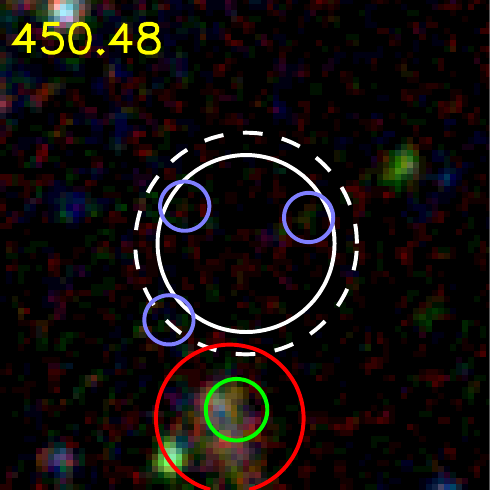}\includegraphics[width=0.32\columnwidth]{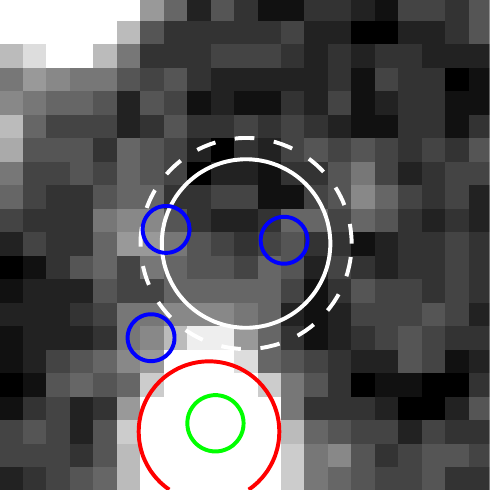}\includegraphics[width=0.32\columnwidth]{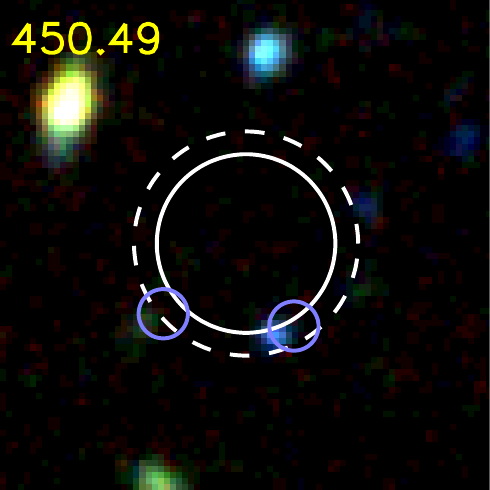}\includegraphics[width=0.32\columnwidth]{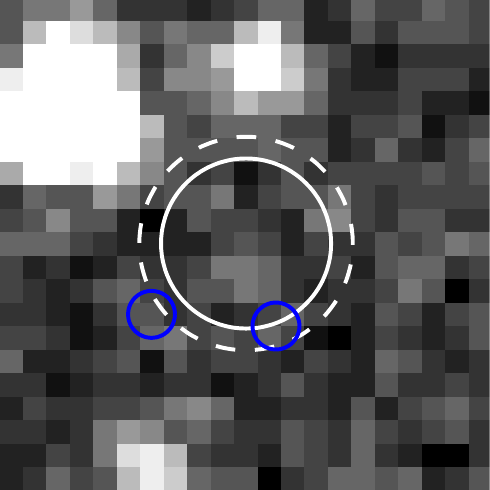}\includegraphics[width=0.32\columnwidth]{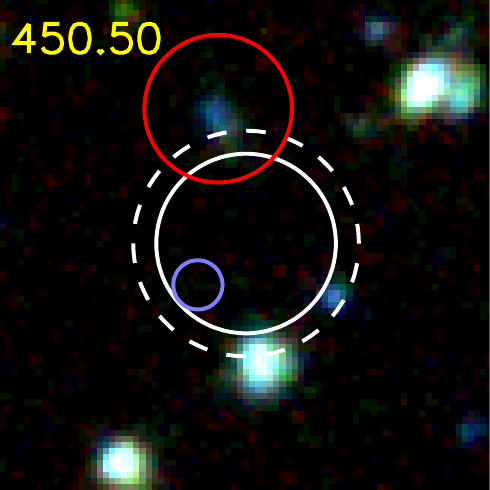}\includegraphics[width=0.32\columnwidth]{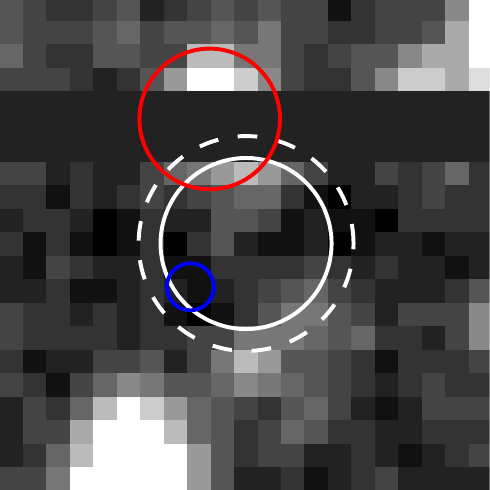}
\includegraphics[width=0.32\columnwidth]{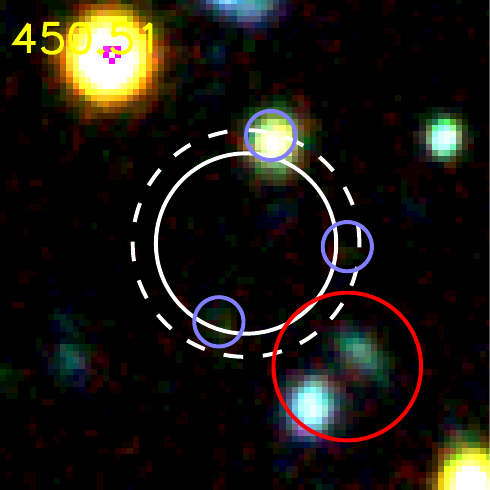}\includegraphics[width=0.32\columnwidth]{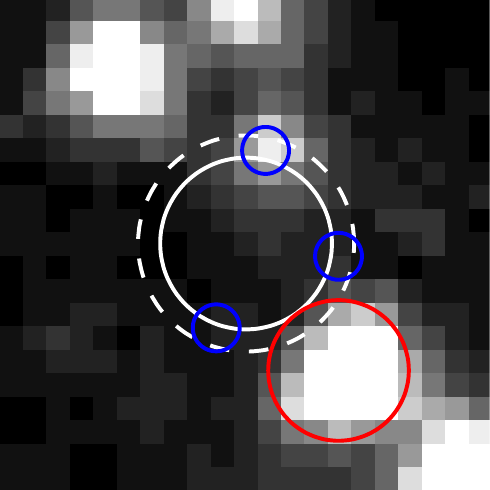}\includegraphics[width=0.32\columnwidth]{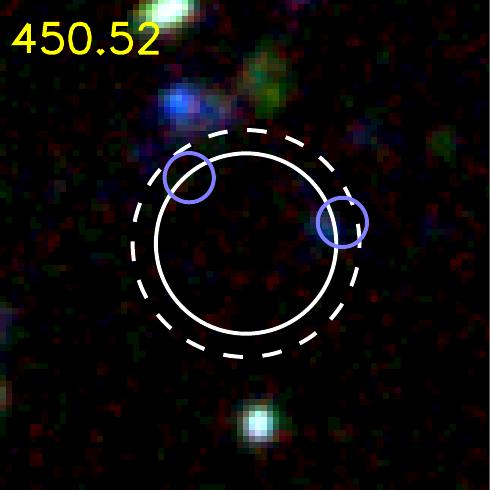}\includegraphics[width=0.32\columnwidth]{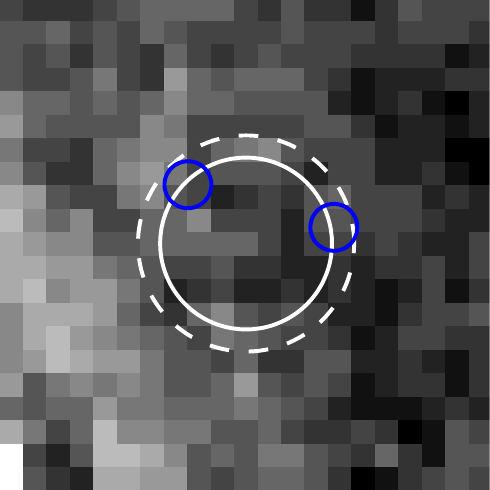}\includegraphics[width=0.32\columnwidth]{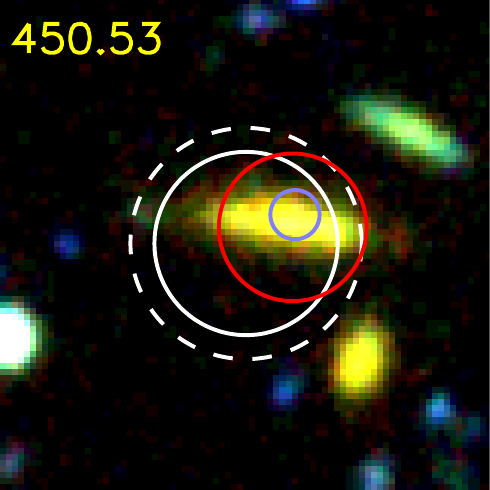}\includegraphics[width=0.32\columnwidth]{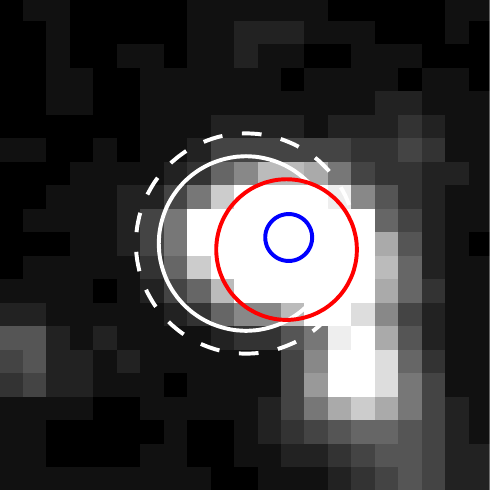}
\includegraphics[width=0.32\columnwidth]{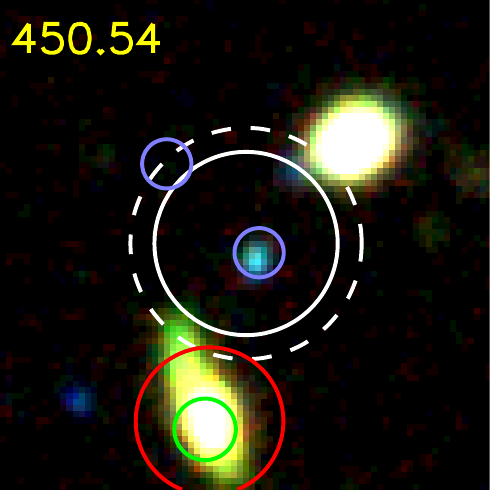}\includegraphics[width=0.32\columnwidth]{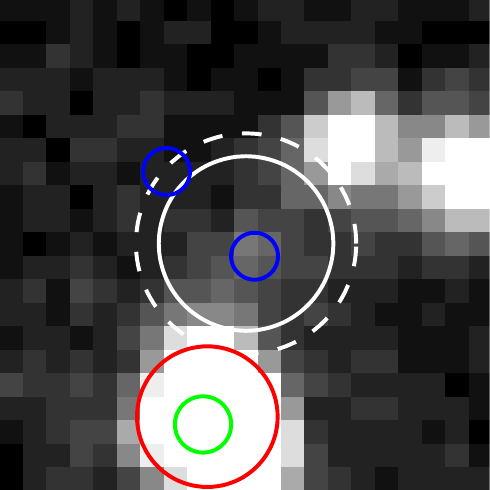}\includegraphics[width=0.32\columnwidth]{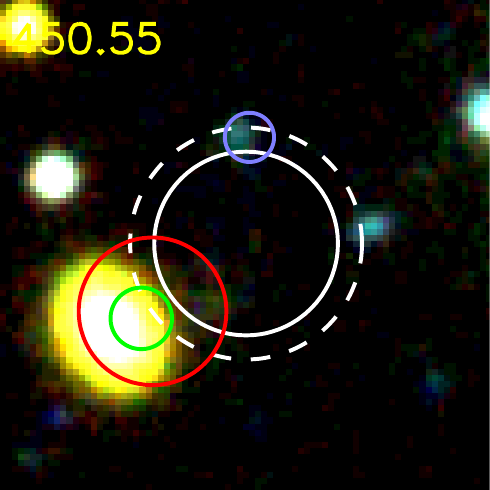}\includegraphics[width=0.32\columnwidth]{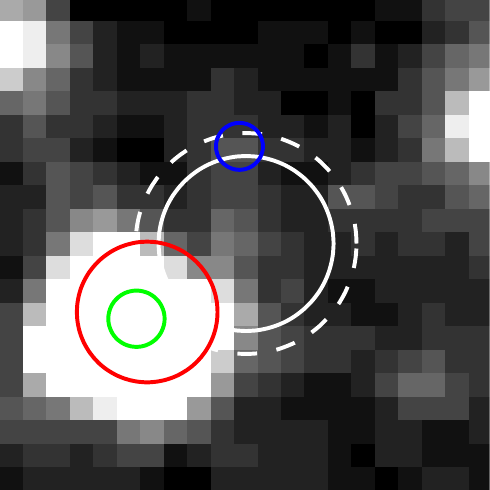}\includegraphics[width=0.32\columnwidth]{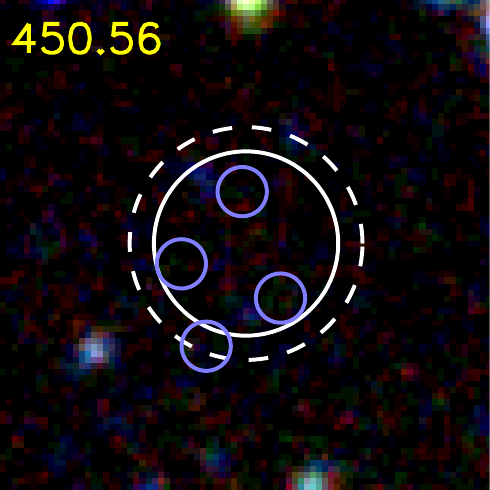}\includegraphics[width=0.32\columnwidth]{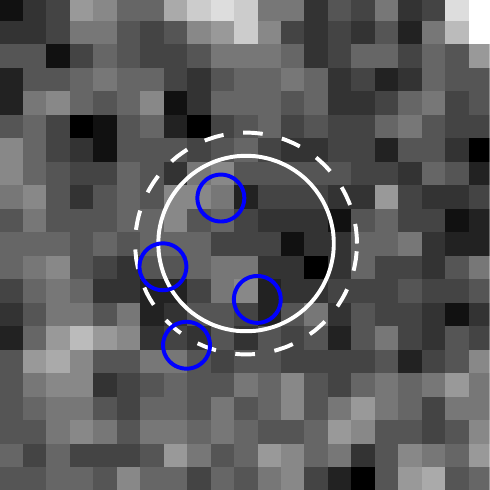}
\includegraphics[width=0.32\columnwidth]{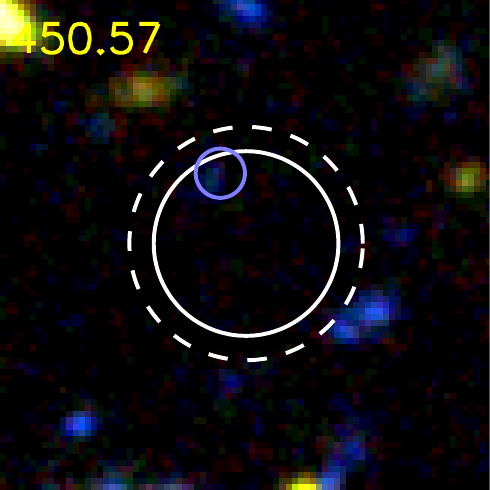}\includegraphics[width=0.32\columnwidth]{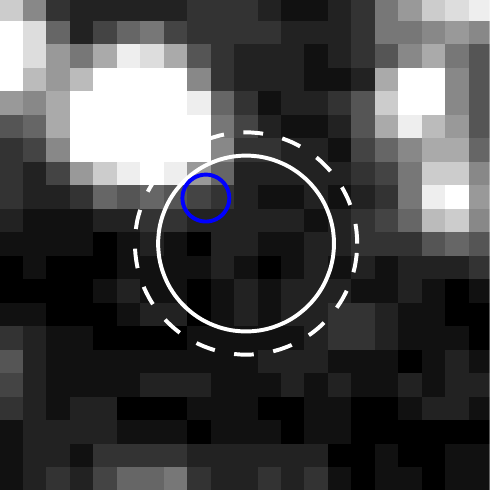}\includegraphics[width=0.32\columnwidth]{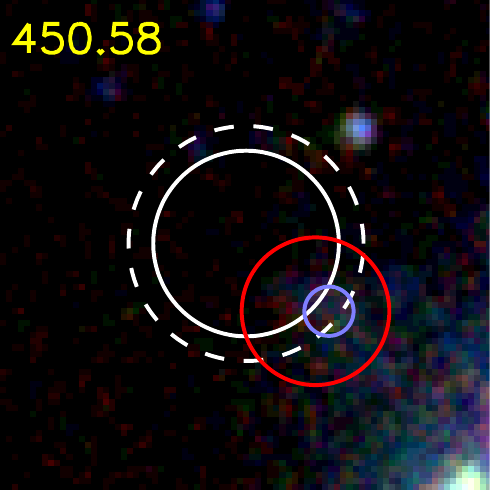}\includegraphics[width=0.32\columnwidth]{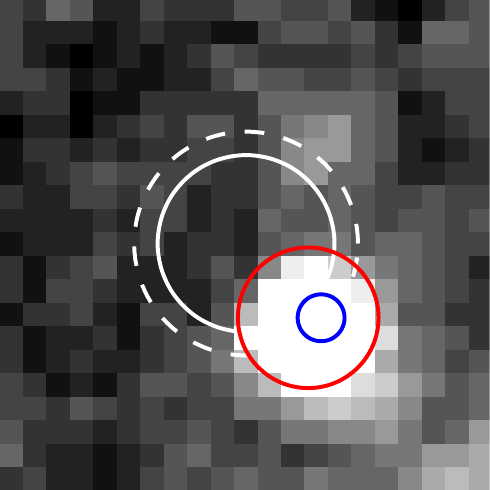}\includegraphics[width=0.32\columnwidth]{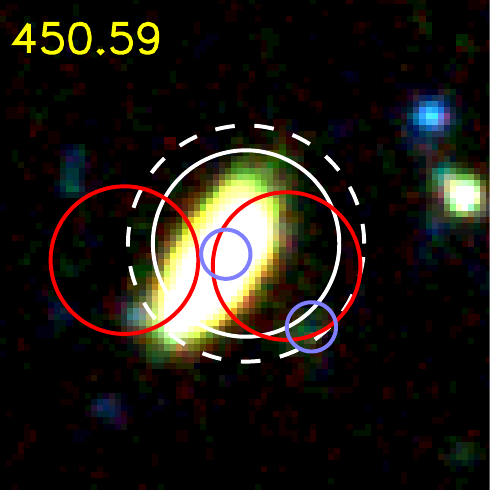}\includegraphics[width=0.32\columnwidth]{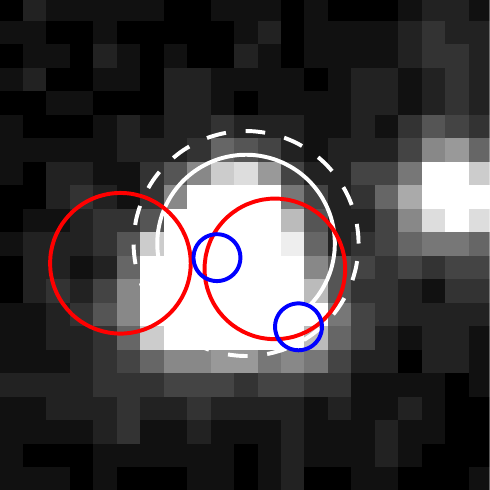}
\includegraphics[width=0.32\columnwidth]{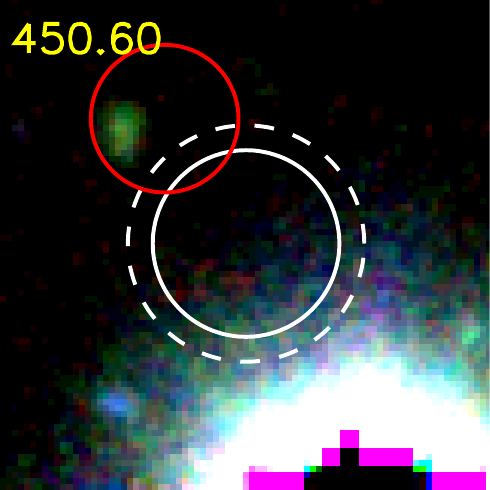}\includegraphics[width=0.32\columnwidth]{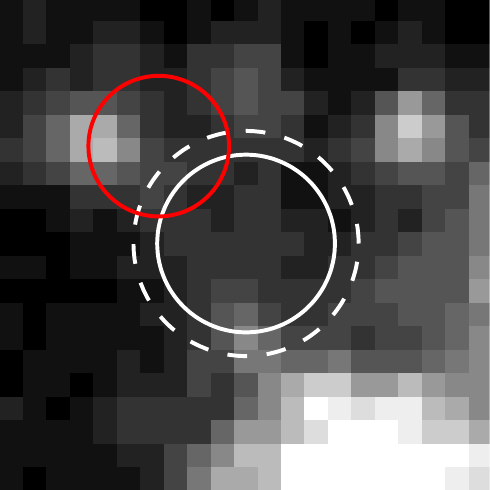}\includegraphics[width=0.32\columnwidth]{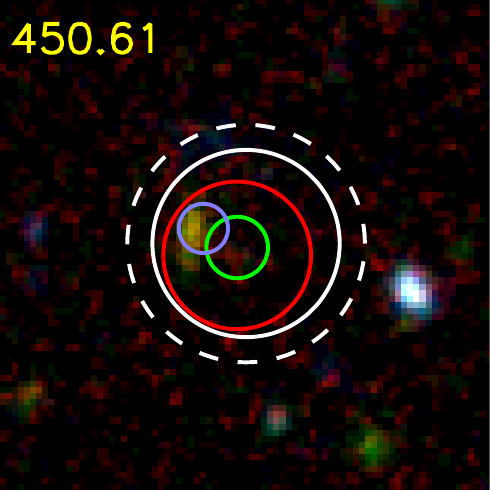}\includegraphics[width=0.32\columnwidth]{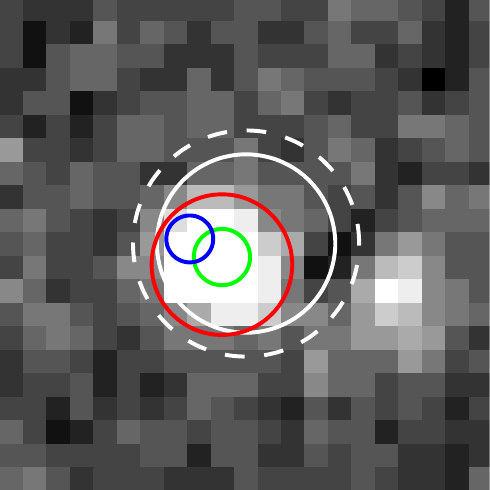}\includegraphics[width=0.32\columnwidth]{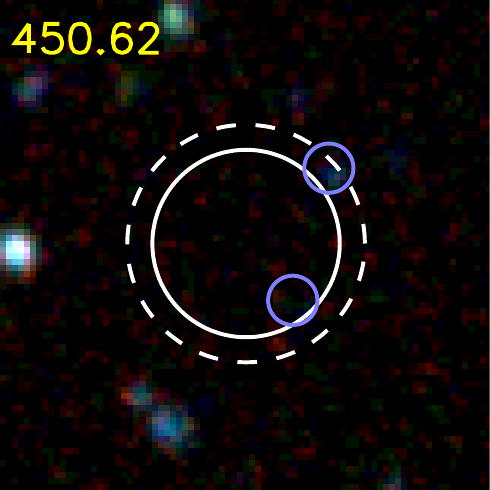}\includegraphics[width=0.32\columnwidth]{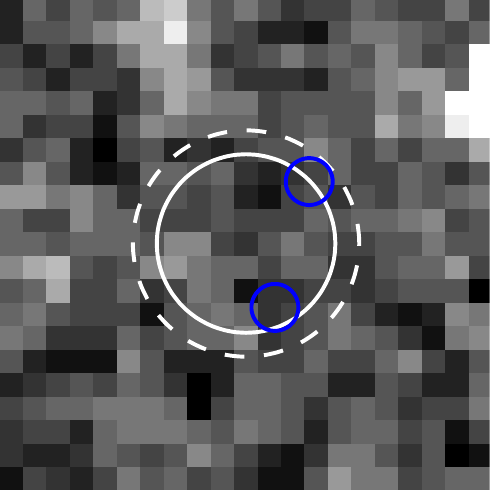}
\centerline{Figure~\ref{fig:cutouts450} -- continued.}\end{center}
\clearpage
\begin{center}
\includegraphics[width=0.32\columnwidth]{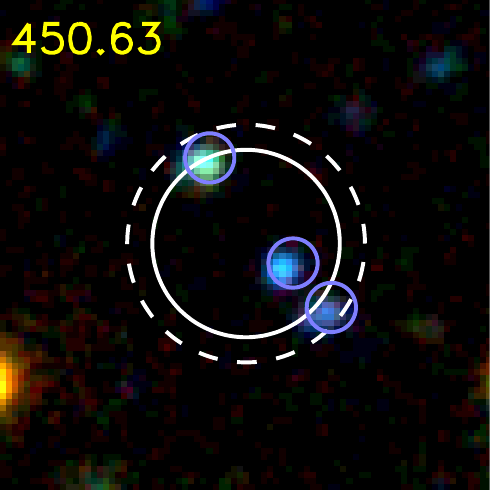}\includegraphics[width=0.32\columnwidth]{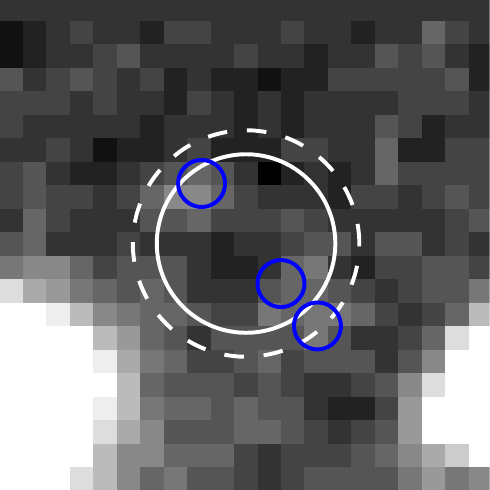}\includegraphics[width=0.32\columnwidth]{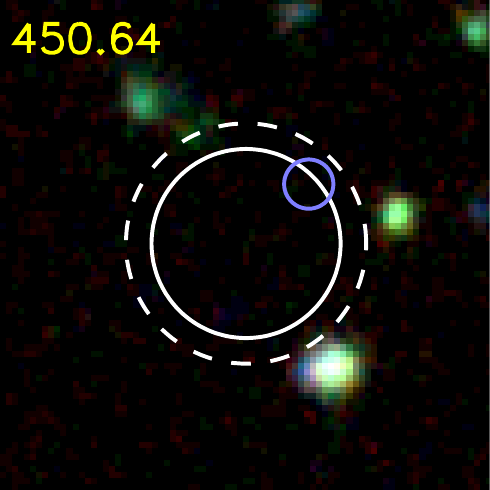}\includegraphics[width=0.32\columnwidth]{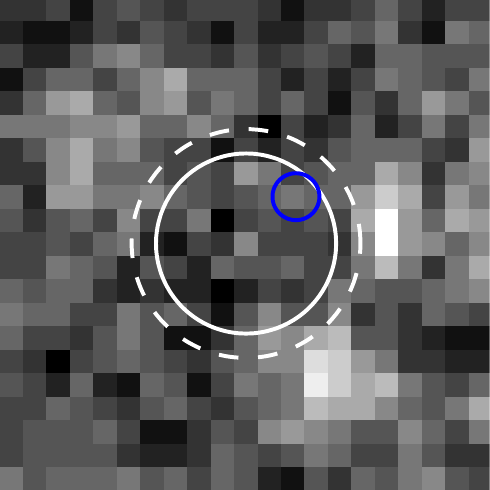}\includegraphics[width=0.32\columnwidth]{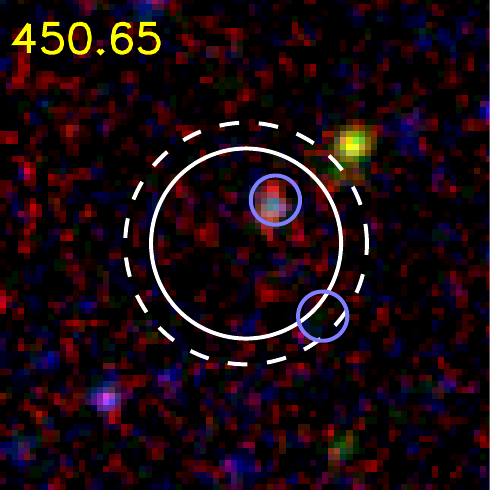}\includegraphics[width=0.32\columnwidth]{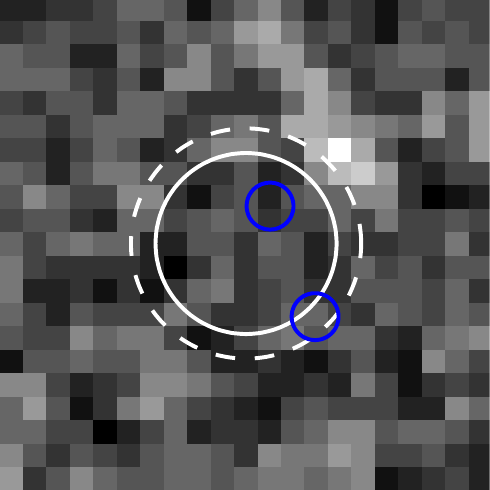}
\includegraphics[width=0.32\columnwidth]{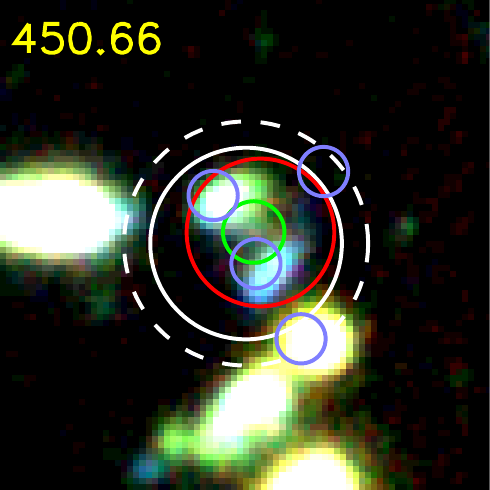}\includegraphics[width=0.32\columnwidth]{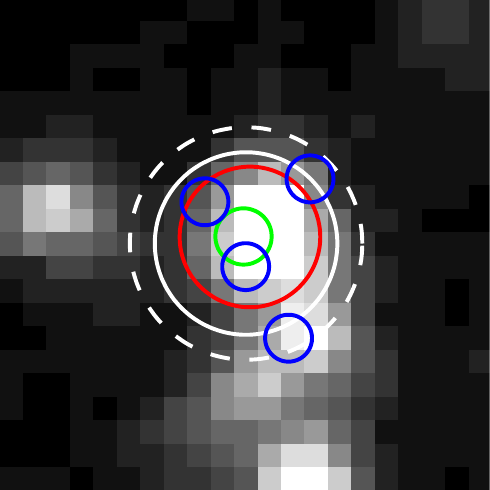}\includegraphics[width=0.32\columnwidth]{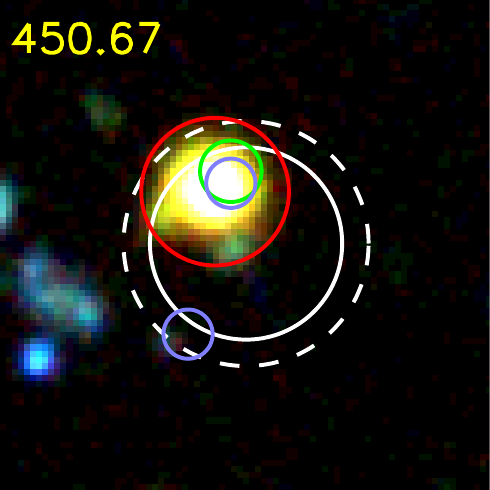}\includegraphics[width=0.32\columnwidth]{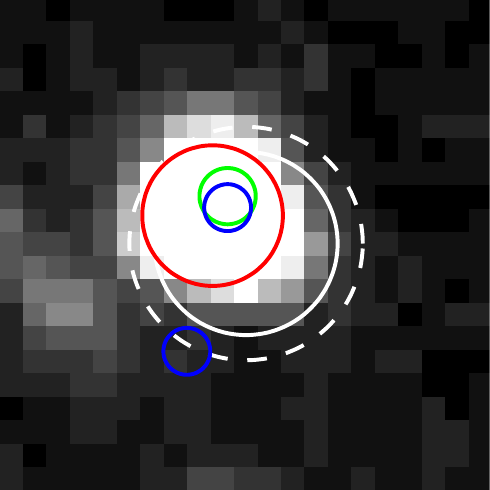}\includegraphics[width=0.32\columnwidth]{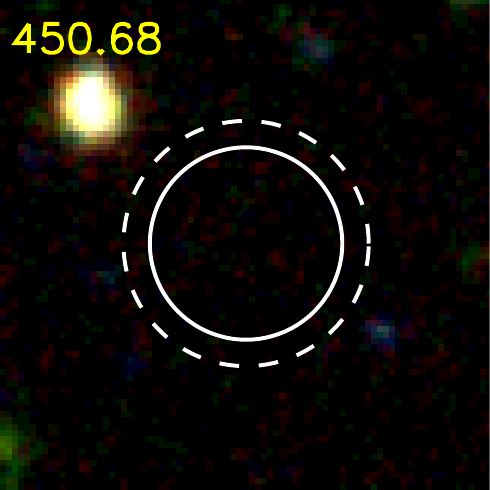}\includegraphics[width=0.32\columnwidth]{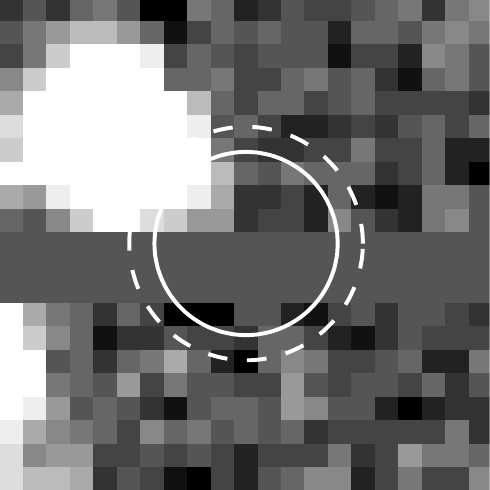}
\includegraphics[width=0.32\columnwidth]{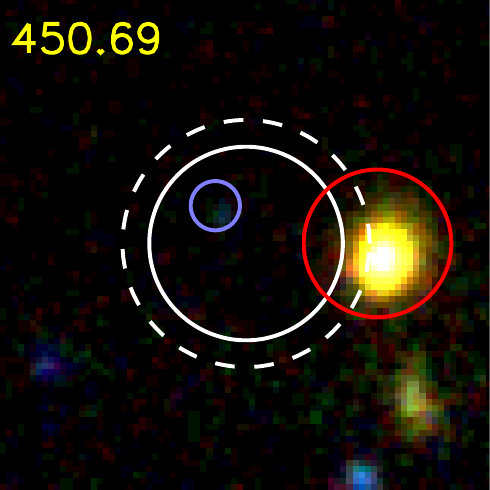}\includegraphics[width=0.32\columnwidth]{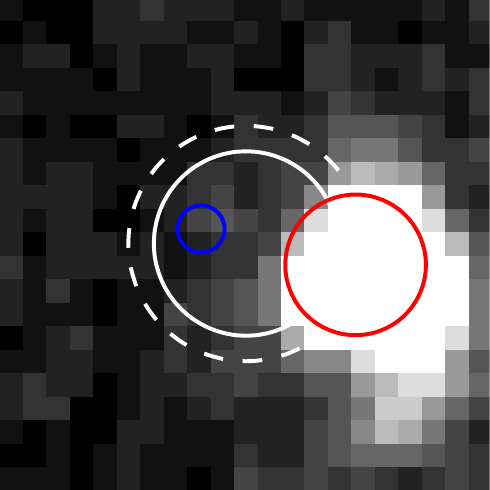}\includegraphics[width=0.32\columnwidth]{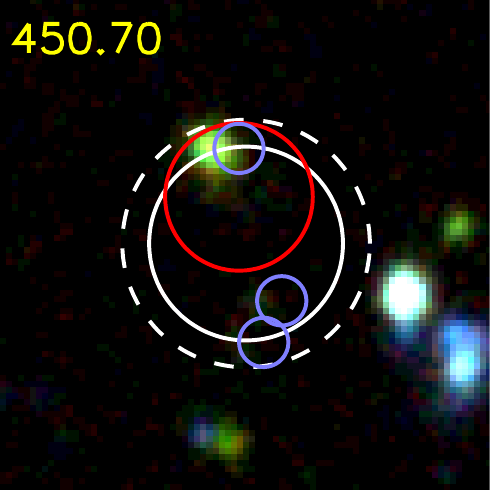}\includegraphics[width=0.32\columnwidth]{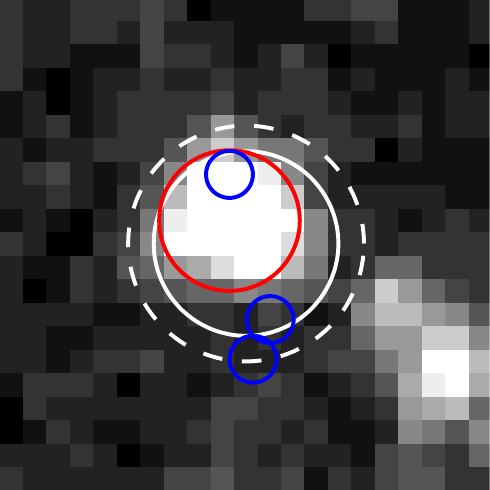}\includegraphics[width=0.32\columnwidth]{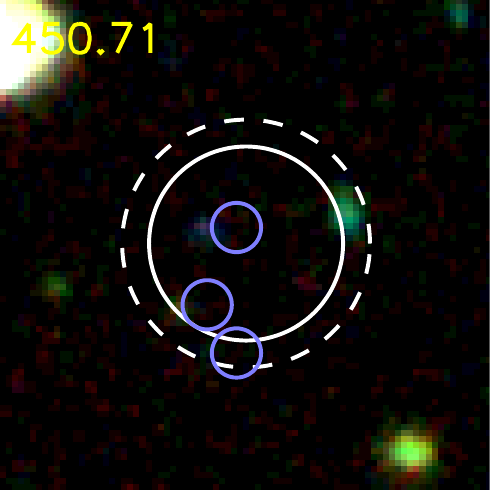}\includegraphics[width=0.32\columnwidth]{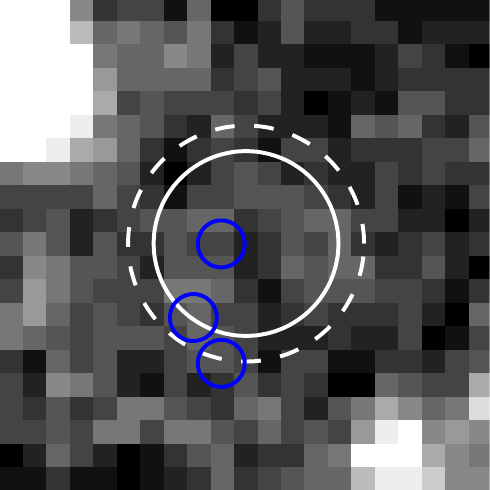}
\includegraphics[width=0.32\columnwidth]{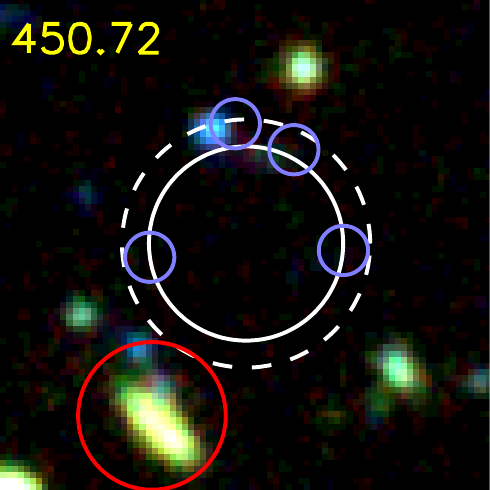}\includegraphics[width=0.32\columnwidth]{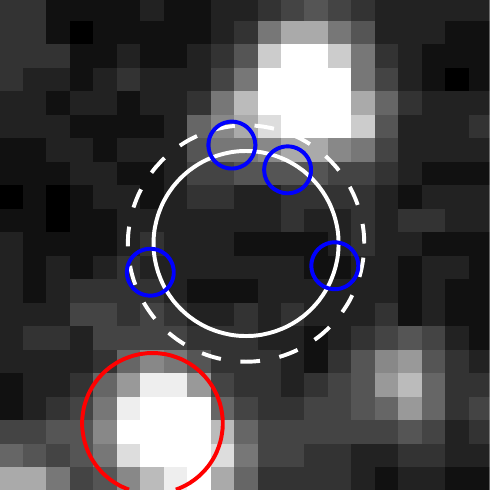}\includegraphics[width=0.32\columnwidth]{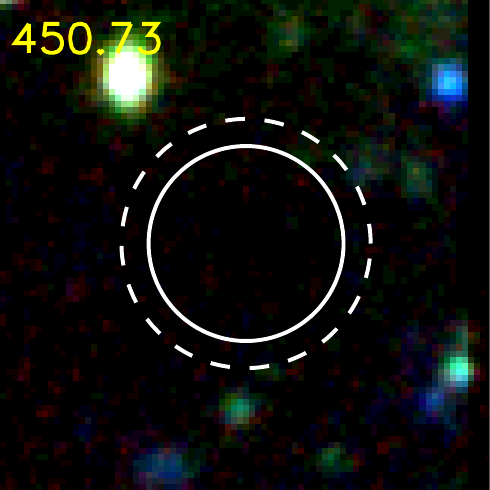}\includegraphics[width=0.32\columnwidth]{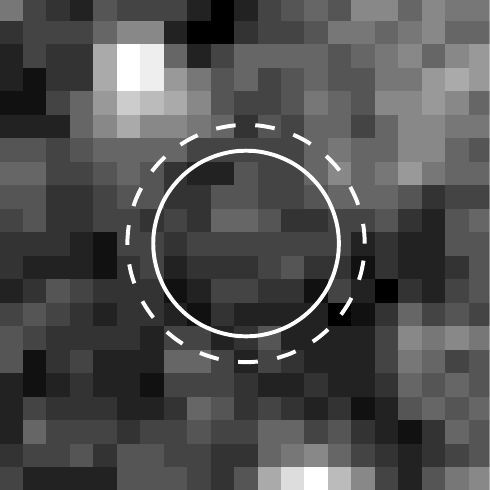}\includegraphics[width=0.32\columnwidth]{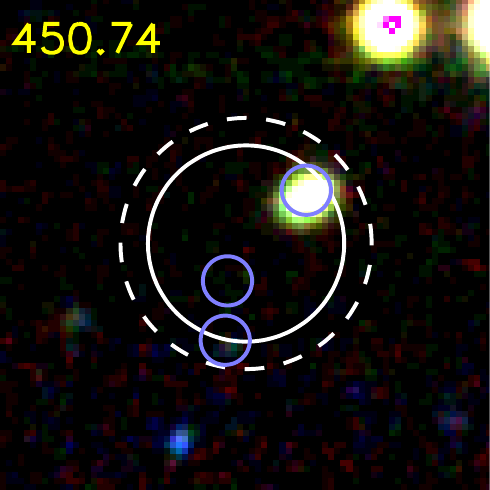}\includegraphics[width=0.32\columnwidth]{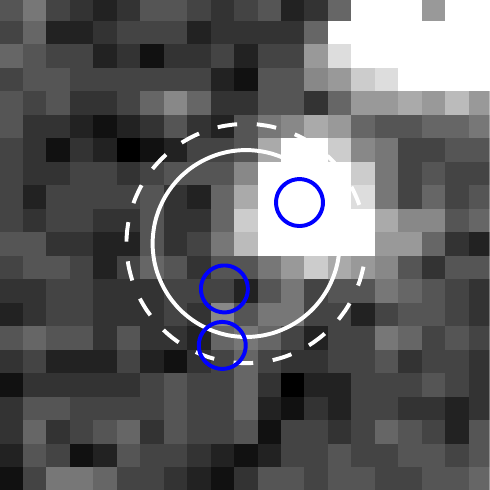}
\includegraphics[width=0.32\columnwidth]{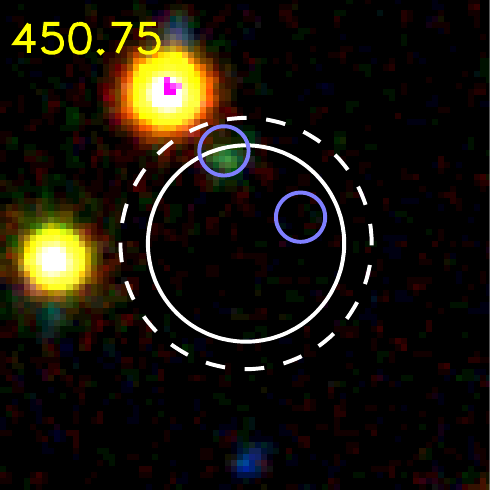}\includegraphics[width=0.32\columnwidth]{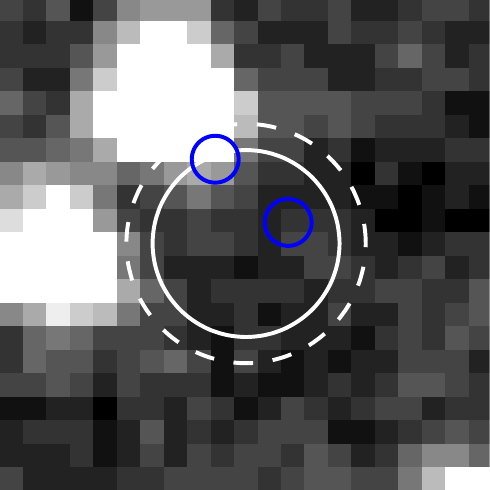}\includegraphics[width=0.32\columnwidth]{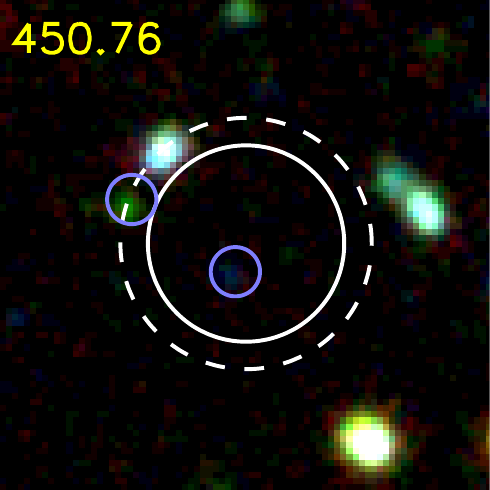}\includegraphics[width=0.32\columnwidth]{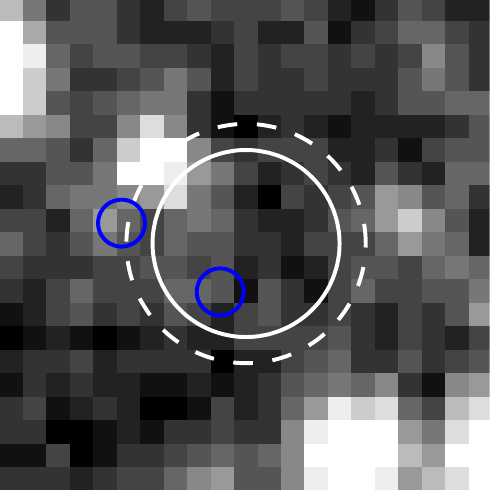}\includegraphics[width=0.32\columnwidth]{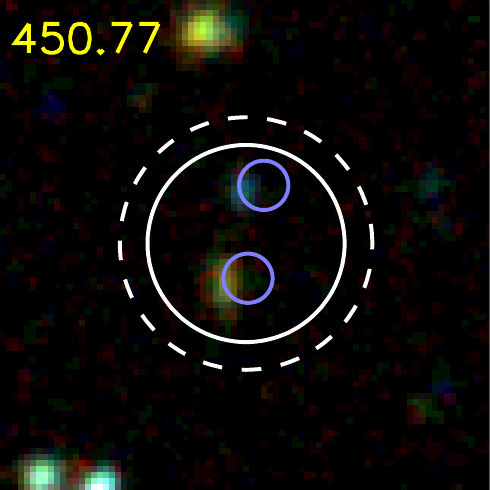}\includegraphics[width=0.32\columnwidth]{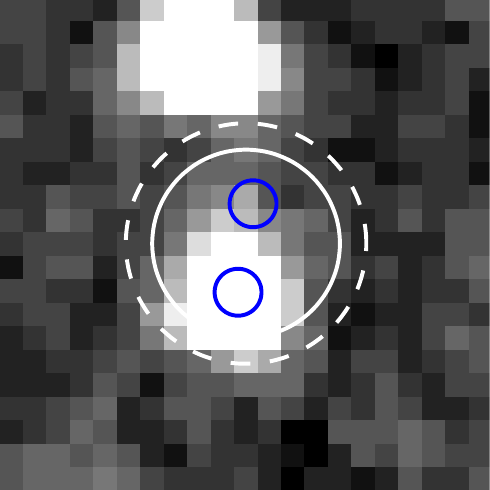}
\centerline{Figure~\ref{fig:cutouts450} -- continued.}\end{center}
\clearpage

\begin{table*}
\centering
\caption{850\um--identified $>$3.6$\sigma$ point sources in COSMOS.}
{\tiny \begin{tabular}{c@{ }c@{ }cccccc@{ }c@{ }c@{ }c@{ }c@{ }c}
\hline\hline
Name & Short Name & RA$_{\rm 850}$ & DEC$_{\rm 850}$ & S/N & S$_{\rm 850}$ & S$_{\rm 850}$ & $\Delta$($\alpha$,$\delta$) & $P_{\rm contam}$ & 450\um-  & Offset & S$_{\rm 450}$ & S$_{\rm 450}$   \\
     &            &                &                 &     &   {\sc Raw} & {\sc Deboosted} &     &                                          & {\sc Source}  &  & {\sc Raw} & {\sc Deboosted} \\
     &            &                &                 &     &   [mJy] & [mJy] &  [\arcsec]   &                                               &               & [\arcsec] & [mJy]     & [mJy] \\
\hline\hline
SMM\,J100008.0+022612... & 850.00 & 10:00:08.0 & 02:26:12 & 20.1 & 16.15$\pm$0.80 & 16.15$\pm$0.80 & 1.19 & $<$0.01 & 450.03 & 0.6  & 23.47$\pm$4.10 & 20.46$\pm$4.78 \\ 
SMM\,J095957.3+022730... & 850.01 & 09:59:57.3 & 02:27:30 & 13.0 & 11.98$\pm$0.92 & 11.49$\pm$1.10 & 1.65 & $<$0.01 & (450.168) &      & 11.02$\pm$4.75 & \\ 
SMM\,J100033.3+022559... & 850.02 & 10:00:33.3 & 02:25:59 & 11.8 & 9.47$\pm$0.80 & 8.97$\pm$1.04 & 1.78 & $<$0.01   & 450.01 & 0.6  & 28.43$\pm$4.13 & 25.62$\pm$4.96 \\ 
SMM\,J100019.7+023204... & 850.03 & 10:00:19.7 & 02:32:04 & 11.7 & 12.11$\pm$1.03 & 11.42$\pm$1.38 & 1.77 & $<$0.01 & 450.04 & 1.5  & 29.15$\pm$5.19 & 25.35$\pm$6.04 \\ 
SMM\,J100015.5+021550... & 850.04 & 10:00:15.5 & 02:15:50 & 10.8 & 11.80$\pm$1.08 & 11.09$\pm$1.56 & 1.85 & $<$0.01 &        &      & 10.67$\pm$5.79 & \\ 
SMM\,J100057.0+022014... & 850.05 & 10:00:57.0 & 02:20:14 & 10.8 & 12.24$\pm$1.13 & 11.49$\pm$1.62 & 1.85 & $<$0.01 & (450.28) &      & 18.48$\pm$5.43 & \\ 
SMM\,J100024.0+021750... & 850.06 & 10:00:24.0 & 02:17:50 & 10.4 & 9.30$\pm$0.89 & 8.64$\pm$1.24 & 1.88 & $<$0.01 & 450.55 & 3.1  & 17.06$\pm$4.57 & 12.70$\pm$5.42 \\ 
SMM\,J100028.6+023202... & 850.07 & 10:00:28.6 & 02:32:02 & 9.41 & 9.94$\pm$1.06 & 9.21$\pm$1.45 & 2.18 & $<$0.01 & 450.00 & 0.6  & 40.58$\pm$5.20 & 37.54$\pm$6.58 \\ 
SMM\,J100023.5+022155... & 850.08 & 10:00:23.5 & 02:21:55 & 8.98 & 7.19$\pm$0.80 & 6.60$\pm$1.12 & 2.10 & $<$0.01 & 450.05 & 1.9  & 23.14$\pm$4.12 & 20.12$\pm$4.80 \\ 
SMM\,J100025.3+021847... & 850.09 & 10:00:25.3 & 02:18:47 & 8.89 & 7.47$\pm$0.84 & 6.83$\pm$1.17 & 2.10 & $<$0.01 & 450.87 & 2.8  & 15.24$\pm$4.30 & 10.74$\pm$5.24 \\ 
SMM\,J100034.4+022121... & 850.10 & 10:00:34.4 & 02:21:21 & 8.61 & 6.90$\pm$0.80 & 6.24$\pm$1.13 & 2.12 & $<$0.01 & 450.81 & 0.8  & 14.71$\pm$4.11 & 10.47$\pm$4.98 \\ 
SMM\,J100049.9+022257... & 850.11 & 10:00:49.9 & 02:22:57 & 8.59 & 7.51$\pm$0.87 & 6.79$\pm$1.24 & 2.12 & $<$0.01 &        &      & 5.94$\pm$4.41 & \\ 
SMM\,J100039.1+022221... & 850.12 & 10:00:39.1 & 02:22:21 & 8.14 & 6.49$\pm$0.80 & 5.83$\pm$1.13 & 2.10 & $<$0.01 & 450.06 & 1.3  & 22.29$\pm$4.11 & 19.32$\pm$4.76 \\ 
SMM\,J100024.1+022005... & 850.13 & 10:00:24.1 & 02:20:05 & 8.00 & 6.44$\pm$0.80 & 5.76$\pm$1.13 & 2.12 & $<$0.01 &        &      & 8.53$\pm$4.13 & \\ 
SMM\,J100010.2+022221... & 850.14 & 10:00:10.2 & 02:22:21 & 7.81 & 6.25$\pm$0.80 & 5.55$\pm$1.11 & 2.17 & $<$0.01 & 450.38 & 2.8  & 16.47$\pm$4.12 & 12.91$\pm$4.73 \\ 
SMM\,J100025.2+022607... & 850.15 & 10:00:25.2 & 02:26:07 & 7.26 & 5.79$\pm$0.80 & 5.08$\pm$1.09 & 2.30 & $<$0.01 &        &      & 11.51$\pm$4.11 &\\ 
SMM\,J100103.9+022447... & 850.16 & 10:01:03.9 & 02:24:47 & 7.12 & 8.94$\pm$1.25 & 7.82$\pm$1.71 & 2.34 & $<$0.01 &        &      & 14.52$\pm$5.92 &\\ 
SMM\,J100000.4+022256... & 850.17 & 10:00:00.4 & 02:22:56 & 7.01 & 5.88$\pm$0.84 & 5.14$\pm$1.14 & 2.39 & $<$0.01 &        &      & 8.96$\pm$4.28 & \\ 
SMM\,J100024.0+022947... & 850.18 & 10:00:24.0 & 02:29:47 & 6.85 & 5.87$\pm$0.86 & 5.13$\pm$1.16 & 2.45 & $<$0.01 &        &      & 6.37$\pm$4.36 & \\ 
SMM\,J100015.7+022445... & 850.19 & 10:00:15.7 & 02:24:45 & 6.74 & 5.40$\pm$0.80 & 4.71$\pm$1.08 & 2.48 & $<$0.01 &        &      & 2.13$\pm$4.13 & \\ 
SMM\,J100027.0+023137... & 850.20 & 10:00:27.0 & 02:31:37 & 6.64 & 6.69$\pm$1.01 & 5.83$\pm$1.36 & 2.51 & $<$0.01 &        &      & 3.93$\pm$5.00 & \\ 
SMM\,J100035.9+022151... & 850.21 & 10:00:35.9 & 02:21:51 & 6.60 & 5.29$\pm$0.80 & 4.60$\pm$1.08 & 2.52 & $<$0.01 & 450.58 & 3.6  & 15.32$\pm$4.12 & 11.35$\pm$4.90 \\ 
SMM\,J100018.7+021655... & 850.22 & 10:00:18.7 & 02:16:55 & 6.25 & 6.05$\pm$0.97 & 5.15$\pm$1.28 & 2.69 & $<$0.01 &        &      & 10.34$\pm$5.10 &\\ 
SMM\,J100022.1+022842... & 850.23 & 10:00:22.1 & 02:28:42 & 6.25 & 5.06$\pm$0.81 & 4.31$\pm$1.07 & 2.69 & $<$0.01 & 450.94 & 0.6  & 14.67$\pm$4.19 & 10.14$\pm$5.17 \\ 
SMM\,J100004.8+023045... & 850.24 & 10:00:04.8 & 02:30:45 & 6.20 & 6.26$\pm$1.01 & 5.32$\pm$1.34 & 2.69 & $<$0.01 & (450.20) &      & 11.37$\pm$5.23 &\\ 
SMM\,J100012.5+021444... & 850.25 & 10:00:12.5 & 02:14:44 & 6.15 & 7.93$\pm$1.29 & 6.72$\pm$1.70 & 2.69 & $<$0.01 &        &      & 14.85$\pm$6.90 &\\ 
SMM\,J100010.3+022627... & 850.26 & 10:00:10.3 & 02:26:27 & 6.09 & 4.87$\pm$0.80 & 4.12$\pm$1.06 & 2.69 & $<$0.01 & 450.32 & 3.0  & 16.86$\pm$4.11 & 13.37$\pm$4.67 \\ 
SMM\,J100023.5+021918... & 850.27 & 10:00:23.5 & 02:19:18 & 6.09 & 5.00$\pm$0.82 & 4.23$\pm$1.08 & 2.69 & $<$0.01 & 450.215 & 2.7 & 13.05$\pm$4.20 & 7.48$\pm$5.28 \\ 
SMM\,J100004.3+022059... & 850.28 & 10:00:04.3 & 02:20:59 & 5.79 & 4.86$\pm$0.84 & 4.07$\pm$1.08 & 2.76 & $<$0.01 & 450.15 & 1.8  & 19.92$\pm$4.33 & 16.53$\pm$4.88 \\ 
SMM\,J100050.1+022615... & 850.29 & 10:00:50.1 & 02:26:15 & 5.70 & 5.13$\pm$0.90 & 4.28$\pm$1.15 & 2.79 & $<$0.01 &        &      & 10.89$\pm$4.50 &\\ 
SMM\,J100006.7+022209... & 850.30 & 10:00:06.7 & 02:22:09 & 5.69 & 4.58$\pm$0.80 & 3.82$\pm$1.03 & 2.79 & $<$0.01 &        &      & 3.90$\pm$4.15 & \\ 
SMM\,J100026.1+021741... & 850.31 & 10:00:26.1 & 02:17:41 & 5.57 & 5.02$\pm$0.90 & 4.19$\pm$1.14 & 2.84 & $<$0.01 &        &      & 12.34$\pm$4.62 &\\ 
SMM\,J100001.5+022429... & 850.32 & 10:00:01.5 & 02:24:29 & 5.56 & 4.56$\pm$0.82 & 3.81$\pm$1.04 & 2.84 & $<$0.01 & (450.13) &      & 13.33$\pm$4.20 &\\ 
SMM\,J095950.7+022827... & 850.33 & 09:59:50.7 & 02:28:27 & 5.55 & 6.20$\pm$1.12 & 5.18$\pm$1.41 & 2.85 & $<$0.01 & 450.99 & 3.7  & 20.15$\pm$5.80 & 13.76$\pm$7.20 \\ 
SMM\,J100016.4+022638... & 850.34 & 10:00:16.4 & 02:26:38 & 5.49 & 4.41$\pm$0.80 & 3.68$\pm$1.01 & 2.86 & $<$0.01 & 450.09 & 2.3  & 20.41$\pm$4.14 & 17.26$\pm$4.68 \\ 
SMM\,J100105.1+022150... & 850.35 & 10:01:05.1 & 02:21:50 & 5.47 & 7.60$\pm$1.39 & 6.35$\pm$1.74 & 2.87 & $<$0.01 &        &      & 0.38$\pm$6.54 & \\ 
SMM\,J100001.3+021745... & 850.36 & 10:00:01.3 & 02:17:45 & 5.39 & 5.60$\pm$1.04 & 4.68$\pm$1.30 & 2.90 & $<$0.01 &        &      & 0.09$\pm$5.62 & \\ 
SMM\,J100035.6+022826... & 850.37 & 10:00:35.6 & 02:28:26 & 5.38 & 4.45$\pm$0.83 & 3.72$\pm$1.03 & 2.91 & $<$0.01 &        &      & 1.74$\pm$4.25 & \\ 
SMM\,J100023.5+021536... & 850.38 & 10:00:23.5 & 02:15:36 & 5.22 & 5.67$\pm$1.09 & 4.72$\pm$1.34 & 2.99 & $<$0.01 &        &      & 13.39$\pm$5.74 &\\ 
SMM\,J100012.1+022310... & 850.39 & 10:00:12.1 & 02:23:10 & 5.21 & 4.17$\pm$0.80 & 3.47$\pm$0.99 & 2.99 & $<$0.01 &        &      & 10.53$\pm$4.13 &\\ 
SMM\,J100013.5+022227... & 850.40 & 10:00:13.5 & 02:22:27 & 5.20 & 4.16$\pm$0.80 & 3.47$\pm$0.99 & 3.00 & $<$0.01 & 450.252 & 3.7 & 12.57$\pm$4.14 & 6.91$\pm$5.15 \\ 
SMM\,J100025.4+022544... & 850.41 & 10:00:25.4 & 02:25:44 & 5.17 & 4.12$\pm$0.80 & 3.43$\pm$0.98 & 3.01 & $<$0.01 & 450.11 & 2.0  & 19.57$\pm$4.11 & 16.40$\pm$4.62 \\ 
SMM\,J100017.2+022519... & 850.42 & 10:00:17.2 & 02:25:19 & 5.14 & 4.12$\pm$0.80 & 3.43$\pm$0.99 & 3.02 & $<$0.01 & 450.17 & 2.6  & 19.01$\pm$4.14 & 15.76$\pm$4.67 \\ 
SMM\,J100026.3+021528... & 850.43 & 10:00:26.3 & 02:15:28 & 5.02 & 5.55$\pm$1.11 & 4.60$\pm$1.35 & 3.10 & $<$0.01 &        &      & 2.32$\pm$5.82 & \\ 
SMM\,J100018.2+022249... & 850.44 & 10:00:18.2 & 02:22:49 & 5.00 & 3.99$\pm$0.80 & 3.31$\pm$0.97 & 3.10 & $<$0.01 &        &      & 3.93$\pm$4.12 & \\ 
SMM\,J100006.9+022047... & 850.45 & 10:00:06.9 & 02:20:47 & 4.99 & 4.12$\pm$0.83 & 3.42$\pm$1.01 & 3.11 & $<$0.01 &        &      & 3.79$\pm$4.25 & \\ 
SMM\,J100025.4+021506... & 850.46 & 10:00:25.4 & 02:15:06 & 4.85 & 5.62$\pm$1.16 & 4.64$\pm$1.40 & 3.20 & $<$0.01 & (450.08)&      & 21.40$\pm$6.11 &\\ 
SMM\,J095952.3+022137... & 850.47 & 09:59:52.3 & 02:21:37 & 4.83 & 4.74$\pm$0.98 & 3.92$\pm$1.19 & 3.21 & $<$0.01 & 450.50 & 3.6  & 19.82$\pm$5.25 & 14.90$\pm$6.19 \\ 
SMM\,J100037.0+021942... & 850.48 & 10:00:37.0 & 02:19:42 & 4.83 & 3.99$\pm$0.83 & 3.30$\pm$1.00 & 3.22 & $<$0.01 & 450.42 & 3.6  & 16.71$\pm$4.24 & 13.01$\pm$4.89 \\ 
SMM\,J100005.0+021719... & 850.49 & 10:00:05.0 & 02:17:19 & 4.80 & 4.91$\pm$1.02 & 4.05$\pm$1.24 & 3.23 & $<$0.01 & 450.193 & 0.6 & 17.40$\pm$5.50 & 10.27$\pm$6.95 \\ 
SMM\,J100009.1+022023... & 850.50 & 10:00:09.1 & 02:20:23 & 4.78 & 3.94$\pm$0.82 & 3.25$\pm$1.00 & 3.25 & $<$0.01 &        &      & 4.68$\pm$4.23 & \\ 
SMM\,J095956.4+021854... & 850.51 & 09:59:56.4 & 02:18:54 & 4.74 & 4.94$\pm$1.04 & 4.06$\pm$1.26 & 3.27 & $<$0.01 &        &      & 9.29$\pm$5.61 & \\ 
SMM\,J100011.9+022937... & 850.52 & 10:00:11.9 & 02:29:37 & 4.74 & 4.09$\pm$0.86 & 3.37$\pm$1.04 & 3.28 & $<$0.01 & 450.134 & 1.7 & 14.45$\pm$4.36 & 9.26$\pm$5.50 \\ 
SMM\,J100011.4+021508... & 850.53 & 10:00:11.4 & 02:15:08 & 4.68 & 5.72$\pm$1.22 & 4.70$\pm$1.47 & 3.32 & $<$0.01 & 450.206 & 4.8 & 20.69$\pm$6.60 & 12.04$\pm$8.32 \\ 
SMM\,J095948.9+022748... & 850.54 & 09:59:48.9 & 02:27:48 & 4.55 & 5.10$\pm$1.12 & 4.17$\pm$1.35 & 3.41 & $<$0.01 &        &      & 0.47$\pm$5.90 & \\ 
SMM\,J095950.9+022742... & 850.55 & 09:59:50.9 & 02:27:42 & 4.52 & 4.77$\pm$1.06 & 3.89$\pm$1.27 & 3.43 & $<$0.01 & 450.86 & 3.6  & 19.80$\pm$5.58 & 13.95$\pm$6.80 \\ 
SMM\,J100005.1+021526... & 850.56 & 10:00:05.1 & 02:15:26 & 4.42 & 5.69$\pm$1.29 & 4.63$\pm$1.54 & 3.58 & 0.02 &        &      & -4.70$\pm$6.84 &\\ 
SMM\,J100006.8+023307... & 850.57 & 10:00:06.8 & 02:33:07 & 4.35 & 5.94$\pm$1.36 & 4.82$\pm$1.63 & 3.68 & 0.02 &        &      & 3.66$\pm$7.17 & \\ 
SMM\,J100101.3+022800... & 850.58 & 10:01:01.3 & 02:28:00 & 4.33 & 5.67$\pm$1.31 & 4.60$\pm$1.57 & 3.71 & 0.03 & 450.36 & 3.8  & 24.27$\pm$6.07 & 19.03$\pm$6.96 \\ 
SMM\,J100000.7+022001... & 850.59 & 10:00:00.7 & 02:20:01 & 4.32 & 3.92$\pm$0.91 & 3.18$\pm$1.08 & 3.72 & 0.03 &        &      & 2.88$\pm$4.76 & \\ 
SMM\,J100000.7+022740... & 850.60 & 10:00:00.7 & 02:27:40 & 4.32 & 3.79$\pm$0.88 & 3.07$\pm$1.05 & 3.73 & 0.03 &        &      & 11.26$\pm$4.49 &\\ 
SMM\,J100002.0+022820... & 850.61 & 10:00:02.0 & 02:28:20 & 4.31 & 3.83$\pm$0.89 & 3.10$\pm$1.06 & 3.73 & 0.03 &        &      & 11.06$\pm$4.53 &\\ 
SMM\,J100031.0+022751... & 850.62 & 10:00:31.0 & 02:27:51 & 4.29 & 3.46$\pm$0.81 & 2.80$\pm$0.96 & 3.76 & 0.03 & 450.78 & 0.9  & 14.94$\pm$4.15 & 10.72$\pm$4.99 \\ 
SMM\,J095953.3+021850... & 850.63 & 09:59:53.3 & 02:18:50 & 4.28 & 4.77$\pm$1.11 & 3.86$\pm$1.33 & 3.78 & 0.03 &        &      & 3.99$\pm$6.07 & \\ 
SMM\,J100000.6+022137... & 850.64 & 10:00:00.6 & 02:21:37 & 4.18 & 3.59$\pm$0.86 & 2.89$\pm$1.02 & 3.95 & 0.04 &        &      & 4.27$\pm$4.41 & \\ 
SMM\,J100059.2+022108... & 850.65 & 10:00:59.2 & 02:21:08 & 4.15 & 4.71$\pm$1.14 & 3.79$\pm$1.35 & 4.04 & 0.04 &        &      & 2.93$\pm$5.46 & \\ 
SMM\,J100024.4+022831... & 850.66 & 10:00:24.4 & 02:28:31 & 4.09 & 3.30$\pm$0.81 & 2.64$\pm$0.96 & 4.19 & 0.05 &        &      & 8.36$\pm$4.17 & \\ 
SMM\,J100032.4+023321... & 850.67 & 10:00:32.4 & 02:33:21 & 4.08 & 5.15$\pm$1.26 & 4.12$\pm$1.50 & 4.21 & 0.05 & 450.240 & 5.9 & 19.01$\pm$6.21 & 10.61$\pm$7.75 \\ 
SMM\,J095955.3+021954... & 850.68 & 09:59:55.3 & 02:19:54 & 4.06 & 4.05$\pm$1.00 & 3.24$\pm$1.19 & 4.27 & 0.05 &        &   & 11.46$\pm$5.35 &\\ 
SMM\,J100043.8+022857... & 850.69 & 10:00:43.8 & 02:28:57 & 4.03 & 3.74$\pm$0.93 & 2.98$\pm$1.10 & 4.34 & 0.06 &        &   & -2.18$\pm$4.64 &\\ 
SMM\,J100020.0+023020... & 850.70 & 10:00:20.0 & 02:30:20 & 4.02 & 3.57$\pm$0.89 & 2.85$\pm$1.06 & 4.36 & 0.06 &        &   & 6.36$\pm$4.49 & \\ 
SMM\,J095944.0+022105... & 850.71 & 09:59:44.0 & 02:21:05 & 4.01 & 5.07$\pm$1.26 & 4.04$\pm$1.50 & 4.38 & 0.06 &        &   & 6.21$\pm$6.92 & \\ 
SMM\,J100013.6+021731... & 850.72 & 10:00:13.6 & 02:17:31 & 3.99 & 3.77$\pm$0.95 & 3.00$\pm$1.13 & 4.45 & 0.06 &        &   & 3.00$\pm$4.96 & \\ 
SMM\,J100020.6+022251... & 850.73 & 10:00:20.6 & 02:22:51 & 3.89 & 3.11$\pm$0.80 & 2.46$\pm$0.95 & 4.69 & 0.07 &        &   & -0.35$\pm$4.11 &\\ 
SMM\,J100005.5+021448... & 850.74 & 10:00:05.5 & 02:14:48 & 3.88 & 5.50$\pm$1.42 & 4.33$\pm$1.69 & 4.77 & 0.08 &        &   & -1.85$\pm$7.61 &\\ 
SMM\,J100026.7+022230... & 850.75 & 10:00:26.7 & 02:22:30 & 3.83 & 3.08$\pm$0.80 & 2.41$\pm$0.96 & 4.95 & 0.09 &        &   & 1.12$\pm$4.14 & \\ 
SMM\,J100048.9+023023... & 850.76 & 10:00:48.9 & 02:30:23 & 3.83 & 4.36$\pm$1.14 & 3.42$\pm$1.36 & 4.95 & 0.09 &        &   & 5.52$\pm$5.43 & \\ 
SMM\,J100005.5+022958... & 850.77 & 10:00:05.5 & 02:29:58 & 3.83 & 3.58$\pm$0.94 & 2.81$\pm$1.12 & 4.97 & 0.09 &        &   & 0.41$\pm$4.78 & \\ 
SMM\,J100008.7+021912... & 850.78 & 10:00:08.7 & 02:19:12 & 3.82 & 3.33$\pm$0.87 & 2.60$\pm$1.04 & 5.00 & 0.09 &        &   & 7.53$\pm$4.52 & \\ 
SMM\,J100102.2+022234... & 850.79 & 10:01:02.2 & 02:22:34 & 3.82 & 4.57$\pm$1.20 & 3.57$\pm$1.43 & 5.04 & 0.10 &        &   & 16.35$\pm$5.70 &\\ 
\hline\hline
\end{tabular}
}

\label{tab:sources850}
\begin{spacing}{0.7}
{\scriptsize {\bf Table Notes.} The $>$3.6$\sigma$ 850\um--detected
  sources we extract within the central 394\arcmin$^2$ of our COSMOS
  map.  The 3.6$\sigma$ detection threshold is chosen based on an
  estimated 3--5\%\ contamination rate. The list is ordered by
  detection signal-to-noise ratio (S/N).  The ``{\sc Raw}'' flux
  densities are those measured directly from our map. The ``{\sc
    Deboosted}'' flux densities are those given after correction for
  confusion and Eddington boosting as a function of detection S/N, as
  described in section~\ref{sec:simulations}.  We also measure a
  90\%\ confidence interval for positional uncertainties and estimated
  probability of contamination, $P_{\rm contam}$, from the results of
  our Monte Carlo tests as functions of detection S/N.  The last four
  columns give details on the corresponding 450\um\ counterparts,
  similar to the 850\um\ counterparts for 450\um\ sources given in
  Table~\ref{tab:sources450}.}
\end{spacing}
\end{table*}
\begin{table*}
\centering
{\tiny \begin{tabular}{c@{ }c@{ }cccccc@{ }c@{ }c@{ }c@{ }c@{ }c}
\hline\hline
Name & Short Name & RA$_{\rm 850}$ & DEC$_{\rm 850}$ & S/N & S$_{\rm 850}$ & S$_{\rm 850}$ & $\Delta$($\alpha$,$\delta$) & $P_{\rm contam}$ & 450\um-  & Offset & S$_{\rm 450}$ & S$_{\rm 450}$   \\
     &            &                &                 &     &   {\sc Raw} & {\sc Deboosted} &     &                  & {\sc Source}  &  & {\sc Raw} & {\sc Deboosted} \\
     &            &                &                 &     &   [mJy] & [mJy] &  [\arcsec]   &                       &               & [\arcsec] & [mJy]     & [mJy] \\
\hline\hline
SMM\,J100014.1+022704... & 850.80 & 10:00:14.1 & 02:27:04 & 3.81 & 3.06$\pm$0.80 & 2.39$\pm$0.96 & 5.06 & 0.10 &        &   & 0.71$\pm$4.13 & $<$12.40 \\ 
SMM\,J095959.0+022441... & 850.81 & 09:59:59.0 & 02:24:41 & 3.80 & 3.21$\pm$0.84 & 2.51$\pm$1.01 & 5.10 & 0.10 &        &   & -1.34$\pm$4.32 & $<$12.97 \\ 
SMM\,J100054.8+021945... & 850.82 & 10:00:54.8 & 02:19:45 & 3.78 & 4.14$\pm$1.09 & 3.22$\pm$1.32 & 5.20 & 0.11 &        &   & 7.10$\pm$5.30 & $<$15.91 \\ 
SMM\,J100010.2+021759... & 850.83 & 10:00:10.2 & 02:17:59 & 3.77 & 3.52$\pm$0.93 & 2.74$\pm$1.12 & 5.22 & 0.11 & 450.166 & 1.8  & 15.87$\pm$4.90 & 9.79$\pm$6.20 \\ 
SMM\,J100041.8+022111... & 850.84 & 10:00:41.8 & 02:21:11 & 3.75 & 3.07$\pm$0.82 & 2.39$\pm$0.99 & 5.32 & 0.11 &        &   & 8.93$\pm$4.19 & $<$12.57 \\ 
SMM\,J100025.0+022753... & 850.85 & 10:00:25.0 & 02:27:53 & 3.73 & 2.99$\pm$0.80 & 2.32$\pm$0.97 & 5.40 & 0.12 & 450.105 & 4.5  & 14.24$\pm$4.14 & 9.60$\pm$5.16 \\ 
SMM\,J100101.3+022439... & 850.86 & 10:01:01.3 & 02:24:39 & 3.73 & 4.23$\pm$1.13 & 3.28$\pm$1.37 & 5.41 & 0.12 &        &   & -0.42$\pm$5.41 & $<$16.22 \\ 
SMM\,J100040.3+021758... & 850.87 & 10:00:40.3 & 02:17:58 & 3.70 & 3.49$\pm$0.94 & 2.70$\pm$1.14 & 5.55 & 0.13 &        &   & 4.66$\pm$4.74 & $<$14.22 \\ 
SMM\,J100022.2+023026... & 850.88 & 10:00:22.2 & 02:30:26 & 3.70 & 3.31$\pm$0.90 & 2.56$\pm$1.08 & 5.56 & 0.13 & 450.179 & 6.8  & 14.51$\pm$4.52 & 8.80$\pm$5.73 \\ 
SMM\,J100020.2+021727... & 850.89 & 10:00:20.2 & 02:17:27 & 3.69 & 3.41$\pm$0.92 & 2.63$\pm$1.11 & 5.60 & 0.13 &        &   & 10.84$\pm$4.77 & $<$14.32 \\ 
SMM\,J100032.8+023049... & 850.90 & 10:00:32.8 & 02:30:49 & 3.68 & 3.52$\pm$0.96 & 2.71$\pm$1.16 & 5.64 & 0.13 & 450.40 & 5.0  & 19.15$\pm$4.81 & 14.98$\pm$5.53 \\ 
SMM\,J100027.8+022554... & 850.91 & 10:00:27.8 & 02:25:54 & 3.67 & 2.93$\pm$0.80 & 2.26$\pm$0.96 & 5.66 & 0.13 &        &   & 6.82$\pm$4.10 & $<$12.30 \\ 
SMM\,J100041.8+022404... & 850.92 & 10:00:41.8 & 02:24:04 & 3.67 & 2.94$\pm$0.80 & 2.27$\pm$0.97 & 5.66 & 0.13 & 450.106 & 5.6  & 14.19$\pm$4.13 & 9.57$\pm$5.15 \\ 
SMM\,J100056.7+022945... & 850.93 & 10:00:56.7 & 02:29:45 & 3.67 & 4.80$\pm$1.31 & 3.69$\pm$1.58 & 5.68 & 0.14 &        &   & 2.09$\pm$6.13 & $<$18.38 \\ 
SMM\,J100002.6+021632... & 850.94 & 10:00:02.6 & 02:16:32 & 3.65 & 4.21$\pm$1.15 & 3.23$\pm$1.39 & 5.77 & 0.14 &        &   & -1.62$\pm$6.22 & $<$18.67 \\ 
SMM\,J095959.9+022705... & 850.95 & 09:59:59.9 & 02:27:05 & 3.65 & 3.18$\pm$0.87 & 2.44$\pm$1.05 & 5.77 & 0.14 &        &   & 6.96$\pm$4.46 & $<$13.38 \\ 
SMM\,J100026.4+022315... & 850.96 & 10:00:26.4 & 02:23:15 & 3.64 & 2.93$\pm$0.80 & 2.24$\pm$0.97 & 5.81 & 0.14 & 450.54 & 8.1  & 15.46$\pm$4.13 & 11.51$\pm$4.90 \\ 
SMM\,J100013.4+021807... & 850.97 & 10:00:13.4 & 02:18:07 & 3.61 & 3.27$\pm$0.90 & 2.50$\pm$1.09 & 5.91 & 0.15 &        &   & 5.64$\pm$4.69 & $<$14.08 \\ 
SMM\,J100014.5+023008... & 850.98 & 10:00:14.5 & 02:30:08 & 3.61 & 3.18$\pm$0.88 & 2.43$\pm$1.07 & 5.95 & 0.15 &        &   & -6.89$\pm$4.43 & $<$13.30 \\
\hline\hline
\end{tabular}
}
\label{tab:sources850_2}
{\small {\sc Table~\ref{tab:sources850} -- Continued.}}
\end{table*}

\begin{table*}	
\centering	
\caption{Counterpart identifications and Multiwavelength Properties for \scubaii\ 450\um-detected Sources}	
{\tiny \begin{tabular}{l@{ }c@{ }cc@{ }c@{ }c@{ }c@{ }c@{ }c@{ }c@{ }c@{ }c@{ }c@{ }c@{ }c@{ }c@{ }c}	
\hline\hline	
NAME & {\sc Counter-} & $p$- & RA$_{\rm 24}$ & DEC$_{\rm 24}$ & $\delta_{24}$ & $S_{\rm 24}$ & RA$_{\rm 1.4}$ & DEC$_{\rm 1.4}$ & $\delta_{1.4}$ & $S_{\rm 1.4}$ & RA$_{\rm opt}$ & DEC$_{\rm opt}$ & $\delta_{\rm opt}$ & $i$       & 3.6\um\ & $z_{\rm phot}$ \\	
     & {\sc part} \#  & $val$ &             &                &    [\arcsec]            &    [\uJy]    &                &                 &       [\arcsec]         &    [\uJy]     &          &           &  [\arcsec]                  & [AB-mag] & [AB-mag] &              \\	
\hline\hline	
450.00 & 1/2 & 0.09 &            &             &      &              &             &             &      &             & 10:00:28.66 & +02:32:03.3 & 1.58 &	26.78 & 23.01 & 2.86 \\ 
       & 2/2 & 0.09 &            &             &      &              &             &             &      &             & 10:00:28.58 & +02:32:02.0 & 1.59 &	24.01 & 22.70 & 0.75 \\ 
450.01 & 1/1 & 0.01 &10:00:33.35 & +02:26:01.6 & 1.87 & 297$\pm$16   &             &             &      &             & {\bf 10:00:33.38} & {\bf +02:26:01.4} & 1.73 & 	26.09 & 21.30 & 2.88 \\ 
       & 2/2 & 0.15 &            &             &      &              &             &             &      &             & 10:00:33.49 & +02:26:00.1 & 2.11 & 	26.94 & 23.84 & 2.10 \\ 
450.02 & 1/3 & 0.03 &            &             &      &              &             &             &      &             & 10:01:09.00 & +02:22:56.1 & 0.87 & 	24.89 & 20.85 & 2.16 \\ 
       & 2/3 & 0.03 &10:01:09.15 & +02:22:55.0 & 1.70 & 1492$\pm$248 &             &             &      &             & 10:01:09.19 & +02:22:54.1 & 2.67 & 	$-$   & $-$   & $-$  \\ 
       & 3/3 & 0.11 &            &             &      &              &             &             &      &             & 10:01:08.95 & +02:22:54.6 & 1.81 & 	25.75 & 20.32 & 2.18 \\ 
450.03 & 1/1 & 0.0009 &10:00:08.01 & +02:26:11.0 & 1.13 & 287$\pm$15   & 10:00:08.02 & +02:26:12.1 & 0.83 & 76$\pm$14   & {\bf 10:00:08.04} & {\bf +02:26:10.7} & 1.29 & 22.09$\star$ & 21.83$\star$ & $-$ \\
450.04 & 1/2 & 0.0009 &10:00:19.76 & +02:32:03.9 & 1.87 & 189$\pm$13   & 10:00:19.75 & +02:32:04.3 & 1.42 & 126$\pm$15  & {\bf 10:00:19.75} & {\bf +02:32:04.5} & 1.24 & 	25.83 & 22.42 & 3.82 \\ 
       & 2/2 & 0.16 &            &             &      &              &             &             &      &             & 10:00:19.73 & +02:32:06.6 & 0.89 & 	26.56 & $-$   & 1.13 \\ 
450.05 & 1/1 & 0.001 &10:00:23.65 & +02:21:55.3 & 1.54 & 236$\pm$16   & 10:00:23.67 & +02:21:55.3 & 1.68 & 43$\pm$11   & {\bf 10:00:23.67} & {\bf +02:21:55.5} & 1.55 & 	27.23 & 21.01 & 3.99 \\ 
450.06 & 1/2 & 0.0004 &10:00:39.25 & +02:22:21.1 & 0.69 & 544$\pm$17   & 10:00:39.24 & +02:22:21.0 & 0.77 & 138$\pm$15  & 10:00:39.23 & +02:22:20.9 & 0.87 & 	23.62 & 20.47 & 2.08 \\ 
       & 2/2 & 0.11 &            &             &      &              &             &             &      &             & 10:00:39.18 & +02:22:23.4 & 1.82 & 	29.93 & 23.20 & 2.27 \\ 
450.07 & 1/1 & 0.002 &10:00:14.25 & +02:30:19.0 & 1.26 & 467$\pm$100  & 10:00:14.20 & +02:30:19.1 & 1.70 & 200$\pm$13  & $-$         & $-$         & $-$  & 	$-$   & $-$   & $-$  \\ 
450.08 & 1/2 & 0.03 &            &             &      &              &             &             &      &             & {\bf 10:00:25.55} & {\bf +02:15:08.4} & 6.76  & 	$-$   & $-$   & $-$  \\ 
       & 2/2 & 0.11 &            &             &      &              &             &             &      &             & 10:00:25.24 & +02:15:09.8 & 1.79 & 	24.69 & 22.85 & 1.86 \\ 
450.09 & 1/3 & 0.0005 &10:00:16.59 & +02:26:38.4 & 0.72 & 890$\pm$17   & 10:00:16.57 & +02:26:38.4 & 0.92 & 5716$\pm$73 & 10:00:16.56 & +02:26:39.8 & 2.25 & 	24.39 & 21.42 & 1.90 \\ 
       & 2/3 & 0.13 &            &             &      &              &             &             &      &             & 10:00:16.70 & +02:26:36.3 & 1.98 & 	25.02 & 22.87 & 1.72 \\ 
       & 2/3 & 0.16 &            &             &      &              &             &             &      &             & 10:00:16.71 & +02:26:39.3 & 2.18 & 	26.47 & 23.47 & 1.98 \\ 
450.10 & 1/1 & 0.01 &10:00:22.27 & +02:23:53.5 & 1.46 & 95$\pm$18    &             &             &      &             & 10:00:22.28 & +02:23:54.1 & 1.13 & 	25.23 & 21.35 & $-$  \\ 
       & 2/2 & 0.14 &            &             &      &              &             &             &      &             & 10:00:22.14 & +02:23:56.5 & 2.06 & 	24.64 & 21.55 & $-$  \\ 
450.11 & 1/2 & 0.01 &10:00:25.45 & +02:25:44.2 & 1.53 & 116$\pm$15   &             &             &      &             & $-$         & $-$         & $-$  & 	$-$   & $-$   & $-$  \\ 
       & 2/2 & 0.19 &            &             &      &              &             &             &      &             & 10:00:25.42 & +02:25:40.5 & 2.37 & 	24.60 & 23.37 & $-$  \\ 
450.12 & 1/1 & 0.13 &            &             &      &              &             &             &      &             & 09:59:57.09 & +02:21:51.9 & 1.91 & 	22.67 & 19.82 & 0.93 \\ 
450.13 & 1/1 & 0.002 &10:00:01.66 & +02:24:27.9 & 1.47 & 287$\pm$12   & 10:00:01.66 & +02:24:28.2 & 1.80 & 90$\pm$14   & 10:00:01.61 & +02:24:28.0 & 2.11 & 	26.39 & 21.32 & 2.87 \\ 
450.14 & 1/2 & 0.02 &10:00:45.01 & +02:19:21.0 & 2.15 & 228$\pm$20   &             &             &      &             & 10:00:45.02 & +02:19:21.1 & 2.37 & 	23.88 & 19.39 & 1.47 \\ 
       & 2/2 & 0.06 &            &             &      &              &             &             &      &             & 10:00:44.93 & +02:19:20.0 & 0.54 & 	$-$   & $-$   & $-$  \\ 
450.15 & 1/1 & 0.001 &10:00:04.36 & +02:20:59.6 & 1.50 & 402$\pm$14   & 10:00:04.37 & +02:20:59.7 & 1.62 & 56$\pm$12   & 10:00:04.33 & +02:20:59.7 & 1.10 & 	26.23 & 22.51 & 2.79 \\ 
       & 2/2 & 0.01 &            &             &      &              &             &             &      &             & 10:00:04.23 & +02:20:59.5 & 0.50 & 	26.97 & 23.18 & 2.62 \\ 
450.16 & 1/2 & 0.004 &10:00:57.28 & +02:20:12.0 & 2.37 & 132$\pm$16   & 10:00:57.26 & +02:20:12.2 & 2.54 & 81$\pm$15   & 10:00:57.46 & +02:20:11.0 & 3.58 & 	24.87 & 20.50 & 2.32 \\ 
       & 2/2 & 0.11 &            &             &      &              &             &             &      &             & 10:00:57.13 & +02:20:08.9 & 1.79 & 	26.44 & 22.76 & 2.22 \\ 
450.17 & 1/4 & 0.05 &            &             &      &              &             &             &      &             & 10:00:17.29 & +02:25:21.4 & 1.23 & 	26.19 & 24.85 & 1.23 \\ 
       & 2/4 & 0.14 &            &             &      &              &             &             &      &             & 10:00:17.24 & +02:25:23.7 & 2.02 & 	25.56 & 24.99 & $-$  \\ 
       & 3/4 & 0.14 &            &             &      &              &             &             &      &             & 10:00:17.16 & +02:25:23.7 & 2.09 & 	25.38 & 24.73 & $-$  \\ 
       & 5/4 & 0.19 &            &             &      &              &             &             &      &             & 10:00:17.10 & +02:25:20.0 & 2.40 & 	24.78 & 22.64 & $-$  \\ 
450.18 & 1/1 & 0.09 &            &             &      &              &             &             &      &             & 10:00:07.22 & +02:18:03.9 & 0.67 & 	27.59 & $-$   & $-$  \\ 
450.19 & 1/2 & 0.14 &            &             &      &              &             &             &      &             & 09:59:48.13 & +02:20:12.6 & 2.06 & 	24.25 & 21.33 & 1.11 \\ 
       & 2/2 & 0.28 &            &             &      &              &             &             &      &             & 09:59:48.08 & +02:20:17.7 & 3.01 & 	23.75 & 20.68 & 1.32 \\ 
450.20 & 1/1 & 0.12 &            &             &      &              &             &             &      &             & 10:00:04.65 & +02:30:42.2 & 1.87 & 	24.31 & 23.33 & 0.76 \\ 
450.21 & 1/1 & 0.005 &10:00:17.23 & +02:21:39.7 & 0.99 & 333$\pm$21   &             &             &      &             & 10:00:17.22 & +02:21:39.8 & 1.08 & 	21.40 & 19.38 & 0.84 \\ 
450.22 & 1/2 & 0.02 &            &             &      &              &             &             &      &             & 10:00:30.80 & +02:31:05.4 & 0.75 & 	26.40 & 24.03 & 2.11 \\ 
       & 2/2 & 0.13 &            &             &      &              &             &             &      &             & 10:00:30.70 & +02:31:03.9 & 1.96 & 	25.68 & 23.85 & 2.11 \\ 
450.23 & 1/1 & 0.15 &            &             &      &              &             &             &      &             & 10:00:04.51 & +02:18:22.3 & 2.09 & 	24.85 & 23.09 & 0.97 \\ 
450.24 & 1/2 & 0.03 &10:00:50.23 & +02:21:18.1 & 2.36 & 1421$\pm$216 &             &             &      &             & 10:00:50.29 & +02:21:19.0 & 3.58 & 	$-$   & 17.50 & 0.16 \\ 
       & 2/2 & 0.09 &            &             &      &              &             &             &      &             & 10:00:50.09 & +02:21:15.1 & 1.64 & 	24.46 & 22.93 & 0.60 \\ 
450.25 & 1/1 & 8$\times$10$^{-5}$ &10:00:28.55 & +02:19:28.2 & 0.49 & 243$\pm$16   & 10:00:28.57 & +02:19:28.0 & 0.39 & 78$\pm$10   & 10:00:28.58 & +02:19:28.3 & 0.76 & 	20.54 & 18.74 & 0.61 \\ 
450.26 & 1/2 & 0.04 &            &             &      &              &             &             &      &             & 10:00:16.59 & +02:20:01.7 & 1.03 & 	24.76 & 23.06 & 2.68 \\ 
       & 2/2 & 0.19 &            &             &      &              &             &             &      &             & 10:00:16.52 & +02:20:02.8 & 2.42 & 	25.26 & 23.17 & 2.37 \\ 
450.27 & 1/1 & 0.0002 &09:59:42.98 & +02:21:44.9 & 0.19 & 401$\pm$13   & 09:59:42.94 & +02:21:45.0 & 0.59 & 153$\pm$12  & 09:59:42.92 & +02:21:45.1 & 0.93 & 	23.42 & 19.20 & $-$  \\ 
450.28 & 1/4 & 0.02 &10:00:56.78 & +02:20:12.9 & 2.17 & 90$\pm$27    &             &             &      &             & 10:00:56.83 & +02:20:13.3 & 2.26 & 	24.03 & 21.39 & 1.98 \\ 
       & 2/4 & 0.14 &            &             &      &              &             &             &      &             & 10:00:56.77 & +02:20:16.5 & 1.99 & 	25.32 & 23.44 & $-$  \\ 
       & 3/4 & 0.19 &            &             &      &              &             &             &      &             & 10:00:56.56 & +02:20:15.7 & 2.40 & 	25.96 & 24.74 & 1.46 \\ 
       & 4/4 & 0.30 &            &             &      &              &             &             &      &             & 10:00:56.65 & +02:20:13.8 & 1.28 & 	26.69 & $-$   & 1.61 \\ 
450.29 & 1/2 & 0.16 &            &             &      &              &             &             &      &             & 10:00:00.58 & +02:25:03.5 & 0.90 & 	28.57 & $-$   & $-$  \\ 
       & 2/2 & 0.34 &            &             &      &              &             &             &      &             & 10:00:00.48 & +02:25:04.9 & 1.40 & 	27.28 & $-$   & 2.49 \\ 
450.30 & 1/1 & 0.002 &10:00:48.33 & +02:29:26.3 & 0.63 & 185$\pm$14   &             &             &      &             & 10:00:48.35 & +02:29:26.0 & 1.04 & 	26.93 & 21.83 & 2.81 \\ 
450.31 & 1/2 & 0.09 &            &             &      &              &             &             &      &             & 10:00:08.36 & +02:22:42.3 & 1.61 & 	25.32 & 21.45 & $-$  \\ 
       & 2/2 & 0.11 &            &             &      &              &             &             &      &             & 10:00:08.58 & +02:22:42.4 & 1.79 & 	25.15 & 23.71 & $-$  \\ 
450.32 & 1/2 & 0.008 &10:00:10.27 & +02:26:25.4 & 1.29 & 143$\pm$13   &             &             &      &             & 10:00:10.22 & +02:26:25.2 & 0.53 & 	26.15 & 20.43 & 1.72 \\ 
       & 2/2 & 0.20 &            &             &      &              &             &             &      &             & 10:00:10.36 & +02:26:24.0 & 2.46 & 	26.53 & 22.62 & 2.71 \\ 
450.33 & 1/3 & 0.02 &10:00:21.19 & +02:30:55.5 & 2.32 & 255$\pm$18   &             &             &      &             & 10:00:21.19 & +02:30:55.4 & 2.38 & 	22.17 & 19.70 & 0.84 \\ 
       & 2/3 & 0.11 &            &             &      &              &             &             &      &             & 10:00:21.46 & +02:30:55.5 & 1.77 & 	26.86 & 22.90 & 2.16 \\ 
       & 3/3 & 0.16 &            &             &      &              &             &             &      &             & 10:00:21.29 & +02:30:57.7 & 2.15 & 	26.74 & 23.23 & 5.62 \\ 
450.34 & 1/2 & 0.004 &09:59:45.11 & +02:22:55.1 & 2.50 & 726$\pm$14   & 09:59:45.10 & +02:22:55.1 & 2.67 & 89$\pm$11   & 09:59:45.06 & +02:22:55.0 & 3.04 & 	$-$   & $-$   & $-$  \\ 
       & 2/2 & 0.02 &            &             &      &              &             &             &      &             & 09:59:45.30 & +02:22:53.6 & 0.76 & 	24.77 & 23.10 & $-$  \\ 
450.35 & NA  & \\
450.36 & 1/2 & 0.04 &10:01:01.24 & +02:28:00.6 & 2.88 & 104$\pm$15   &             &             &      &             & 10:01:01.26 & +02:28:00.5 & 3.21 & 	$-$   & $-$   & $-$  \\ 
       & 2/2 & 0.10 &            &             &      &              &             &             &      &             & 10:01:01.00 & +02:28:01.0 & 0.70 & 	27.10 & $-$   & $-$  \\ 
450.37 & 1/1 & 0.002 &            &             &      &              &             &             &      &             & 10:00:01.54 & +02:19:39.9 & 0.22 & 	24.21 & 22.97 & 2.17 \\ 
450.38 & 1/4 & 0.0007 &10:00:10.38 & +02:22:24.4 & 0.98 & 381$\pm$14   & 10:00:10.36 & +02:22:24.8 & 1.14 & 100$\pm$13  & 10:00:10.38 & +02:22:24.4 & 1.03 & 	25.78 & 20.61 & $-$  \\ 
       & 2/4 & 0.04 &            &             &      &              &             &             &      &             & 10:00:10.31 & +02:22:22.7 & 1.06 & 	26.59 & 22.22 & $-$  \\ 
       & 3/4 & 0.19 &            &             &      &              &             &             &      &             & 10:00:10.28 & +02:22:21.5 & 2.41 & 	26.43 & 23.81 & $-$  \\ 
       & 4/4 & 0.21 &            &             &      &              &             &             &      &             & 10:00:10.43 & +02:22:21.7 & 2.53 & 	26.84 & 23.73 & $-$  \\ 
450.39 & 1/1 & 0.22 &            &             &      &              &             &             &      &             & 10:00:11.37 & +02:15:55.3 & 2.58 & 	23.29 & 22.29 & 0.61 \\ 
450.40 & NA & \\
450.41 & 1/1 & 0.20 &            &             &      &              &             &             &      &             & 10:00:06.77 & +02:15:29.7 & 2.43 & 	27.31 & 24.76 & 3.46 \\ 
450.42 & 1/3 & 0.04 &10:00:36.88 & +02:19:41.3 & 2.85 & 87$\pm$15    &             &             &      &             & 10:00:36.89 & +02:19:40.5 & 2.07 & 	26.89 & 22.03 & 4.81 \\ 
       & 2/3 & 0.09 &            &             &      &              &             &             &      &             & 10:00:37.06 & +02:19:39.0 & 1.60 & 	26.23 & 24.17 & 2.72 \\ 
       & 3/3 & 0.19 &            &             &      &              &             &             &      &             & 10:00:36.88 & +02:19:36.7 & 2.36 & 	23.91 & 23.35 & 0.44 \\ 
450.43 & 1/2 & 0.16 &            &             &      &              &             &             &      &             & 10:00:28.79 & +02:33:38.3 & 2.18 & 	24.91 & 23.92 & 0.36 \\ 
       & 2/2 & 0.23 &            &             &      &              &             &             &      &             & 10:00:29.03 & +02:33:35.2 & 2.64 & 	25.39 & 22.99 & 2.19 \\ 
450.44 & 1/3 & 0.02 &            &             &      &              &             &             &      &             & 10:00:54.54 & +02:19:19.2 & 0.70 & 	26.38 & 24.91 & 1.75 \\ 
       & 2/3 & 0.08 &            &             &      &              &             &             &      &             & 10:00:54.43 & +02:19:20.7 & 1.55 & 	26.06 & 24.33 & 2.25 \\ 
       & 3/3 & 0.16 &            &             &      &              &             &             &      &             & 10:00:54.44 & +02:19:17.9 & 2.14 & 	25.15 & 24.11 & 0.92 \\ 
450.45 & 1/1 & 0.18 &            &             &      &              &             &             &      &             & 10:00:56.16 & +02:18:29.4 & 2.29 & 	26.97 & 24.17 & 2.01 \\ 
\hline\hline	
\end{tabular}	
}	
\label{tab:counterparts450}	
\begin{spacing}{0.7}	
{\scriptsize {\bf Table Notes.}  	
All possible counterparts (24\um, radio and optical/near-IR) for
\scubaii\ 450\um-identified $>$3.6$\sigma$ sources.  The best-guess
counterpart is the source with the lowest $p$-value of any within
the search radius (see text for details).  Sources with a
$p$-value $>$0.50 are excluded from this table since they are more
likely a random coincident source than a true counterpart.
Counterparts with optical/near-IR positions in {\bf bold} are
submillimeter sources which have been observed interferometrically in
the literature, therefore the correct counterpart is known as the
source at the given position.  See the text (\S~\ref{sec:purity}) for
details on reliability of counterpart identifications.
All positional offsets, $\delta_{\rm 24}$, $\delta_{\rm 1.4}$ and	
$\delta_{\rm opt}$ are measured with respect to the	
\scubaii\ 450\um\ position.	
[$\star$] The $i$-band and 3.6\um\ magnitudes for 450.03/850.00 is	
measured from the HST imaging, where it can be decoupled from the	
bright elliptical nearby (1.6\arcsec).	
[$\ddag$] The 24\um\ flux density for 450.07 is originally measured as	
  1.3\,mJy; this is an obvious blend with a nearby bright source.  We	
  have approximated the contribution from 450.07 to be $\sim$38\%,	
  thus we arrive at 467\uJy.	
Sources whose optical magnitudes are given as ``$-$'' are dropouts and the	
position is measured from the {\it Spitzer}-IRAC 3.6\um.	
  }	
\end{spacing}	
\end{table*}

\begin{table*}	
\centering	
{\tiny \begin{tabular}{l@{ }c@{ }cc@{ }c@{ }c@{ }c@{ }c@{ }c@{ }c@{ }c@{ }c@{ }c@{ }c@{ }c@{ }c@{ }c}	
\hline\hline	
NAME & {\sc Counter-} & $p$- & RA$_{\rm 24}$ & DEC$_{\rm 24}$ & $\delta_{24}$ & $S_{\rm 24}$ & RA$_{\rm 1.4}$ & DEC$_{\rm 1.4}$ & $\delta_{1.4}$ & $S_{\rm 1.4}$ & RA$_{\rm opt}$ & DEC$_{\rm opt}$ & $\delta_{\rm opt}$ & $i$       & 3.6\um\ & $z_{\rm phot}$ \\	
     & {\sc part} \#  & $val$ &              &                &    [\arcsec]            &    [\uJy]    &                &                 &       [\arcsec]         &    [\uJy]     &          &           &  [\arcsec]                  & [AB-mag] & [AB-mag] &              \\	
\hline\hline	
450.46 & 1/2 & 0.02 &            &             &      &              &             &             &      &             & 10:00:32.46 & +02:21:49.1 & 0.74 & 	24.77 & 22.16 & 3.06 \\ 
       & 2/2 & 0.14 &            &             &      &              &             &             &      &             & 10:00:32.29 & +02:21:49.0 & 2.01 & 	24.69 & 22.49 & 2.73 \\ 
450.47 & 1/3 & 0.02 &10:00:25.24 & +02:19:32.9 & 2.24 & 84$\pm$15    &             &             &      &             & 10:00:25.26 & +02:19:33.1 & 2.47 & 	$-$   & $-$   & $-$  \\ 
       & 2/3 & 0.02 &            &             &      &              &             &             &      &             & 10:00:25.21 & +02:19:30.1 & 0.68 & 	28.22 & 24.39 & 2.77 \\ 
       & 3/3 & 0.19 &            &             &      &              &             &             &      &             & 10:00:25.07 & +02:19:29.7 & 2.36 & 	25.87 & 24.60 & 3.03 \\ 
450.48 & 1/3 & 0.09 &            &             &      &              &             &             &      &             & 10:00:09.37 & +02:22:24.3 & 1.62 & 	$-$   & 24.43 & $-$  \\ 
       & 2/3 & 0.11 &            &             &      &              &             &             &      &             & 10:00:09.57 & +02:22:24.6 & 1.75 & 	27.38 & 23.47 & $-$  \\ 
       & 3/3 & 0.23 &            &             &      &              &             &             &      &             & 10:00:09.60 & +02:22:21.8 & 2.68 & 	27.17 & 23.17 & $-$  \\ 
450.49 & 1/2 & 0.17 &            &             &      &              &             &             &      &             & 09:59:59.98 & +02:25:22.7 & 2.28 & 	25.41 & 24.16 & 1.70 \\ 
       & 2/2 & 0.23 &            &             &      &              &             &             &      &             & 10:00:00.20 & +02:25:23.0 & 2.68 & 	26.21 & $-$   & 2.55 \\ 
450.50 & 1/2 & 0.40 &            &             &      &              &             &             &      &             & 09:59:52.33 & +02:21:32.7 & 1.53 & 	27.79 & $-$   & $-$  \\ 
       & 2/2 & 0.48 &            &             &      &              &             &             &      &             & 09:59:52.21 & +02:21:35.3 & 1.75 & 	$-$   & $-$   & $-$  \\ 
450.51 & 1/3 & 0.13 &            &             &      &              &             &             &      &             & 10:00:44.41 & +02:23:11.9 & 1.95 & 	25.56 & 23.87 & 1.58 \\ 
       & 2/3 & 0.21 &            &             &      &              &             &             &      &             & 10:00:44.20 & +02:23:13.7 & 2.50 & 	26.22 & 22.57 & 2.58 \\ 
       & 3/3 & 0.25 &            &             &      &              &             &             &      &             & 10:00:44.32 & +02:23:16.4 & 2.77 & 	23.29 & 21.72 & 0.97 \\ 
450.52 & NA & \\
450.53 & 1/1 & 0.007 &10:00:10.33 & +02:20:27.1 & 1.19 & 413$\pm$15   &             &             &      &             & 10:00:10.32 & +02:20:27.4 & 1.36 & 	22.95 & 19.03 & 1.01 \\ 
450.54 & 1/2 & 0.006 &            &             &      &              &             &             &      &             & 10:00:26.86 & +02:23:18.5 & 0.40 & 	24.82 & 22.74 & 1.90 \\ 
       & 2/2 & 0.24 &            &             &      &              &             &             &      &             & 10:00:27.01 & +02:23:20.7 & 2.73 & 	27.04 & 24.52 & 2.01 \\ 
450.55 & 1/2 & 0.04 &10:00:23.96 & +02:17:50.1 & 2.78 & 162$\pm$17   &             &             &      &             & 10:00:24.01 & +02:17:50.3 & 3.26 & 	$-$   & $-$   & $-$  \\ 
       & 2/2 & 0.22 &            &             &      &              &             &             &      &             & 10:00:23.81 & +02:17:54.3 & 2.60 & 	24.73 & 22.64 & 2.60 \\ 
450.56 & 1/3 & 0.30 &            &             &      &              &             &             &      &             & 09:59:59.40 & +02:30:02.0 & 1.29 & 	26.49 & $-$   & 2.82 \\ 
       & 2/3 & 0.40 &            &             &      &              &             &             &      &             & 09:59:59.34 & +02:29:59.4 & 1.54 & 	26.06 & $-$   & 3.18 \\ 
       & 3/3 & 0.46 &            &             &      &              &             &             &      &             & 09:59:59.50 & +02:30:00.2 & 1.69 & 	26.82 & $-$   & 1.94 \\ 
450.57 & 1/1 & 0.11 &            &             &      &              &             &             &      &             & 10:00:32.46 & +02:18:04.4 & 1.79 & 	28.59 & 21.85 & 3.48 \\ 
450.58 & 1/1 & 0.02 &10:00:36.05 & +02:21:51.1 & 2.28 & 194$\pm$17   &             &             &      &             & 10:00:36.03 & +02:21:51.1 & 2.53 & 	24.47 & 20.79 & $-$  \\ 
450.59 & 1/2 & 0.005 &10:00:19.35 & +02:20:24.2 & 1.08 & 197$\pm$52   &             &             &      &             & 10:00:19.44 & +02:20:24.5 & 0.61 & 	20.61 & 19.60 & 0.47 \\ 
       & 2/2 & 0.04 &10:00:19.61 & +02:20:24.3 & 3.05 & 86$\pm$15    &             &             &      &             & $-$         & $-$         & $-$  & 	$-$   & $-$   & $-$  \\ 
450.60 & NA \\
450.61 & 1/1 & 2$\times$10$^{-5}$  &10:00:18.76 & +02:28:13.5 & 0.32 & 136$\pm$13   & 10:00:18.76 & +02:28:13.7 & 0.20 & 92$\pm$11   & 10:00:18.81 & +02:28:14.1 & 1.11 & 	25.04 & 21.88 & 4.58 \\ 
450.62 & 1/1 & 0.47 &            &             &      &              &             &             &      &             & 10:00:47.23 & +02:20:48.4 & 1.71 & 	27.78 & $-$   & 1.95 \\ 
450.63 & 1/3 & 0.05 &            &             &      &              &             &             &      &             & 10:00:19.94 & +02:21:29.2 & 1.20 & 	25.09 & 24.02 & 2.36 \\ 
       & 2/3 & 0.17 &            &             &      &              &             &             &      &             & 10:00:20.07 & +02:21:31.8 & 2.25 & 	24.40 & 23.74 & 2.84 \\ 
       & 3/3 & 0.22 &            &             &      &              &             &             &      &             & 10:00:19.87 & +02:21:28.1 & 2.58 & 	25.19 & 24.42 & 1.56 \\ 
450.64 & NA \\
450.65 & 1/1 & 0.30 &            &             &      &              &             &             &      &             & 10:00:27.23 & +02:24:49.8 & 1.29 & 	25.42 & $-$   & 0.70 \\ 
450.66 & 1/4 & 8$\times$10$^{-5}$ &10:01:04.63 & +02:26:34.0 & 0.47 & 586$\pm$24   & 10:01:04.64 & +02:26:34.0 & 0.38 & 86$\pm$11   & 10:01:04.63 & +02:26:33.2 & 0.51 & 	23.20 & 20.54 & $-$  \\ 
       & 2/4 & 0.07 &            &             &      &              &             &             &      &             & 10:01:04.70 & +02:26:34.9 & 1.45 & 	22.59 & 20.44 & $-$  \\ 
       & 3/4 & 0.22 &            &             &      &              &             &             &      &             & 10:01:04.53 & +02:26:35.5 & 2.60 & 	25.94 & 22.14 & $-$  \\ 
       & 4/4 & 0.23 &            &             &      &              &             &             &      &             & 10:01:04.56 & +02:26:31.4 & 2.64 & 	21.86 & 20.50 & $-$  \\ 
450.67 & 1/2 & 0.002 &10:00:00.37 & +02:29:03.9 & 1.43 & 538$\pm$15   & 10:00:00.35 & +02:29:04.4 & 1.76 & 114$\pm$14  & 10:00:00.35 & +02:29:04.1 & 1.47 & 	21.37 & 19.06 & 0.93 \\ 
       & 2/2 & 0.23 &            &             &      &              &             &             &      &             & 10:00:00.42 & +02:29:00.5 & 2.64 & 	25.83 & 23.47 & 1.67 \\ 
450.68 & NA & \\
450.69 & 1/2 & 0.05 &            &             &      &              &             &             &      &             & 10:00:47.55 & +02:25:21.6 & 1.16 & 	26.82 & 23.67 & 2.04 \\ 
       & 2/2 & 0.05 &10:00:47.29 & +02:25:20.7 & 3.20 & 251$\pm$14   &             &             &      &             & 10:00:47.27 & +02:25:20.3 & 3.45 & 	22.28 & 19.36 & 0.95 \\ 
450.70 & 1/2 & 0.006 &09:59:52.66 & +02:27:12.8 & 1.13 & 263$\pm$15   &             &             &      &             & 09:59:52.66 & +02:27:14.0 & 2.30 & 	23.77 & 21.06 & 3.03 \\ 
       & 2/2 & 0.10 &            &             &      &              &             &             &      &             & 09:59:52.59 & +02:27:10.3 & 1.67 & 	26.17 & 22.47 & 2.89 \\ 
450.71 & 1/1 & 0.006 &            &             &      &              &             &             &      &             & 09:59:46.40 & +02:29:32.0 & 0.42 & 	26.81 & 24.22 & 1.95 \\ 
450.72 & 1/2 & 0.20 &            &             &      &              &             &             &      &             & 09:59:50.44 & +02:20:18.9 & 2.48 & 	25.61 & 22.92 & 1.43 \\ 
       & 2/2 & 0.26 &            &             &      &              &             &             &      &             & 09:59:50.54 & +02:20:19.6 & 2.87 & 	24.92 & 23.11 & 1.72 \\ 
450.73 & NA  \\
450.74 & 1/2 & 0.13 &            &             &      &              &             &             &      &             & 10:00:41.26 & +02:16:42.1 & 1.96 & 	22.79 & 21.36 & 0.75 \\ 
       & 2/2 & 0.19 &            &             &      &              &             &             &      &             & 10:00:41.39 & +02:16:39.9 & 0.99 & 	26.45 & $-$   & 0.96 \\ 
450.75 & 1/2 & 0.08 &            &             &      &              &             &             &      &             & 10:00:34.20 & +02:34:22.4 & 1.48 & 	26.81 & 24.79 & 2.26 \\ 
       & 2/2 & 0.18 &            &             &      &              &             &             &      &             & 10:00:34.33 & +02:34:24.0 & 2.32 & 	$-$   & 23.01 & $-$  \\ 
450.76 & 1/2 & 0.02 &            &             &      &              &             &             &      &             & 10:00:21.76 & +02:31:14.0 & 0.77 & 	31.02 & 24.83 & 1.96 \\ 
       & 2/2 & 0.27 &            &             &      &              &             &             &      &             & 10:00:21.93 & +02:31:15.7 & 2.92 & 	25.27 & 23.61 & 0.73 \\ 
450.77 & 1/2 & 0.03 &            &             &      &              &             &             &      &             & 09:59:52.65 & +02:22:57.8 & 0.92 & 	25.12 & 21.05 & 1.43 \\ 
       & 2/2 & 0.07 &            &             &      &              &             &             &      &             & 09:59:52.62 & +02:23:00.0 & 1.41 & 	25.64 & 22.35 & 1.65 \\ 
\hline\hline	
\end{tabular}	
}	
	
{\small	
{\sc Table~\ref{tab:counterparts450} Continued.}}	
\end{table*}	

\begin{table*}	
\centering	
\caption{Counterpart identifications and Multiwavelength Properties for marginal \scubaii\ 450\um- and 850\um-detected Sources}	
{\tiny \begin{tabular}{l@{ }c@{ }c@{ }c@{ }c@{ }c@{ }c@{ }c@{ }c@{ }c@{ }c@{ }c@{ }c@{ }c@{ }c@{ }c@{ }c}	
\hline\hline	
NAME & {\sc Counter-} & $p$- & RA$_{\rm 24}$ & DEC$_{\rm 24}$ & $\delta_{24}$ & $S_{\rm 24}$ & RA$_{\rm 1.4}$ & DEC$_{\rm 1.4}$ & $\delta_{1.4}$ & $S_{\rm 1.4}$ & RA$_{\rm opt}$ & DEC$_{\rm opt}$ & $\delta_{\rm opt}$ & $i$       & 3.6\um\ & $z_{\rm phot}$ \\	
     & {\sc part} \#  & $val$ &              &                &    [\arcsec]            &    [\uJy]    &                &                 &       [\arcsec]         &    [\uJy]     &          &           &  [\arcsec]                  & [AB-mag] & [AB-mag] & \\	
\hline\hline	
  450.78/850.62 & 1/2 & 0.006 &            &             &      &              &             &             &      &             & 10:00:31.05 & +02:27:51.8 & 0.40 &  	27.35 & 22.40 & $-$  \\ 
                & 2/2 & 0.21 &            &             &      &              &             &             &      &             & 10:00:31.05 & +02:27:54.2 & 2.55 &  	26.26 & 24.53 & $-$  \\ 
  450.81/850.10 & 1/3 & 2$\times$10$^{-5}$ &10:00:34.33 & +02:21:21.4 & 0.53 & 130$\pm$17   & 10:00:34.37 & +02:21:21.6 & 0.21 & 517$\pm$23  & 10:00:34.33 & +02:21:21.0 & 0.76 &  	25.34 & 21.78 & 1.75 \\ 
                & 2/3 & 0.19 &            &             &      &              &             &             &      &             & 10:00:34.43 & +02:21:19.6 & 2.36 &  	25.18 & 23.21 & 1.66 \\ 
                & 3/3 & 0.27 &            &             &      &              &             &             &      &             & 10:00:34.53 & +02:21:20.2 & 2.94 &  	25.61 & 23.80 & 1.95 \\ 
  450.86/850.55 & 1/2 & 0.06 &            &             &      &              &             &             &      &             & 09:59:50.90 & +02:27:44.6 & 1.34 &  	25.86 & 23.56 & 3.59 \\ 
                & 2/2 & 0.15 &            &             &      &              &             &             &      &             & 09:59:50.97 & +02:27:46.7 & 2.07 &  	26.80 & 24.16 & 3.69 \\ 
  450.87/850.09 & 1/2 & 0.004 &10:00:25.29 & +02:18:45.9 & 2.76 & 204$\pm$29   & 10:00:25.28 & +02:18:46.2 & 2.53 & 58$\pm$12   & 10:00:25.32 & +02:18:46.3 & 2.98 &  	26.22 & 21.45 & 2.71 \\ 
                & 2/2 & 0.12 &            &             &      &              &             &             &      &             & 10:00:25.20 & +02:18:46.0 & 1.90 &  	26.82 & 21.87 & 4.85 \\ 
  450.94/850.23 & 1/3 & 0.03 &            &             &      &              &             &             &      &             & 10:00:22.26 & +02:28:43.2 & 0.83 &  	21.82 & 20.20 & 0.71 \\ 
                & 2/3 & 0.03 &10:00:22.12 & +02:28:44.8 & 2.50 & 128$\pm$15   &             &             &      &             & 10:00:22.04 & +02:28:43.9 & 2.92 &  	24.18 & 20.44 & 3.34 \\ 
                & 3/3 & 0.27 &            &             &      &              &             &             &      &             & 10:00:22.32 & +02:28:45.2 & 2.95 &  	24.83 & 24.95 & 0.52 \\ 
 450.96/850.133 & 1/1 & 0.26 &            &             &      &              &             &             &      &             & 10:00:49.95 & +02:24:51.0 &  2.85 & 	26.67 & 22.93 & 2.34 \\ 
  450.99/850.33 & 1/4 & 0.02 &09:59:50.90 & +02:28:22.5 & 2.19 & 217$\pm$15   &             &             &      &             & 09:59:50.86 & +02:28:22.4 & 1.77 &  	26.18 & 21.21 & 2.16 \\ 
                & 2/4 & 0.06 &            &             &      &              &             &             &      &             & 09:59:50.69 & +02:28:23.8 & 1.30 &  	27.43 & 22.86 & 2.15 \\ 
                & 3/4 & 0.19 &            &             &      &              &             &             &      &             & 09:59:50.64 & +02:28:22.5 & 2.38 &  	26.51 & 23.82 & 1.82 \\ 
                & 4/4 & 0.29 &            &             &      &              &             &             &      &             & 09:59:50.78 & +02:28:26.7 & 3.03 &  	26.75 & 24.02 & 2.33 \\ 
450.103/850.128 & 1/1 & 0.42 &            &             &      &              &             &             &      &             & 10:00:18.42 & +02:23:59.3 &  3.87 & 	23.82 & 21.86 & 1.54 \\ 
 450.105/850.85 & 1/1 & 0.03 &10:00:25.19 & +02:27:55.8 & 2.58 & 173$\pm$13   &             &             &      &             & 10:00:25.17 & +02:27:56.0 & 2.20 &  	22.82 & 19.21 & $-$  \\ 
 450.106/850.92 & 1/1 & 0.44 &            &             &      &              &             &             &      &             & 10:00:42.08 & +02:23:59.8 & 3.95 &  	23.33 & 20.67 & 1.00 \\ 
450.126/850.159 & 1/1 & 0.32 &            &             &      &              &             &             &      &             & 10:00:25.70 & +02:30:51.2 &  1.33 & 	26.35 & $-$   & 2.04 \\ 
450.133/850.131 & 1/4 & 0.03 &10:00:29.67 & +02:21:29.6 & 2.74 & 518$\pm$16   &             &             &      &             & 10:00:29.68 & +02:21:29.8 & 2.68 &  	20.30 & 18.51 & 0.93 \\ 
                & 2/4 & 0.16 &            &             &      &              &             &             &      &             & 10:00:29.60 & +02:21:33.8 & 2.14 &  	26.21 & 23.39 & 2.61 \\ 
                & 3/4 & 0.22 &            &             &      &              &             &             &      &             & 10:00:29.68 & +02:21:33.5 & 2.61 &  	26.63 & 23.34 & 1.43 \\ 
                & 4/4 & 0.36 &            &             &      &              &             &             &      &             & 10:00:29.46 & +02:21:35.0 & 3.49 &  	25.41 & 24.20 & 0.46 \\ 
 450.134/850.52 & 1/1 & 0.0002 &10:00:11.87 & +02:29:35.9 & 0.20 & 110$\pm$16   &             &             &      &             & 10:00:11.82 & +02:29:35.3 & 0.77 &  	$-$   & $-$   & $-$  \\ 
                & 2/4 & 0.19 &            &             &      &              &             &             &      &             & 10:00:11.82 & +02:29:33.5 & 2.36 &  	25.99 & 22.11 & 2.89 \\ 
                & 3/4 & 0.21 &            &             &      &              &             &             &      &             & 10:00:12.02 & +02:29:34.7 & 2.51 &  	27.82 & 23.50 & 2.54 \\ 
                & 4/4 & 0.30 &            &             &      &              &             &             &      &             & 10:00:11.86 & +02:29:38.9 & 3.14 &  	28.49 & 23.70 & 4.60 \\ 
450.135/850.163 & 1/4 & 0.01 &09:59:52.96 & +02:26:42.7 & 1.64 & 340$\pm$14   &             &             &      &             & 09:59:52.95 & +02:26:42.8 & 1.83 &  	24.45 & 21.04 & 1.68 \\ 
                & 2/4 & 0.05 &            &             &      &              &             &             &      &             & 09:59:52.92 & +02:26:40.3 & 2.46 &  	25.88 & 23.47 & 3.06 \\ 
                & 3/4 & 0.08 &            &             &      &              &             &             &      &             & 09:59:53.12 & +02:26:40.7 & 1.46 &  	26.10 & 24.50 & 1.11 \\ 
                & 4/4 & 0.20 &            &             &      &              &             &             &      &             & 09:59:53.18 & +02:26:42.6 & 2.10 &  	25.67 & 24.00 & 0.66 \\ 
 450.166/850.83 & 1/2 & 0.14 &            &             &      &              &             &             &      &             & 10:00:10.34 & +02:17:57.3 & 3.31 &  	23.38 & 22.20 & 0.93 \\ 
                & 2/2 & 0.33 &            &             &      &              &             &             &      &             & 10:00:10.17 & +02:17:59.7 & 1.12 &  	26.71 & 21.35 & 3.25 \\ 
450.173/850.104 & 1/1 & 0.002 &10:00:05.38 & +02:25:16.1 & 0.67 & 260$\pm$20   &             &             &      &             & 10:00:05.31 & +02:25:15.7 & 1.67 &  	21.55 & 19.92 & 1.01 \\ 
                & 2/3 & 0.27 &            &             &      &              &             &             &      &             & 10:00:05.23 & +02:25:18.2 & 2.93 &  	24.67 & 22.75 & 1.09 \\ 
                & 3/3 & 0.33 &            &             &      &              &             &             &      &             & 10:00:05.61 & +02:25:17.4 & 3.28 &  	23.95 & 20.04 & 1.19 \\ 
 450.179/850.88 & 1/4 & 0.02 &10:00:22.77 & +02:30:25.3 & 2.02 & 103$\pm$14   &             &             &      &             & 10:00:22.78 & +02:30:25.2 & 2.06 &  	22.06 & 20.69 & 0.66 \\ 
                & 2/4 & 0.06 &            &             &      &              &             &             &      &             & 10:00:22.65 & +02:30:22.5 & 1.33 &  	25.78 & 24.99 & 0.55 \\ 
                & 3/4 & 0.13 &            &             &      &              &             &             &      &             & 10:00:22.76 & +02:30:22.2 & 1.93 &  	25.06 & 24.54 & 0.65 \\ 
                & 4/4 & 0.13 &            &             &      &              &             &             &      &             & 10:00:22.55 & +02:30:24.0 & 1.96 &  	25.68 & 23.31 & 2.48 \\ 
450.189/850.114 & 1/2 & 0.003 &10:00:14.10 & +02:28:38.6 & 1.94 & 1345$\pm$17  & 10:00:14.07 & +02:28:38.8 &  2.30 & 146$\pm$11 & 10:00:14.08 & +02:28:38.7 &  2.16 & 	$-$   & 18.34 & $-$  \\ 
                & 2/2 & 0.14 &            &             &      &              &             &             &      &             & 10:00:14.25 & +02:28:35.6 &  1.99 & 	26.30 & 22.83 & $-$  \\ 
 450.193/850.49 & 1/6 & 0.05 &            &             &      &              &             &             &      &             & 10:00:05.13 & +02:17:19.4 & 1.14 &  	24.62 & 21.16 & 3.08 \\ 
                & 2/6 & 0.08 &10:00:05.09 & +02:17:14.6 & 4.08 & 159$\pm$28   &             &             &      &             & 10:00:05.06 & +02:17:14.1 & 4.63 &  	22.61 & 20.15 & 1.15 \\ 
                & 3/6 & 0.14 &            &             &      &              &             &             &      &             & 10:00:04.94 & +02:17:19.3 & 2.03 &  	26.59 & 22.68 & 3.06 \\ 
                & 4/6 & 0.29 &            &             &      &              &             &             &      &             & 10:00:04.91 & +02:17:16.7 & 3.06 &  	24.95 & 22.08 & 2.00 \\ 
                & 5/6 & 0.30 &            &             &      &              &             &             &      &             & 10:00:04.96 & +02:17:21.4 & 3.13 &  	26.57 & 23.26 & 3.15 \\ 
                & 6/6 & 0.33 &            &             &      &              &             &             &      &             & 10:00:05.26 & +02:17:17.2 & 3.31 &  	23.99 & 22.50 & 3.09 \\ 
 450.206/850.53 & 1/4 & 0.08 &            &             &      &              &             &             &      &             & 10:00:11.10 & +02:15:06.4 & 1.47 &  	25.86 & 24.67 & 1.70 \\ 
                & 2/4 & 0.22 &            &             &      &              &             &             &      &             & 10:00:11.24 & +02:15:05.6 & 2.58 &  	25.27 & 23.60 & 1.65 \\ 
                & 3/4 & 0.31 &            &             &      &              &             &             &      &             & 10:00:11.07 & +02:15:04.7 & 3.15 &  	25.49 & 23.45 & 1.88 \\ 
                & 4/4 & 0.41 &            &             &      &              &             &             &      &             & 10:00:10.89 & +02:15:07.1 & 3.80 &  	26.08 & 22.93 & 1.44 \\ 
 450.215/850.27 & 1/2 & 0.01 &10:00:23.69 & +02:19:15.0 & 1.69 & 119$\pm$17   &             &             &      &             & 10:00:23.67 & +02:19:15.1 & 1.67 &  	25.19 & 21.73 & 2.76 \\ 
                & 3/3 & 0.38 &            &             &      &              &             &             &      &             & 10:00:23.66 & +02:19:13.1 & 3.62 &  	26.62 & 22.74 & 2.87 \\ 
450.240/850.67  & 1/2 & 0.37 &            &             &      &              &             &             &      &             & 10:00:32.03 & +02:33:28.1 & 3.53 &  	28.88 & 23.08 & 2.73 \\ 
                & 2/2 & 0.40 &            &             &      &              &             &             &      &             & 10:00:32.23 & +02:33:27.8 & 3.72 &  	25.63 & 22.48 & 2.64 \\ 
450.247/850.151 & 1/2 & 0.15 &            &             &      &              &             &             &      &             & 10:00:41.42 & +02:25:33.7 &  2.10 & 	26.92 & 25.04 & 1.78 \\ 
                & 2/2 & 0.18 &            &             &      &              &             &             &      &             & 10:00:41.35 & +02:25:34.4 &  0.95 & 	26.95 & $-$   & 1.74 \\ 
 450.252/850.40 & 1/5 & 0.003 &10:00:13.58 & +02:22:25.6 & 2.78 & 605$\pm$14   & 10:00:13.55 & +02:22:25.7 & 2.43 & 88$\pm$15   & 10:00:13.57 & +02:22:25.4 & 2.52 &  	23.10 & 19.95 & $-$  \\ 
                & 2/5 & 0.30 &            &             &      &              &             &             &      &             & 10:00:13.21 & +02:22:23.6 & 3.11 &  	25.70 & 23.68 & $-$  \\ 
                & 3/5 & 0.34 &            &             &      &              &             &             &      &             & 10:00:13.41 & +02:22:21.4 & 3.33 &  	23.81 & 23.66 & $-$  \\ 
                & 4/5 & 0.37 &            &             &      &              &             &             &      &             & 10:00:13.21 & +02:22:26.7 & 3.55 &  	24.31 & 23.57 & $-$  \\ 
                & 5/5 & 0.44 &            &             &      &              &             &             &      &             & 10:00:13.31 & +02:22:21.1 & 3.94 &  	26.19 & 24.18 & $-$  \\ 
\hline\hline	
\end{tabular}	
}	
\label{tab:counterpartsextra450}	
	
{\small {\bf Table Notes.}  	
All quantities are measured and given as in Table~\ref{tab:counterparts450}.	
  }	
\end{table*}

\begin{table*}	
\centering	
\caption{Counterpart identifications and Multiwavelength Properties for \scubaii\ 850\um-detected Sources}	
{\tiny \begin{tabular}{l@{ }c@{ }c@{ }c@{ }c@{ }c@{ }c@{ }c@{ }c@{ }c@{ }c@{ }c@{ }c@{ }c@{ }c@{ }c@{ }c@{ }c}	
\hline\hline	
NAME & {\sc Counter-} & $p$- & RA$_{\rm 24}$ & DEC$_{\rm 24}$ & $\delta_{24}$ & $S_{\rm 24}$ & RA$_{\rm 1.4}$ & DEC$_{\rm 1.4}$ & $\delta_{1.4}$ & $S_{\rm 1.4}$ & RA$_{\rm opt}$ & DEC$_{\rm opt}$ & $\delta_{\rm opt}$ & $i$       & 3.6\um\ & $z_{\rm phot}$ \\	
     & {\sc part} \#  & $val$ &               &                &    [\arcsec]            &    [\uJy]    &                &                 &       [\arcsec]         &    [\uJy]     &          &           &  [\arcsec]                  & [AB-mag] & [AB-mag] & \\	
\hline\hline	
850.01 & 1/1 & 0.0008 &           &             &      &              & 09:59:57.30 & +02:27:30.5 & 1.20 & 68$\pm$13   & {\bf 09:59:57.29} & {\bf +02:27:30.1} & 1.23 &     25.67 & 23.22 & 1.45 \\
850.04 & 1/1 & 0.01 & 10:00:15.47 & +02:15:50.5 & 1.78 & 123$\pm$18   &             &             &      &             & 10:00:15.49 & +02:15:50.2 & 1.43 & 	23.83 & 20.72 & 1.39 \\ 
850.05 & 1/2 & 0.02 &             &             &      &              &             &             &      &             & 10:00:57.02 & +02:20:14.0 & 0.31 &     24.59 & $-$   & 2.76 \\
       & 2/2 & 0.19 &             &             &      &              &             &             &      &             & 10:00:57.01 & +02:20:12.0 & 2.32 &     27.03 & 22.18 & 2.48 \\
850.11 & 1/2 & 0.10 &             &             &      &              &             &             &      &             & 10:00:49.90 & +02:22:58.8 & 1.71 & 	25.83 & 22.08 & 2.77 \\ 
       & 2/2 & 0.11 &             &             &      &              &             &             &      &             & 10:00:49.94 & +02:22:55.4 & 1.82 & 	27.02 & 24.07 & 2.68 \\ 
850.13 & 1/3 & 0.002 &             &             &      &              &             &             &      &             & 10:00:24.12 & +02:20:05.0 & 0.28 & 	$-$   & 24.09 & 0.86 \\ 
       & 2/3 & 0.21 &             &             &      &              &             &             &      &             & 10:00:24.27 & +02:20:03.8 & 2.55 & 	26.14 & 24.71 & 1.03 \\ 
       & 3/3 & 0.22 &             &             &      &              &             &             &      &             & 10:00:24.08 & +02:20:02.8 & 2.58 & 	24.63 & 23.81 & 0.94 \\ 
850.15 & 1/1 & 0.08 &             &             &      &              &             &             &      &             & 10:00:25.25 & +02:26:05.9 & 1.50 & 	21.70 & 20.25 & $-$  \\ 
850.16 & 1/1 & 0.002 & 10:01:03.92 & +02:24:48.2 & 1.08 & 229$\pm$18   & 10:01:03.91 & +02:24:48.9 & 1.69 & 57$\pm$10   & 10:01:03.92 & +02:24:48.0 & 0.89 & 	24.95 & 20.82 & $-$  \\ 
       & 2/2 & 0.18 &             &             &      &              &             &             &      &             & 10:01:03.83 & +02:24:45.1 & 2.32 & 	23.15 & 20.70 & $-$  \\ 
850.17 & 1/1 & 0.29 &             &             &      &              &             &             &      &             & 10:00:00.36 & +02:22:56.3 & 1.26 & 	26.09 & $-$   & $-$  \\ 
850.18 & 1/2 & 0.01 &             &             &      &              &             &             &      &             & 10:00:24.02 & +02:29:47.3 & 0.62 & 	25.67 & 23.71 & 3.01 \\ 
       & 2/2 & 0.27 &             &             &      &              &             &             &      &             & 10:00:24.04 & +02:29:50.1 & 2.90 & 	26.81 & 23.61 & 2.83 \\ 
850.19 & 1/2 & 0.11 &             &             &      &              &             &             &      &             & 10:00:15.67 & +02:24:45.5 & 1.76 & 	26.87 & 24.99 & 2.24 \\ 
       & 2/2 & 0.21 &             &             &      &              &             &             &      &             & 10:00:15.92 & +02:24:46.8 & 2.51 & 	27.43 & 23.17 & 3.61 \\ 
850.20 & 1/1 & 0.18 &             &             &      &              &             &             &      &             & 10:00:27.13 & +02:31:38.5 & 2.32 & 	24.77 & 22.22 & 2.80 \\ 
{\bf 850.22} & 1/2 & 0.05 & 10:00:18.69 & +02:16:51.9 & 3.38 & 300$\pm$17 &         &             &      &             & 10:00:18.70 & +02:16:52.5 & 2.72 & 	25.50 & 20.40 & 2.06 \\ 
       & 2/2 & 0.26 &             &             &      &              &             &             &      &             & 10:00:18.77 & +02:16:58.0 & 2.83 & 	27.05 & 24.11 & 2.29 \\ 
850.24 & 1/1 & 0.0002 & 10:00:04.82 & +02:30:45.3 & 0.51 & 321$\pm$14 & 10:00:04.81 & +02:30:45.2 & 0.55 & 96$\pm$15   & 10:00:04.78 & +02:30:45.2 & 1.00 &     27.21 & 20.82 & 3.51 \\
850.25 & 1/5 & 0.0002 & 10:00:12.59 & +02:14:44.1 & 0.20 & 331$\pm$15   &             &             &      &             & {\bf 10:00:12.58} & {\bf +02:14:44.0} & 0.20 & 	24.09 & 20.31 & 2.33 \\ 
       & 2/5 & 0.08 &             &             &      &              &             &             &      &             & 10:00:12.57 & +02:14:42.7 & 1.52 & 	23.67 & 20.79 & 2.78 \\ 
       & 3/5 & 0.18 &             &             &      &              &             &             &      &             & 10:00:12.71 & +02:14:45.6 & 2.30 & 	26.28 & 22.99 & 2.61 \\ 
       & 4/5 & 0.20 &             &             &      &              &             &             &      &             & 10:00:12.75 & +02:14:43.9 & 2.44 & 	26.27 & 23.01 & 2.32 \\ 
       & 5/5 & 0.20 &             &             &      &              &             &             &      &             & 10:00:12.44 & +02:14:43.1 & 2.45 & 	25.83 & 22.14 & 2.75 \\ 
850.29 & 1/2 & 0.03 & 10:00:50.18 & +02:26:17.8 & 2.57 & 155$\pm$18   &             &             &      &             & 10:00:50.15 & +02:26:18.3 & 3.10 & 	$-$   & 21.12 & $-$  \\ 
       & 2/2 & 0.02 &             &             &      &              &             &             &      &             & 10:00:50.15 & +02:26:14.4 & 0.77 & 	25.86 & 23.42 & $-$  \\ 
850.30 & 1/1 & 0.02 & 10:00:06.82 & +02:22:07.4 & 1.89 & 186$\pm$14   &             &             &      &             & 10:00:06.80 & +02:22:07.5 & 1.71 & 	20.94 & 20.61 & 0.23 \\ 
       & 2/2 & 0.10 &             &             &      &              &             &             &      &             & 10:00:06.86 & +02:22:11.6 & 2.64 & 	25.36 & 23.80 & $-$  \\ 
850.31 & 1/1 & 0.02 & 10:00:26.17 & +02:17:39.0 & 2.27 & 296$\pm$44   &             &             &      &             & 10:00:26.18 & +02:17:39.5 & 1.89 & 	24.43 & 20.46 & 1.80 \\ 
850.32 & 1/2 & 0.004 & 10:00:01.67 & +02:24:27.9 & 2.73 & 287$\pm$12  & 10:00:01.67 & +02:24:28.3 & 2.52 & 90$\pm$14   & 10:00:01.62 & +02:24:28.0 & 2.02 &     27.53 & 21.71 & 1.95 \\
       & 2/2 & 0.34 &             &            &       &              &             &             &      &             & 10:00:01.44 & +02:24:28.4 & 1.40 &     25.07 & $-$   & 1.58 \\
850.35 & 1/2 & 0.06 &             &             &      &              &             &             &      &             & 10:01:05.07 & +02:21:51.5 & 1.35 & 	25.59 & 23.02 & 2.36 \\ 
       & 2/2 & 0.23 &             &             &      &              &             &             &      &             & 10:01:05.15 & +02:21:52.7 & 2.65 & 	25.08 & 22.90 & 2.63 \\ 
850.36 & 1/3 & 0.04 & 10:00:01.23 & +02:17:42.4 & 3.08 & 127$\pm$49   &             &             &      &             & 10:00:01.38 & +02:17:42.4 & 2.96 & 	24.20 & 20.42 & 2.05 \\ 
       & 2/3 & 0.13 &             &             &      &              &             &             &      &             & 10:00:01.21 & +02:17:43.9 & 1.94 & 	20.88 & 21.44 & 0.22 \\ 
       & 3/3 & 0.13 &             &             &      &              &             &             &      &             & 10:00:01.44 & +02:17:45.2 & 1.97 & 	24.09 & 21.70 & 2.04 \\ 
850.37 & 1/3 & 0.12 &             &             &      &              &             &             &      &             & 10:00:35.70 & +02:28:27.4 & 1.84 & 	28.03 & 23.31 & 4.57 \\ 
       & 2/3 & 0.15 &             &             &      &              &             &             &      &             & 10:00:35.68 & +02:28:28.0 & 2.08 & 	27.96 & 23.31 & 3.99 \\ 
       & 3/3 & 0.27 &             &             &      &              &             &             &      &             & 10:00:35.54 & +02:28:29.0 & 2.92 & 	28.41 & 23.98 & 3.53 \\ 
850.38 & 1/4 & 0.07 &             &             &      &              &             &             &      &             & 10:00:23.45 & +02:15:35.5 & 1.39 & 	24.90 & 22.75 & 2.04 \\ 
       & 2/4 & 0.08 &             &             &      &              &             &             &      &             & 10:00:23.58 & +02:15:37.5 & 1.50 & 	26.39 & 23.01 & 2.72 \\ 
       & 3/4 & 0.14 &             &             &      &              &             &             &      &             & 10:00:23.63 & +02:15:35.0 & 1.99 & 	27.04 & 24.92 & 2.58 \\ 
       & 4/4 & 0.22 &             &             &      &              &             &             &      &             & 10:00:23.36 & +02:15:36.4 & 2.60 & 	26.60 & 23.31 & 2.43 \\ 
850.39 & 1/2 & 0.01 &             &             &      &              &             &             &      &             & 10:00:12.21 & +02:23:09.8 & 0.62 & 	26.48 & 24.32 & $-$  \\ 
       & 2/2 & 0.18 &             &             &      &              &             &             &      &             & 10:00:12.16 & +02:23:07.9 & 2.31 & 	24.76 & 23.20 & $-$  \\ 
850.43 & 1/1 & 0.12 &             &             &      &              &             &             &      &             & 10:00:26.31 & +02:15:29.0 & 0.77 & 	$-$   & $-$   & 0.84 \\ 
850.44 & 1/3 & 0.007 & 10:00:18.32 & +02:22:50.0 & 1.24 & 128$\pm$16   &             &             &      &             & 10:00:18.29 & +02:22:51.0 & 1.81 & 	25.03 & 22.78 & 2.48 \\ 
       & 2/3 & 0.11 &             &             &      &              &             &             &      &             & 10:00:18.16 & +02:22:50.3 & 1.79 & 	23.27 & 21.41 & 2.56 \\ 
       & 3/3 & 0.14 &             &             &      &              &             &             &      &             & 10:00:18.33 & +02:22:47.5 & 2.06 & 	27.58 & 23.56 & 2.03 \\ 
850.45 & 1/1 & 0.06 & 10:00:06.75 & +02:20:44.7 & 3.52 & 105$\pm$33   &             &             &      &             & 10:00:06.78 & +02:20:44.8 & 3.07 & 	24.22 & 21.03 & 1.32 \\ 
850.46 & 1/1 & 0.0006 & 10:00:25.54 & +02:15:05.9 & 1.12 & 560$\pm$17 & 10:00:25.52 & +02:15:05.8 & 1.01 & 112$\pm$10  & 10:00:25.51 & +02:15:06.1 & 0.65 &     26.03 & 20.78 & 3.04 \\
850.50 & 1/3 & 0.009 &             &             &      &              & 10:00:08.97 & +02:20:26.5 & 3.92 & 62$\pm$12   & 10:00:09.11 & +02:20:26.4 & 3.20 & 	26.07 & 23.73 & 2.38 \\ 
       & 2/3 & 0.02 &             &             &      &              &             &             &      &             & 10:00:09.17 & +02:20:23.2 & 0.78 & 	25.06 & 22.90 & 1.97 \\ 
       & 3/3 & 0.24 &             &             &      &              &             &             &      &             & 10:00:08.96 & +02:20:24.7 & 2.71 & 	26.43 & 23.47 & 2.87 \\ 
850.51 & 1/4 & 0.007 & 09:59:56.41 & +02:18:53.0 & 1.23 & 193$\pm$15   &             &             &      &             & 09:59:56.39 & +02:18:52.7 & 1.66 & 	24.39 & 20.82 & 1.11 \\ 
       & 2/4 & 0.23 &             &             &      &              &             &             &      &             & 09:59:56.32 & +02:18:56.2 & 2.64 & 	25.18 & 22.04 & 1.09 \\ 
       & 3/4 & 0.26 &             &             &      &              &             &             &      &             & 09:59:56.56 & +02:18:56.5 & 2.89 & 	21.06 & 19.07 & 0.66 \\ 
       & 4/4 & 0.29 &             &             &      &              &             &             &      &             & 09:59:56.63 & +02:18:53.3 & 3.04 & 	25.46 & 22.37 & 1.02 \\ 
850.54 & 1/1 & 0.02 &             &             &      &              &             &             &      &             & 09:59:48.89 & +02:27:48.9 & 0.71 & 	26.82 & 23.91 & 3.07 \\ 
850.56 & NA  \\
850.57 & 1/4 & 0.0003 & 10:00:06.90 & +02:33:08.1 & 1.20 & 592$\pm$184  & 10:00:06.82 & +02:33:07.9 & 0.75 & 82$\pm$14   & $-$         & $-$         & $-$  & 	$-$   & $-$   & $-$  \\ 
       & 2/4 & 0.13 &             &             &      &              &             &             &      &             & 10:00:06.81 & +02:33:05.4 & 1.94 & 	23.93 & 22.89 & 0.50 \\ 
       & 3/4 & 0.20 &             &             &      &              &             &             &      &             & 10:00:06.68 & +02:33:07.2 & 2.49 & 	20.13 & 18.28 & 0.74 \\ 
       & 4/4 & 0.21 &             &             &      &              &             &             &      &             & 10:00:07.01 & +02:33:07.2 & 2.51 & 	24.78 & 21.95 & 1.51 \\ 
850.59 & 1/4 & 0.007 & 10:00:00.80 & +02:20:02.4 & 1.21 & 85$\pm$13    &             &             &      &             & 10:00:00.84 & +02:20:02.4 & 1.60 & 	26.08 & 24.06 & 3.15 \\ 
       & 2/4 & 0.15 &             &             &      &              &             &             &      &             & 10:00:00.92 & +02:20:01.2 & 2.13 & 	26.18 & 24.01 & 2.64 \\ 
       & 3/4 & 0.16 &             &             &      &              &             &             &      &             & 10:00:00.77 & +02:20:03.4 & 2.15 & 	24.22 & 21.79 & 2.97 \\ 
       & 4/4 & 0.23 &             &             &      &              &             &             &      &             & 10:00:00.60 & +02:20:00.8 & 2.69 & 	24.59 & 22.85 & 0.97 \\ 
850.60 & 1/2 & 0.03 & 10:00:00.86 & +02:27:41.2 & 2.54 & 239$\pm$14   &             &             &      &             & 10:00:00.82 & +02:27:41.3 & 2.03 & 	24.38 & 19.74 & 1.44 \\ 
       & 2/2 & 0.10 &             &             &      &              &             &             &      &             & 10:00:00.81 & +02:27:39.6 & 1.69 & 	26.20 & 22.12 & 2.75 \\ 
850.61 & 1/1 & 0.06 &             &             &      &              &             &             &      &             & 10:00:01.96 & +02:28:20.8 & 1.34 & 	23.88 & 22.16 & 0.53 \\ 
850.63 & 1/3 & 0.08 &             &             &      &              &             &             &      &             & 09:59:53.47 & +02:18:50.6 & 1.55 & 	25.87 & 23.90 & 1.89 \\ 
       & 2/3 & 0.30 &             &             &      &              &             &             &      &             & 09:59:53.36 & +02:18:48.9 & 1.27 & 	25.40 & $-$   & 2.35 \\ 
       & 3/3 & 0.36 &             &             &      &              &             &             &      &             & 09:59:53.32 & +02:18:51.4 & 1.43 & 	$-$   & $-$   & $-$  \\ 
850.64 & 1/3 & 0.04 &             &             &      &              &             &             &      &             & 10:00:00.72 & +02:21:37.5 & 1.10 & 	26.44 & 24.01 & 3.39 \\ 
       & 2/3 & 0.05 &             &             &      &              &             &             &      &             & 10:00:00.66 & +02:21:36.1 & 1.19 & 	24.90 & 23.58 & 2.53 \\ 
       & 3/3 & 0.28 &             &             &      &              &             &             &      &             & 10:00:00.84 & +02:21:36.6 & 3.01 & 	27.94 & 23.18 & 4.12 \\ 
850.65 & 1/2 & 0.06 & 10:00:59.10 & +02:21:11.4 & 3.68 & 172$\pm$15   &             &             &      &             & 10:00:59.04 & +02:21:11.1 & 4.09 & 	23.33 & 21.76 & 0.48 \\ 
       & 2/2 & 0.13 &             &             &      &              &             &             &      &             & 10:00:59.13 & +02:21:09.6 & 1.97 & 	23.02 & 21.30 & 1.27 \\ 
850.66 & 1/5 & 0.09 &             &             &      &              &             &             &      &             & 10:00:24.46 & +02:28:30.0 & 1.60 & 	24.98 & 23.96 & 0.94 \\ 
       & 2/5 & 0.19 &             &             &      &              &             &             &      &             & 10:00:24.44 & +02:28:33.5 & 2.36 & 	25.76 & 23.26 & 2.47 \\ 
       & 3/5 & 0.21 &             &             &      &              &             &             &      &             & 10:00:24.56 & +02:28:30.6 & 2.56 & 	23.58 & 23.62 & 0.36 \\ 
       & 4/5 & 0.21 &             &             &      &              &             &             &      &             & 10:00:24.23 & +02:28:31.7 & 2.54 & 	26.76 & 24.77 & 1.52 \\ 
       & 5/5 & 0.31 &             &             &      &              &             &             &      &             & 10:00:24.32 & +02:28:34.2 & 3.16 & 	26.58 & 23.31 & 2.26 \\ 
850.68 & 1/4 & 0.05 & 09:59:55.30 & +02:19:51.1 & 3.13 & 218$\pm$14   &             &             &      &             & 09:59:55.26 & +02:19:51.2 & 3.09 & 	25.31 & 20.47 & 1.95 \\ 
       & 2/4 & 0.02 &             &             &      &              &             &             &      &             & 09:59:55.31 & +02:19:53.5 & 0.71 & 	26.22 & 21.78 & 2.04 \\ 
       & 3/4 & 0.11 &             &             &      &              &             &             &      &             & 09:59:55.31 & +02:19:56.0 & 1.78 & 	24.09 & 22.27 & 0.93 \\ 
       & 4/4 & 0.30 &             &             &      &              &             &             &      &             & 09:59:55.10 & +02:19:53.4 & 3.14 & 	27.59 & 23.11 & 2.30 \\ 
850.69 & 1/1 & 0.06 & 10:00:43.66 & +02:28:55.6 & 3.66 & 142$\pm$15   &             &             &      &             & 10:00:43.64 & +02:28:55.3 & 4.15 & 	25.60 & 21.28 & 1.80 \\ 
850.70 & 1/1 & 0.005 & 10:00:20.01 & +02:30:19.5 & 1.00 & 119$\pm$15   &             &             &      &             & 10:00:20.07 & +02:30:19.3 & 0.96 & 	23.46 & 20.56 & 0.78 \\ 
       & 2/2 & 0.29 &             &             &      &              &             &             &      &             & 10:00:20.24 & +02:30:21.5 & 3.03 & 	26.66 & 23.75 & 2.46 \\ 
\hline\hline	
\end{tabular}	
}	
\label{tab:counterparts850}	

{\small {\bf Table Notes.}  	
All quantities are measured and given as in
Table~\ref{tab:counterparts450}.  Sources which have
450\um\ counterparts are excluded from this table as their
450\um\ positional uncertainty is smaller than their
850\um\ uncertainty; the former is used to determine their
counterparts which are given in either Table~\ref{tab:counterparts450}
or Table~\ref{tab:counterpartsextra450}.  Source 850.22 is in bold
text in this table to indicate it was interferometrically observed by
\citep{smolcic12a} however no source was detected; it is likely that
the reason for this is that the submillimeter emission eminates from
multiple possible counterparts.  
%
  }	
\end{table*}	
	
\begin{table*}	
\centering	
{\tiny \begin{tabular}{l@{ }c@{ }c@{ }c@{ }c@{ }c@{ }c@{ }c@{ }c@{ }c@{ }c@{ }c@{ }c@{ }c@{ }c@{ }c@{ }c@{ }c}	
\hline\hline	
NAME & {\sc Counter-} & $p$- & RA$_{\rm 24}$ & DEC$_{\rm 24}$ & $\delta_{24}$ & $S_{\rm 24}$ & RA$_{\rm 1.4}$ & DEC$_{\rm 1.4}$ & $\delta_{1.4}$ & $S_{\rm 1.4}$ & RA$_{\rm opt}$ & DEC$_{\rm opt}$ & $\delta_{\rm opt}$ & $i$       & 3.6\um\ & $z_{\rm phot}$ \\	
     & {\sc part} \#  & $val$ &              &                &    [\arcsec]            &    [\uJy]    &                &                 &       [\arcsec]         &    [\uJy]     &          &           &  [\arcsec]                  & [AB-mag] & [AB-mag] & \\	
\hline\hline	
850.71 & 1/2 & 0.24 &             &             &      &              &             &             &      &             & 09:59:44.05 & +02:21:06.3 & 1.12 & 	27.01 & $-$   & 2.17 \\ 
       & 2/2 & 0.08 &             &             &      &              &             &             &      &             & 09:59:43.99 & +02:21:03.8 & 1.51 & 	24.55 & 24.13 & 0.90 \\ 
850.72 & 1/4 & 0.04 & 10:00:13.69 & +02:17:34.2 & 3.06 & 150$\pm$13   &             &             &      &             & 10:00:13.70 & +02:17:34.8 & 3.62 & 	$-$   & $-$   & $-$  \\ 
       & 2/4 & 0.21 &             &             &      &              &             &             &      &             & 10:00:13.73 & +02:17:28.9 & 2.55 & 	$-$   & 24.84 & 0.93 \\ 
       & 3/4 & 0.23 &             &             &      &              &             &             &      &             & 10:00:13.81 & +02:17:32.6 & 2.67 & 	25.64 & 23.40 & 2.08 \\ 
       & 4/4 & 0.26 &             &             &      &              &             &             &      &             & 10:00:13.47 & +02:17:32.2 & 2.88 & 	25.48 & 24.22 & 1.82 \\ 
850.73 & 1/5 & 0.04 & 10:00:20.52 & +02:22:49.2 & 2.93 & 112$\pm$26   &             &             &      &             & 10:00:20.48 & +02:22:48.6 & 3.80 & 	26.48 & 22.48 & 3.91 \\ 
       & 2/5 & 0.005 &             &             &      &              &             &             &      &             & 10:00:20.63 & +02:22:51.2 & 0.38 & 	25.75 & 22.90 & 1.82 \\ 
       & 3/5 & 0.02 &             &             &      &              &             &             &      &             & 10:00:20.70 & +02:22:51.3 & 0.67 & 	25.07 & 23.14 & 1.65 \\ 
       & 4/5 & 0.14 &             &             &      &              &             &             &      &             & 10:00:20.69 & +02:22:53.2 & 1.99 & 	25.59 & 22.43 & 2.50 \\ 
       & 5/5 & 0.15 &             &             &      &              &             &             &      &             & 10:00:20.69 & +02:22:49.1 & 2.16 & 	25.13 & 23.59 & 1.67 \\ 
850.74 & 1/1 & 0.25 &             &             &      &              &             &             &      &             & 10:00:05.41 & +02:14:47.1 & 2.78 & 	27.42 & 23.87 & 2.76 \\ 
850.75 & 1/4 & 0.007 &             &             &      &              & 10:00:26.97 & +02:22:30.5 & 3.64 & 82$\pm$16   & 10:00:26.96 & +02:22:30.8 & 3.40 & 	25.86 & 21.91 & $-$  \\ 
       & 2/4 & 0.05 &             &             &      &              &             &             &      &             & 10:00:26.79 & +02:22:31.1 & 1.20 & 	26.69 & 23.28 & $-$  \\ 
       & 3/4 & 0.25 &             &             &      &              &             &             &      &             & 10:00:26.91 & +02:22:29.3 & 2.81 & 	25.99 & 23.05 & $-$  \\ 
       & 4/4 & 0.34 &             &             &      &              &             &             &      &             & 10:00:26.52 & +02:22:29.4 & 3.34 & 	25.59 & 23.43 & $-$  \\ 
850.76 & 1/2 & 0.08 &             &             &       &             &             &              &     &             & 10:00:48.95 & +02:30:24.88 & 1.62 &    24.31 & 22.94 & 1.06 \\
       & 2/2 & 0.31 &             &             &       &             &             &              &     &             & 10:00:48.82 & +02:30:21.00 & 3.08 &    23.22 & 22.61 & 0.86 \\
850.77 & 1/1 & 0.12 & 10:00:05.47 & +02:29:53.1 & 5.17 & 441$\pm$178  &             &             &      &             & $-$         & $-$         & $-$  & 	$-$   & $-$   & $-$  \\ 
850.78 & 1/6 & 0.08 & 10:00:08.77 & +02:19:16.2 & 4.11 & 207$\pm$14   &             &             &      &             & 10:00:08.93 & +02:19:16.2 & 5.07 & 	22.89 & 21.18 & 0.94 \\ 
       & 2/6 & 0.11 &             &             &      &              &             &             &      &             & 10:00:08.78 & +02:19:13.8 & 1.82 & 	25.98 & 24.27 & 2.39 \\ 
       & 3/6 & 0.22 &             &             &      &              &             &             &      &             & 10:00:08.88 & +02:19:11.4 & 2.58 & 	29.41 & $-$   & 3.24 \\ 
       & 4/6 & 0.29 &             &             &      &              &             &             &      &             & 10:00:08.66 & +02:19:13.2 & 1.26 & 	$-$   & $-$   & 0.71 \\ 
       & 5/6 & 0.31 &             &             &      &              &             &             &      &             & 10:00:08.73 & +02:19:15.4 & 3.19 & 	26.46 & 22.75 & 4.35 \\ 
       & 6/6 & 0.32 &             &             &      &              &             &             &      &             & 10:00:08.67 & +02:19:09.0 & 3.26 & 	27.65 & 23.83 & 2.94 \\ 
850.79 & 1/3 & 0.001 &             &             &      &              &             &             &      &             & 10:01:02.28 & +02:22:34.1 & 0.20 & 	21.39 & 20.55 & $-$  \\ 
       & 2/3 & 0.18 &             &             &      &              &             &             &      &             & 10:01:02.14 & +02:22:34.1 & 2.29 & 	23.34 & 22.20 & $-$  \\ 
       & 3/3 & 0.26 &             &             &      &              &             &             &      &             & 10:01:02.27 & +02:22:37.0 & 2.83 & 	23.99 & 23.40 & $-$  \\ 
850.80 & 1/3 & 0.16 &             &             &      &              &             &             &      &             & 10:00:14.13 & +02:27:02.1 & 2.15 & 	25.23 & 22.72 & 3.54 \\ 
       & 2/3 & 0.17 &             &             &      &              &             &             &      &             & 10:00:14.26 & +02:27:03.3 & 2.28 & 	24.07 & 23.74 & 2.40 \\ 
       & 3/3 & 0.31 &             &             &      &              &             &             &      &             & 10:00:13.99 & +02:27:01.7 & 3.19 & 	26.74 & 23.94 & 3.87 \\ 
850.81 & 1/2 & 0.06 &             &             &      &              &             &             &      &             & 09:59:59.12 & +02:24:41.9 & 1.35 & 	24.11 & 23.58 & 0.40 \\ 
       & 2/2 & 0.47 &             &             &      &              &             &             &      &             & 09:59:58.97 & +02:24:39.8 & 1.73 & 	26.81 & $-$   & 2.81 \\ 
850.82 & 1/1 & 0.03 & 10:00:54.96 & +02:19:47.4 & 2.46 & 203$\pm$17   &             &             &      &             & 10:00:54.95 & +02:19:47.1 & 2.09 & 	26.03 & 21.11 & 2.87 \\ 
850.84 & 1/1 & 0.22 &             &             &      &              &             &             &      &             & 10:00:41.67 & +02:21:12.7 & 2.58 & 	26.85 & 24.03 & 2.42 \\ 
850.86 & 1/2 & 0.005 & 10:01:01.30 & +02:24:39.4 & 1.04 & 299$\pm$19   &             &             &      &             & 10:01:01.27 & +02:24:39.9 & 1.51 & 	23.28 & 20.03 & $-$  \\ 
       & 2/2 & 0.10 &             &             &      &              &             &             &      &             & 10:01:01.31 & +02:24:37.6 & 1.73 & 	21.73 & 19.55 & $-$  \\ 
850.87 & 1/2 & 0.05 & 10:00:40.23 & +02:17:55.5 & 3.18 & 284$\pm$15   &             &             &      &             & 10:00:40.26 & +02:17:55.7 & 2.85 & 	26.99 & 20.78 & 2.24 \\ 
       & 2/2 & 0.11 &             &             &      &              &             &             &      &             & 10:00:40.33 & +02:18:00.0 & 1.75 & 	26.48 & 22.91 & 1.45 \\ 
850.89 & 1/4 & 0.02 & 10:00:20.26 & +02:17:25.1 & 2.12 & 401$\pm$16   &             &             &      &             & 10:00:20.25 & +02:17:25.7 & 1.53 & 	22.53 & 19.94 & $-$  \\ 
       & 2/4 & 0.04 &             &             &      &              &             &             &      &             & 10:00:20.19 & +02:17:27.0 & 1.00 & 	25.42 & 22.16 & 1.87 \\ 
       & 3/4 & 0.11 &             &             &      &              &             &             &      &             & 10:00:20.24 & +02:17:29.0 & 1.79 & 	25.72 & 23.02 & 1.72 \\ 
       & 4/4 & 0.20 &             &             &      &              &             &             &      &             & 10:00:20.37 & +02:17:28.9 & 2.43 & 	24.88 & 23.21 & 0.92 \\ 
850.91 & 1/3 & 0.18 &             &             &      &              &             &             &      &             & 10:00:27.85 & +02:25:51.9 & 2.30 & 	26.95 & 23.91 & 2.71 \\ 
       & 2/3 & 0.26 &             &             &      &              &             &             &      &             & 10:00:27.94 & +02:25:51.6 & 2.84 & 	26.73 & 24.91 & 2.21 \\ 
       & 3/3 & 0.46 &             &             &      &              &             &             &      &             & 10:00:27.76 & +02:25:55.1 & 1.69 & 	25.95 & $-$   & 2.68 \\ 
850.93 & 1/3 & 0.01 &             &             &      &              & 10:00:57.06 & +02:29:42.9 & 4.97 & 123$\pm$10  & 10:00:57.06 & +02:29:42.8 & 5.08 & 	24.47 & 20.88 & $-$  \\ 
       & 2/3 & 0.09 &             &             &      &              &             &             &      &             & 10:00:56.66 & +02:29:44.7 & 1.59 & 	25.82 & 24.76 & $-$  \\ 
       & 3/3 & 0.32 &             &             &      &              &             &             &      &             & 10:00:56.69 & +02:29:48.3 & 3.23 & 	21.25 & 20.79 & 0.37 \\ 
850.94 & 1/2 & 0.002 & 10:00:02.62 & +02:16:34.0 & 1.84 & 393$\pm$16   & 10:00:02.61 & +02:16:33.8 & 1.65 & 72$\pm$11   & 10:00:02.61 & +02:16:34.2 & 2.03 & 	24.62 & 20.29 & 2.04 \\ 
       & 2/2 & 0.006 &             &             &      &              &             &             &      &             & 10:00:02.62 & +02:16:32.1 & 0.39 & 	25.70 & 22.38 & 4.16 \\ 
850.95 & 1/3 & 0.02 & 09:59:59.82 & +02:27:06.6 & 1.88 & 387$\pm$67   &             &             &      &             & 09:59:59.80 & +02:27:07.4 & 2.74 & 	22.78 & 19.60 & 1.55 \\ 
       & 2/3 & 0.13 &             &             &      &              &             &             &      &             & 09:59:59.82 & +02:27:03.7 & 1.94 & 	$-$   & 19.82 & 0.74 \\ 
       & 3/3 & 0.32 &             &             &      &              &             &             &      &             & 09:59:59.93 & +02:27:02.0 & 3.21 & 	$-$   & 21.05 & $-$  \\ 
850.97 & 1/3 & 0.005 &             &             &      &              & 10:00:13.57 & +02:18:05.0 & 2.87 & 148$\pm$15  & 10:00:13.59 & +02:18:05.1 & 3.00 & 	20.46 & 18.63 & 0.73 \\ 
       & 2/3 & 0.03 & 10:00:13.60 & +02:18:08.6 & 2.62 & 349$\pm$15   &             &             &      &             & 10:00:13.59 & +02:18:08.5 & 2.40 & 	24.02 & 21.12 & 1.48 \\ 
       & 3/3 & 0.25 &             &             &      &              &             &             &      &             & 10:00:13.29 & +02:18:08.7 & 2.82 & 	25.22 & 22.91 & 1.80 \\ 
850.98 & 1/1 & 0.19 &             &             &       &             &             &             &      &             & 10:00:14.49 & +02:30:10.1 & 2.37 &     $-$   & 19.02 & $-$  \\
\hline\hline	
\end{tabular}	
}	
	
{\small	
{\sc Table~\ref{tab:counterparts850} Continued.}}	
\end{table*}

\begin{table*}
\centering
\caption{Physical Characteristics of \scubaii\ Selected Galaxies}
\begin{spacing}{0.7}
{\tiny
\begin{tabular}{cccc@{ }c@{ }c@{ }c@{ }cc@{ }c}
\hline\hline
{\sc Short}  & {\sc Short}  & $z_{\rm phot}$ & \multicolumn{5}{c}{\underline{Estimates from Infrared SED Fitting (\S \ref{sec:sed})}}                  & \multicolumn{2}{c}{\underline{{\sc Le Phare} Properties}} \\
{\sc 450}\um & {\sc 850}\um &                & $\chi^2_{\rm fit}$ & $L_{\rm IR}$  & $\log$SFR$_{\rm IR}$   & $\log\lambda_{\rm peak}$ & $M_{\rm dust}$ & $\log$SFR$_{\rm UV}$   & $M_\star$ \\ 
{\sc Name}   & {\sc Name}   &                &                    & ($L_{\odot}$) & ($M_\odot$\,yr$^{-1}$) & (\um)                    & ($M_\odot$)    & ($M_\odot$\,yr$^{-1}$) & ($M_\odot$) \\ 
\hline\hline
450.00 & 850.07 & 2.86$^{+0.21}_{-0.26}$ & 0.4 & (5.1$^{+7.4}_{-3.5}$)$\times10^{13}$  & 3.9$\pm$0.2 & 1.6$\pm$0.4 & (5$^{+6}_{-4}$)$\times10^{8}$   &  1.71$^{+0.13}_{-0.33}$ & (2$^{+3}_{-1}$)$\times10^{10}$ \\
450.01 & 850.02 & 2.88$^{+0.09}_{-0.18}$ & 0.4 & (1.1$^{+1.7}_{-0.7}$)$\times10^{13}$  & 3.3$\pm$0.2 & 1.9$\pm$0.4 & (1$^{+2}_{-1}$)$\times10^{9}$   &  1.86$^{+0.29}_{-0.15}$ & (2$^{+2}_{-1}$)$\times10^{11}$ \\
450.02 &        & 2.16$^{+0.04}_{-0.06}$ & 2.4 & (8.6$^{+16.6}_{-4.5}$)$\times10^{12}$ & 3.2$\pm$0.3 & 2.0$\pm$0.5 & (1$^{+3}_{-1}$)$\times10^{9}$   &  1.10$^{+0.07}_{-0.07}$ & (9$^{+10}_{-8}$)$\times10^{10}$ \\
450.03 & 850.00 & 0.34$^{+0.01}_{-0.01}$ & 4.7 & (4.3$^{+5.1}_{-3.6}$)$\times10^{10}$  & 0.9$\pm$0.1 & $\equiv$2.0 & $-$                             & -0.50$^{+0.07}_{-0.07}$ & (6$^{+8}_{-5}$)$\times10^{10}$ \\
450.04 & 850.03 & 3.82$^{+0.44}_{-0.69}$ & 1.2 & (1.9$^{+3.3}_{-1.1}$)$\times10^{13}$  & 3.5$\pm$0.3 & 1.8$\pm$0.5 & (1$^{+2}_{-1}$)$\times10^{9}$   &  2.45$^{+0.11}_{-0.32}$ & (7$^{+10}_{-5}$)$\times10^{10}$ \\
450.05 & 850.08 & 3.99$^{+0.08}_{-0.85}$ & 2.4 & (1.9$^{+3.4}_{-1.1}$)$\times10^{13}$  & 3.5$\pm$0.2 & 1.8$\pm$0.6 & (7$^{+9}_{-6}$)$\times10^{8}$   &  2.16$^{+0.10}_{-0.09}$ & (7$^{+8}_{-6}$)$\times10^{11}$ \\
450.06 & 850.12 & 2.08$^{+0.08}_{-0.07}$ & 0.3 & (7.2$^{+9.2}_{-5.7}$)$\times10^{12}$  & 3.1$\pm$0.1 & 1.9$\pm$0.1 & (6$^{+7}_{-5}$)$\times10^{8}$   &  2.00$^{+0.07}_{-0.07}$ & (1$^{+1}_{-1}$)$\times10^{11}$ \\
450.07 &        & 4.70$^{+0.07}_{-0.08}$ & 4.3 & (7.2$^{+8.7}_{-6.0}$)$\times10^{13}$  & 4.1$\pm$0.1 & $\equiv$2.0 & $-$                             &  2.53$^{+0.14}_{-0.47}$ & (4$^{+6}_{-3}$)$\times10^{11}$ \\
450.08 &        & 5.15$^{+0.20}_{-0.33}$ & 3.7 & (8.7$^{+10.4}_{-7.2}$)$\times10^{13}$ & 4.2$\pm$0.1 & $\equiv$2.0 & $-$                             &  2.51$^{+0.25}_{-0.59}$ & (2$^{+3}_{-2}$)$\times10^{11}$ \\
450.09 & 850.34 & 1.90$^{+0.03}_{-0.04}$ & 0.1 & (7.7$^{+10.3}_{-5.7}$)$\times10^{12}$ & 3.1$\pm$0.1 & 1.8$\pm$0.1 & (3$^{+4}_{-2}$)$\times10^{8}$   &  2.26$^{+0.09}_{-0.08}$ & (8$^{+8}_{-7}$)$\times10^{9}$ \\
450.12 &        & 0.93$^{+0.01}_{-0.01}$ & 0.1 & (5.8$^{+18.6}_{-1.8}$)$\times10^{11}$ & 2.0$\pm$0.5 & $\equiv$2.0 & $-$                             &  1.61$^{+0.07}_{-0.07}$ & (3$^{+3}_{-3}$)$\times10^{10}$ \\
450.13 &        & 2.87$^{+0.13}_{-0.26}$ & 0.1 & (8.7$^{+11.5}_{-6.6}$)$\times10^{12}$ & 3.2$\pm$0.1 & 1.8$\pm$0.1 & (4$^{+5}_{-3}$)$\times10^{8}$   &  2.01$^{+0.08}_{-0.09}$ & (2$^{+2}_{-2}$)$\times10^{11}$ \\
450.14 &        & 1.47$^{+0.01}_{-0.02}$ & 3.2 & (1.6$^{+2.8}_{-0.9}$)$\times10^{12}$  & 2.4$\pm$0.2 & 2.0$\pm$0.2 & (6$^{+10}_{-4}$)$\times10^{8}$  &  1.81$^{+0.08}_{-0.10}$ & (2$^{+2}_{-2}$)$\times10^{11}$ \\
450.15 & 850.28 & 2.79$^{+0.28}_{-0.29}$ & 0.5 & (1.1$^{+1.4}_{-0.8}$)$\times10^{13}$  & 3.3$\pm$0.1 & 1.8$\pm$0.1 & (4$^{+4}_{-3}$)$\times10^{8}$   &  1.62$^{+0.33}_{-0.19}$ & (4$^{+5}_{-3}$)$\times10^{10}$ \\
450.16 &        & 2.32$^{+0.02}_{-0.03}$ & 2.1 & (2.0$^{+2.8}_{-1.4}$)$\times10^{12}$  & 2.5$\pm$0.2 & 2.2$\pm$0.2 & (9$^{+40}_{-2}$)$\times10^{9}$  &  1.20$^{+0.07}_{-0.07}$ & (1$^{+1}_{-1}$)$\times10^{11}$ \\
450.17 & 850.42 & 1.23$^{+0.21}_{-0.11}$ & 0.1 & (7.9$^{+18.9}_{-3.3}$)$\times10^{11}$ & 2.1$\pm$0.4 & 2.1$\pm$0.5 & (6$^{+20}_{-2}$)$\times10^{8}$  & -0.35$^{+0.40}_{-0.31}$ & (6$^{+9}_{-3}$)$\times10^{8}$ \\
450.19 &        & 1.11$^{+0.06}_{-0.02}$ & 9.7 & (9.7$^{+32.0}_{-3.0}$)$\times10^{11}$ & 2.2$\pm$0.5 & $\equiv$2.0 & $-$                             &  0.56$^{+0.09}_{-0.09}$ & (8$^{+10}_{-7}$)$\times10^{9}$ \\
450.20 &        & 0.76$^{+0.04}_{-0.03}$ & 0.8 & (2.6$^{+5.7}_{-1.2}$)$\times10^{11}$  & 1.6$\pm$0.3 & 2.2$\pm$0.5 & (7$^{+40}_{-2}$)$\times10^{8}$  & -0.05$^{+0.37}_{-0.28}$ & (8$^{+10}_{-6}$)$\times10^{8}$ \\
450.21 &        & 0.84$^{+0.01}_{-0.01}$ & 0.1 & (4.7$^{+7.9}_{-2.8}$)$\times10^{11}$  & 1.9$\pm$0.2 & 2.1$\pm$0.2 & (2$^{+4}_{-1}$)$\times10^{8}$   &  1.98$^{+0.08}_{-0.07}$ & (4$^{+4}_{-3}$)$\times10^{10}$ \\
450.22 & 850.149& 2.11$^{+0.20}_{-0.28}$ & 5.5 & (3.2$^{+10.6}_{-1.}$)$\times10^{12}$  & 2.7$\pm$0.5 & $\equiv$2.0 & $-$                             &  0.31$^{+0.13}_{-0.36}$ & (4$^{+5}_{-3}$)$\times10^{9}$ \\
450.23 &        & 0.97$^{+0.06}_{-0.04}$ & 0.6 & (5.4$^{+17.9}_{-1.6}$)$\times10^{11}$ & 2.0$\pm$0.5 & $\equiv$2.0 & $-$                             &  0.46$^{+0.34}_{-0.16}$ & (8$^{+10}_{-6}$)$\times10^{8}$ \\
450.24 &        & 0.16$^{+0.01}_{-0.01}$ & 3.5 & (3.2$^{+10.6}_{-1.}$)$\times10^{10}$  & 0.7$\pm$0.5 & $\equiv$2.0 & $-$                             &  0.55$^{+0.09}_{-0.11}$ & (6$^{+7}_{-5}$)$\times10^{10}$ \\
450.25 &        & 0.61$^{+0.01}_{-0.01}$ & 2.0 & (1.7$^{+3.1}_{-1.}$)$\times10^{11}$   & 1.5$\pm$0.3 & 2.2$\pm$0.2 & (2$^{+10}_{-1}$)$\times10^{8}$  &  0.80$^{+0.07}_{-0.07}$ & (1$^{+2}_{-1}$)$\times10^{11}$ \\
450.26 &        & 2.68$^{+0.11}_{-0.13}$ & 4.6 & (5.3$^{+17.7}_{-1.6}$)$\times10^{12}$ & 3.0$\pm$0.5 & $\equiv$2.0 & $-$                             &  1.89$^{+0.63}_{-0.12}$ & (7$^{+9}_{-3}$)$\times10^{9}$ \\
450.28 &        & 1.98$^{+0.03}_{-0.07}$ & 0.7 & (1.2$^{+1.7}_{-0.8}$)$\times10^{12}$  & 2.3$\pm$0.2 & 2.3$\pm$0.2 & (1$^{+10}_{-1}$)$\times10^{10}$ &  2.10$^{+0.07}_{-0.07}$ & (1$^{+1}_{-1}$)$\times10^{10}$ \\
450.30 &        & 2.81$^{+0.56}_{-0.56}$ & 3.5 & (6.0$^{+10.}_{-3.6}$)$\times10^{12}$  & 3.0$\pm$0.2 & 1.9$\pm$0.2 & (4$^{+5}_{-3}$)$\times10^{8}$   &  1.81$^{+0.12}_{-0.18}$ & (1$^{+1}_{-1}$)$\times10^{11}$ \\
450.32 & 850.26 & 1.72$^{+0.15}_{-0.09}$ & 5.0 & (1.6$^{+2.3}_{-1.1}$)$\times10^{12}$  & 2.4$\pm$0.2 & 2.1$\pm$0.2 & (7$^{+10}_{-5}$)$\times10^{8}$  & -0.27$^{+0.22}_{-0.52}$ & (4$^{+20}_{-1}$)$\times10^{8}$ \\
450.33 &        & 0.84$^{+0.01}_{-0.01}$ & 0.1 & (3.8$^{+7.2}_{-2.1}$)$\times10^{11}$  & 1.8$\pm$0.3 & 2.1$\pm$0.2 & (3$^{+6}_{-1}$)$\times10^{8}$   &  1.30$^{+0.08}_{-0.09}$ & (4$^{+4}_{-4}$)$\times10^{10}$ \\
450.34 &        & 1.83$^{+0.03}_{-0.03}$ & 3.9 & (2.4$^{+6.5}_{-0.9}$)$\times10^{12}$  & 2.6$\pm$0.4 & 2.0$\pm$0.7 & (6$^{+10}_{-3}$)$\times10^{8}$  &  1.89$^{+0.08}_{-0.07}$ & (8$^{+9}_{-7}$)$\times10^{10}$ \\
450.36 & 850.58 & 3.04$^{+0.16}_{-0.12}$ & 0.1 & (6.9$^{+18.8}_{-2.6}$)$\times10^{12}$ & 3.1$\pm$0.4 & 1.9$\pm$0.7 & (7$^{+10}_{-5}$)$\times10^{8}$  &  2.39$^{+0.13}_{-0.41}$ & (3$^{+6}_{-3}$)$\times10^{10}$ \\
450.37 &        & 2.17$^{+0.08}_{-0.08}$ & 1.9 & (3.1$^{+10.6}_{-0.9}$)$\times10^{12}$ & 2.7$\pm$0.5 & $\equiv$2.0 & $-$                             &  1.26$^{+0.37}_{-0.15}$ & (4$^{+6}_{-2}$)$\times10^{9}$ \\
450.39 &        & 0.61$^{+0.02}_{-0.02}$ & 0.1 & (1.6$^{+5.4}_{-0.5}$)$\times10^{11}$  & 1.4$\pm$0.5 & $\equiv$2.0 & $-$                             & -2.37$^{+0.67}_{-3.00}$ & (3$^{+4}_{-3}$)$\times10^{9}$ \\
450.41 &        & 3.46$^{+1.74}_{-1.90}$ & 0.1 & (8.8$^{+30.3}_{-2.5}$)$\times10^{12}$ & 3.2$\pm$0.5 & $\equiv$2.0 & $-$                             &  1.14$^{+0.39}_{-0.56}$ & (5$^{+10}_{-2}$)$\times10^{9}$ \\
450.42 & 850.48 & 4.81$^{+0.14}_{-0.21}$ & 3.1 & (1.6$^{+4.0}_{-0.6}$)$\times10^{13}$  & 3.4$\pm$0.4 & 1.7$\pm$0.8 & (3$^{+4}_{-2}$)$\times10^{8}$   &  2.60$^{+0.12}_{-0.34}$ & (3$^{+4}_{-3}$)$\times10^{11}$ \\
450.43 &        & 0.36$^{+0.06}_{-0.07}$ & 1.6 & (3.8$^{+13.3}_{-1.1}$)$\times10^{10}$ & 0.8$\pm$0.5 & $\equiv$2.0 & $-$                             & -1.11$^{+0.48}_{-0.55}$ & (2$^{+3}_{-1}$)$\times10^{8}$ \\
450.44 &        & 1.75$^{+0.50}_{-0.51}$ & 0.1 & (1.8$^{+6.3}_{-0.5}$)$\times10^{12}$  & 2.5$\pm$0.5 & $\equiv$2.0 & $-$                             &  0.33$^{+0.41}_{-0.39}$ & (8$^{+20}_{-4}$)$\times10^{8}$ \\
450.45 &        & 2.01$^{+0.93}_{-0.71}$ & 1.0 & (2.5$^{+8.9}_{-0.7}$)$\times10^{12}$  & 2.6$\pm$0.5 & $\equiv$2.0 & $-$                             &  0.61$^{+0.43}_{-0.37}$ & (3$^{+4}_{-1}$)$\times10^{9}$ \\
450.46 &        & 3.06$^{+0.08}_{-0.06}$ & 4.3 & (5.5$^{+19.6}_{-1.6}$)$\times10^{12}$ & 3.0$\pm$0.5 & $\equiv$2.0 & $-$                             &  2.17$^{+0.12}_{-0.36}$ & (3$^{+5}_{-3}$)$\times10^{10}$ \\
450.47 &        & 1.77$^{+0.07}_{-0.24}$ & 0.1 & (1.1$^{+4.0}_{-0.3}$)$\times10^{12}$  & 2.3$\pm$0.5 & $\equiv$2.0 & $-$                             &  1.64$^{+0.35}_{-0.10}$ & (6$^{+8}_{-4}$)$\times10^{9}$ \\
450.49 &        & 1.70$^{+0.16}_{-0.10}$ & 4.1 & (1.4$^{+5.1}_{-0.4}$)$\times10^{12}$  & 2.4$\pm$0.6 & $\equiv$2.0 & $-$                             &  0.48$^{+0.28}_{-0.34}$ & (2$^{+3}_{-1}$)$\times10^{9}$ \\
450.51 &        & 1.58$^{+0.22}_{-0.45}$ & 0.1 & (1.1$^{+3.8}_{-0.3}$)$\times10^{12}$  & 2.3$\pm$0.6 & $\equiv$2.0 & $-$                             &  0.79$^{+0.37}_{-0.28}$ & (1$^{+2}_{-0.7}$)$\times10^{9}$ \\
450.53 & 850.109& 1.01$^{+0.01}_{-0.01}$ & 1.8 & (8.6$^{+14.6}_{-5.1}$)$\times10^{11}$ & 2.2$\pm$0.2 & 2.0$\pm$0.2 & (2$^{+3}_{-1}$)$\times10^{8}$   &  1.40$^{+0.08}_{-0.08}$ & (8$^{+9}_{-7}$)$\times10^{10}$ \\
450.54 & 850.96 & 1.90$^{+0.14}_{-0.12}$ & 1.1 & (1.5$^{+4.7}_{-0.5}$)$\times10^{12}$  & 2.4$\pm$0.5 & 2.0$\pm$0.8 & (4$^{+7}_{-2}$)$\times10^{8}$   &  1.73$^{+0.13}_{-0.36}$ & (3$^{+5}_{-3}$)$\times10^{9}$ \\
450.55 & 850.06 & 0.37$^{+0.01}_{-0.01}$ & 1.6 & (2.8$^{+3.4}_{-2.3}$)$\times10^{10}$  & 0.7$\pm$0.1 & $\equiv$2.0 & $-$                             & -2.12$^{+0.11}_{-3.40}$ & (3$^{+4}_{-3}$)$\times10^{10}$ \\
450.56 &        & 2.82$^{+1.39}_{-1.20}$ & 3.6 & (5.0$^{+17.8}_{-1.4}$)$\times10^{12}$ & 2.9$\pm$0.6 & $\equiv$2.0 & $-$                             &  0.94$^{+0.74}_{-0.74}$ & (3$^{+20}_{-1}$)$\times10^{9}$ \\
450.57 &        & 3.48$^{+1.33}_{-1.60}$ & 3.6 & (7.4$^{+26.5}_{-2.1}$)$\times10^{12}$ & 3.1$\pm$0.6 & $\equiv$2.0 & $-$                             & -3.61$^{+2.43}_{-0.32}$ & (1$^{+1}_{-0.9}$)$\times10^{11}$ \\
450.59 & 850.101& 0.47$^{+0.01}_{-0.01}$ & 4.0 & (8.3$^{+18.6}_{-3.7}$)$\times10^{10}$ & 1.2$\pm$0.3 & 2.2$\pm$0.4 & (2$^{+50}_{-1}$)$\times10^{8}$  &  0.91$^{+0.08}_{-0.08}$ & (3$^{+3}_{-2}$)$\times10^{10}$ \\
450.61 &        & 4.58$^{+0.11}_{-0.06}$ & 1.4 & (1.3$^{+2.6}_{-0.7}$)$\times10^{13}$  & 3.3$\pm$0.3 & 1.7$\pm$0.3 & (2$^{+3}_{-2}$)$\times10^{8}$   &  2.69$^{+0.12}_{-0.36}$ & (1$^{+20}_{-7}$)$\times10^{10}$ \\
450.62 &        & 1.95$^{+1.32}_{-1.10}$ & 0.1 & (1.9$^{+6.7}_{-0.5}$)$\times10^{12}$  & 2.5$\pm$0.6 & $\equiv$2.0 & $-$                             &  0.40$^{+0.56}_{-0.74}$ & (1$^{+7}_{-0.2}$)$\times10^{9}$ \\
450.63 &        & 2.36$^{+0.13}_{-0.12}$ & 3.4 & (2.8$^{+10.2}_{-0.8}$)$\times10^{12}$ & 2.7$\pm$0.6 & $\equiv$2.0 & $-$                             &  0.53$^{+0.35}_{-0.11}$ & (3$^{+5}_{-2}$)$\times10^{9}$ \\
450.65 &        & 0.70$^{+2.57}_{-0.49}$ & 0.1 & (1.5$^{+5.5}_{-0.4}$)$\times10^{11}$  & 1.4$\pm$0.6 & $\equiv$2.0 & $-$                             & -0.30$^{+0.50}_{-0.57}$ & (2$^{+4}_{-1}$)$\times10^{8}$ \\
450.67 &        & 0.93$^{+0.01}_{-0.01}$ & 1.4 & (8.7$^{+18.1}_{-4.2}$)$\times10^{11}$ & 2.2$\pm$0.3 & 2.0$\pm$0.2 & (2$^{+4}_{-1}$)$\times10^{8}$   &  2.23$^{+0.09}_{-0.10}$ & (5$^{+5}_{-4}$)$\times10^{10}$ \\
450.69 &        & 2.04$^{+0.49}_{-0.48}$ & 0.1 & (2.0$^{+7.3}_{-0.5}$)$\times10^{12}$  & 2.5$\pm$0.6 & $\equiv$2.0 & $-$                             &  0.71$^{+0.41}_{-0.38}$ & (3$^{+5}_{-2}$)$\times10^{9}$ \\
\hline\hline  
\end{tabular}  
} 
\end{spacing}
\label{tab:physical}

{\small Some basic derived properties of 450\um\ sources using the
  infrared SED fitting described in \S~\ref{sec:sed}. The sources'
  optical/near-infrared photometric redshifts are given by $z_{\rm
    p}$. Their integrated 8--1000\um\ infrared luminosity, $L_{\rm
    IR}$, infrared star formation rates $SFR_{\rm IR}$
  \citep[calculated via the relation given in][]{kennicutt98b}, SED
  peak wavelength, and approximated dust masses $M_{\rm dust}$ are
  calculated assuming SEDs as described in section~\ref{sec:sed}.
  Sources with fewer than three IR photometric points have the SED
  peak wavelength (or dust temperature) fixed to the mean of the
  remainder of the sample, measured to be $\langle \log(\lambda_{\rm
    peak}/$\um$)\rangle=2.05\pm0.04$ for 450\um-detected galaxies and
  $\langle \log(\lambda_{\rm peak}/$\um$)\rangle=2.12\pm0.04$ for
  850\um-only detected galaxies.  }
\end{table*}
\begin{table*}
\centering
\begin{spacing}{0.7}
{\tiny
\begin{tabular}{cccc@{ }c@{ }c@{ }c@{ }cc@{ }c}
\hline\hline
{\sc Short}  & {\sc Short}  & $z_{\rm phot}$ & \multicolumn{5}{c}{\underline{Estimates from Infrared SED Fitting (\S \ref{sec:sed})}}                  & \multicolumn{2}{c}{\underline{{\sc Le Phare} Properties}} \\
{\sc 450}\um & {\sc 850}\um &                & $\chi^2_{\rm fit}$ & $L_{\rm IR}$  & $\log$SFR$_{\rm IR}$   & $\log\lambda_{\rm peak}$ & $M_{\rm dust}$ & $\log$SFR$_{\rm UV}$   & $M_\star$ \\ 
{\sc Name}   & {\sc Name}   &                &                    & ($L_{\odot}$) & ($M_\odot$\,yr$^{-1}$) & (\um)                    & ($M_\odot$)    & ($M_\odot$\,yr$^{-1}$) & ($M_\odot$) \\ 
\hline\hline
450.70 &        & 3.03$^{+0.04}_{-0.03}$  & 0.1 & (8.6$^{+17.7}_{-4.1}$)$\times10^{12}$ & 3.2$\pm$0.3 & 1.8$\pm$0.3 & (3$^{+4}_{-2}$)$\times10^{8}$   &  3.06$^{+0.09}_{-0.36}$ & (6$^{+8}_{-5}$)$\times10^{10}$ \\
450.71 &        & 1.95$^{+0.61}_{-0.42}$  & 0.6 & (2.6$^{+9.5}_{-0.7}$)$\times10^{12}$  & 2.6$\pm$0.6 & $\equiv$2.0 & $-$                             &  0.67$^{+0.41}_{-0.38}$ & (2$^{+4}_{-1}$)$\times10^{9}$ \\
450.72 &        & 1.43$^{+0.31}_{-0.11}$  & 2.8 & (9.9$^{+36.1}_{-2.7}$)$\times10^{11}$ & 2.2$\pm$0.6 & $\equiv$2.0 & $-$                             &  0.91$^{+0.24}_{-0.51}$ & (2$^{+4}_{-1}$)$\times10^{9}$ \\
450.74 &        & 0.75$^{+0.01}_{-0.01}$  & 3.8 & (1.9$^{+7.0}_{-0.5}$)$\times10^{11}$  & 1.5$\pm$0.6 & $\equiv$2.0 & $-$                             &  1.04$^{+0.09}_{-0.10}$ & (4$^{+5}_{-3}$)$\times10^{9}$ \\
450.75 &        & 2.26$^{+2.30}_{-1.30}$  & 3.9 & (3.2$^{+11.9}_{-0.9}$)$\times10^{12}$ & 2.7$\pm$0.6 & $\equiv$2.0 & $-$                             &  0.56$^{+0.44}_{-0.39}$ & (2$^{+3}_{-1}$)$\times10^{9}$ \\
450.76 &        & 1.96$^{+0.45}_{-0.51}$  & 0.1 & (1.9$^{+6.8}_{-0.5}$)$\times10^{12}$  & 2.5$\pm$0.6 & $\equiv$2.0 & $-$                             & -0.03$^{+0.43}_{-0.38}$ & (1$^{+2}_{-0.5}$)$\times10^{9}$ \\
450.77 &        & 1.43$^{+0.03}_{-0.04}$  & 0.1 & (8.8$^{+32.3}_{-2.4}$)$\times10^{11}$ & 2.2$\pm$0.6 & $\equiv$2.0 & $-$                             &  1.27$^{+0.63}_{-0.48}$ & (3$^{+4}_{-1}$)$\times10^{10}$ \\
450.81 & 850.10 & 1.75$^{+0.08}_{-0.12}$  & 0.1 & (1.6$^{+2.3}_{-1.1}$)$\times10^{12}$  & 2.4$\pm$0.2 & 2.1$\pm$0.2 & (1$^{+2}_{-1}$)$\times10^{9}$   &  1.58$^{+0.14}_{-0.15}$ & (2$^{+2}_{-1}$)$\times10^{10}$ \\
450.86 & 850.55 & 3.59$^{+0.45}_{-0.81}$  & 1.8 & (6.8$^{+25.4}_{-1.8}$)$\times10^{12}$ & 3.1$\pm$0.6 & $\equiv$2.0 & $-$                             &  1.69$^{+0.41}_{-0.29}$ & (1$^{+2}_{-0.9}$)$\times10^{10}$ \\
450.87 & 850.09 & 2.71$^{+0.19}_{-0.17}$  & 3.2 & (6.3$^{+9.0}_{-4.4}$)$\times10^{12}$  & 3.0$\pm$0.2 & 1.9$\pm$0.2 & (1$^{+10}_{-8}$)$\times10^{8}$  &  2.04$^{+0.09}_{-0.10}$ & (1$^{+2}_{-1}$)$\times10^{11}$ \\
450.94 & 850.23 & 0.71$^{+0.01}_{-0.01}$  & 3.4 & (8.2$^{+30.9}_{-2.1}$)$\times10^{10}$ & 1.1$\pm$0.6 & $\equiv$2.0 & $-$                             & -4.41$^{+3.35}_{-0.33}$ & (2$^{+3}_{-2}$)$\times10^{10}$ \\
450.96 & 850.133& 2.34$^{+0.40}_{-0.41}$  & 0.1 & (2.1$^{+8.0}_{-0.6}$)$\times10^{12}$  & 2.6$\pm$0.6 & $\equiv$2.0 & $-$                             &  1.09$^{+0.42}_{-0.22}$ & (2$^{+2}_{-1}$)$\times10^{10}$ \\
450.99 & 850.33 & 2.16$^{+0.08}_{-0.13}$  & 2.9 & (1.9$^{+5.0}_{-0.7}$)$\times10^{12}$  & 2.5$\pm$0.4 & 2.1$\pm$0.6 & (2$^{+4}_{-1}$)$\times10^{9}$   &  1.74$^{+0.11}_{-0.32}$ & (9$^{+10}_{-7}$)$\times10^{10}$ \\
450.106 & 850.92 & 1.0$^{+0.01}_{-0.01}$  & 0.1 & (2.3$^{+8.8}_{-0.6}$)$\times10^{11}$  & 1.6$\pm$0.6 & $\equiv$2.0 & $-$                             &  0.96$^{+0.27}_{-0.11}$ & (2$^{+2}_{-1}$)$\times10^{10}$ \\
450.126 & 850.159& 2.04$^{+0.92}_{-1.30}$ & 0.4 & (1.4$^{+5.5}_{-0.4}$)$\times10^{12}$  & 2.4$\pm$0.6 & $\equiv$2.0 & $-$                             &  1.27$^{+0.27}_{-0.35}$ & (1$^{+3}_{-0.5}$)$\times10^{10}$ \\
450.133 & 850.131& 0.93$^{+0.01}_{-0.01}$ & 0.1 & (1.8$^{+7.2}_{-0.4}$)$\times10^{11}$  & 1.5$\pm$0.6 & $\equiv$2.0 & $-$                             &  2.80$^{+0.07}_{-0.07}$ & (3$^{+3}_{-2}$)$\times10^{10}$ \\
450.134 & 850.52 & 2.89$^{+0.14}_{-0.24}$ & 2.8 & (4.1$^{+7.0}_{-2.5}$)$\times10^{12}$  & 2.8$\pm$0.2 & 1.9$\pm$0.3 & (4$^{+6}_{-4}$)$\times10^{8}$   &  1.73$^{+0.14}_{-0.32}$ & (6$^{+7}_{-4}$)$\times10^{10}$ \\
450.135 & 850.163& 1.68$^{+0.08}_{-0.06}$ & 1.0 & (2.5$^{+9.9}_{-0.6}$)$\times10^{12}$  & 2.6$\pm$0.6 & $\equiv$2.0 & $-$                             &  1.88$^{+0.08}_{-0.08}$ & (2$^{+3}_{-2}$)$\times10^{10}$ \\
450.166 & 850.83 & 0.93$^{+0.01}_{-0.01}$ & 3.2 & (1.7$^{+7.1}_{-0.4}$)$\times10^{11}$  & 1.5$\pm$0.6 & $\equiv$2.0 & $-$                             &  0.54$^{+0.12}_{-0.12}$ & (2$^{+3}_{-2}$)$\times10^{9}$ \\
450.173 & 850.104& 1.01$^{+0.01}_{-0.01}$ & 3.3 & (5.8$^{+23.9}_{-1.4}$)$\times10^{11}$ & 2.0$\pm$0.6 & $\equiv$2.0 & $-$                             &  1.83$^{+0.36}_{-0.12}$ & (2$^{+3}_{-2}$)$\times10^{10}$ \\
450.179 & 850.88 & 0.66$^{+0.01}_{-0.01}$ & 0.1 & (6.9$^{+28.8}_{-1.7}$)$\times10^{10}$ & 1.1$\pm$0.6 & $\equiv$2.0 & $-$                             &  1.62$^{+0.11}_{-0.42}$ & (5$^{+6}_{-4}$)$\times10^{9}$ \\
450.193 & 850.49 & 3.08$^{+0.13}_{-0.16}$ & 2.9 & (3.2$^{+13.3}_{-0.8}$)$\times10^{12}$ & 2.7$\pm$0.6 & $\equiv$2.0 & $-$                             &  2.28$^{+0.10}_{-0.29}$ & (1$^{+2}_{-1}$)$\times10^{11}$ \\
450.206 & 850.53 & 1.70$^{+0.20}_{-0.20}$ & 3.4 & (7.9$^{+33.5}_{-1.9}$)$\times10^{11}$ & 2.1$\pm$0.6 & $\equiv$2.0 & $-$                             & -0.16$^{+0.15}_{-0.09}$ & (1$^{+2}_{-0.6}$)$\times10^{9}$ \\
450.215 & 850.27 & 2.76$^{+0.10}_{-0.12}$ & 3.0 & (4.1$^{+6.5}_{-2.6}$)$\times10^{12}$  & 2.8$\pm$0.2 & 1.9$\pm$0.3 & (6$^{+8}_{-5}$)$\times10^{8}$   &  1.66$^{+0.25}_{-0.10}$ & (8$^{+9}_{-6}$)$\times10^{10}$ \\
450.240 & 850.67 & 2.73$^{+2.39}_{-1.20}$ & 3.0 & (2.4$^{+10.3}_{-0.5}$)$\times10^{12}$ & 2.6$\pm$0.6 & $\equiv$2.0 & $-$                             &  1.12$^{+0.28}_{-2.30}$ & (2$^{+3}_{-1}$)$\times10^{10}$ \\
450.247 & 850.151& 1.78$^{+0.93}_{-0.71}$ & 0.1 & (7.1$^{+31.1}_{-1.6}$)$\times10^{11}$ & 2.1$\pm$0.6 & $\equiv$2.0 & $-$                             &  0.07$^{+0.44}_{-0.42}$ & (5$^{+10}_{-2}$)$\times10^{8}$ \\
        & 850.01 & 1.37$^{+0.44}_{-0.25}$ & 1.9 & (4.7$^{+8.6}_{-2.6}$)$\times10^{11}$  & 1.9$\pm$0.3 & 2.3$\pm$0.3 & (6$^{+20}_{-2}$)$\times10^{9}$  &  0.98$^{+0.29}_{-0.25}$ & (2$^{+2}_{-1}$)$\times10^{9}$ \\
        & 850.04 & 1.39$^{+0.02}_{-0.02}$ & 1.3 & (4.7$^{+7.9}_{-2.8}$)$\times10^{11}$  & 1.9$\pm$0.2 & 2.4$\pm$0.2 & (1$^{+2}_{-1}$)$\times10^{10}$  &  1.37$^{+0.09}_{-0.07}$ & (3$^{+4}_{-3}$)$\times10^{10}$ \\
        & 850.05 & 2.55$^{+0.12}_{-0.15}$ & 0.1 & (2.8$^{+5.1}_{-1.5}$)$\times10^{12}$  & 2.7$\pm$0.3 & 2.1$\pm$0.3 & (5$^{+10}_{-2}$)$\times10^{9}$  &  1.47$^{+0.10}_{-0.08}$ & (2$^{+2}_{-2}$)$\times10^{10}$ \\
        & 850.11 & 2.77$^{+0.16}_{-0.19}$ & 0.3 & (1.5$^{+4.4}_{-0.5}$)$\times10^{12}$  & 2.4$\pm$0.5 & $\equiv$2.1 & $-$                             &  1.36$^{+0.10}_{-0.34}$ & (4$^{+5}_{-4}$)$\times10^{10}$ \\
        & 850.13 & 0.86$^{+0.48}_{-0.55}$ & 0.2 & (9.0$^{+17.5}_{-4.7}$)$\times10^{10}$ & 1.2$\pm$0.3 & 2.4$\pm$0.3 & (2$^{+10}_{-1}$)$\times10^{9}$  & -0.13$^{+0.51}_{-0.38}$ & (8$^{+10}_{-3}$)$\times10^{8}$ \\
        & 850.18 & 3.01$^{+0.17}_{-0.21}$ & 0.1 & (2.3$^{+5.6}_{-1.}$)$\times10^{12}$   & 2.6$\pm$0.4 & 2.0$\pm$0.6 & (1$^{+2}_{-1}$)$\times10^{9}$   &  1.08$^{+0.35}_{-0.30}$ & (9$^{+10}_{-6}$)$\times10^{9}$ \\
        & 850.19 & 2.24$^{+2.19}_{-1.00}$ & 0.1 & (4.5$^{+14.0}_{-1.5}$)$\times10^{11}$ & 1.9$\pm$0.5 & $\equiv$2.1 & $-$                             &  0.48$^{+0.43}_{-0.39}$ & (1$^{+2}_{-0.7}$)$\times10^{9}$ \\
        & 850.20 & 2.80$^{+0.05}_{-0.07}$ & 0.1 & (1.4$^{+3.4}_{-0.5}$)$\times10^{12}$  & 2.4$\pm$0.4 & 2.1$\pm$0.5 & (2$^{+4}_{-1}$)$\times10^{9}$   &  1.08$^{+0.08}_{-0.07}$ & (3$^{+4}_{-3}$)$\times10^{10}$ \\
        & 850.22 & 2.06$^{+0.10}_{-0.10}$ & 0.4 & (1.2$^{+2.6}_{-0.6}$)$\times10^{12}$  & 2.3$\pm$0.3 & 2.3$\pm$0.3 & (5$^{+20}_{-1}$)$\times10^{9}$  &  1.83$^{+0.11}_{-0.35}$ & (2$^{+2}_{-2}$)$\times10^{11}$ \\
        & 850.24 & 2.94$^{+0.96}_{-0.36}$ & 1.2 & (1.0$^{+1.5}_{-0.7}$)$\times10^{13}$  & 3.2$\pm$0.2 & 1.8$\pm$0.1 & (5$^{+6}_{-4}$)$\times10^{8}$   &  0.21$^{+0.14}_{-0.35}$ & (1$^{+10}_{-0.2}$)$\times10^{9}$ \\
        & 850.25 & 2.33$^{+0.05}_{-0.05}$ & 1.3 & (1.9$^{+3.5}_{-1.1}$)$\times10^{12}$  & 2.5$\pm$0.2 & 2.3$\pm$0.3 & (8$^{+30}_{-2}$)$\times10^{9}$  &  2.50$^{+0.13}_{-0.50}$ & (2$^{+2}_{-2}$)$\times10^{11}$ \\
        & 850.30 & 0.23$^{+0.01}_{-0.01}$ & 1.9 & (1.4$^{+2.2}_{-1.}$)$\times10^{10}$   & 0.4$\pm$0.2 & 2.3$\pm$0.1 & (1$^{+2}_{-1}$)$\times10^{8}$   &  0.15$^{+0.10}_{-0.10}$ & (1$^{+2}_{-1}$)$\times10^{9}$ \\
        & 850.31 & 1.80$^{+0.04}_{-0.05}$ & 0.6 & (2.9$^{+4.5}_{-1.9}$)$\times10^{12}$  & 2.7$\pm$0.2 & 2.0$\pm$0.3 & (5$^{+6}_{-4}$)$\times10^{8}$   &  2.05$^{+0.10}_{-0.12}$ & (7$^{+8}_{-5}$)$\times10^{10}$ \\
        & 850.32 & 2.87$^{+0.13}_{-0.26}$ & 13. & (8.5$^{+11.9}_{-6.1}$)$\times10^{12}$ & 3.2$\pm$0.1 & 1.8$\pm$0.1 & (4$^{+4}_{-3}$)$\times10^{8}$   &  2.01$^{+0.08}_{-0.09}$ & (2$^{+2}_{-2}$)$\times10^{11}$ \\
        & 850.35 & 2.36$^{+0.20}_{-0.33}$ & 0.1 & (1.1$^{+3.4}_{-0.4}$)$\times10^{12}$  & 2.3$\pm$0.5 & 2.2$\pm$0.6 & (3$^{+10}_{-1}$)$\times10^{9}$  &  1.00$^{+0.08}_{-0.09}$ & (1$^{+2}_{-0.9}$)$\times10^{10}$ \\
        & 850.36 & 2.05$^{+0.02}_{-0.03}$ & 2.9 & (2.2$^{+7.0}_{-0.7}$)$\times10^{12}$  & 2.6$\pm$0.5 & $\equiv$2.1 & $-$                             &  1.72$^{+0.10}_{-0.30}$ & (1$^{+2}_{-1}$)$\times10^{11}$ \\
        & 850.37 & 4.57$^{+1.07}_{-1.60}$ & 0.3 & (3.3$^{+10.6}_{-1.0}$)$\times10^{12}$ & 2.8$\pm$0.5 & $\equiv$2.1 & $-$                             &  2.10$^{+0.19}_{-0.33}$ & (7$^{+10}_{-4}$)$\times10^{10}$ \\
        & 850.38 & 2.04$^{+0.14}_{-0.09}$ & 0.1 & (1.5$^{+4.0}_{-0.6}$)$\times10^{12}$  & 2.4$\pm$0.4 & 2.1$\pm$0.6 & (1$^{+2}_{-1}$)$\times10^{9}$   &  1.79$^{+0.07}_{-0.07}$ & (3$^{+3}_{-2}$)$\times10^{9}$ \\
        & 850.43 & 0.84$^{+0.43}_{-0.42}$ & 1.3 & (5.1$^{+16.4}_{-1.6}$)$\times10^{10}$ & 0.9$\pm$0.5 & $\equiv$2.1 & $-$                             & -0.07$^{+0.52}_{-0.38}$ & (8$^{+20}_{-4}$)$\times10^{8}$ \\
        & 850.44 & 2.48$^{+0.13}_{-0.10}$ & 1.5 & (3.0$^{+4.9}_{-1.9}$)$\times10^{12}$  & 2.7$\pm$0.2 & 1.9$\pm$0.2 & (4$^{+5}_{-3}$)$\times10^{8}$   &  1.61$^{+0.11}_{-0.34}$ & (1$^{+2}_{-0.7}$)$\times10^{10}$ \\
        & 850.45 & 1.32$^{+0.03}_{-0.05}$ & 1.1 & (2.3$^{+7.4}_{-0.7}$)$\times10^{11}$  & 1.6$\pm$0.5 & $\equiv$2.1 & $-$                             &  1.56$^{+0.11}_{-0.27}$ & (3$^{+3}_{-2}$)$\times10^{10}$ \\
        & 850.46 & 2.82$^{+0.13}_{-0.19}$ & 0.1 & (1.5$^{+2.0}_{-1.0}$)$\times10^{13}$  & 3.4$\pm$0.1 & 1.8$\pm$0.1 & (4$^{+5}_{-4}$)$\times10^{8}$   &  1.90$^{+0.16}_{-0.24}$ & (2$^{+2}_{-2}$)$\times10^{11}$ \\
        & 850.50 & 2.38$^{+0.37}_{-0.51}$ & 0.1 & (3.2$^{+10.6}_{-1.}$)$\times10^{11}$  & 1.7$\pm$0.5 & 2.4$\pm$0.4 & (2$^{+70}_{-1}$)$\times10^{10}$ &  0.90$^{+0.40}_{-0.19}$ & (5$^{+8}_{-2}$)$\times10^{9}$ \\
        & 850.51 & 1.11$^{+0.07}_{-0.04}$ & 1.0 & (6.6$^{+10.5}_{-4.1}$)$\times10^{11}$ & 2.1$\pm$0.2 & 2.1$\pm$0.2 & (6$^{+10}_{-3}$)$\times10^{8}$  &  1.29$^{+0.08}_{-0.07}$ & (8$^{+10}_{-7}$)$\times10^{9}$ \\
        & 850.54 & 3.07$^{+1.19}_{-1.20}$ & 1.6 & (2.3$^{+7.5}_{-0.7}$)$\times10^{12}$  & 2.6$\pm$0.5 & $\equiv$2.1 & $-$                             &  1.25$^{+0.40}_{-0.48}$ & (8$^{+10}_{-4}$)$\times10^{9}$ \\
        & 850.57 & 0.74$^{+0.01}_{-0.01}$ & 1.2 & (9.4$^{+21.3}_{-4.1}$)$\times10^{10}$ & 1.2$\pm$0.4 & 2.4$\pm$0.4 & (2$^{+7}_{-1}$)$\times10^{9}$   & -1.83$^{+1.35}_{-0.14}$ & (2$^{+3}_{-2}$)$\times10^{11}$ \\
        & 850.59 & 3.15$^{+0.13}_{-0.17}$ & 0.3 & (3.8$^{+12.9}_{-1.1}$)$\times10^{12}$ & 2.8$\pm$0.5 & $\equiv$2.1 & $-$                             &  1.05$^{+0.38}_{-0.13}$ & (8$^{+10}_{-4}$)$\times10^{9}$ \\
\hline\hline  
\end{tabular}  
} 
\end{spacing}
{\small {\sc Table~\ref{tab:physical} -- Continued.}}
\end{table*}
\begin{table*}
\centering
\begin{spacing}{0.7}
{\tiny
\begin{tabular}{cccc@{ }c@{ }c@{ }c@{ }cc@{ }c}
\hline\hline
{\sc Short}  & {\sc Short}  & $z_{\rm phot}$ & \multicolumn{5}{c}{\underline{Estimates from Infrared SED Fitting (\S \ref{sec:sed})}}                  & \multicolumn{2}{c}{\underline{{\sc Le Phare} Properties}} \\
{\sc 450}\um & {\sc 850}\um &                & $\chi^2_{\rm fit}$ & $L_{\rm IR}$  & $\log$SFR$_{\rm IR}$   & $\log\lambda_{\rm peak}$ & $M_{\rm dust}$ & $\log$SFR$_{\rm UV}$   & $M_\star$ \\ 
{\sc Name}   & {\sc Name}   &                &                    & ($L_{\odot}$) & ($M_\odot$\,yr$^{-1}$) & (\um)                    & ($M_\odot$)    & ($M_\odot$\,yr$^{-1}$) & ($M_\odot$) \\ 
\hline\hline
        & 850.60 & 1.44$^{+0.02}_{-0.02}$ & 4.5 & (1.4$^{+2.2}_{-0.9}$)$\times10^{12}$  & 2.4$\pm$0.2 & 2.0$\pm$0.2 & (4$^{+5}_{-3}$)$\times10^{8}$   &  1.43$^{+0.08}_{-0.09}$ & (1$^{+1}_{-1}$)$\times10^{11}$ \\
        & 850.61 & 0.53$^{+0.02}_{-0.02}$ & 0.4 & (5.4$^{+14.4}_{-2.0}$)$\times10^{10}$ & 1.0$\pm$0.4 & 2.3$\pm$0.5 & (4$^{+50}_{-1}$)$\times10^{8}$  & -0.31$^{+0.23}_{-0.15}$ & (6$^{+8}_{-6}$)$\times10^{8}$ \\
        & 850.63 & 1.89$^{+0.23}_{-0.18}$ & 1.2 & (3.4$^{+4.1}_{-2.8}$)$\times10^{11}$  & 1.8$\pm$0.1 & $\equiv$2.1 & $-$                             &  0.23$^{+0.14}_{-0.33}$ & (3$^{+4}_{-2}$)$\times10^{9}$ \\
        & 850.64 & 3.39$^{+1.16}_{-1.50}$ & 0.1 & (2.1$^{+7.1}_{-0.6}$)$\times10^{12}$  & 2.6$\pm$0.5 & $\equiv$2.1 & $-$                             &  1.61$^{+0.30}_{-0.40}$ & (1$^{+2}_{-0.7}$)$\times10^{10}$ \\
        & 850.65 & 0.48$^{+0.01}_{-0.01}$ & 5.3 & (7.6$^{+13.1}_{-4.4}$)$\times10^{10}$ & 1.1$\pm$0.2 & 2.3$\pm$0.2 & (3$^{+30}_{-1}$)$\times10^{8}$  &  0.42$^{+0.80}_{-0.09}$ & (4$^{+4}_{-2}$)$\times10^{8}$ \\
        & 850.66 & 0.94$^{+0.14}_{-0.16}$ & 1.2 & (1.6$^{+5.6}_{-0.5}$)$\times10^{11}$  & 1.4$\pm$0.5 & $\equiv$2.1 & $-$                             &  0.14$^{+0.37}_{-0.37}$ & (5$^{+8}_{-4}$)$\times10^{8}$ \\
        & 850.68 & 1.95$^{+0.04}_{-0.04}$ & 0.1 & (2.7$^{+4.4}_{-1.7}$)$\times10^{12}$  & 2.7$\pm$0.2 & 2.0$\pm$0.2 & (4$^{+5}_{-3}$)$\times10^{8}$   &  1.76$^{+0.10}_{-0.30}$ & (2$^{+2}_{-1}$)$\times10^{11}$ \\
        & 850.69 & 1.80$^{+0.15}_{-0.11}$ & 0.1 & (1.6$^{+2.9}_{-0.9}$)$\times10^{12}$  & 2.4$\pm$0.3 & 2.0$\pm$0.2 & (4$^{+6}_{-3}$)$\times10^{8}$   &  1.60$^{+0.07}_{-0.07}$ & (4$^{+5}_{-3}$)$\times10^{10}$ \\
        & 850.70 & 0.78$^{+0.01}_{-0.03}$ & 2.3 & (1.8$^{+3.2}_{-1.}$)$\times10^{11}$   & 1.5$\pm$0.3 & 2.2$\pm$0.2 & (4$^{+10}_{-1}$)$\times10^{8}$  &  1.40$^{+0.07}_{-0.07}$ & (2$^{+3}_{-2}$)$\times10^{9}$ \\
        & 850.71 & 2.17$^{+1.36}_{-1.40}$ & 0.3 & (8.7$^{+30.0}_{-2.5}$)$\times10^{11}$ & 2.2$\pm$0.5 & $\equiv$2.1 & $-$                             &  0.61$^{+0.60}_{-0.78}$ & (2$^{+10}_{-1}$)$\times10^{9}$ \\
        & 850.72 & 2.80$^{+0.15}_{-0.16}$ & 2.0 & (4.5$^{+7.9}_{-2.6}$)$\times10^{12}$  & 2.9$\pm$0.2 & 1.9$\pm$0.2 & (3$^{+4}_{-2}$)$\times10^{8}$   &  1.78$^{+0.10}_{-0.09}$ & (2$^{+2}_{-2}$)$\times10^{11}$ \\
        & 850.73 & 3.91$^{+0.94}_{-1.20}$ & 0.1 & (7.2$^{+25.2}_{-2.1}$)$\times10^{12}$ & 3.1$\pm$0.5 & $\equiv$2.1 & $-$                             &  2.32$^{+0.13}_{-0.19}$ & (6$^{+9}_{-4}$)$\times10^{10}$ \\
        & 850.74 & 2.76$^{+2.78}_{-1.50}$ & 0.1 & (1.8$^{+6.4}_{-0.5}$)$\times10^{12}$  & 2.5$\pm$0.5 & $\equiv$2.1 & $-$                             &  1.21$^{+0.36}_{-0.42}$ & (8$^{+10}_{-4}$)$\times10^{9}$ \\
        & 850.76 & 1.99$^{+0.07}_{-0.22}$ & 1.4 & (9.1$^{+32.0}_{-2.6}$)$\times10^{11}$ & 2.2$\pm$0.5 & $\equiv$2.1 & $-$                             &  2.69$^{+0.07}_{-0.07}$ & (5$^{+5}_{-5}$)$\times10^{9}$ \\
        & 850.78 & 0.94$^{+0.01}_{-0.01}$ & 0.3 & (4.2$^{+7.3}_{-2.4}$)$\times10^{11}$  & 1.9$\pm$0.2 & 2.1$\pm$0.2 & (3$^{+5}_{-2}$)$\times10^{8}$   &  0.79$^{+0.08}_{-0.07}$ & (1$^{+2}_{-1}$)$\times10^{10}$ \\
        & 850.80 & 3.54$^{+0.12}_{-0.15}$ & 0.2 & (2.8$^{+9.8}_{-0.8}$)$\times10^{12}$  & 2.7$\pm$0.5 & $\equiv$2.1 & $-$                             &  1.77$^{+0.12}_{-0.36}$ & (4$^{+6}_{-3}$)$\times10^{10}$ \\
        & 850.81 & 0.40$^{+0.03}_{-0.03}$ & 1.3 & (1.4$^{+4.9}_{-0.4}$)$\times10^{10}$  & 0.4$\pm$0.5 & $\equiv$2.1 & $-$                             & -1.45$^{+0.35}_{-0.23}$ & (5$^{+7}_{-3}$)$\times10^{8}$ \\
        & 850.82 & 2.87$^{+0.17}_{-0.25}$ & 1.3 & (6.2$^{+11.3}_{-3.4}$)$\times10^{12}$ & 3.0$\pm$0.3 & 1.8$\pm$0.2 & (3$^{+4}_{-2}$)$\times10^{8}$   &  2.10$^{+0.08}_{-0.10}$ & (2$^{+3}_{-2}$)$\times10^{11}$ \\
        & 850.84 & 2.42$^{+0.38}_{-0.48}$ & 1.1 & (2.0$^{+7.1}_{-0.6}$)$\times10^{12}$  & 2.5$\pm$0.6 & $\equiv$2.1 & $-$                             &  0.68$^{+0.40}_{-0.17}$ & (4$^{+6}_{-2}$)$\times10^{9}$ \\
        & 850.87 & 2.24$^{+0.42}_{-0.19}$ & 0.1 & (4.2$^{+7.3}_{-2.4}$)$\times10^{12}$  & 2.9$\pm$0.2 & 1.8$\pm$0.2 & (2$^{+2}_{-1}$)$\times10^{8}$   &  1.07$^{+0.62}_{-0.70}$ & (6$^{+40}_{-1}$)$\times10^{9}$ \\
        & 850.91 & 2.71$^{+0.46}_{-0.55}$ & 0.1 & (2.0$^{+7.1}_{-0.6}$)$\times10^{12}$  & 2.5$\pm$0.6 & $\equiv$2.1 & $-$                             &  1.03$^{+0.41}_{-0.36}$ & (8$^{+10}_{-5}$)$\times10^{9}$ \\
        & 850.94 & 2.04$^{+0.10}_{-0.21}$ & 0.1 & (7.3$^{+9.1}_{-5.8}$)$\times10^{11}$  & 2.1$\pm$0.1 & $\equiv$2.1 & $-$                             &  2.10$^{+0.07}_{-0.07}$ & (1$^{+1}_{-1}$)$\times10^{11}$ \\
        & 850.95 & 1.55$^{+0.05}_{-0.08}$ & 0.2 & (2.2$^{+7.9}_{-0.6}$)$\times10^{12}$  & 2.6$\pm$0.6 & $\equiv$2.1 & $-$                             &  2.48$^{+0.09}_{-0.28}$ & (8$^{+10}_{-7}$)$\times10^{10}$ \\
        & 850.97 & 0.73$^{+0.01}_{-0.01}$ & 0.1 & (3.4$^{+6.4}_{-1.8}$)$\times10^{11}$  & 1.8$\pm$0.3 & 2.1$\pm$0.2 & (2$^{+4}_{-1}$)$\times10^{8}$   & -3.90$^{+1.94}_{-0.88}$ & (2$^{+2}_{-2}$)$\times10^{11}$ \\
\hline\hline  
\end{tabular}  
} 
\end{spacing}
{\small {\sc Table~\ref{tab:physical} -- Continued.}}
\end{table*}

\end{document}